\documentclass[12pt,a4paper]{report}
\usepackage[]{graphicx}
\usepackage{amssymb}
\usepackage[brazilian]{babel}
\usepackage[T1]{fontenc}
\usepackage{latexsym}
\usepackage{dsfont} % Fonte para ''1'' = Matrix identidade
\usepackage{amsbsy}

\usepackage{cjhebrew}
\usepackage{amsmath}
\usepackage{hyperref}
\usepackage{indentfirst}
\usepackage{bm}
\usepackage{makeidx}
\usepackage{pdfpages}
\usepackage{fancyhdr}
\usepackage{setspace} % \begin\end{doublespacing} & \begin\end{spacing}{1.5}
\newtheorem{theorem}{\rmfamily\bfseries{Teorema}}[section]

\newtheorem{lemma}[theorem]{\rmfamily\bfseries{Lema}}

\newtheorem{definition}[theorem]{\rmfamily\bfseries{Definição}}%[section]

\newcounter{fig}[chapter]
\begin{document}
 %\bibliographystyle{unsrt}

%\pagenumbering{roman} \thispagestyle{empty}

%-----------------------------------------------
%%%%%%%%%%%%%%%%%%    CAPA    %%%%%%%%%%%%%%%%%%
%-----------------------------------------------
%\thispagestyle{empty}
%
\begin{spacing}{1.0}
\begin{center}
%\textbf{André Herkenhoff Gomes} \\
\textbf{BRUNO CARVALHO NEVES} \\

\end{center}
\vfill
\begin{center}
\begin{large}\textbf{UM MODELO DE GRAVITAÇÃO TIPO-TOPOLÓGICO EM UM ESPAÇO-TEMPO $4$D}\end{large} \\
\end{center}
\vfill
\begin{flushright}
\begin{minipage}[b][1.4cm][t]{8.5cm}
\textbf{Tese apresentada à Universidade Federal de Viçosa, como parte das exigências do Programa de Pós-Graduação em Física Aplicada, para obtenção do título de \textit{Doctor Scientiae}}.
\end{minipage}

\end{flushright}

\vfill
\begin{center}
\textbf{{VIÇOSA}\\
{MINAS GERAIS -  BRASIL}\\%\vspace{-0.15cm}
{2016}}
\end{center}
\end{spacing}
\pagenumbering{roman}\thispagestyle{empty}
%\pagebreak
\newpage
%\topmargin10cm
%\vspace*{12cm}
%

%-----------------------------------------------
%%%%%%%%%%%%%%%%%%    CAPA    %%%%%%%%%%%%%%%%%%

\includepdf[pages=-]{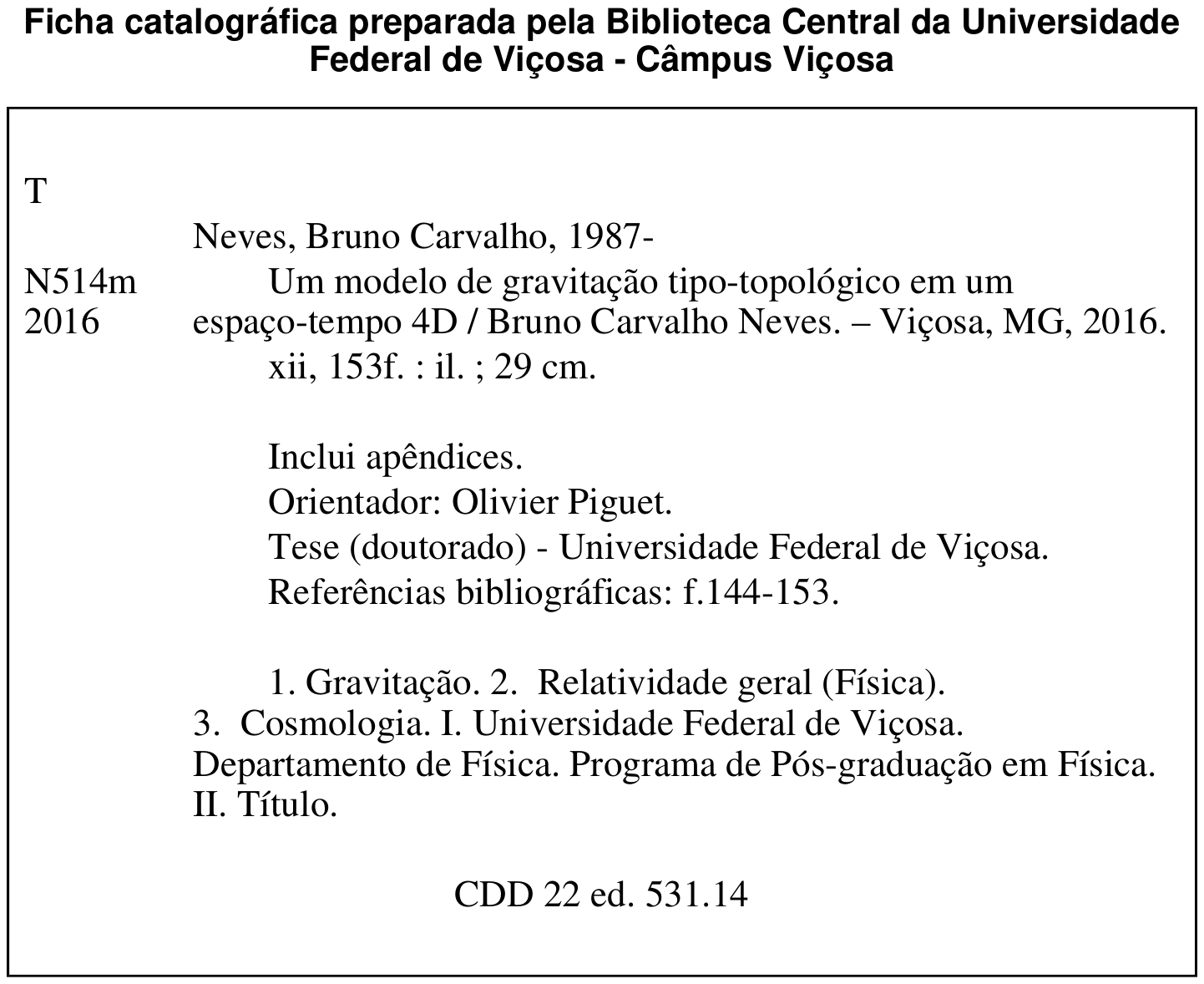}

\pagenumbering{roman}\thispagestyle{empty}

\includepdf[pages=-]{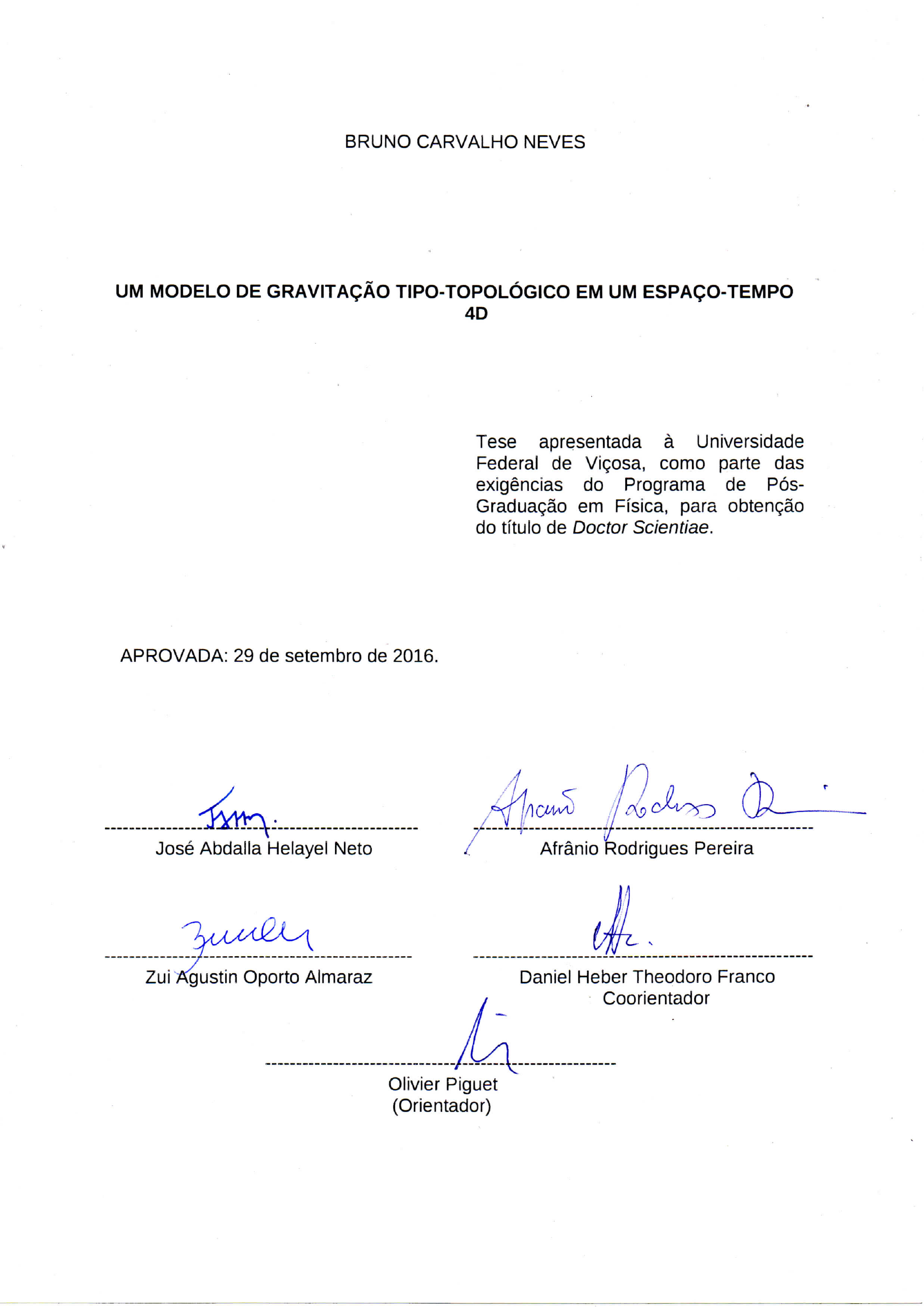}

%-----------------------------------------------
%\newpage

%-----------------------------------------------
%%%%%%%%%%%%%%    DEDICATÓRIA    %%%%%%%%%%%%%%%
%-----------------------------------------------

\topmargin0cm
\linespread{1.3}
\vspace*{19cm}
 \begin{flushright}
 %------------------------------------------ Dedicatoria
 À minha família e amigos.
 \end{flushright}
\pagebreak
%-----------------------------------------------
%%%%%%%%%%%%%%    DEDICATÓRIA    %%%%%%%%%%%%%%%
%-----------------------------------------------

%-----------------------------------------------
%%%%%%%%%%%%%%%%%    FRASE    %%%%%%%%%%%%%%%%%%
%-----------------------------------------------
\begin{flushright}
\cjRL{h}" \cjRL{b}
\end{flushright}

\vspace*{19cm} \begin{center}
\cjRL{hw' hyh 'wmr ; 'M 'yN 'ny ly my ly wk+s'ny l`.smy mh 'ny w'M l' `k+sw 'ymty} \\
``Hilel dizia: Se eu não for por mim, quem será por mim? Mas se eu for só por mim, o que sou eu? E se não agora, quando?''
\\
\textit{Pirkê Avót, Capítulo 1, Mishná 14.}\end{center}
\pagebreak
%-----------------------------------------------
%%%%%%%%%%%%%%%%%    FRASE    %%%%%%%%%%%%%%%%%%
%-----------------------------------------------

%-----------------------------------------------
%%%%%%%%%%%%    AGRADECIMENTOS    %%%%%%%%%%%%%%
%-----------------------------------------------
\chapter*{\centering Agradecimentos}
\addcontentsline{toc}{chapter}{Agradecimentos}
\begin{flushright}
	\parbox[t]{15.0cm}{{\setlength{\baselineskip}%
			{0.8\baselineskip}
Agradeço, primeira e fundamentalmente, a D-us (\cjRL{yhwh}) o Deus de Avraham, Yitzchak e Ya'akov, que nos mantiveste vivos, preservaste-nos e nos fizeste chegar a este momento. À  minha mãe Aparecida, a quem tenho a honra de dedicar esse trabalho, ao meu irmão Thiago e meu pai José. 
%\medskip

Ao professor Olivier Piguet, pela amizade ao longo destes anos de trabalho em conjunto, por fazer um centro de reunião onde pude assentar-me aos seus pés e de frente a sua erudição e sabedoria a fim de beber e absorver avidamente de suas palavras. 

%\medskip

Ao professor Daniel por sua confiança desde a graduação e por ter sido essa ponte fundamental para esse trabalho. Ao professor José Helayël pelas incontáveis referências bem como por sua disposição e carisma, sendo um referencial privilegiado no ensino e pesquisa no Brasil.

%\medskip

Aos meus amigos da pós-graduação, aos grandes amigos do grupo de pesquisa Ivan Morales e Zui Oporto pela amizade, pelos cafés filosóficos e companheirismo nesses anos de trabalho. Agradeço ao Fabiano e Carlos pela grande amizade e pelas inúmeras horas de almoço e conversas. Ao meu grande amigo Felipe Apolônio que mesmo estando tão distante quanto o pré-sal se faz presente com sua disposição carisma e amizade. 

%\medskip

À Camila Moraes por sua benevolência, carinho e dedicação para comigo em todas as situações que temos vivido nesse um ano de adorável convivência que tem sido uma verdadeira médica na nossa linha de mundo.

%\medskip

Aos meus familiares que de maneira direta e indireta contribuiram para esse trabalho. À Gabriela que fez parte integrante desse trabalho com muito carinho. Ao meu filho (\emph{in memoriam}) que pelo pouco que viveu me transformou em uma nova pessoa e retirou-me de toda visão míope e superficial sobre a vida.

%\medskip

Agradeço a todos brasileiros que contribuem ao desenvolvimento da Ciência em nosso país e também àqueles que, através da CAPES, concedem, a nós, estudantes, o necessários apoio financeiro.

%\medskip

A todos, meus mais sinceros agradecimentos.\par}}
\end{flushright}
\pagebreak
%-----------------------------------------------
%%%%%%%%%%%%    AGRADECIMENTOS    %%%%%%%%%%%%%%
%-----------------------------------------------

%-----------------------------------------------
%%%%%%%%%%%    TABLE OF CONTENTS    %%%%%%%%%%%%
%-----------------------------------------------
\tableofcontents
%\addcontentsline{toc}{chapter}{Sumário}
\vfill
\parindent3em
\pagebreak
%-----------------------------------------------
%%%%%%%%%%%    TABLE OF CONTENTS    %%%%%%%%%%%%
%-----------------------------------------------

%-----------------------------------------------
%%%%%%    LISTA DE FIGURAS E TABELAS    %%%%%%%%
%-----------------------------------------------
%\listoffigures\addcontentsline{toc}{chapter}{Lista de Figuras}
%\listoftables\addcontentsline{toc}{chapter}{Lista de Tabelas}

%-----------------------------------------------
%%%%%%%%%    NOTAÇÕES E CONVENÇÕES    %%%%%%%%%%
%-----------------------------------------------
\chapter*{Notações e Convenções}
\addcontentsline{toc}{chapter}{Notações e Convenções}
\section*{Notações e Convenções}
%=======
\subsection*{Convenções} 

\begin{itemize}

\item Os índices do espaço-tempo $4$D e $5$D : $\mu, \cdots=0, \cdots, 3 $ e
$\alpha, \beta, ... = 0,...,4 $\vspace{0.1cm}

 \item Os índices espaciais em $3$D e $4$D são : $a, b, ... = 1, ..., 3 $ e
$m, n, \cdots = 1, \cdots, 4 $. \vspace{0.1cm}
 \item Os grupos de de Sitter e anti-de Sitter SO$(n, N - n)$ serão denotados de maneira compacta como: (A)dS$_{N}$.
 \end{itemize}
\begin{itemize} 
  \item Os índices e as correspondentes métricas serão denotadas por:
  \end{itemize}
\begin{eqnarray} %\[\ba{ll}
\label{metricADS6}&&\textrm{(A)dS}_{6}:\quad  M,N, \cdots = 0, \cdots, 5\,,\quad
\eta_{MN} = \textrm{diag}(-1,1,1,1,1,s )\,,\\[0.1cm]
\label{metricADS5}&&\textrm{(A)dS}_{5}:\quad   A,B, \cdots = 0, \cdots, 4\,,\quad 
\eta_{AB} = \textrm{diag}(-1,1,1,1,s)\,,
\end{eqnarray} %\ea\]
onde $s$ assume os valores $\pm1$ para dS ou AdS, respectivamente. Os índices de Lorentz SO$(1,3)$ $4$D  serão denotados por $(I, J\cdots = 0,\cdots,3)$, $(i,j\cdots = 1, 2, 3)$
a métrica correspondente sendo $\eta_{IJ} = \textrm{diag}(-1,1,1,1)$.
Essas métricas e suas inversas permitem abaixar e subir os vários índices do grupo.
  \begin{itemize}
\item  Os respectivos símbolos de Levi-Civita são definidos como
\end{itemize}
%========
\begin{eqnarray*}
\varepsilon_{MNPQRS}&=&\left\{
	\begin{array}{l}
	 \varepsilon_{012345}:=1\hspace{30pt} \\
	 \varepsilon_{ABCDE5}:= \varepsilon_{ABCDE}
	 \end{array}
	      \right.\\
\varepsilon_{ABCDE}&=&\left\{
	\begin{array}{c}
	 \varepsilon_{01234}:=1\hspace{30pt} \\
	 \varepsilon_{IJKL4}:= \varepsilon_{IJKL}
	 \end{array}
	      \right.\\
\varepsilon_{IJKL}&=&\left\{
	\begin{array}{c}
	 \varepsilon_{0123}:=1\hspace{10pt} \\
	 \varepsilon_{0ijk}:=\varepsilon_{ijk}
	 \end{array}
	      \right.
\end{eqnarray*}
para o espaço interno, ou seja, do grupo de simetria, e 
\begin{eqnarray*}
\varepsilon^{\alpha\beta\gamma\delta\varepsilon}&=&\left\{
	\begin{array}{l}
	 \varepsilon^{01234}:=1\hspace{10pt} \\
	 \varepsilon^{\mu\nu\rho\sigma 4}:=\varepsilon^{\mu\nu\rho\sigma}
	 \end{array}
	      \right.\\
\varepsilon^{\mu\nu\rho\sigma}&=&\left\{
	\begin{array}{c}
	 \varepsilon^{0123}:=1\hspace{10pt} \\
	 \varepsilon^{0abc}:=\varepsilon^{abc}
	 \end{array}
	      \right.
\end{eqnarray*}
para  o espaços-tempo $5$D e $4$D.

%===========

\subsection*{Base da álgebra de Lie}\label{A - Lie algebra basis}

Uma base para álgebra de Lie (a)ds$_{6}$ do grupo (A)dS$_{6}$ pode ser dada por $15$ matrizes $M_{PQ}=-M_{QP}$:
\begin{eqnarray*}
(M_{PQ})^{M}{}_{N}:=-(\delta_{P}^{M}\eta_{NQ} -\eta_{PN}\delta{_Q}^{M})
\end{eqnarray*}
satisfazendo as relações de comutação de (a)ds$_{6}$
\begin{eqnarray}
\lbrack M_{MN}, M_{PQ}\rbrack = -\eta_{MQ}M_{NP}-\eta_{NP}M_{MQ}+\eta_{MP}M_{NQ}+\eta_{NQ}M_{MP}.
\label{alg-so6}\end{eqnarray}
Pode-se decompor essa base de acordo com representações do grupo de Lorentz $5$D SO$(1,4)$ como
\begin{eqnarray*}
M_{MN}=\left\{
	\begin{array}{l}
	 M_{AB}\hspace{50pt} \\
	 P_{A} := \frac{1}{l} M_{A5}
	 \end{array}
	      \right.
\end{eqnarray*}
onde o parâmetro $l$ positivo é introduzido relacionando-se com a constante cosmológica $\Lambda \sim \dfrac{s}{l^{2}}$ ($s=\eta_{55}$)  de uma teoria de gravitação em $5$D.
Dessa forma, as relações de comutação leem-se
\begin{eqnarray}
\lbrack M_{AB}, M_{CD}\rbrack &=& 
-\tilde{\eta}_{AD}M_{BC}-\tilde{\eta}_{BC}M_{AD}
+\tilde{\eta}_{AC}M_{BD}+\tilde{\eta}_{BD}M_{AC}\,,\nonumber\\
\lbrack M_{AB}, P_{C}\rbrack &=& \tilde{\eta}_{AC}P_{B}-\tilde{\eta}_{BC}P_{A}\,,\label{alg-(A)dS_6->Lorentz}\\
\lbrack P_{A}, P_{B}\rbrack &=& \Lambda M_{AB}\,,\nonumber
\end{eqnarray}
com $\tilde\eta_{AB} = \textrm{diag}(-1,1,1,1,1)$.
Os dez geradores $M_{AB}$ são os geradores do grupo de Lorentz $5$D, e juntos com os $5$ geradores $P_{A}$, geram o grupo (A)dS$_{6}$  para o espaço-tempo $5$D. Os geradores $M_{AB}$ podem ser representados por matrizes $5\times5$  
%  UTIL?? 
\begin{eqnarray*}
(M_{CD})^{A}\,_{B}:=-(\delta_{C}^{A}\tilde\eta_{BD}-\tilde\eta_{CB}\delta_{D}^{A})
\end{eqnarray*}
A primeira linha das relações de comutação acima, a saber
\begin{equation}
\lbrack M_{AB}, M_{CD}\rbrack = 
-{\eta}_{AD}M_{BC}-{\eta}_{BC}M_{AD}
+{\eta}_{AC}M_{BD}+{\eta}_{BD}M_{AC}\,,
\end{equation}
mas dessa vez com a métrica
$\eta_{AB} = \textrm{diag}(-1,1,1,1,s)$, nos fornece as regras de comutação da álgebra de Lie de (A)dS$_{5}$. Sua decomposição de acordo com as representações do grupo de Lorentz $4$D são
\begin{eqnarray*}
M_{AB}=\left\{
	\begin{array}{l}
	 M_{IJ}\hspace{50pt} \\
	 P_{I} := \frac{1}{l} M_{I4}
	 \end{array}
	      \right.
\end{eqnarray*}
Analogamente introduzimos um parâmetro positivo para dimensionalizar corretamente o gerador $P_{I}$ $l$ associado a constante cosmológica em uma teoria de gravitação em $4$D. Assim
\begin{eqnarray*}
\lbrack M_{IJ}, M_{KL}\rbrack &=& -\eta_{IL}M_{JK}-\eta_{JK}M_{IL}+\eta_{IK}M_{JL}+\eta_{JL}M_{IK}\,,\\
\lbrack M_{IJ}, P_{K}\rbrack &=& \eta_{IK}P_{J}-\eta_{JK}P_{I}\,,\\
\lbrack P_{I}, P_{J}\rbrack &=& \Lambda M_{AB}\,.
\end{eqnarray*}
Além disso, estaremos interessados na decomposição completa de (A)dS$_{6}$ de acordo com as representações do grupo de Lorentz SO$(1,3)$
\begin{equation}
	 M_{IJ}\,,\quad 
	 P_{I} := \frac{1}{l} M_{I5}\,,\quad 
	 Q_{I} := \frac{1}{l} M_{I4}\,,\quad R: = M_{45}\,,
\end{equation}

\begin{eqnarray*}
\lbrack M_{IJ}, M_{KL}\rbrack & = & -\eta_{IL}M_{JK}-\eta_{JK}M_{IL}+\eta_{IK}M_{JL}+\eta_{JL}M_{IK}\,,\\[0.1cm]
\lbrack M_{IJ}, P_{K}\rbrack & = & \eta_{IK}P_{J}-\eta_{JK}P_{I},\\[0.1cm] 
\lbrack M_{IJ}, Q_{K}\rbrack & = & \eta_{IK}Q_{J}-\eta_{JK}Q_{I},\\[0.1cm]
\lbrack M_{IJ},R \rbrack & = & 0\,,\\[0.1cm]
\lbrack P_{I}, P_{J}\rbrack & = & \frac{s}{l^{2}} M_{IJ}\,,\quad
\lbrack Q_{I}, Q_{J}\rbrack = \frac{1}{l^{2}} M_{IJ}\,,\\[0.1cm]
\lbrack P_{I}, Q_{J}\rbrack & = & \frac{1}{l^{2}} \eta_{IJ} R\,,\\[0.1cm]
\lbrack P_{I}, R\rbrack & = & s Q_{I} \,,\quad
\lbrack Q_{I}, R\rbrack = - P_{I} \,.
\end{eqnarray*}

%=================

\subsection*{Dimensões}
%  IN THE TEXT?? 
As dimensões dos campos e dos parâmetros da teoria e dos geradores do grupo, dados em unidade de massa, são:
  \[
 \left|\begin{array}{c|c|c|c|c|c|c|c}
 \hline
  & ds & \omega^{IJ} & e^{I} & l & \Lambda & M_{IJ} & P_{I} \\
 \hline 
 \mbox{dim} & -1 & 1 & 0 & -1 & 2 & 0 & 1\\
 \hline
 \end{array}\right|
\]
%\left|\begin{array}{c}0 \\0\end{array}\right|

\pagebreak
%-----------------------------------------------
%%%%%%%%%    NOTAÇÕES E CONVENÇÕES    %%%%%%%%%%
%-----------------------------------------------

%-----------------------------------------------
%%%%%%%%%%%%%%%%    RESUMO    %%%%%%%%%%%%%%%%%%
%-----------------------------------------------
\chapter*{\centering Resumo}
\addcontentsline{toc}{chapter}{Resumo}
\begin{flushright}
\parbox[t]{15.0cm}{{\setlength{\baselineskip}%
{0.6\baselineskip}
%\hspace{-0.30cm} GOMES, André Herkenhoff, M. Sc., Universidade Federal de Viçosa, Julho de $2010$
%\textbf{Testes em Guias de Onda de uma Eletrodinâmica com Violação das Simetrias de Lorentz e CPT}. Orientador: Winder Alexander de Moura Melo. Co-orientadores: Afrânio Rodrigues Pereira e Daniel Heber Theodoro Franco\par}}.
%\end{flushright}
 NEVES, Bruno Carvalho, D. Sc., Universidade Federal de Viçosa, setembro de $2016$.
\textbf{Um modelo de gravitação tipo-topológico em um espaço-tempo $4$D}. Orientador: Olivier Piguet. Coorientadores: Daniel Heber Theodoro Franco e Oswaldo Monteiro Del Cima.\par}}
\end{flushright}

%~\\
\bigskip

\bigskip

\bigskip
\begin{flushright}
	\parbox[t]{15.0cm}{{\setlength{\baselineskip}%
			{1.1\baselineskip}
Nesse trabalho consideramos um modelo para gravitação em um espaço-tempo quadridimensional, originalmente proposto por Chamseddine, que pode ser obtido por uma redução dimensional e truncação de uma teoria de Chern-Simons pentadimensional. Sua origem topológica, torna-o um candidato interessante para uma quantização mais fácil, por exemplo, na abordagem da quantização de laços. O presente trabalho é dedicado a análise clássica das propriedades do modelo. Soluções cosmológicas, bem como soluções de onda, foram encontradas e comparadas com as soluções correspondentes da relatividade geral de Einstein com constante cosmológica.\par}} 
\end{flushright}
%\parindent3em
%\linespread{1.3}
\vfill
\pagebreak
%-----------------------------------------------
%%%%%%%%%%%%%%%%    RESUMO    %%%%%%%%%%%%%%%%%%
%-----------------------------------------------

%-----------------------------------------------
%%%%%%%%%%%%%%%    ABSTRACT    %%%%%%%%%%%%%%%%%
%-----------------------------------------------
\chapter*{\centering Abstract}
\addcontentsline{toc}{chapter}{Abstract}
\begin{flushright}
\parbox[t]{15.0cm}{{\setlength{\baselineskip}%
{0.6\baselineskip}
%\hspace{-0.30cm} GOMES, André Herkenhoff, M. Sc., Universidade Federal de Viçosa, July, $2010$
%\textbf{Testing a Electrodynamics with Broken Lorentz and CPT Symmetries with Waveguides}. Adviser: Winder Alexander de Moura Melo. Co-advisers: Afrânio Rodrigues Pereira e Daniel Heber Theodoro Franco\par}}.
%\end{flushright}
 NEVES, Bruno Carvalho, D. Sc., Universidade Federal de Viçosa, September, $2016$. \textbf{A topological-like model for gravity in 4D space-time}. Adviser: Olivier Piguet. Co-advisers: Daniel Heber Theodoro Franco and Oswaldo Monteiro Del Cima.\par}}
\end{flushright}

%~\\
\bigskip

\bigskip

\bigskip
\begin{flushright}
	\parbox[t]{15.0cm}{{\setlength{\baselineskip}%
			{1.1\baselineskip}
In this work we consider a model for gravity in 
$4$-dimensional space-time originally proposed by Chamseddine, 
  which may be  derived by  dimensional reduction and truncation from a 
$5$-dimensional Chern-Simons theory. Its topological origin makes it an interesting 
candidate for an easier quantization, \textit{e.g.} in the Loop Quantization framework.
The present work is dedicated to a classical analysis of the model's properties. 
Cosmological solutions as well as   
wave solutions are found and compared with the corresponding solutions 
of Einstein's General Relativity with cosmological constant.
\par}}
\end{flushright}
\vfill
\pagebreak
%-----------------------------------------------
%%%%%%%%%%%%%%%    ABSTRACT    %%%%%%%%%%%%%%%%%
%-----------------------------------------------

%\vfill
%\pagebreak
%-----------------------------------------------
%%%%%%    LISTA DE FIGURAS E TABELAS    %%%%%%%%
%-----------------------------------------------

%-----------------------------------------------
%%%%%%%%%%%    ESTILO DA PÁGINA    %%%%%%%%%%%%%
%-----------------------------------------------
\renewcommand{\chaptermark}[1]{\markboth{#1}{}}
\renewcommand{\sectionmark}[1]{\markright{\thesection\ #1}}
\setcounter{chapter}{0} \pagenumbering{arabic}
%-----------------------------------------------
%%%%%%%%%%%    ESTILO DA PÁGINA    %%%%%%%%%%%%%
%-----------------------------------------------

%-----------------------------------------------
%%%%%%%%%%%%    INÍCIO DA TESE    %%%%%%%%%%%%%%
%-----------------------------------------------

%
\chapter*{Introdução e Motivação}
\addcontentsline{toc}{chapter}{Introdução e Motivação}
\pagestyle{fancy}
\lhead{\bfseries Introdução e Motivação}
\rhead{}
\parindent3em
\label{intro}

Usualmente apresenta-se a relatividade geral (RG) como uma teoria métrica, onde o campo dinâmico da teoria é codificado exclusivamente na métrica do espaço-tempo. Essa descrição é conhecida como formalismo de segunda-ordem. Contudo, ela pode ser reformulada como uma teoria dinâmica de conexões no chamado formalismo de primeira-ordem. Essa formulação coloca a relatividade geral mais próxima às teorias de gauge que descrevem as demais interações da Natureza. No entanto, com uma diferença sutíl na sua dinâmica. Em particular, enquanto a dinâmica das teorias de gauge do Modelo Padrão exigem uma geometria de fundo fixa, a saber espaço de Minkowski, a interação gravitacional é fundamentalmente diferente. A grande lição da relatividade geral é que os graus de liberdade do campo gravitacional são codificados na geometria do espaço-tempo. O espaço-tempo é completamente dinâmico: a noção de um fundo fixo sobre o qual as ``coisas acontecem'' perde o sentido. O campo gravitacional define a geometria sobre a qual seus graus de liberdade bem como dos campos de matéria propagam-se. Consequentemente, essa perda da noção familiar de um espaço-tempo como uma espécie de ``palco'' onde a dinâmica dos demais campos se desdobra é conhecida como uma teoria independente de fundo (\textit{background independence}).

 A gravitação, como descrita pela RG de Einstein, não é uma teoria de campos em um fundo curvo, mais do que isso, aquilo que chamamos de espaço-tempo é uma entidade física dinâmica. Uma consequência imediata é que um \emph{quanta} do campo gravitacional não pode estar no espaço-tempo: eles devem construir o espaço-tempo. Com efeito, existem propostas para deteção de um comportamento anômalo na propagação da luz, em um espaço-tempo que emerge de uma teoria de gravitação quântica de laços (LQG) \cite{Ash1, Rov1, Rov2}, devido a sua estrutura tipo-polímero ou granular, que deveria produzir modificações nas equações de Maxwell \cite{Pullin}. Essa é uma tarefa necessária se desejamos compreender a estrutura do espaço-tempo em escalas próximas ao comprimento de Planck $l_{P} \sim 10^{-33}$ cm. Além disso, há a esperança de que uma teoria quântica da gravitação irá curar as singularidades da teoria no nível clássico, tais como \emph{Big Bang} e Buracos-Negros. Por outro lado, a ausência de uma noção preferencial de tempo implica que a hamiltoniana da teoria seja uma combinação linear de vínculos \cite{Boj1,Wal}. Consequentemente, as equações de movimento de Hamilton não podem mais ser interpretadas como uma evolução temporal, pelo contrário, correspondem a um ``movimento'' ao longo das órbitas de gauge da RG. Nessa perspectiva, a noção de espaço-tempo torna-se secundária e a interpretação dinâmica da teoria parece ser problemática.  

A independência de fundo da teoria implica que a sua formulação canônica traz à tona o grupo de difeomorfismos como parte do grupo de gauge da RG. A LQG é construída como uma formulação  de quantização canônica do espaço de fase da RG em termos das conexões tipo Yang-Mills de SU$(2)$, o que nos introduz um grupo de simetria extra na formulção via a representação de laços \cite{Ash1, Rov1, Thi}. A presença de simetrias de gauge nos leva, naturalmente, a existência de relações entre as variáveis do espaço de fase - definidas em uma superfície tipo-espaço - conhecidas como vínculos. Esses vínculos, por sua vez, definiem uma álgebra através dos parêntes de Poisson, e são caracterizados como sendo os geradores infinitesimais das transformações de gauge. Existem três tipo de vínculos locais nessa formulação: $G^{i}$ - chamado de vínculo de Gauss - geradores das transformações de gauge de SU$(2)$, três vínculos locais $\hat{V}_{a}$ - vínculos vetoriais associados aos geradores dos difeomorfismos espaciais, e finalmente o vínculo escalar $S$ relacionado as simetrias de gauge remanescentes.

A quantização canônica de sistemas com simetrias de gauge é conhecida como programa de Dirac-Bergmann \cite{Hans, Dir}. Esse algorítimo aplicado à quantização da RG na representação de holonomias é conhecido como LQG. Portanto, seguindo esse programa, devemos identificar todos os vínculos associados as variáveis do espaço de fase e promovê-los a operadores auto-adjuntos que irão atuar sobre o espaço de Hilbert, satisfazendo as relações de comutação e, finalmente, resolvê-los. A dificuldade em se realizar esse programa tem trazido problemas de cunho matemático tanto quanto de interpretação física. Esse desafio levou a novas ideias de se abordar o problema tais como a formulação dos \emph{spin-foams} \cite{Perez}, que foi construída como uma alternativa para superar os problemas dinâmicos da LQG bem como da definição de quantidades observáveis. Portanto, na tentativa de se contornar essas dificuldades, repensa-se o problema através da perspectiva de uma quantização via integral de caminho.

Por outro lado, essas dificuldades são evitadas em teorias de cunho topológico. Na série de trabalhos \cite{Piguet1, Piguet2, Piguet3}, estudou-se a quantização de laços em teorias topológicas em dimensões mais baixas. O interesse em tais modelos reside no fato de eles compartilharem a propriedade de ser independente de fundo como na RG em $4$D e poderem, em certos casos, representarem teorias de gravitação com resultados interessantes \cite{Witten}. Além disso, é sabido que em teorias topológicas genéricas \cite{Hen2} o difeomofismo temporal não é independente dos difeomorfismos espaciais e das demais simetrias de gauge. Em outras palavras, o difeomorfismo temporal pode ser escrito em termos dos difeomorfismos espaciais e das transformações de gauge internas,  o que facilita muito a aplicação das técnicas de quantização via laços pois nesse caso o vínculo hamiltoniano ou escalar é consequência dos demais.

 Em nosso trabalho, propomos a investigação de problemas mais atuais da gravitação. A RG em um espaço-tempo $3$-dimensional com ou sem constante cosmológica, na ausência de matéria, pode ser descrita como uma teoria de Chern-Simons tendo como grupo de calibre, ou seja, como simetria local o grupo de Poincaré ou (anti-) de Sitter. Essa teoria de cunho topológico, isto é, sem uma estrutura métrica dada \textit{a priori}, mostrou-se ser ausente de graus de liberdade locais. No entanto, M. Bañados e seus colaboradores \cite{Hen2} mostraram que as teorias de Chern-Simons em dimensões mais altas, mesmo sendo construídas via o mesmo padrão topológico que em $2 \,+\, 1$ dimensões, possuem, em geral, graus de liberdades locais não-nulos. Apresentamos como motivação do trabalho um modelo, devido a Chamseddine \cite{Cham, Cham2}, de gravitação em $4$ dimensões obtida de uma teoria de Chern-Simons em $5$ dimensões, via o processo de Kaluza-Klein de redução dimensional,  tendo como grupo de calibre o grupo de (A)dS SO$(1,5)$ ou SO$(2,4)$. Mais especificamente, a proposta do presente trabalho é investigar a dinâmica do modelo de Chamseddine e comparar suas soluções com as soluções da teoria de Einstein convencionais. 
 
 O presente trabalho começa no Capítulo $1$ com uma breve revisão sobre o formalismo de Einstein-Cartan bem como elementos do cálculo exterior. No Capítulo $2$ fazemos a derivação do modelo de Chamseddine em $4$D - cuja invariância de gauge é a de de Sitter SO$(1,4)$ ou anti-de Sitter SO$(2,3)$ - de uma teoria de Chern-Simons em $5$D, sob o grupo de simetria local SO$(1,5)$ ou SO$(2,4)$, via uma redução dimensional e truncação de alguns campos. Mostramos que a teoria, através de uma boa escolha de fixação de gauge, se reduz a uma teoria de gravitação com torção interagindo com um campo escalar tipo-dilaton. Mostramos que as equações de campo da teoria de Chamseddine são soluções especiais da teoria de Chern-Simons completa reduzida a $4$D. As aproximações lineares são estudadas no Capítulo $3$, o que nos leva ao limite newtoniano e a presença de soluções de ondas gravitacionais. Além disso, fizemos um estudo sobre as soluções cosmológicas da teoria e sua comparação com o modelo $\Lambda$CDM. Conclusões e Perspectivas são apresentadas no Capítulo $4$.  Três apêndices apresentando alguns detalhes operacionais foram anexados. Esse trabalho rendeu a publicação de um artigo \cite{Top-Grav} que é de acesso aberto ao público.

\chapter{Formalismo de Einstein-Cartan}
\pagestyle{fancy}
\lhead{\bfseries 1. O Formalismo de Einstein-Cartan}
\rhead{}
\parindent3em
\label{chap.sme}
\section{O Formalismo de Gauge da Gravitação}
A relatividade geral é, antes de tudo, uma teoria que descreve com grande acuidade a gravitação. Desde sua gênese, por volta do ano de $1915$, ela vem sendo testada e corroborada por inúmeros experimentos terrestres bem como observações astronômicas \cite{Will}. Em $1916$, um ano após a formulação das equações de campo da relatividade geral, Einstein previu que no limite de linearização de suas equações, essas apresentavam soluções que previam a existência de ondas gravitacionais. Um século após essas predições de Einstein, a equipe do LIGO (\textit{Laser Interferometer Gravitational-Wave Observatory}) anunciou a primeira detecção direta de ondas gravitacionais \cite{ligo1,ligo2}. Mais uma vez os experimentos corroboram que, de fato, a descrição da gravitação einsteiniana nos concede um esquadrinhamento preciso da natureza. Lev Landau considerou, acredito que com muita razão, a relatividade geral como: ``the most beautiful of the scientific theories''. Entretanto, a relatividade geral é muito mais do que isso. Ela é uma modificação na nossa compreensão da natureza do espaço-tempo cujo  conteúdo ainda possui consequências insondadas. Essa seção não tem a pretensão de uma introdução a relatividade, muito menos uma descrição exaustiva de toda amplitude da teoria. Para isso, convido ao leitor aos livros textos clássicos de referência \cite{Trautman1,Trautman2,Misner, Gasperini, Wal, Weinberg}. Com efeito, darei apenas uma curta apresentação do formalismo em sua versão mais moderna, enfatizando as características fundamentais, bem como o verdadeiro conteúdo físico por tras do fenômeno gravitacional que é sua invariância de gauge.

A relatividade geral é usualmente apresentada como uma teoria métrica, onde o campo dinâmico da teoria é codificado exclusivamente na métrica do espaço-tempo. Essa descrição é conhecida como formalismo de segunda-ordem. Contudo, ela pode ser reformulada como uma teoria dinâmica de conexões no chamado formalismo de primeira-ordem. Essa formulação coloca a relatividade geral mais próxima às teorias de gauge que descrevem as demais interações da Natureza, no entanto, com uma diferença sutíl na sua dinâmica. Em particular, enquanto a dinâmica das teorias de gauge do Modelo Padrão exigem uma geometria de fundo fixa, a saber espaço de Minkowski, a interação gravitacional é fundamentalmente diferente. A grande lição da relatividade geral é que os graus de liberdade do campo gravitacional são codificados na geometria do espaço-tempo. O espaço-tempo é completamente dinâmico: a noção de um fundo fixo sobre o qual as ``coisas acontecem'' perde o sentido. O campo gravitacional define a geometria sobre a qual seus graus de liberdade bem como dos campos de matéria propagam-se. A relatividade geral não é uma teoria de campos em um fundo curvo, mais do que isso, aquilo que chamamos de espaço-tempo é uma entidade física dinâmica. Uma consequência imediata é que um quanta do campo gravitacional não pode estar no espaço-tempo: eles devem construir o espaço-tempo. Essa perda da noção familiar de um espaço-tempo como uma espécie de ``palco'' onde a dinâmica dos demais campos se desdobra é conhecida como uma teoria independente de fundo (\textit{background independence}). 

A fim de identificar a invariância de gauge associada a interação gravitacional, isto é, a possibilidade de escrever as equações de Einstein num formalismo tipo Yang-Mills, tendo o grupo de Lorentz (não compacto) como grupo de simetria local \cite{Blago,report,Kibble}, precisamos fazer algumas considerações sobre um princípio básico da relatividade geral, a saber, princípio da equivalência.

Pouco tempo depois da sua descoberta da relatividade especial, Einstein observou, em seus famosos experimentos mentais, que o efeito da gravitação pode ser neutralizado. Portanto, ele percebeu que um observador que estivesse em queda livre não seria capaz de sentir seu próprio peso. Em outras palavras, em um elevador em queda livre, o efeito da gravitação pode ser eliminado. No entanto, esse ``truque'' é muito restrito, ou seja, funciona apenas localmente: o laboratório ou o elevador tem de ser pequeno o suficiente e o tempo do experimento curto o suficiente para que nenhuma inomogeneidade do campo gravitacional seja percebida. Sob tais condições, experimentos realizados em queda livre são indistinguíveis daqueles realizados na ausência de gravitação. Naturalmente, as leis da física estarão sob o regime de validade do espaço de Minkowski. Portanto, isso significa que, em uma vizinhança local, o espaço-tempo possui invariância de Lorentz. Em cada ponto do espaço-tempo, podemos obter um referencial cujo movimento é inercial. Para que essa invariânica seja manifestada, é necessário fazermos uma transformação de coordenadas apropriada a um sistema de referência inercial particular. Denotemos por $X^{I}$ as coordenadas locais definidas pelo referencial inercial e seja $x = x^{\mu}$ as coordenadas arbitrárias não necessariamente inerciais. As coordenadas $X^{I}$ podem ser expressadas como funções
\begin{equation*}
X^{I} = X^{I}(x)
\end{equation*} 
das coordenadas arbitrárias $x$. Nas coordenadas $x$, a não linearidade do movimento é interpretada como o efeito de um campo gravitacional. Portanto, gravitação é a informação codificada na mudança de coordenadas que nos leva de um sistema arbitrário a um inercial. Essa informação está contida na função que relaciona os sistemas de coordenadas $X(x)$. Mas como discutido acima, apenas valores dessas funcões em uma vizinhança pequena o suficiente é relevante, pois se nos afastarmos muito dessa região, o sistema local de referência inercial irá mudar. Podemos fazer uma expansão em Taylor, se consideramos que a origem $x = 0$ seja associada ao evento a ser analisado, e que $X^{I}(x = 0) = 0$; a única contribuição que devemos guardar seria em primeira ordem. Daí,
\begin{equation}
X^{I} = \frac{\partial X^{I}}{\partial x^{\mu}}\bigg|_{x = 0} x^{\mu},
\end{equation}
define como podemos passar de um conjunto de coordenadas arbitrárias para um inercial. A quantidade $\dfrac{\partial X^{I}}{\partial x^{\mu}}\bigg|_{x = 0}$ nada mais é que o jacobiano dessa transformação em particular que guarda em si a informação do campo gravitacional presente na região. Definimos o a quantidade

\begin{equation*}
e^{I}_{\mu}(x = 0) = \frac{\partial X^{I}}{\partial x^{\mu}}\bigg|_{x = 0},
\end{equation*}  
como vierbein\footnote{\textit{Vierbein} é uma palavra do alemão que é a conjunção de \textit{vier} que sigifica quatro com \textit{bein} que é traduzido como perna, logo \textit{vierbein} igual a quatro pernas.} ou tetrada, responsável por nos informar sobre a presença local do campo gravitacional. Naturalmente essa construção não se restringe ao nosso ponto em particular e pode ser concebida em cada ponto $x$. E, portanto, a quantidade 
\begin{equation}
e^{I}_{\mu}(x) = \frac{\partial X^{I}}{\partial x^{\mu}}\bigg|_{x}
\end{equation} 
é o campo gravitacional no ponto $x$.

O campo gravitacional $e^{I}_{\mu}(x)$ é, portanto, representado pela matriz jacobiana da transformação de coordenadas de $x$ para as coordenadas localmente inerciais $X^{I}$. De maneira mais precisa, $e^{I}_{\mu}(x)$ nos informa que sendo espaço-tempo uma variedade diferenciável $\mathcal{M}$, em cada ponto $x \in \mathcal{M}$ existe um espaço tangente $T_{x}$, que é uma boa aproximação de $\mathcal{M}$ nas vizinhanças de $x$. Esse espaço tangente é um sistema de referência local e inercial, ou seja, em queda livre como indicado pelo princípio da equivalência. O fato de medições poderem ser feitas independentes da escolha do referencial e poderem ser traduzidas a um referencial inercial, significa que existe um isomorfismo entre tensores em $\mathcal{M}$ e tensores em $T_{x}$, de maneira que esses são definidos em $T_{x}$, representado através de um mapeamento linear chamado de \textit{vierbein}, que de fato faz uma troca de base em $T_{x}$.

Podemos definir a separação entre coordenadas de dois pontos infinitesimalmente próximos em $\mathcal{M}$. Sua correspondência com a separação no referencial em queda livre será 
\begin{equation*}
dX^{I} = e^{I}_{\mu}(x)dx^{\mu}.
\end{equation*}  
Como o espaço tangente é um espaço de Minkowski local, consequência do princípio da equivalência, ele possui naturalmente uma métrica $\eta_{IJ} = \textrm{diag}(-1, 1, 1, 1)$ que define a métrica em $\mathcal{M}$ através do isomorfismo (campo gravitacional) $e^{I}_{\mu}(x)$. De fato, o elemento de linha
\begin{eqnarray*}
ds^{2} & = & \eta_{IJ}dX^{I}dX^{J} \\[0.1cm]
       & = & \eta_{IJ}e^{I}_{\mu}(x)e^{J}_{\nu}(x)dx^{\mu}dx^{\nu}\\[0.1cm]
       & = & g_{\mu\nu}(x)dx^{\mu}dx^{\nu},
\end{eqnarray*}
onde
\begin{equation}
e^{I}_{\mu}(x)\eta_{IJ}e^{J}_{\nu}(x) = g_{\mu\nu}(x),
\end{equation}
é a métrica em $\mathcal{M}$, induzida pelo \textit{vierbein} $e^{I}_{\mu}(x)$ e a métrica do espaço tangente local $\eta_{IJ}$.

Essa relação nos mostra que dado $e^{I}_{\mu}(x)$ podemos derivar a métrica do espaço-tempo e, portanto, todas as propriedades da métrica $g_{\mu\nu}(x)$ estão codificadas no \textit{vierbein}. Muitas vezes pensa-se em $e^{I}_{\mu}(x)$ como a ``raíz quadrada'' da métrica. Se sabemos $e^{I}_{\mu}(x)$ podemos facilmente calcular $g_{\mu\nu}(x)$. Portanto, podemos considerar $e^{I}_{\mu}(x)$ como sendo o verdadeiro campo fundamental da teoria capaz de revelar a verdadeira simetria da relatividade geral, invariância de Lorentz local, e a métrica como um campo secundário. A recíproca, no entanto, não é verdadeira: dada uma métrica $g_{\mu\nu}(x)$, existe uma infinidade de escolhas possíveis de $e^{I}_{\mu}(x)$ que reproduzem a mesma métrica. Essa perda da unicidade na definição do \textit{vierbein}, dada uma métrica \textit{a priori}, é facilmente verificada. O fato de irmos da descrição $g_{\mu\nu} \mapsto e^{I}_{\mu}(x)$ estamos ganhando uma simetria extra, pois podemos fazer ``rotações'' de Lorentz sobre essa nova base de modo que a métrica não consiga perceber essa simetria local. Em outras palavras, é possível fazer transformações de Lorentz sobre o \textit{vierbein} de maneira a serem indetectáveis do ponto de vista da variedade $\mathcal{M}$. Sob uma transformação de Lorentz, o  \textit{vierbein} transforma-se como
\begin{equation}
e^{I}_{\mu}(x) \longmapsto e'^{I}_{\mu}(x) = \Lambda^{I}_{J}(x) e^{J}_{\mu}(x) 
\end{equation}
onde a matriz $\Lambda(x) \in$ SO$(1,3)$. Pela definição do grupo de Lorentz SO$(1,3)$, a transformação $\Lambda(x)$ deixa a métrica do espaço tangente invariante,
\begin{equation*}
\Lambda_{K}^{I}\eta_{IJ}\Lambda^{J}_{L} = \eta_{KL},
\end{equation*}
de maneira mais compacta temos 
\begin{equation*}
\Lambda^{T}\eta\Lambda = \eta.
\end{equation*}
A métrica $g_{\mu\nu}$ claramente não percebe as transformações de simetria feitas em $T_{x}$. Isso significa, em particular, que existem mais componentes independentes, ou melhor, mais graus de liberdade em $e^{I}_{\mu}(x)$ do que em $g_{\mu\nu}$. De fato o \textit{vierbein} possui $16$ componentes independentes enquanto a métrica, por ser simétrica, apenas $10$. A pergunta que nos fazemos é: se as descrições são equivalentes como é possível haver mais graus de liberdade nos \textit{vierbein}? A resposta reside justamente, no conteúdo de \textit{gauge} que se revela através das transformações de Lorentz locais. Temos exatamente, $6$ possíveis transformações de Lorentz independentes ($3$ rotações e  $3$ \textit{boosts}) que nos conecta a essa simetria local associada ao \textit{vierbein}. Assim, a observação de Einstein de que o princípio de equivalência é uma característica central na relatividade geral, implica que as transformações de SO$(1,3)$ aparecem como a verdadeira simetria de \textit{gauge} local da gravitação.

Naturalmente, uma mudança de base ou de coordenadas deve ser inversível. Assim, deve existir uma matriz $e^{\mu}_{J}$ de modo que 
\begin{eqnarray*}
e^{I}_{\mu}e^{\mu}_{J} = \delta^{I}_{J}, & & e^{I}_{\mu}e^{\nu}_{I} = \delta^{\mu}_{\nu}.
\end{eqnarray*}

Vimos que o campo gravitacional $e^{I}_{\mu}(x)$ age como uma espécie de projetor, isto é, via essa mudança particular de coordenadas podemos trazer todos os campos da variedade a um espaço tangente local. Em outras palavras, existe uma coleção de espaços tangentes que a cada ponto da variedade $\mathcal{M}$ definem a ação do grupo de simetria local (Lorentz, SO$(1,3)$), e matematicamente isso nos leva a uma estrutura de fibrados \cite{Isham,Nakahara,report}. Portanto, os campos de \textit{gauge} do espaço-tempo serão representações locais do grupo interno. De fato, o grupo de Lorentz restrito, isto é, \emph{próprio} e \emph{ortóchrono} \cite{Barut,Kirsten} ($\bigr(\Lambda^{0}\,_{0}\bigl)^{2} \geq 1$ e $\Lambda^{0}\,_{0} \geq 1$), isto é, conexo com a identidade,  é um grupo de Lie contínuo com $6$ parâmetros, e uma transformação geral pode ser representada pela exponencial dos geradores da álgebra so$(1,3)$, como segue
\begin{equation*}
\Lambda = e^{\frac{i}{2}\omega^{IJ}M_{IJ}}.
\end{equation*} 
A matriz $\omega^{IJ} = - \omega^{JI}$ é anti-simétrica e contém $6$ parâmetros reais, enquanto os geradores $M_{IJ} = - M_{JI}$ satisfazem a álgebra de Lie do grupo SO$(1,3)$:
\begin{equation}
\lbrack M_{IJ},M_{KL}\rbrack = (-i) \bigl(\eta_{IK}M_{JL} -\eta_{JK}M_{IL} + \eta_{JK}M_{IL} - \eta_{IK}M_{JL} \bigr).
\end{equation}

Sabemos das teorias de \textit{gauge}\cite{Abers} que para levarmos um grupo de simetria global ao nível local, devemos introduzir uma derivação covariante, construída com os campos de \textit{gauge} associados a simetria. Portanto, se $\varphi \in \textrm{irrep}(G)$ e, $\varphi \longmapsto \varphi' = e^{i\alpha(x)} \varphi$ vemos imediatamente que a derivada parcial não é mais covariante devido ao parâmetro $\alpha = \alpha(x)$,
\begin{eqnarray*}
(\partial_{\mu}\varphi)' & = & \partial_{\mu}\varphi' = \partial_{\mu}(e^{i\alpha}\varphi)\\[0.1cm]
                         & = & i\partial_{\mu}\alpha e^{i\alpha}\varphi + e^{i\alpha}\partial_{\mu}\varphi.
\end{eqnarray*}  
A covariância sob transformações locais pode ser recuperada através da introdução de uma conexão de \textit{gauge}, de maneira a obtermos uma derivada que seja, de fato, covariante. No caso mais simples de uma teoria abeliana, como a eletrodinâmica, o potencial vetor $A_{\mu}$ faz esse papel. Assim,
\begin{equation}
D_{\mu} = \partial_{\mu} + igA_{\mu},
\end{equation}
onde $g$ seria a carga (gerador) associado a simetria, na eletrodinâmica, seria a carga elétrica $e$. Nos casos não-abelianos \cite{Gelfand, Fuchs, Ramond}, como veremos em particular no SO$(1,3)$, essa carga será obtida através dos geradores, produzindo o que se chama de álgebra de Lie valorada

\begin{equation}
\bigr(D_{\mu}\varphi\bigl)' = e^{i\alpha}D_{\mu}\varphi.
\end{equation}
Dessa forma $A_{\mu}$ deve se transformar como
\begin{equation*}
A'_{\mu} = e^{i\alpha}A_{\mu}e^{-i\alpha} - \frac{i}{g}(\partial_{\mu}e^{i\alpha})e^{-i\alpha},
\end{equation*}
ou de forma mais compacta, considerando $U := e^{i\alpha}$
\begin{equation}
A'_{\mu} = UA_{\mu}U^{-1} - \frac{i}{g}(\partial_{\mu}U)U^{-1}.
\end{equation}
Portanto, no caso de Lorentz local, os parâmetros passam a depender das coordenadas do espaço-tempo $\omega_{IJ} = \omega_{IJ}(x)$. Precisamos associar aos geradores do grupo, campos de \textit{gauge} $\omega_{\mu}$, onde
\begin{equation}
\omega_{\mu} = \frac{1}{2}\omega_{\mu}^{IJ}M_{IJ}
\end{equation}
representando as componentes da chamada conexão de Lorenz, ou conexão de \textit{spin}, e introduzindo a derivada covariante, definida por:
\begin{equation}
D_{\mu} = \partial_{\mu} - \frac{i}{2}\omega_{\mu}^{IJ}M_{IJ}. 
\end{equation}
Naturalmente, para que a derivada covariante se transforme como o próprio campo sob SO$(1,3)$ local, a conexão de spin deve obedecer a seguinte lei de transformação:
\begin{equation}
\omega'_{\mu} = \Lambda\omega_{\mu}\Lambda^{-1} - i\Lambda\bigr(\partial_{\mu}\Lambda^{-1}\bigl)
\end{equation}

Um exemplo de aplicação desse conceito de derivada covariante associa-se com a tentativa de introdução de matéria fermiônica na gravitação. A representação espinorial irá nos permitir acoplar fémions de Dirac com o campo gravitacional. Com efeito, seja $\psi_{\alpha} \in (\frac{1}{2},0)\oplus (0,\frac{1}{2})\, (\alpha,\beta,.., = 1,2,3,4)$, onde os geradores assumem a seguinte forma:
\begin{equation*}
M^{IJ} = \frac{i}{4}\lbrack\gamma^{I},\gamma^{J}\rbrack.
\end{equation*}

Nesse caso, $\psi$ é um campo com quatro componentes complexas, transformando-se como um espinor da representação $(\frac{1}{2},0)\oplus (0,\frac{1}{2})$ do grupo de Lorentz, e $\gamma^{I}$ são as matrizes de Dirac $4\times 4$ satisfazendo a chamada álgebra de Clifford \cite{Todorov,Rey},
\begin{equation*}
\bigl\{\gamma^{I},\gamma^{J}\bigr\} = 2\eta^{IJ}\mathds{1}_{4\times 4}.
\end{equation*}
Com nossa convenção de assinatura de métrica temos em particular,
\begin{eqnarray*}
(\gamma^{0})^{2} = \mathds{1}, & & (\gamma^{0})^{\dagger} = \gamma^{0}\\[0.1cm]
(\gamma^{i})^{2} = -\mathds{1}, & & (\gamma^{i})^{\dagger} = -\gamma^{i},\, i = 1,2,3.
\end{eqnarray*} 
A forma explícita das matrizes de Dirac depende da escolha da sua representação. Para o propósito do nosso exemplo de aplicação não será necessário abordarmos esse contexto. Segue que a derivada covariante $D_{\mu}$ para um espinor de Dirac assume a forma:
\begin{equation}
D_{\mu}\psi_{\alpha} = \partial_{\mu}\psi_{\alpha} -\frac{1}{8}\omega_{\mu}^{IJ}\lbrack\gamma_{I},\gamma_{J}\rbrack_{\alpha\beta}\psi_{\beta}.
\end{equation}

Nesse ponto vale a pena notarmos que essa simples introdução dos férmions de Dirac na gravitação, via o formalismo de primeira ordem, não se processa na formulação métrica usual einsteiniana da relatividade geral. Pois o grupo de simetria é o grupo de difeomorfismos e não possui representação finita e/ou espinorial. De fato, o grupo $GL(4,\mathbb{R})$  de simetria da relatividade geral não apresenta representação espinorial \cite{Siegel}. Portanto, a formulação do campo gravitacional como uma métrica pseudo-riemanniana não pode estar fundamentalmente correta \cite{Rov1,Rov2}, justamente pelo fato de não permitir acoplamento fermiônico, e os férmions existem no universo. Assim, precisamos do formalismo do vierbein para acoplar a dinâmica de férmions regida pela equação de Dirac
\begin{equation*}
(i\hslash\gamma^{I}\partial_{I} - mc)\psi = 0.
\end{equation*} 

Consideremos agora a aplicação da derivada covariante a um campo vetorial $A^{I}(x) \in (\frac{1}{2},\frac{1}{2})$. Nesse caso, os geradores são da forma
\begin{equation*}
\bigl(M_{IJ}\bigr)^{K}\,_{L} = -i(\eta_{IL}\delta^{K}_{J} - \eta_{JL}\delta^{K}_{I}).
\end{equation*}
Implicando que a derivada covariante assuma a forma
\begin{eqnarray*}
D_{\mu}A^{I} & = & \partial_{\mu}A^{I} - \frac{i}{2}\omega_{\mu}^{IJ}\bigl(M_{IJ}\bigr)^{K}\,_{L}A^{L}\\[0.1cm]
             & = & \partial_{\mu}A^{I} +\omega_{\mu}^{I}\,_{J}A^{J}, \quad \omega_{\mu}^{I}\,_{J} = - \omega_{\mu\, J}\,^{I},
\end{eqnarray*}
e assim por diante. Sabendo em qual representação do grupo local o campo se situa podemos construir a derivada covariante. Para obtermos a derivada covariante de um vetor covariante $A_{I}$, podemos partir de considerações do produto escalar $A_{I}X^{I}$, e
\begin{equation*}
D_{\mu}(A_{I}X^{I}) = X^{I}D_{\mu}A_{I} + A_{I}(\partial_{\mu}X^{I} + \omega_{\mu}^{I}\,_{J}X^{J}).
\end{equation*}
Portanto, lembrando que a derivada covariante de um escalar sob o grupo é a própria derivação parcial sem termos de conexão, segue-se que
\begin{equation}
D_{\mu}A_{I} = \partial_{\mu}A_{I} - \omega_{\mu}^{J}\,_{I}A_{J}.
\end{equation}
E assim sucessivamente para tensores de rank arbitrário.

Seguindo as nossas convenções de regra de diferenciação covariante, vemos que a ação da conexão de spin segue um padrão específico nos índices de Lorentz dos campos, e deve possuir o sinal positivo quando agir sobre vetores contravariantes, e um sinal negativo para os covariantes. Para índices mistos, temos
\begin{equation}
D_{\mu}T^{I}\,_{J} = \partial_{\mu}T^{I}\,_{J} + \omega_{\mu}^{I}\,_{K}T^{K}\,_{J} - \omega_{\mu}^{K}\,_{J}T^{I}\,_{K}.
\end{equation}
Assim, vemos a analogia com a derivada covariante $\nabla_{\mu} = \partial_{\mu}\, +\, \Gamma_{\mu}$ para vetores da variedade $\mathcal{M}$, utilizada no formalismo métrico da relativiade geral \cite{Zee,Bryce, Poisson, Inverno, Padmanabhan}.
\section{Metricidade, Torção e Curvatura}

Vimos na seção anterior que o formalismo de vierbein nos possibilita escrever todas as quantidades geométricas da variedade $\mathcal{M}$ no espaço tangente $T_{x}$. Em outras palavras, os vierbein nos fornecem uma transformação de coordenadas específica, baseada no princípio da equivalência, onde as quantidades da variedade podem ser projetadas ponto a ponto no espaço tangente. Além do mais, para as quantidades projetadas podemos introduzir o conceito de uma derivada covariante, justamente pelo fato da simetria ser local, baseada na conexão de Lorentz que seja capaz de preservar a invariância local do espaço tangente de Minkowski. Todo esse formalismo mostra-se consistente com a covariância sob difeomorfismos associada a variedade (pseudo)riemanniana.

A pergunta que se segue é: seria possível conectar ou estabeler uma correspondência entre os formalismo de primeira e segunda ordem? Em particular, entre as derivadas de Lorentz e de Riemann, entre a conexão de Christoffel ($\Gamma$)\footnote{Em geral, tanto a métrica $g$ quanto a conexão afim $\Gamma$ são necessárias para descrição completa da geometria do espaço-tempo. Se $g$ e $\Gamma$ são quantidades completamente independentes então temos uma variedade equipada com uma estrutura geométrica e afim. Por outro lado, se $\Gamma$ for determinada completamente por $g$, implica que a métrica, por si só, é capaz de dar uma descrição completa e suficiente da geometria da variedade. De fato, Einstein em sua formulação da gravitação faz uso da variedade assumindo que está contendo apenas uma estrutura métrica, isto é, ele assume a conexão de Levi-Civita conhecida como o símbolo de Christoffel: $\Gamma^{\alpha}_{\mu\nu}\, =\, \frac{1}{2}g^{\alpha\beta}(\partial_{\mu}g_{\nu\beta}\,+\, \partial_{\nu}g_{\beta\mu}\,-\, \partial_{\beta}g_{\mu\nu}).$} e a conexão de spin ($\omega$). Sendo possível essa correspondência, assim como $\Gamma$ pode ser escrito em termos da métrica $g$, e $g$ como uma função do vierbein $e$, poderíamos esperar a existência de uma relação precisa que nos dê a conexão de spin como uma função do vierbein $\omega(e)$. Portanto, a pergunta que também iremos respoder é: haveria, de fato, alguma vantagem na descrição do formalismo de primeira ordem? Existe algum ganho de informação na descrição de gauge da relativade geral, ou são completamente equivalentes?

Para começarmos a responder essas perguntas, notemos que o vierbein $e^{I}_{\mu}$ carrega tanto índice de grupo local quanto de variedade. Logo, ele deve sentir uma variação tanto do fibrado \cite{Nakahara, report} tangente de $\mathcal{M}$, $T(\mathcal{M}) = \cup_{p}(p,T_{p}(\mathcal{M}))$, quanto do fibrado de Lorentz $F = (\mathcal{M}, SO(1,3))$. Com efeito, podemos definir uma derivada covariante para objetos com ambos os índices, tais como o vierbein, 
\begin{equation}
\mathcal{D}_{\mu}e^{I}_{\nu} = \partial_{\mu}e^{I}_{\nu} + \omega_{\mu}^{I}\,_{J}e^{J}_{\nu} - \Gamma^{\rho}_{\mu\nu}e^{I}_{\rho}.
\end{equation} 
Assim como a conexão de Levi-Civita $\Gamma(g)$ é compatível com a métrica (metricidade), isto é, $\nabla_{\mu}g_{\nu\rho} = 0$, iremos demandar que $\omega_{\mu}$ seja compatível com o vierbein, isto é, $\mathcal{D}_{\mu}e^{I}_{\nu} = 0$. Isso implica que
\begin{equation}
\partial_{(\mu}e^{I}_{\nu)} + \omega_{(\mu}\,^{I}\,_{J}e^{J}_{\nu)} = \Gamma^{\rho}_{(\mu\nu)}e^{I}_{\rho}, \quad \partial_{[\mu}e^{I}_{\nu]} + \omega_{[\mu}\,^{I}\,_{J}e^{J}_{\nu]} = \Gamma^{\rho}_{[\nu\mu]}e^{I}_{\rho} = 0, \label{metricidade}
\end{equation}
onde separamos os índices do espaço-tempo em sua combinação simétrica e anti-simétrica, e usamos o fato  de que a conexão de Levi-Civita $\Gamma(g)$ não possui parte anti-simétrica.
Dessas equações vemos imediatamente a existência da seguinte relação entre a conexão de spin  e a conexão de Levi-Civita,
\begin{equation}
\omega_{\mu}\,^{I}\,_{J} = e^{I}_{\nu}\nabla_{\mu}e^{\nu}_{J}.
\end{equation}
 
 Naturalmente, podemos generalizar o caso para uma relação entre a conexão de spin e uma conexão que não seja necessariamente de Levi-Civita. Da condição de metricidade para o vierbein e, usando-os para projetar índices entre o epaço tangente e os índices da variedade, temos para a parte anti-simétrica, segundo (\ref{metricidade})
 \begin{equation}
 \partial_{[\mu}e^{I}_{\nu]} + \omega_{[\mu}\,^{I}\,_{J}e^{J}_{\nu]} - T^{I}_{[\nu\mu]} = 0,\label{definição-torção}
 \end{equation}
 onde $T^{I}_{[\mu\nu]} := \Gamma^{\rho}_{[\nu\mu]}e^{I}_{\rho}$.
 É interessante notar que a presença da parte anti-simétrica na conexão $\Gamma$ não é necessariamente excluída pela condição de metricidade, e podemos calcular $\omega$ levando em consideração contribuições não-nulas da parte anti-simétrica de $\Gamma$ chamadas de Torção \cite{Moshe, Simone}. Assim, podemos obter a conexão de spin mais geral que seja compatível com a condição de metricidade do veirbein.
 
 Para facilitar as contas iremos projetar todos os índices da relação acima no espaço tangente (contraindo-os com o uso do vierbein $e^{\mu}_{J} e^{\nu}_{K}$): obtemos
 \begin{equation}
 \xi^{I}\,_{JK} + \frac{1}{2}(\omega_{J}\,^{I}\,_{K} - \omega_{K}\,^{I}\,_{J}) - T^{I}\,_{JK} = 0,
 \end{equation}
onde 
\begin{equation}
\xi^{I}\,_{JK} := (\partial_{\mu}e^{I}_{\nu} - \partial_{\nu}e^{I}_{\mu})e^{\mu}_{J}e^{\nu}_{K} = \xi^{I}\,_{[JK]} \label{xi-eqn}
\end{equation}
são chamados de coeficientes de rotação de Ricci. Escrevendo a relação acima com suas permutações cíclicas dos índices $I,\, J,\, K$, temos:
\begin{eqnarray*}
\xi_{IJK} + \frac{1}{2}(\omega_{JIK} - \omega_{KIJ}) - T_{IJK} & = & 0, \\[0.1cm]
\xi_{JKI} + \frac{1}{2}(\omega_{IKJ} - \omega_{IJK}) - T_{JKI} & = & 0, \\[0.1cm]
\xi_{KIJ} + \frac{1}{2}(\omega_{KJI} - \omega_{JKI}) - T_{KIJ} & = & 0.
\end{eqnarray*}
Somando as duas primeiras equações e subtraindo a terceira, e usando as propriedades de simetria $\omega_{IJK} = -\omega_{I[JK]}$, encontramos a seguinte relação para a conexão de spin
\begin{equation}
\omega_{IJK} = (\xi_{IJK} + \xi_{JKI} - \xi_{KIJ}) + (T_{KIJ} - T_{IJK} - T_{JKI}).
\end{equation}
Finalmente, podemos colocar o resultado acima na forma usualmente conhecida, subindo os índices $J$ e $K$, e projetando o índice $I$ na variedade: chega-se assim a seguinte forma
\begin{equation}
\omega_{\mu}\,^{JK} = \bar{\omega}_{\mu}\,^{JK} + \mathcal{C}_{\mu}\,^{JK},\label{conexão-completa}
\end{equation}
onde
\begin{equation}
\bar{\omega}_{\mu}\,^{IJ} = e^{K}_{\mu}(\xi_{K}\,^{IJ} + \xi^{J}\,_{K}\,^{I} - \xi^{IJ}\,_{K}) \label{conexão-sem torção}
\end{equation}
é a chamada conexão de Levi-Civita (conexão com torção nula), e 
\begin{equation}
\mathcal{C}_{\mu}\,^{IJ} = -e^{K}_{\mu}(T_{K}\,^{IJ} + T^{J}\,_{K}\,^{I} - T^{IJ}\,_{K})\label{contorção}
\end{equation}
é o tensor de contorção que é formado por essa combinação dos termos da torção. Enfim, das equa\c c\~oes (\ref{definição-torção}-\ref{contorção}) segue a seguinte express\~ao para a 
tors\~ao:
\begin{equation}
T^{I}\,_{\mu\nu} = D_{\mu}e^{I}_{\nu} - D_{\nu}e^{I}_{\mu}.
\end{equation}

A fim de completarmos nossa pequena apresentação desse formalismo geométrico baseado no vierbein bem como nas conexões de gauge, ainda precisamos de uma expressão para a curvatura. Depois de escrevermos o tensor de Riemann em termos dessas novas variáveis dinâmicas seremos capazes de expressar as equações de Einstein usando essa nova abordagem. Veremos pelo princípio da ação que os dois formalismos se equivalem, porém, somos capazes de mostrar que a condição de torção nula não é uma necessidade \textit{a priori}, mas um resultado das equações de movimento. Através do formalismo de primeira ordem fomos capazes de descortinar e trazer à tona a verdadeira invariância por trás da gravitação, a saber, a gravitação é uma teoria de gauge sob o grupo de Lorentz local como vimos\footnote{Alguns grupos de pesquisa não concordam que o grupo de simetria local da gravitação seja Lorentz. Assumem como grupo local o grupo de \emph{Poincaré} o que leva a novos \emph{insights} interpetativos. Nessa formulação, o papel da Torção torna-se ainda mais relevante, como é o caso da abordagem do Teleparalelismo\cite{Pereira, Blago2}}. 

Todavia, existe uma diferença de suma importância entre a relatividade geral e as teorias de gauge convencionais, devido ao fato de que a ação de Einstein-Hilbert é linear, ao invés de quadrática, no campo de Yang-Mills, isto é, na curvatura. A razão física para esse fato é que  no caso da gravitação, a conexão de gauge apresente-se como uma função de outro campo, a saber, o vierbein - que é de fato a variável dinâmica fundamental. Entretanto, isso não exclui a possibilidade de considerar modelos de gravitação contendo potências mais altas na curvatura \cite{Felice, Stelle, Capo}.

\section{Elementos do Cálculo Exterior}
Essa seção não contém nenhuma noção física a mais, mas tem o propósito de reescrever em uma linguagem mais moderna os resultados anteriores. Essa é a linguagem do formalismo das formas diferenciais ou cálculo exterior. Com esse formalismo poderemos reescrever as equações anteriores de uma maneira mais compacta, onde os índices tensorias do espaço-tempo estarão ``escondidos'', de maneira precisa, nas próprias variáveis dinâmicas. Portanto, as formas diferenciais são objetos extremamente práticos, primeiro porque essa notação nos permite escrever quantidades tensoriais de maneira independente da escolha particular de coordenadas, segundo, todo o formalismo matemático, tais como a ideia de derivação, covariância, etc., torna-se bem simples com o uso do cálculo exterior. A ideia dessa seção está longe de fornecer uma introdução rigorosa ou exaurir o conteúdo desse assunto. Todavia, será uma seção pedagógica e com a finalidade prática e instrumental do assunto. Para uma busca de verticalidade e precisão matemática do assunto o leitor poderá consultar as referências \cite{Schutz,Felsager, Hull, Lovelock2, Tu, Crampin, Schreiber, Darling, Walter}.

Uma $p$-forma é definida como uma quantidade tensorial com a propriedade de ser completamente anti-simétrica, o que reduz drásticamente as possíveis construções desses objetos. Uma $p$-forma escreve-se, num sistema de coordenadas $x^{\mu}$, como
\begin{equation*}
T_{p} = \frac{1}{p!}T_{\mu_{1}...\mu_{p}}dx^{\mu_{1}}\wedge dx^{\mu_{2}}\wedge...\wedge dx^{\mu_{p}},
\end{equation*}
onde  $T_{\mu_{1}...\mu_{p}} = T_{[\mu_{1}...\mu_{p}]}$ denota um campo tensorial covariante completamente anti-simétrico de rank $p \leq D$, onde $D$ é a dimensão da variedade considerada $\textrm{D} \mathcal{M}$ $=$ $D$. Com efeito, o requerimento de ser completamente anti-simétrica víncula o rank das formas com a dimensão da variedade. Consequentemente, o rank máximo que podemos obter para uma $p$-forma será sempre menor ou igual a dimensão da variedade considerada.  $T_{p} = 0$ sempre que $p > D$, e $\wedge$ denota o produto exterior tal que
\begin{equation*}
dx^{\mu}\wedge dx^{\nu} = \frac{1}{2}\bigl(dx^{\mu}\otimes dx^{\nu} - dx^{\nu}\otimes dx^{\mu}\bigr).
\end{equation*}
Os objetos $dx^{\mu}$ e seus produtos exteriores formam uma base para as formas. Isto é bem simples de se ver, pois os diferenciais $dx^{\mu}$ transformam-se como vetores contravariantes: 

\begin{equation*}
dx^{\mu} \longmapsto dx'^{\mu} = \frac{\partial x'^{\mu}}{\partial x^{\nu}}dx^{\nu}.
\end{equation*}

Denota-se o conjunto de todas as $p$-formas por $\Lambda^{p}$, que é um espaço vetorial, ou seja, se $T_{p}$ e $F_{p}$ são $p$-formas, então $T_{p}\,+\,F_{p}$ e $aF_{p}$, $a\in \mathbb{R}$ também o são. Novamente, se $p > D$, então $\Lambda^{p}$ irá conter apenas o elemento nulo:
\begin{equation*}
\Lambda^{p} = \{0\}, \quad p > D.
\end{equation*} 
 Para vermos isso, seja $F_{p}$ uma $p$-forma, logo $F$ é caracterizado por suas componentes
 \begin{equation*}
 F_{\mu_{1}...\mu_{p}},
 \end{equation*}
 mas se $p > D$, dois dos índices devem sempre coincidir (pois cada índice $\mu_{i}$ pode assumir valores apenas até a dimensão da variedade $1,..., D$), daí segue que $F_{\mu_{1}...\mu_{p}}$ se anula. Portanto, no caso de $p$-formas, não podemos contruir uma infinidade de família de tensores, isto é, nosso espaço vetorial irá possuir dimensão
\begin{equation*}
\textrm{dim}\Lambda^{p} = \binom{D}{p} = \frac{D!}{p!(D-p)!}.
\end{equation*}
Devemos notar que $\Lambda^{p}$ e $\Lambda^{D-p}$ possuem a mesma dimensão, ou seja, uma $p$-forma possui o mesmo número de componentes independentes que uma $(D-p)$-forma, ou seja, dim $\Lambda^{p}$ $=$ dim $\Lambda^{D-p}$. Essa propriedade estabelece o conceito de dualidade entre esses dois espaços, um isomorfismo que é obtido através da operação Hodge\,$^{\star}$, muitas vezes chamado de operação de dualidade.
\begin{equation*}
\Lambda^{p} \stackrel{\star}{\longmapsto} \Lambda^{D-p}.
\end{equation*}
A operação de dualidade $\star$ transforma uma $p$-forma em uma $(D-p)$-forma e sua ação é definida por:
\begin{equation}
^{\star}T_{p} = \frac{1}{(D-p)!}\varepsilon^{\mu_{1}...\mu_{p}}\,_{\mu_{p +1}...\mu_{D}}T_{\mu_{1}...\mu_{p}}dx^{\mu_{p+1}}\wedge...\wedge dx^{\mu_{D}},
\end{equation}
onde 
\begin{equation*}
\varepsilon^{\mu_{1}...\mu_{p}}\,_{\mu_{p +1}...\mu_{D}} = g^{\mu_{1}\nu_{1}}...g^{\mu_{p}\nu_{p}}\varepsilon_{\nu_{1}...\nu_{p}\mu_{p+1}...\mu_{D}}
\end{equation*}
$g_{\mu\nu}$ é o tensor métrico e $g = \textrm{det}(g_{\mu\nu})$. Em uma variedade curva com métrica $g_{\mu\nu}$ temos
\begin{equation*}
\varepsilon_{123...D} = \sqrt{\mid g\mid}.
\end{equation*}
Em geral tomamos o valor absoluto do determinante da métrica $\sqrt{\mid g\mid}$ e não apenas $\sqrt{-g}$, pois dependendo da dimensão e da assinatura da métrica está poderá produzir um determinante negativo. Em outras palavras, dependendo do número de componentes  $D - 1$ tipo-espaço (par ou ímpar) teremos $g > 0$ ou $g < 0$. Podemos normalizar o símbolo de Levi-Civita a unidade (convenção), isto é, $\varepsilon_{123...D} = 1$ mas devemos tomar cuidado com seu correspondente covariante. Lembrando que levantamos e abaixamos os índices via métria ou sua inversa assim,
\begin{equation*}
\varepsilon^{\mu_{1}...\mu_{D}} = g^{\mu_{1}\nu_{1}}...g^{\mu_{D}\nu_{D}}\varepsilon_{\nu_{1}...\nu_{D}},
\end{equation*}
fazendo-se uso da fórmula de Caley para o determinante
\begin{equation*}
\varepsilon^{12...D} = g^{-1} \varepsilon_{12...D}
\end{equation*}

Uma propriedade interessante sobre a operação de dualidade é que em geral o dual do dual  de uma $p$-forma, não necessariamente, retorna o valor original da $p$-forma
\begin{eqnarray*}
^{\star}(^{\star}T) & = & \frac{1}{p!(D-p)!}T_{\mu_{1}...\mu_{p}}\varepsilon^{\mu_{1}...\mu_{D}}\varepsilon_{\mu_{p + 1}...\mu_{D}\nu_{1}...\nu_{p}}dx^{\nu_{1}}\wedge...\wedge dx^{\nu_{p}}\\[0.1cm]
 & = & (-1)^{p(D-p)} (-1)^{D-1}\frac{1}{p!}\delta^{\mu_{1}...\mu_{p}}_{\nu_{1}...\nu_{p}}T_{\mu_{1}...\mu_{p}}dx^{\nu_{1}}\wedge...\wedge dx^{\nu_{p}}\\[0.1cm]
 & = & (-1)^{p(D-p) + D - 1}T.
\end{eqnarray*}
O fator $(-1)^{D-1}$ vem das regras de produto de tensores completamente anti-simétricos, em $D-1$ dimensões espaciais (e com as nossas convenções),
\begin{equation*}
\varepsilon_{12...D-1} = (-1)^{D-1} g \varepsilon^{12...D-1} = (-1)^{D-1}.
\end{equation*}
As leis do produto tornam-se, em geral, 
\begin{equation*}
\varepsilon_{\nu_{1}...\nu_{p}\mu_{p + 1}...\mu_{D}}\varepsilon^{\mu_{1}...\mu_{D}} = (-1)^{D-1}(D-p)!\delta^{\mu_{1}...\mu_{p}}_{\nu_{1}...\nu_{p}},
\end{equation*}
onde $\delta^{\mu_{1}...\mu_{p}}_{\nu_{1}...\nu_{p}}$ é  determinante dos deltas de Kronecker. O fator $(-1)^{p(D-p)}$, por outro lado, vem dos rearranjamentos de $p$ índices de $T$ com $D-p$ índices do seu dual (esse rearranjamento é necessário para que os índices de $\varepsilon$ possam se encaixar com a definição das regras de produto definidas acima.) Para uma prova dessas proposições vide \cite{Felsager, Bertlmann}.

Assim, se considerarmos o espaço euclidiano $3$-dimensional, podemos construir as seguintes formas
\begin{eqnarray}
V = V_{i}dx^{i}, & & T_{2} = \frac{1}{2}(T_{ij} -T_{ji})dx^{i}\wedge dx^{j}
\end{eqnarray}
que são uma $1$-forma e uma $2$-forma respectivamente. A próxima quantidade completamente anti-simétrica possível de se construir em um espaço $3$-dimensional seria uma $3$-forma
\begin{equation}
W_{3} = W_{ijk}dx^{i}\wedge dx^{j} \wedge dx^{k},
\end{equation}
onde
\begin{equation*}
W_{ijk} = W_{[ijk]} = \frac{1}{3!}\bigr(W_{ijk} + W_{jki} + W_{kij} -W_{ikj} - W_{kji} - W_{jik}\bigl).
\end{equation*}
Com a propriedade de anti-simetria na ordem do produto $dx^{i}\wedge dx^{j} \wedge dx^{k} = - dx^{i}\wedge dx^{k} \wedge dx^{j} = dx^{k}\wedge dx^{i} \wedge dx^{j}$. 

Vejamos um exemplo sobre a dualidade no espaço euclidiano $3$-dimensional. Nesse caso, uma $2$-forma guarda o mesmo conteúdo de informação de uma $1$-forma. Portanto, seja $v_{i}$ uma $1$-forma (com $i = 1, 2, 3$), o dual será $^{*}v_{i} := V_{ij} = \varepsilon_{ijk}v_{k}$. Em termos matriciais lê-se, considerando $\varepsilon_{123} = 1$:
$$ V_{2} = (V_{ij})_{3 \times 3} =  \left(
\begin{array}{c|c|c}
0 & v_{3} & -v_{2}\\ \hline
-v_{3} & 0 & v_{1} \\ \hline
v_{2} & -v_{1} & 0\\
\end{array}
\right) $$
assim, vemos que a $2$-forma que é o dual de um vetor possui o mesmo conteúdo de informação que um vetor em $3$D. Poderíamos tomar o dual de $V_{ij}$, onde vemos que nesse caso particulas de espaço euclidiano retornamos ao vetor original
\begin{eqnarray*}
V_{ij} & = & \varepsilon_{ijk}v_{k}\\[0.1cm]
\varepsilon_{mij}V_{ij} & = & \varepsilon_{mij}\varepsilon_{ijk}v_{k}\\[0.1cm]
\varepsilon_{mij}V_{ij} & = & 2\delta_{mk}v_{k}\\[0.1cm]
\end{eqnarray*}
portanto,
\begin{equation*}
v_{i} = \frac{1}{2}\varepsilon_{ijk}V_{jk}.
\end{equation*}

De modo geral o produto exterior de duas formas $p$ e $q$ segue a regra
\begin{equation}
T_{p}\wedge T_{q} = (-1)^{pq}T_{q}\wedge T_{p},
\end{equation}
que é uma $(p + q)$-forma, com $p + q$ índices completamente anti-simétricos. Para vermos como esse conceito de produto exterior generaliza a ideia de produto vetorial do cálculo vetorial usual, vamos novamente conseiderar o espaço euclidiano $3$D. Seja, $u$ e $v$ $1$-formas, tal que 
\begin{eqnarray*}
v \wedge u & = &  v_{i}dx^{i}\wedge u_{j}dx^{j}\\[0.1cm]
v\wedge u & = & \frac{1}{2}(v_{i}u_{j} - v_{j}u_{i})dx^{i}\wedge dx^{j}\\[0.1cm]
          & = & \frac{1}{2}\varepsilon_{ijk}(\vec{v}\times\vec{u})_{k}dx^{i}\wedge dx^{j}
\end{eqnarray*}
matricialmente, vemos 
$$ W_{2} = v\wedge u = (W_{ij})_{3 \times 3} =   \left(
\begin{array}{ccc}
0 & (\vec{u}\times \vec{v})_{z} & -(\vec{u}\times \vec{v})_{y}\\ 
-(\vec{u}\times \vec{v})_{z} & 0 & (\vec{u}\times \vec{v})_{x} \\
(\vec{u}\times \vec{v})_{y} & -(\vec{u}\times \vec{v})_{x} & 0\\
\end{array}
\right) $$
Com efeito, a definição de ``produto vetorial'' em dimensões mais altas é na verdade o produto exterior de duas $1$-formas e todo o cálculo vetorial em $3$D pode ser recuperado através desse formalismo exterior. Assim, em $D = 3$ uma $3$-forma é o dual de uma $0$-forma portanto, podemos construir uma $3$-forma através do produto de $3$ $1$-formas, digamos $u$, $v$ e $w$:
\begin{eqnarray*}
u\wedge(v\wedge w) & = & u_{i}dx^{i}\wedge \frac{1}{2}(v_{j}w_{k} - v_{k}w_{j})dx^{j}\wedge dx^{k}\\[0.1cm]
      & = & \frac{1}{2}\varepsilon_{jkl}u_{i}(\vec{v}\times\vec{w})_{l}dx^{i}\wedge dx^{j}\wedge dx^{k}
\end{eqnarray*}
como estamos em $3$D, a base $dx^{i}\wedge dx^{j}\wedge dx^{k}$ é uma $3$-forma que é na verdade uma combinação do elemento $dx^{1}\wedge dx^{2}\wedge dx^{3}$. Assim, $dx^{i}\wedge dx^{j}\wedge dx^{k} = \varepsilon^{ijk}dx^{1}\wedge dx^{2}\wedge dx^{3}$, onde o termo $dx^{1}\wedge dx^{2}\wedge dx^{3}$ é chamado de elemento de volume e denotado simplesmente por $d^{3}x$. Substituindo essa condição para a base $3$-forma, temos:
\begin{eqnarray*}
u \wedge (v \wedge w) & = & \frac{1}{2}\varepsilon^{ijk}\varepsilon_{jkl}u_{i}(\vec{v}\times\vec{w})_{l}d^{3}x\\[0.1cm]
& = & \vec{u}\cdot(\vec{v}\times\vec{w})d^{3}x,
\end{eqnarray*}
onde usamos o fato de que $\varepsilon^{ijk}\varepsilon_{jkl} = 2\delta_{il}$.

 Podemos definir, ainda, uma operação de derivação exterior, que eleva o rank da forma em uma unidade, da seguinte forma. 
\begin{definition} 
 Seja $T_{p}$ uma $p$-forma, a derivada exterior de $T_{p}$ é uma $(p + 1)$-forma, denotada por $\bm{d}T_{p}$,  definida num sistema de coordenadas $x^{\mu}$ por
\begin{equation}
\bm{d}T_{p} = \frac{1}{p!}\partial_{\nu}T_{\mu_{1}...\mu_{p}}dx^{\nu}\wedge dx^{\mu_{1}}\wedge...\wedge dx^{\mu_{p}}.
\end{equation}
\end{definition}
 A derivada exterior satisfazendo as propriedades do chamado Lema de Poincaré, isto é, a aplicação sucessiva da derivada exterior é nula
\begin{equation}
\bm{d}^{2}T_{p} = 0, \quad \forall T_{p},
\end{equation}
e obedecendo à regra de Leibniz do cálculo exterior
\begin{equation}
\bm{d}(T_{p}\wedge T_{q}) = \bm{d}T_{p}\wedge T_{q} + (-1)^{p}T_{p}\wedge \bm{d}T_{q}.\label{Leibniz}
\end{equation}

Essa definição de derivação exterior possibilita a generalização dos operadores gradiente e rotacional para quaisquer dimensões. Para vermos que de fato essa definição abarca o cálculo vetorial usual, consideremos novamente o espaço euclidiano $3$-dimensional. Seja $v = v_{i}dx^{i}$ uma $1$-forma, a derivada exterior de $v$ será
\begin{eqnarray*}
\bm{d}v & = & \partial_{j}v_{i}dx^{j}\wedge dx^{i} = \frac{1}{2}T_{[ji]}dx^{j}\wedge dx^{i}\\[0.1cm]
& = & \frac{1}{2}(\partial_{j}v_{i} - \partial_{i}v_{j})dx^{j}\wedge dx^{i}\\[0.1cm]
& = & \frac{1}{2}\varepsilon_{ijk}(\nabla\times\vec{v})_{k}dx^{i}\wedge dx^{j},
\end{eqnarray*}
que contém toda a informação sobre o rotacional de um vetor em $3$D. Portanto, se desejamos calcular o rotacional de uma $1$-forma em dimensões mais altas este será dado pela derivada exterior. Dessa forma, construímos um operador diferencial $\bm{d}\,: \,\Lambda^{p}\,\rightarrow\,\Lambda^{p + 1}$, que leva uma $p$-forma a uma $(p+1)$-forma.

Outra quantidade importante no cálculo exterior é o conceito de produto interior que é uma operação de contração. Dessa forma, temos a seguinte
\begin{definition}
 O produto interior de uma $p$-forma $\omega \in \Lambda^{p}$ com o vetor $V \in T_{x}$, denotado por $i_{V}\omega$ é um mapeamento $i_{V} : \Lambda^{p} \to \Lambda^{p -1}$ de maneira que, em um sistema de coordenadas $x^{\mu}$, tem a propriedade
 \begin{equation}
 i_{V} \omega = \frac{1}{(p - 1)!} V^{\nu}\omega_{\nu\mu_{2}...\mu_{p}}dx^{\mu_{2}}\wedge ...\wedge dx^{\mu_{p}}. \label{produto interior}
 \end{equation}
  
\end{definition}
Se aplicarmos o produto interior duas vezes em uma $2$-forma, obtemos um escalar

\begin{eqnarray*}
i_{U}i_{V} T & = & \frac{1}{2}T_{\mu\nu}i_{U}i_{V}dx^{\mu}\wedge dx^{\nu}\\[0.1cm]
             & = & \frac{1}{2}T_{\mu\nu}\bigl(U^{\mu}V^{\nu} - U^{\nu}V^{\mu}\bigr),
\end{eqnarray*}
 onde $i_{U}i_{V}dx^{\mu}\wedge dx^{\nu} = U^{\mu}V^{\nu} - U^{\nu}V^{\mu}$. A derivada de Lie $\pounds$ (vide Apêndice $1$) ao longo de um vetor $\xi$, nesse formalismo, lê-se
 \begin{equation}
\pounds_{\xi} = i_{\xi}\bm{d} + \bm{d}i_{\xi}. \label{Lie-formas} 
 \end{equation}
 
Uma vez que temos o mapa de dualidade $\star$ a nossa disposição, podemos construir um novo operador diferencial que, a exemplo de $\bm{d}$, irá generalizar o conceito de divergência. 
\begin{definition}
Considere uma $p$-forma $T_{p}$ em $D$ dimensões, a coderivada ou codiferencial de $T_{p}$, denotada por $\bm{\delta} T_{p}$, é uma $(p-1)$-forma definida por
\begin{eqnarray}
\bm{\delta} T & := & g (-1)^{p(D - p + 1)}\,\star \textbf{d} \star T, 
\end{eqnarray}
\end{definition}
onde $g$ denota o determinante da métrica. Dessa forma, nosso operador $\bm{\delta}\,:\,\Lambda^{p}\,\rightarrow\,\Lambda^{p-1}$, ou seja, a ação da coderivata em uma $p$-forma atua como uma espécie de contração que reduz o rank da forma em uma unidade. Esse operador é linear por construção, pois é composto de mapas lineares $\star$ e $\bm{d}$. Observe que $\bm{\delta}$ possui uma propriedade em comum com a derivada exterior $\bm{d}$:
\begin{equation*}
\bm{\delta}^{2} = 0.
\end{equation*}
ou seja, a composição de dois $\bm{\delta}$ leva a zero. Novamente, como exemplo de aplicação no espaço euclidiano $3$D, consideremos uma $1$-forma $u = u_{i}dx^{i}$ e calculemos 
\begin{eqnarray*}
\bm{\delta}u & = & \star \bm{d} (\star u).\\[0.1cm] 
\end{eqnarray*} 
Primeiro que a operação de dualidade de $u$ 
\begin{equation*}
\star u = \frac{1}{2}\varepsilon_{mni}u_{i}dx^{m}\wedge dx^{n},
\end{equation*}
e tomando-se a derivada exterior de $\star u$,
\begin{eqnarray*}
d(\star u) & = & \frac{1}{2}\varepsilon_{mni}(\partial_{j}u_{i})dx^{j}\wedge dx^{m} \wedge dx^{n}\\[0.1cm]
& = & \frac{1}{2}\varepsilon_{mni}\varepsilon^{jmn}(\partial_{j}u_{i})dx^{1}\wedge dx^{2}\wedge dx^{3}\\[0.1cm]
& = & (\vec{\nabla}\cdot\vec{u})dx^{1}\wedge dx^{2}\wedge dx^{3}.
\end{eqnarray*}
lembrando que $dx^{1}\wedge dx^{2}\wedge dx^{3} = \dfrac{1}{3!}\varepsilon_{ijk}dx^{i}\wedge dx^{j}\wedge dx^{k}$, vemos que $\bm{d}(\star u)$ é uma $3$-forma em $D = 3$, portanto o dual será uma $0$-forma. Donde obtemos,
\begin{equation*}
\bm{\delta} u = \star(\bm{d}\star u) = \vec{\nabla}\cdot \vec{u}
\end{equation*}

Finalmente, podemos introduzir o operador Laplaciano, que é definido por
\begin{equation}
\vartriangle : = -(\bm{d}\bm{\delta} + \bm{\delta d}) = - \{\bm{d},\bm{\delta} \}.
\end{equation}
Temos que $\vartriangle \,:\, \Lambda^{p}\,\rightarrow\,\Lambda^{p}$ é um mapa de $p$-formas em $p$-formas. Seguindo os mesmos passos calculados anteriormente para o espaço euclidiano $3$-dimensional é bem direto mostrar que a ação de $\vartriangle$ sobre uma $1$-forma $u$ é:
\begin{equation*}
\vartriangle u = - (\bm{d \delta} + \bm{\delta d}) u = \nabla^{2}u,
\end{equation*}
que é o Laplaciano usual do cálculo vetorial. Se considerarmos uma variedade pseudo-riemanniana, então as considerações acima continuam valendo, exceto que, nesse caso, $-\{\bm{d},\bm{\delta}\}$ representa o operador d'Alembertiano ($\square : = g^{\mu\nu}\partial_{\mu}\partial_{\nu}$).  Temos ainda duas definições importantíssimas que diz respeito a diferenciabilidade de formas que são os conceitos de forma \emph{exata} e \emph{fechada}. Assim, temos a seguinte:
\begin{definition}
 a) Uma forma diferencial $F$ é fechada se sua derivada exterior se anula, isto é, $\bm{d}F = 0.$ \\[0.1cm]
 b) Uma forma diferencial $F$ é exata se puder ser escrita como a derivada exterior de uma forma $A$, isto é, $F = \bm{d}A$.
\end{definition}
Com essas definições podemos enunciar o seguinte
\begin{lemma}
Toda forma fechada, $\bm{d}F = 0$, pode ser escrita localmente como uma forma exata, $F = \bm{d}A$.
\end{lemma}

Naturalmente, podemos escrever o formalismo da eletrodinâmica através dessa nova roupagem geométrica e independente de coordenadas. Temos que o \textit{field strength} $F_{\mu\nu}$ da eletrodinâmica é dado pelo ``rotacional'' do $4$-potencial $A_{\mu}$, ou seja,
\begin{equation*}
F_{\mu\nu} = \partial_{\mu}A_{\nu} - \partial_{\nu}A_{\mu}, \quad A_{\mu} = (\Phi,\vec{A}).
\end{equation*}
$F_{\mu\nu}$ é uma matriz anti-simétrica $4\times 4$ que acomoda os campos elétricos e magnéticos
$$ (F_{\mu\nu}) = \left(
\begin{array}{cccc}
0 & -E_{x} & - E_{y} & - E_{z}\\
E_{x} & 0 & B_{z} & -B_{y}\\
E_{y} & -B_{z} & 0 & B_{x}\\
E_{z} & B_{y} & -B_{x} & 0\\
\end{array}
\right)
$$ 
Portanto, $F_{\mu\nu}$ é uma $2$-forma que pode ser representada sob a base de formas como:
\begin{equation*}
\bm{F} = \frac{1}{2}F_{\mu\nu}dx^{\mu}\wedge dx^{\nu}.
\end{equation*}
O campo de gauge $A_{\mu}$ é representado por uma $1$-forma de maneira que
\begin{equation*}
\bm{A} = A_{\mu}dx^{\mu} = \Phi dt - A_{x}dx - A_{y}dy - A_{z}dz.
\end{equation*}
Contudo, podemos reescrever as equações de Maxwell de uma forma que seja independente do sistema de coordenadas, ou seja, podemos geometrizar a eletrodinâmica com
\begin{equation*}
\bm{F} = \bm{dA}.
\end{equation*}
Dessa maneira, a eletrodinâmica pode ser descrita em quaisquer sistemas de coordenadas o que facilita e muito quando se pretende acoplar com a geometria como no caso da RG. Além disso, podemos ver facilmente que sob uma transformação de gauge em $\bm{A}$, ou seja, $\bm{A}\,\rightarrow\,\bm{A}\,+\,\bm{d}\chi$, $\bm{F}$ é invariante.

Além do mais, $\bm{dF}$ é uma identidade pois segundo o lemma de Poincaré $\bm{d}^{2} = 0$, logo
\begin{equation*}
\bm{dF} = 0
\end{equation*}
é uma identidade de Bianchi, e mostra que $\bm{F}$ é uma forma fechada. Em componentes num sistema de coordenadas lê-se
\begin{equation*}
\partial_{\kappa}F_{\mu\nu} + \partial_{\mu}F_{\nu\kappa} + \partial_{\nu}F_{\kappa\mu} = 0.
\end{equation*}

Para finalizar esse exemplo sobre o campo eletromagnético, deveríamos tentar dar uma interpretação geométrica da equação de Maxwell com fontes, a saber
\begin{equation*}
\partial_{\mu}F^{\mu\nu} = j^{\nu}.
\end{equation*}
Como ela está associada a divergência de $\bm{F}$, deveríamos olhar para a coderivada $\bm{\delta F}$. Pelas definições de coderivação vemos que sua aplicação a uma $2$-forma deverá produzir uma $1$-forma. Consequentemente, se a $4$-corrente se escreve como uma $1$-forma $\bm{j} = j_{\mu}dx^{\mu}$, a equação de Maxwell lê-se
\begin{equation*}
\bm{\delta F} = \bm{j}.
\end{equation*}
A lei de conservação da carga é automaticamente recuperada lembrando-se que $\bm{\delta}^{2} = 0$, portanto 
\begin{equation*}
0 = \bm{\delta j}.
\end{equation*}
\subsection{Integração de Formas Diferenciais}

 Além dos conceitos de derivação, podemos integrar formas diferenciais. Esses objetos já são candidatos naturais para integração por serem, por construção, objetos expandidos na base dos diferenciais. Além disso, para a formulação variacional é necessário termos uma noção básica de como podemos integrar formas e quais formas são passíveis de integração. A integração é,  a grosso modo, a operação inversa da derivada exterior. Para integrarmos em uma variedade $\mathcal{M}$ de dimensão $D$ é necessário introduzir uma medida de integração ou um elemento de volume. Uma $D$-forma diferencial, naturalmente, é um objeto que possui as propriedades necessárias. Iremos assumir todas as condições básicas para uma variedade e sua consequente integrabilidade tais como: compacidade, orientabilidade etc. Para uma abordagem matematicamente mais precisa vide \cite{Felsager, Tu, Rudin}.

Seja $\mathcal{M}$ uma variedade diferencial $D$-dimensional, orientada e compacta. Desejamos definir integrais 
\begin{equation*}
\int_{\mathcal{M}} F, \quad F \in \Lambda^{D}
\end{equation*} 
Seja $ x = (x^{1},...x^{D})\in U$, onde $U$ é a imagem do mapeamento de uma região da variedade em $\mathbb{R}^{D}$, ou seja $U \subset \mathbb{R}^{D}$, um sistema de coordenadas local. Se $F = f(x)dx^{1}\wedge...\wedge dx^{D}$, defini-se
\begin{equation}
\int_{\mathcal{M}}F = \int_{U} f(x^{1}, ...,x^{D})dx^{1}\wedge...\wedge dx^{D} = \int_{U} f(x)d^{D}x,\label{int}
\end{equation}
onde o lado direito de (\ref{int})  é a integral de Riemann-Lebesgue, com o elemento de volume $d^{D}x = dx^{1}\wedge...\wedge dx^{D}$.

Essa definição faz sentido, desde que $F$ possua propriedades de transfomação corretas. Em outras palavras se consideramos um outro sistema de coordenadas $(y^{1},...,y^{D})$ a $D$-forma com suas propriedades de anti-simetria produza o determinante do Jacobiano da mudança de coordenadas. Com efeito, se $F = g(y)dy^{1}\wedge...\wedge dy^{D}$, temos
\begin{equation*}
\int_{\phi(U)} g\, dy^{1}\wedge...\wedge dy^{D} = \int_{U} g \circ \phi\, \textrm{det}\bigl(\frac{\partial y^{i}}{\partial x^{j}}\bigr)dx^{1}\wedge...\wedge dx^{D},
\end{equation*}
onde $\phi$ denota a mudança de coordenadas. Por outro lado, temos que 
\begin{equation}
f(x^{1},...,x^{D}) = (g\circ \phi)(x^{1},...,x^{D})\textrm{det}
\bigl(\frac{\partial y^{i}}{\partial x^{j}}\bigr).\label{volume-transf}
\end{equation}
Por exemplo, em duas dimensões considerando a função $f(x)$ como sendo a unidade temos
\begin{eqnarray*}
dx^{1}\wedge dx^{2} & = & \textrm{det}\bigl(\frac{\partial x^{i}}{\partial y^{j}}\bigr)dy^{1}\wedge dy^{2}. 
\end{eqnarray*}
Em geral, para uma $D$-forma qualquer obtemos (\ref{volume-transf}) para o elemento de volume
\begin{equation*}
F = dx^{1}\wedge...\wedge dx^{D} = \textrm{det}\bigl(\frac{\partial x^{\mu}}{\partial y^{\nu}}\bigr)dy^{1}\wedge...\wedge dy^{D},
\end{equation*}
que é a propriedade de transformação correta para o elemento de volume.

Uma conexão importante entre uma informação local de uma determinada quantidade em uma variedade $\mathcal{M}$ com a respectiva informação global sob a borda da variedade $\partial\mathcal{M}$ é obtida através do teorema de Stokes
\begin{theorem}
Seja $\mathcal{M}$ uma variedade compacta e orientada de dimensão $D$ e com uma borda $\partial\mathcal{M}$; $F \in \Lambda^{D - 1}$ é uma $(D-1)$-forma, então
\begin{equation}
\int_{\mathcal{M}}\bm{d}F = \int_{\partial\mathcal{M}}F.
\end{equation}
\end{theorem}
A prova completa desse teorema pode ser encontrada em \cite{Tu,Spivak}. O teorema de Stokes para formas diferenciais generaliza, de forma mais compacta, todos os teoremas de integração do cálculo vetorial usual. Vejamos um exemplo de aplicação, vamos considerar o caso bidimensional, ou seja, dim $\mathcal{M}$ $=$ $2$, onde a variedade pode ser definida como
\begin{equation*}
\mathcal{M} = \{(x_{1},x_{2}), x_{1}^{2} + x_{2}^{2} \leq 1\}, \quad \partial\mathcal{M} = \{(x_{1},x_{2}), x_{1}^{2} + x_{2}^{2} = 1\}.
\end{equation*} 
e seja $A$ nossa $1$-forma
\begin{equation*}
A = A_{1}dx^{1} + A_{2}dx^{2} = A_{i}dx^{i},
\end{equation*}
a derivada exterior de $A$ assume a forma
\begin{eqnarray*}
\bm{d}A & = & \partial_{j}A_{i}dx^{j}\wedge dx^{i} = \frac{1}{2}(\partial_{i}A_{j} - \partial_{j}A_{i})dx^{i}\wedge dx^{j}\\[0.1cm]
& = & \frac{1}{2}\varepsilon_{ijk}B_{k}dx^{i}\wedge dx^{j} = \vec{B}\cdot \bm{d}\vec{\Sigma}, 
\end{eqnarray*}
onde o rotacional $B_{k} = \varepsilon_{kij}\partial_{i}A_{j}$, ou ainda $\vec{B} = \vec{\nabla}\times \vec{A}$, está multiplicando escalarmente um elemento de área orientado $\bm{d}\vec{\Sigma}$ ou em componentes $\bm{d}\Sigma_{k} = \dfrac{1}{2}\varepsilon_{kij}dx^{i}\wedge dx^{j}$. Portanto, recuperamos o teorema de Stokes na notação familiar do cálculo vetorial
\begin{equation*}
\int_{\mathcal{M}}\,\bm{d}A = \int_{x_{1}^{2} + x_{2}^{2} < 1} \,(\vec{\nabla}\times\vec{A})\cdot\bm{d}\vec{\Sigma} = \int_{x_{1}^{2} + x_{2}^{2} = 1}\,\vec{A}\cdot d\vec{x} = \int_{\partial\mathcal{M}}\,A.
\end{equation*}
Físicamente, $\vec{A}$ pode ser o potencial vetor e $\vec{B}$ o campo magnético. A integral do fluxo do campo magnético através de um disco de raio unitário é igual a integral do potencial vetor ao longo da borda, no caso, o círculo de raio unitário.

Finalizando, temos a regra de integração por partes. Vimos anteriormente a generalização da regra de Leibniz (\ref{Leibniz}). Sejam $F$ e $G$ uma  $p$ e uma $q$-forma respectivamente. Considere $\Omega$ um domínio regular orientado de dimensão $(p + q -1)$. Assim,
\begin{equation}
\int_{\Omega}\,\bm{d}(F\wedge G) = \int_{\Omega}\,\bm{d}F \wedge G + (-1)^{p}\int_{\Omega}\,F\wedge\bm{d}G.\label{partes}
\end{equation}
Note que podemos usar o teorema de Stokes do lado esquerdo da equação (\ref{partes}), de maneira que
\begin{equation*}
\int_{\partial\Omega}\,F\wedge G = \int_{\Omega}\,\bm{d}F\wedge G + (-1)^{p}\int_{\Omega}\,F \wedge \bm{d}G.
\end{equation*}
Com efeito, somos levados ao seguinte
\begin{theorem}(Integração por partes de formas diferenciais)
\begin{equation}
\int_{\Omega}\,\bm{d}F\wedge G = \int_{\partial\Omega}\,F\wedge G -\, (-1)^{p}\int_{\Omega}\,F\wedge \bm{d}G.
\end{equation}
\end{theorem}
Essa é, naturalmente, uma generalização direta da regra de integração por partes usual:
\begin{equation*}
\int_{a}^{b}\,\frac{d f}{dx}g(x)dx = f(x)g(x)\bigg{|}^{b}_{a} - \int_{a}^{b}f(x)\frac{d g}{dx}dx
\end{equation*}

 Em várias aplicações somos capazes de eliminar o termo de superfície, seja porque $\Omega$ não tenha borda, isto é $\partial\Omega = \emptyset$, ou porque desejamos que um dos campos dinâmicos, no caso uma das formas diferenciais, se anulem na superfície.
\section{Relatividade Geral no Formalismo de Einstein-Cartan}

Feito essas observações de cunho estritamente matemático, voltemo-nos para suas aplicações no caso da RG. Podemos observar que tanto o vierbein quanto a conexão de spin, que são as variáveis dinâmicas dessa nova descrição geométrica\cite{Trautman1, Trautman2, Hehl}, podem ser colocadas sobre a forma de $1$-formas
\begin{eqnarray}
e^{I} := e^{I}_{\mu}dx^{\mu}, & & \omega^{IJ}:= \omega_{\mu}^{IJ}dx^{\mu}.
\end{eqnarray}
O efeito de usarmos $1$-formas locais não é acidental, ou seja, vemos que todas as propriedades geométricas da variedade podem ser expressas através dessas formas, seus produtos exteriores e suas derivadas exteriores. Como $e^{I}$ e $\omega^{IJ}$ não carregam índices de coordenadas ($\mu,\nu,$ etc.), eles se transformam sob difeomorfismos em $\mathcal{M}$. Assim, seja $\xi$ uma campo vetorial de componentes em um dado sistema de coorenadas $\xi^{\mu}$, a variação de uma forma $\omega$ sob difeomorfismos é dada por
\begin{eqnarray*}
\delta_{\textrm{diff}}\omega = \pounds_{\xi}\omega, & & \pounds_{\xi}\omega = (i_{\xi} d + di_{\xi})\omega.
\end{eqnarray*}
onde $\pounds_{\xi}$ é a derivada de Lie na direção do vetor $\xi$. A grosso modo, difeomorfismo é uma transformação que de certa forma arrasta suavemente todos os campos dinâmicos de uma região da variedade para outra. Assim, a descrição da geometria usando apenas essas formas, seus produtos e suas derivadas exteriores são naturalmente independentes de coordenadas.

Utilizando o formalismo das formas diferenciais vemos que deixamos manifestada apenas a simetria de gauge da teoria, isto é, $e^{I}$ é considerado como um vetor sob Lorentz local, cujas componentes são $1$-fomas. A pergunta que surge naturalmente é: seria a derivada exterior de um vetor também um vetor sob Lorentz? Sabemos que
\begin{eqnarray*}
dv^{I}  =  \partial_{\mu}v^{I}_{\nu}dx^{\mu}\wedge dx^{\nu} = 
         \frac{1}{2}(\partial_{\mu}v^{I}_{\nu} -\partial_{\nu}v^{I}_{\mu})dx^{\mu}\wedge dx^{\nu},
\end{eqnarray*}
e que 
\begin{equation*}
v^{I} \longmapsto v'^{I} = \Lambda^{I}\,_{J}(x)v^{J}.
\end{equation*}
Então, $dv^{I}\,\longmapsto\,(dv^{I})'\,=\,d(\Lambda^{I}\,_{J}v^{J})\,=\, d\Lambda^{I}\,_{J}v^{J}\,+\,\Lambda^{I}\,_{J}dv^{J}$. Novamente, vemos que essa é uma  lei de transformação não-homogênea, isto é, $dv^{I}$ não se transforma como um vetor de Lorentz, pois nosso grupo de simetria é local. Para tanto, introduzimos a conexão $1$-forma $\omega^{IJ}$ e definimos a derivada covariante exterior de Lorentz:
\begin{equation}
Dv^{I} = dv^{I} + \omega^{I}\,_{J}\wedge v^{J}.
\end{equation}
Demandamos que $De^{I}$ se transforme como um vetor de Lorentz, ou seja,
\begin{equation*}
De^{I} \longmapsto (De^{I})' = \Lambda^{I}\,_{J}De^{J},
\end{equation*}
o que está assegurado pela lei de transformação de potenciais de gauge não-abelianos
\begin{equation*}
\omega \longmapsto \omega' = \Lambda\omega\Lambda^{-1} - (d\Lambda)\Lambda^{-1}.
\end{equation*}
A ação da derivada covariante exterior nas diversas representações do grupo de simetria segue as regras de aplicação em índices covariantes e contravariantes
\begin{eqnarray*}
Dv^{I} = dv^{I} + \omega^{I}\,_{J}\wedge v^{J}, && DV^{I}\,_{J} = dV^{I}\,_{J} + \omega^{I}\,_{K}\wedge V^{K}\,_{J} - \omega^{K}\,_{J}\wedge V^{I}\,_{K}.
\end{eqnarray*}
\begin{definition} A Torção $2$-forma é a derivada exterior covariante do vierbein
\begin{equation}
T^{I}:= De^{I} = \frac{1}{2}T^{I}\,_{\mu\nu}dx^{\mu}\wedge dx^{\nu}. \label{Torção-cap1}
\end{equation}
\end{definition}
Poderímos ser levados a pensar em um Lema de Poincaré para derivada covariante exterior, já que está é construída através da derivada covariante. Assim, e se aplicarmos a derivada covariante exterior na Torção $2$-forma, $DT^{I}$? O que vemos é que, apesar a derivada covariante satisfazer o Lema de Poincaré $d^{2} = 0$ não implica que $D^{2} = 0$. Por exemplo, sobre um vetor $V$ teríamos: $D^{2}V^{I} = R^{I}\,_{J}\wedge V^{J}$, onde 
\begin{equation}
R^{IJ} := d\omega^{IJ} + \omega^{I}\,_{K}\wedge\omega^{KJ} 
\end{equation}
é a curvatura $2$-forma de Yang-Mills que no caso gravitacional em questão coincide com a curvatura de Riemann. Com efeito, a ação de $D$ sob a Torção $2$-forma
\begin{eqnarray}
DT^{I} & = & R^{I}\,_{K}\wedge e^{K}.
\end{eqnarray}
Como a derivada covariante exterior age de maneira covariante e, como $e^{K}$ é um tensor sob Lorentz, a quantidade $d\omega^{I}\,_{K} + \omega^{I}\,_{J}\wedge\omega^{J}\,_{K}$ deve ser um tensor sob Lorentz. Essa grandeza é justamente a curvatura $2$-forma
\begin{equation}
R^{IJ} = d\omega^{IJ} + \omega^{I}\,_{K}\wedge\omega^{KJ} = \frac{1}{2}R^{IJ}\,_{\mu\nu}dx^{\mu}\wedge dx^{\nu}.
\end{equation}
Em resumo, temos $e^{I}$, $T^{I} = De^{I}$ e $DT^{I} = R^{I}\,_{J}\wedge e^{J}$. É bem simples de mostrar que a derivada covariante exterior satisfaz a mesma regra de Leibniz que a derivada exterior $d$,

\begin{equation*}
D(T_{p}\wedge T_{q}) = DT_{p}\wedge T_{q} + (-1)^{p}T_{p}\wedge DT_{q}.
\end{equation*} 

De maneira análoga poderíamos nos perguntar se poderíamos obter alguma nova quantidade geométrica ao aplicarmos mais uma vez a derivada covariante exterior, isto é,
\begin{equation*}
D^{2}T^{I} = DR^{I}\,_{J}\wedge e^{J} + R^{I}\,_{J}\wedge T^{J}.
\end{equation*}
O único termo que teria a capacidade de produzir alguma quantidade tensorial nova seria $DR^{I}\,_{J}$, no entanto, vemos que
\begin{eqnarray*}
DR^{I}\,_{J} & = & d(d\omega^{I}\,_{J} + \omega^{I}\,_{K}\wedge\omega^{K}\,_{J}) + \omega^{I}\,_{K}\wedge(d\omega^{K}\,_{J} + \omega^{K}\,_{L}\wedge\omega^{L}\,_{J}) + \nonumber\\[0.1cm]
& & - \omega^{K}\,_{J}\wedge(d\omega^{I}\,_{K} + \omega^{I}\,_{L}\wedge\omega^{L}\,_{K}).
\end{eqnarray*}
Usando-se das propriedades de anti-simetria de $\omega^{IJ} = -\omega^{JI}$ e do produto exterior entre formas, segue-se que 
\begin{equation}
DR^{IJ} = 0.\label{bianchi}
\end{equation}
Essas equações $DR^{IJ} = 0$ e $DT^{I} = R^{I}\,_{J}\wedge e^{J}$ são, na verdade,  identidades. Elas são chamadas de identidades de Bianchi. Portanto, o conjunto $\bigl\{e^{I},\,\omega^{IJ},\,T^{I},\,R^{IJ}\bigr\}$ forma um conjunto de formas diferenciais fechadas sob o produto exterior $\wedge$ e sob a derivada exterior $d$. Em outras palavras, eles formam um conjunto completo para a descrição da geometria diferencial.

A pergunta que se faz é: seria possível escrever a relatividade geral nesse formalismo? A resposta é afirmativa e não somente isso, o formalismo das formas diferenciais nos levam a uma abordagem muito mais elegante e concisa onde a independência das coordenadas, ou melhor, as simetrias da teoria tornam-se manifestas. Portanto, precisamos de uma lagrangiana
 
\begin{equation*}
S = \int_{\mathcal{M}}d^{4}x \mathcal{L}(\phi,\partial_{\mu}\phi),
\end{equation*}
 e o fato das formas serem objetos completamente antisimétricos simplifica muito a construção de uma ação. Portanto, o grau da forma que necessitamos para a lagrangiana está intimamente ligado a dimensão da variedade considerada. A lagrangiana em $4$D deve ser uma $4$-forma, que pode ser integrada facilmente no espaço-tempo $4$D. Além disso, nossa ação deve possuir invariância de Lorentz. Os ingredientess fundamentais de que dispomos para a construção da teoria são: $\bigl\{e^{I},\,\omega^{IJ},\,T^{I},\,R^{IJ},\,\eta_{IJ},\,\varepsilon_{IJKL}\bigr\}$\footnote{ onde $\varepsilon_{IJKL}$ é o tensor de Levi-Civita completamente anti-simétrico com a condição de normalização $\varepsilon_{0123} = 1$.}. Com efeito, devemos tentar selecionar, dentre os elementos desse conjunto, possíveis $4$-formas SO$(1,3)$-invariantes que possam ter uma interpretação dinâmica precisa. Existem algumas possibilidades
 \begin{eqnarray}
 T^{I}\wedge T_{I}, \nonumber\\[0.1cm]  R^{I}\,_{J}\wedge R^{J}\,_{I}, \nonumber\\[0.1cm]
 \varepsilon_{IJKL}e^{I}\wedge e^{J}\wedge R^{KL},\label{L0}\\[0.1cm] \varepsilon_{IJKL}e^{I}\wedge e^{J}\wedge e^{K}\wedge e^{L}, \nonumber\\[0.1cm]
 \varepsilon_{IJKL}R^{IJ}\wedge R^{KL}\nonumber
 \end{eqnarray}
 
 Dessas possibilidades de $4$-formas para construção de uma lagrangiana para a gravitação, três dessas são componentes de uma família específica de lagrangianas chamadas de família Lovelock em $4$D \cite{Lovelock} e são denotadas por
\begin{equation*}
L_{0} = \varepsilon_{IJKL}e^{I}\wedge e^{J}\wedge e^{K}\wedge e^{L}, \, L_{1} = \varepsilon_{IJKL}e^{I}\wedge e^{J}\wedge R^{KL}, \, L_{2} = \varepsilon_{IJKL}R^{IJ}\wedge R^{KL}.
\end{equation*}  
 Na verdade, Lovelock fornece as possíveis extensões naturais da ação de Einstein-Hilbert, para quaisquer dimensão, que não envolve torção e produz equações de campo de segunda ordem para a métrica. Portanto, a ação pode ser expandida em uma série cujos elementos são os integrantes das famílias de Lovelock,
 
 \begin{equation*}
 L = \sum_{p = 0}^{n = [D/2]}\alpha_{p}L_{p}, 
 \end{equation*}
 com
 \begin{equation}
 L_{p} = \varepsilon_{a_{1}....a_{D}}\underbrace{R^{a_{1}a_{2}}\wedge....\wedge R^{a_{2p-1 a_{2p}}}}_{p \, \textrm{vezes}}\wedge \underbrace{e^{a_{2p + 1}}\wedge...\wedge e^{a_{D}}}_{D - 2p \,\textrm{vezes}}. \label{lovelock}
 \end{equation}
  Uma excelente referência sobre as categorizações das famílias de Lovelock está em \cite{Mardones, Bruno, Za1}. Além disso, a determinação dos coeficientes $\alpha_{p}$ pode ser achada em \cite{Banados2}.
 
Lembrando que a ação de Einstein-Hilbert, que é linear nos termos de curvatura, é da forma
\begin{equation*}
S_{EH} = \int_{\mathcal{M}}d^{4}x \sqrt{-g} R,
\end{equation*}
onde $R$ é o escalar de curvatura que é a contração do tensor de Ricci com a métrica $R = g_{\mu\nu}R^{\mu\nu}$. Vamos tomar $L_{1}$ e abrir em componentes
\begin{eqnarray*}
L_{1} & = & \varepsilon_{IJKL}e^{I}\wedge e^{J}\wedge R^{KL},\\[0.1cm]
      & = & \frac{1}{2}\varepsilon_{IJKL}R^{IJ}\,_{\mu\nu}e^{K}_{\rho}e^{L}_{\sigma}dx^{\mu}\wedge dx^{\nu}\wedge dx^{\rho}\wedge dx^{\sigma},\\[0.1cm]
      & = & \frac{1}{2}\varepsilon_{IJKL}\varepsilon^{\mu \nu \rho \sigma} R^{IJ}\,_{\mu\nu}e^{K}_{\rho}e^{L}_{\sigma}dx^{0}\wedge dx^{1}\wedge dx^{2}\wedge dx^{3}
\end{eqnarray*}
o termo $dx^{0}\wedge dx^{1}\wedge dx^{2}\wedge dx^{3}$ é o elemento de volume em $4$D denotado simplesmente por $d^{4}x$. Usando-se do vierbein para colocar os índices de Lorentz da curvatura $2$-forma na variedade, obtemos
\begin{eqnarray*}
L_{1} & = & \frac{1}{2}\varepsilon_{IJKL}\varepsilon^{\mu \nu \rho \sigma}e^{K}_{\rho}e^{L}_{\sigma}e^{I}_{\kappa}e^{J}_{\lambda}  R^{\kappa\lambda}\,_{\mu\nu} d^{4}x.
\end{eqnarray*}
Lembrando ainda da fórmula de Caley para o determinante de uma matriz,
\begin{equation*}
\varepsilon_{IJKL}e^{I}_{\kappa}e^{J}_{\lambda}e^{K}_{\rho}e^{L}_{\sigma} = \varepsilon_{\kappa\lambda\rho\sigma} \textrm{det}(e),
\end{equation*}
o elemento da família de Lovelock torna-se
\begin{eqnarray*}
L_{1} & = & \frac{1}{2}\varepsilon_{\kappa\lambda\rho\sigma}\varepsilon^{\mu\nu\rho\sigma}R^{\kappa\lambda}\,_{\mu\nu}
 \,d^{4}x \textrm{det}e\\[0.1cm]
    & = & \frac{1}{2}(\delta^{\mu}_{\kappa}\delta^{\nu}_{\lambda} - \delta^{\mu}_{\lambda}\delta^{\nu}_{\kappa})R^{\kappa\lambda}\,_{\mu\nu}\,d^{4}x \textrm{det}e \\[0.1cm]
    & = &  \frac{1}{2}(R^{\mu\nu}\,_{\mu\nu} - R^{\nu\mu}\,_{\mu\nu})d^{4}x e.
\end{eqnarray*}
Usando-se das propriedades de anti-simetria do tensor de curvatura de Riemann, $R^{\mu\nu}\,_{\rho\sigma} = - R^{\nu\mu}\,_{\rho\sigma}$, e denotando por $e = \sqrt{-g}$ o determinante do vierbein\footnote{O determinante do vierbein $e^{I}_{\mu}$ em função do determinante da métrica $g_{\mu\nu}$ é obtido de maneira bem direta ao tomarmos o determinante $\textrm{det}(e^{I}_{\mu}\eta_{IJ}e^{J}_{\nu}) = \textrm{det}(g_{\mu\nu})$ $\Rightarrow$ $\textrm{det}(e^{I}_{\mu}) = \sqrt{-g}$.}, $L_{1}$ assume a seguinte forma:
\begin{equation}
L_{1} = d^{4}x\sqrt{-g}R.
\end{equation}
Portanto, o termo $L_{1}$ da família de Lovelock reproduz a lagrangiana de Einstein-Hilbert da relatividade geral. Analogamente, o termo $L_{0}$ é proporcional ao elemento de volume, ou seja, $L_{0} = 4!\sqrt{-g}d^{4}x$ e portanto é um candidato a um termo que representaria a inserção da constante cosmológica ($\Lambda$) na relatividade geral. Lembrando, que a ação que nos leva as equações de Einstein com constante cosmológica é da forma
\begin{equation*}
S = \int_{\mathcal{M}}d^{4}x\sqrt{-g} (R - 2\Lambda).
\end{equation*}

Para $L_{2}$ encontramos que
\begin{eqnarray*}
L_{2} & = & \frac{1}{4}\varepsilon_{IJKL}\varepsilon^{\mu\nu\rho\sigma}e^{I}_{\alpha}e^{J}_{\beta}e^{K}_{\gamma}
e^{L}_{\delta} R^{\alpha\beta}\,_{\mu\nu}R^{\gamma\delta}\,_{\rho\sigma} d^{4}x\\[0.1cm]
      & = & \frac{1}{4}\delta^{\mu\nu\rho\sigma}_{\alpha\beta\gamma\delta}R^{\alpha\beta}\,_{\mu\nu}R^{\gamma\delta}
\,_{\rho\sigma}\sqrt{-g}d^{4}x.
\end{eqnarray*}
onde $\delta^{\mu\nu\rho\sigma}_{\alpha\beta\gamma\delta}$ é um determinante de uma matriz $4\,\times\,4$ formada pelos deltas de Kronecker( $ \varepsilon_{\alpha\beta\gamma\delta}\varepsilon^{\mu\nu\rho\sigma} = \delta_{\alpha\beta\gamma\delta}^{\mu\nu\rho\sigma}$), de modo que depois de alguma manipulação algébrica obtemos
\begin{equation}
L_{2} = \bigl(R^{2} - 4R^{\mu\nu}R_{\mu\nu} + R^{\mu\nu\rho\sigma}R_{\mu\nu\rho\sigma}\bigr)d^{4}x \sqrt{-g}.
\end{equation} 
Essa densidade é  conhecida como Gauss-Bonnet e representa um invariante topológico em $D = 4$ que de certa forma está associado a diferenças na topologia da variedade. Gauss-Bonnet não produz nenhuma diferença classicamente mas pode ter importância em uma teoria quântica. Ainda considerando $L_{2}$, podemos nos perguntar o que uma variação nos vierbein $e^{I}\,\longmapsto\,e'^{I} = e^{I} + \delta e^{I}$ poderia produzir em $L_{2}$? Naturalmente, $\delta_{e}L_{2} = 0$ pois é função apenas das curvaturas, por outro lado, uma variação na conexão $\omega^{IJ}\, \longmapsto\,\omega'^{IJ} = \omega^{IJ} + \delta\omega^{IJ}$ poderia produzir alguma variação na densidade $L_{2}$? Como a curvatura $2$-forma é definida por $R^{IJ} = d\omega^{IJ}\, + \,\omega^{I}\,_{K}\wedge \omega^{KJ}$, donde uma variação na conexão irá produzir uma variação na curvatura 
\begin{eqnarray*}
\delta R^{IJ} & = & d\delta\omega^{IJ} + \omega^{I}\,_{K}\wedge\delta\omega^{KJ} + \omega^{I}\,_{K}\wedge\delta\omega^{KJ},\\[0.1cm]
              & = & D(\delta\omega^{IJ}).
\end{eqnarray*} 
Introduzindo em $\delta L_{2}$:
\begin{eqnarray*}
\delta L_{2} & = & 2\varepsilon_{IJKL}R^{IJ}\wedge \delta R^{KL}\\[0.1cm]
            & = &  2\varepsilon_{IJKL}R^{IJ}\wedge D\delta\omega^{KL}\\[0.1cm]
            & = & D(2\varepsilon_{IJKL}R^{IJ}\wedge \delta\omega^{KL})\\[0.1cm]
            & = & d\Omega.
\end{eqnarray*}
Onde definimos $\Omega := 2\varepsilon_{IJKL}R^{IJ}\wedge \delta\omega^{KL}$ que é um escalar de SO$(1,3)$ e fizemos uso da identidade de Bianchi ($DR^{IJ} = 0$). Assim, vemos que $L_{2}$ não produz nenhuma equação de campo, pois sua variação $\delta L_{2}$ se resume em uma derivada total $d\Omega$. Portanto, adicionar $L_{2}$ produz apenas uma mudança nas condições de contorno.

De acordo com o princípio variacional junto com a abordagem  desenvolvida por Palatini, a ação que descreve os fenômenos gravitacionais parte do princípio de que tanto a conexão de spin quanto o vierbein são campos independentes. Considerando por simplicidade uma variedade espaço-temporal de dimensão $D \,=\, 4$, na ausência de matéria e sem constante cosmológica, a ação de Einstein-Hilbert pode ser reescrita fazendo-se uso das formas diferenciais na ação proposta por Palatini. 
Notemos, primeiramente, que a ação gravitacional que corresponde a integral da densidade de escalar de curvatura ao longo de uma região espaço-temporal - pode ser escrita como uma integral de uma $4$-forma como se segue:

\begin{equation}
S_{\textrm{Palatini}}[\omega,e] = \frac{1}{2\kappa}\varepsilon_{IJKL}\int_{\mathcal{M}_{4}}e^{I}\wedge e^{J}\wedge R^{KL} - \frac{\Lambda}{12}e^{I}\wedge e^{J}\wedge e^{K}\wedge e^{L}
\end{equation} 
onde $\kappa \,=\, \dfrac{8\pi G}{c^{4}}$, $G$ é a constante gravitacional de Newton e $\Lambda$ a constante cosmológica. Como nesse formalismo temos dois campos independentes, a princípio, é de se esperar alguma contribuição a mais nas equações de movimento. Com efeito, fazendo-se as variações funcionais obtemos as seguintes equações\footnote{Lembrando das definições de Curvatura e Torção $2$-forma: $R^{IJ} \,=\, \dfrac{1}{2}R^{IJ} \,_{\mu\nu}dx^{\mu}\wedge dx^{\nu}\,=\, d\omega^{IJ} + \omega^{I}_{K}\wedge \omega^{KJ}$ e $T^{I}\,=\, de^{I} + \omega^{I}_{J}\wedge e^{J}\,\equiv\,De^{I}$. Temos ainda que a variação da ação em relação a conexão de spin terá contribuição apenas no termo de curvatura, isto é, $\delta R^{IJ} = d\delta\omega^{IJ} + \omega^{I}_{K}\wedge \delta\omega^{KJ} + \omega^{J}_{K}\wedge\delta\omega^{IK}\,\equiv\, D\delta\omega^{IJ}$. De modo que, via uma integração por partes, chega-se a Eq. (\ref{Palatine2})} de movimento:
\begin{eqnarray}
\frac{\delta S}{\delta e^{I}} = 0 & : & \varepsilon_{IJKL}(R^{KL}\wedge e^{J} - \frac{\Lambda}{3}e^{J}\wedge e^{K}\wedge e^{L}) = 0 \label{Palatini1}\\ [0.1cm]
\frac{\delta S}{\delta \omega^{IJ}} = 0 & : & \varepsilon_{IJKL}T^{K}\wedge e^{L} = 0\label{Palatine2}
\end{eqnarray}
onde a equação (\ref{Palatini1}) são as equações de campo de Einstein e (\ref{Palatine2}) implicam que a torção, na ausência de férmions, anula-se via equação de movimento. 

Podíamos nos perguntar quais outros termos satisfazendo as exigências de invariância sob Lorentz local e que seja uma $4$-forma poderiam ser adicionadosa à ação de Palatini. De fato, a menos de termos de superfície, como no caso de Gauss-Bonnet, existe essencialmente um único termo que poderia ser adicionado, a saber o termo de Holst. Portanto, a ação de Palatini com a modificação de Holst lê-se
\begin{equation*}
S[\omega,e]_{\textrm{Holst}} = S_{\textrm{Palatini}} + \frac{1}{2\kappa\gamma}\int_{\mathcal{M}_{4}}e^{I}\wedge e^{J}\wedge R_{IJ}.
\end{equation*} 
As equações de movimento dessa ação, com a modificação de Holst, ficam inalteradas e, consequentemente, reproduzem no nível clássico as mesmas equações de campo da relatividade geral, ou seja, o segundo termo não produz efeito sobre as equações de movimento. O que acontece é que a variação em relação a conexão produz novamente a condição de torção nula. A constante de acoplamento $\gamma$ é chamada de parâmetro de \emph{Barbero-Immirzi}. Para aplicações na gravitação quântica o parâmetro $\gamma$ assume um papel importante e não pode ser zero. O termo de Holst possui um papel  em análogia com as teorias de Yang-Mills, onde podemos adicionar um termo topológico a ação que não modifica as equações clássicas de movimento pois o integrando pode ser reexpresso como uma derivada total. No caso presente, mesmo a modificação de Holst não sendo de origem topológica, devido a primeira identidade de Bianchi ele se anula identicamente:
\begin{equation*}
\frac{\delta S_{\gamma}}{\delta e^{I}} := 0 \quad \Rightarrow \quad \frac{1}{\kappa\gamma} e^{J}\wedge R_{IJ} = 0.
\end{equation*}
Na ausência de torção a última expressão é precisamente a identidade de Bianchi. Consequentemente, o parâmetro $\gamma$ é de certa forma análogo ao conhecido parâmetro $\theta_{\textrm{QCD}}$ na Cromodinâmica Quântica \cite{Perez2,Rov4, Pullin2, Noui}.

\chapter{Modelo de Gravitação tipo-Topológico em 4D}
\pagestyle{fancy}
\lhead{\bfseries 2. Chern-Simons 5D e o modelo tipo-topológico}
\rhead{}
\section{Extensão do Grupo de Lorentz}
Vimos que a série de Lovelock, que é a extensão mais geral para dimensões mais altas da relatividade geral, produz equações de campo de segunda-ordem na métrica com torção nula. Em $D = 4$ ela contém a densidade de lagrangiana que descreve as equações de campo da relatividade geral, a saber Einstein-Hilbert. Embora seja bem similar em estrutura e conteúdo da teoria usual, a teoria de Lovelock possui características singulares. Além da constante de Newton ($G$) e da constante cosmológica ($\Lambda$), a ação possui uma coleção de parâmetros dimensionais e arbitrários ($\alpha_{p}$) que torna a análise das propriedades físicas da teoria muito complexa. Nesse formalismo, a dimensão canônica do vierbein é $[e^{I}] = l^{0}$, enquanto a conexão de spin $\omega^{IJ}$ não possui dimensão, isto é, $[\omega^{IJ}] = l^{1}$, como se espera de um verdadeiro campo de gauge. Isso é um reflexo do fato de que a gravitação einsteineana é naturalmente uma teoria de gauge para o grupo de Lorentz, onde $e^{I}$ assume o papel do campo gravitacional, ou seja, um campo vetorial sob Lorentz e não uma conexão. A pergunta que poderíamos fazer é: será que seria possível acomodar $e^{I}$ e $\omega^{IJ}$ como componentes de uma única conexão? A resposta a essa pergunta acabou sendo na afirmativa, porém, com um preço a se pagar: devemos estender o grupo de simetria de gauge \cite{Derek, Mansouri}.

Portanto, sendo $A$ a conexão de Yang-Mills, que toma valores na álgebra de Lie de um grupo $G$
\begin{equation}
A = \frac{1}{2}A^{AB}M_{AB}
\end{equation}
e, os geradores satisfazendo as relações 
\begin{equation*}
\lbrack M_{AB}, M_{CD}\rbrack = \eta_{CB}M_{AD} - \eta_{CA}M_{BD} + \eta_{DB}M_{CA} - \eta_{DA}M_{CB},
\end{equation*}
a construção da curvatura de Yang-Mills é dada por
\begin{eqnarray*}
F & = & dA + A\wedge A \\[0.1cm]
  & = & \frac{1}{2}dA^{AB}M_{AB} + \frac{1}{8}\lbrack A^{AB}M_{AB}, A^{CD}M_{CD}\rbrack\\[0.1cm]
  & = & \frac{1}{2}\bigl(dA^{AB} + A^{A}\,_{C}\wedge A^{CB}\bigr)M_{AB}\\[0.1cm]
  & = & \frac{1}{2}F^{AB}M_{AB},
\end{eqnarray*}
onde $F^{AB} = dA^{AB} + A^{A}\,_{C}\wedge A^{CB}$ é a curvatura $2$-forma associada ao grupo de simetria $G$.

 As transformações de gauge infinitesimais podem ser escritas como
\begin{equation*}
\delta_{\epsilon}A = D \epsilon,
\end{equation*}
onde $D \epsilon = d \epsilon + \lbrack A,\epsilon \rbrack$, e $\epsilon$ é um parâmetro infinitesimal que é uma zero-forma local valorado na álgebra de Lie
\begin{equation*}
\epsilon = \frac{1}{2}\epsilon^{AB}M_{AB}.
\end{equation*} 

A fim de incluírmos o vierbein nessa abordagem de Yang-Mills necessitamos de um grupo $G$ que seja uma extensão do grupo de Lorentz, ou seja,

\begin{equation}
A \stackrel{!}{=} \frac{1}{2}\omega^{IJ}M_{IJ} + e^{I}P_{I}.
\end{equation}
Uma transformação de gauge infinitesimal sob a conexão lê-se $\delta \omega = D\epsilon$, no entanto, sob o vierbein teríamos $\delta e^{I} = \epsilon^{I}\,_{J}e^{J}$. Para unificarmos essas duas transformações em uma única, iremos precisar de termos na álgebra que dêem conta dessa transformação da componente vetorial da conexão estendida, assim

\begin{equation}
\lbrack M_{IJ},P_{K}\rbrack = \eta_{KJ}P_{I} - \eta_{KA}P_{B}
\end{equation} 
para que em uma álgebra estendida a variação da conexão que compõe $\{\omega^{IJ},e^{I}\}$ seja equivalente as transformações individuais, ou seja, iremos alargar o grupo de simetria de tal forma a garantir que
\begin{eqnarray*}
\delta_{\epsilon} A  =  D \epsilon &\Longrightarrow  & \delta_{\epsilon}\omega^{IJ} = D\epsilon^{IJ} \,\,\textrm{e}\,\, \delta_{\epsilon}e^{I} = \epsilon^{I}\,_{J}e^{J},
\end{eqnarray*}
no entanto, iremos ganhar uma simetria nova associada a ``translações'' como veremos a seguir.

As escolhas mais simples para uma álgebra estendida estão contidas em uma imersão da álgebra de Lie do grupo de Lorentz em um grupo maior. Naturalmente, tais escolhas são associadas a Poincaré que acrescenta translações espaço-temporais ISO$(1,3)$. No caso da presença de uma constante cosmológica $\Lambda \neq 0$, espaço-tempo plano não é mais solução das equações de Einstein, assim é possível estender o grupo de Lorentz ao grupo --  de tipo minkowskiano --  de de Sitter ou anti-de Sitter, para $\Lambda > 0$ ou $\Lambda < 0$, respectivamente. As transformações de gauge serão aquelas que deixam invariantes a métrica $\eta_{MN}\,=\, \textrm{diag}(-1,1,1,1,1,s)$, no caso $5$D, onde $M,N,...\,=\, 0,...,5$ e $s$ assume os valores $\pm\,1$. As assinaturas $(-1,1,1,1,1,1)$ e $(-1,1,1,1,1,-1)$, correspondem, respectivamente, ao grupo minkowskiano de de Sitter SO$(1,5)$ e anti-de Sitter SO$(2,4)$ para espaço-tempo $5$D. Esses grupos serão denotados de maneira mais compacta por (A)dS$_{6}$.

Uma base para a álgebra de Lie (a)ds$_{6}$ de AdS$_{6}$ é dada pelos geradores $M_{MN} = - M_{NM}$ que são, de fato, matrizes $6\times 6$ assumindo a forma $\bigl(M_{MN}\bigr)^{P}\,_{Q} := -\delta^{P}_{M}\eta_{NQ} + \delta^{P}_{N}\eta_{MQ} $, obedecendo a relação de comutação
\begin{equation}
\lbrack M_{MN},M_{PQ}\rbrack = \eta_{MP}M_{NQ} - \eta_{MQ}M_{NP} - \eta_{NP}M_{MQ} + \eta_{NQ}M_{MP}.
\end{equation}
Os campos dinâmicos\footnote{Campos e formas no espaço-tempo penta-dimensional $\mathcal{M}_{5}$ serão escritos com um ``chapéu'', índices do espaço-tempo são denotados por $\alpha, \beta,... = 0,..., 4$.} da teoria são componentes de uma conexão $\hat{A} = \hat{A}_{\alpha}dx^{\alpha}$, valoradas na álgebra de Lie (a)ds$_{6}$. Em respeito a base ($M_{MN}$), a conexão lê-se
\begin{equation*}
\hat{A} = \frac{1}{2}\hat{A}^{MN}M_{MN} = \frac{1}{2}\hat{A}_{\alpha}^{MN}dx^{\alpha}M_{MN},
\end{equation*} 
cuja transformação de gauge é da forma
\begin{equation*}
\delta \hat{A}^{MN} = D\epsilon^{MN} = d\hat{A}^{MN} + \hat{A}^{M}\,_{P}\epsilon^{PN} - \hat{A}^{N}\,_{P}\epsilon^{PM}
\end{equation*}

Portanto, a nossa conexão associada ao grupo (A)dS$_{6}$ produz a curvatura de Yang-Mills $\hat{F}^{MN} = d\hat{A}^{MN} + (\hat{A}^{2})^{MN}$. Consequentemente, se desejamos acomodar tanto o fünfbein quanto a conexão de spin em uma única conexão $\hat{A}$, podemos fazer a seguinte identificação dos geradores $M_{MN}$ com os geradores de Lorentz $5$-dimensional $M_{AB}$ e os geradores de ``translações'' $P_{A}$, com $A, B,... = 0,...,4$:
\begin{equation*}
M_{AB} = M_{AB},\quad P_{A} = \frac{1}{l}M_{A5},
\end{equation*}
onde $l > 0$ é um parâmetro com dimensão de comprimento no sistema de unidades onde $c = \hslash = 1$. Com efeito, nossa conexão desacopla-se da seguinte forma

\begin{eqnarray*}
\hat{A}^{MN}=\left\{
	\begin{array}{c}
	\hat{A}^{AB}= \hat{\omega}^{AB},\quad \mbox{  conexão de spin}\\
	\hat{A}^{A5}=\dfrac{1}{l}\hat{e}^{A},\quad \mbox{fünfbein} \hspace{30pt}
	 \end{array}
	      \right.
\end{eqnarray*}
 As relações de comutação assumem a forma explícita como se segue
\begin{eqnarray}
\lbrack M_{AB}, M_{CD}\rbrack &=& 
-\tilde{\eta}_{AD}M_{BC}-\tilde{\eta}_{BC}M_{AD}
+\tilde{\eta}_{AC}M_{BD}+\tilde{\eta}_{BD}M_{AC}\,,\nonumber\\
\lbrack M_{AB}, P_{C}\rbrack &=& \tilde{\eta}_{AC}P_{B}-\tilde{\eta}_{BC}P_{A}\,,\label{alg-(A)dS_6->Lorentz}\\
\lbrack P_{A}, P_{B}\rbrack &=& \frac{s}{l^{2}} M_{AB}\,,\nonumber
\end{eqnarray}
com $\tilde\eta_{AB} = \textrm{diag}(-1,1,1,1,1)$.
Os dez geradores $M_{AB}$ geram o grupo de Lorentz $5$D, e junto com os $5$ geradores $P_{A}$, geram o grupo AdS$_{6}$ para o espaço-tempo $5$D.
Com essa identificação dos geradores a curvatura de Yang-Mills assume a seguinte forma:
\begin{eqnarray}
\hat{F} & = & \frac{1}{2}d\hat{\omega}^{AB}M_{AB} + \frac{1}{l}d\hat{e}^{A}P_{A} + \frac{1}{2}\lbrack\frac{1}{2}\hat{\omega}^{AB}M_{AB} + \frac{1}{l}\hat{e}^{A}P_{A}, \frac{1}{2}\hat{\omega}^{CD}M_{CD} + \frac{1}{l}\hat{e}^{C}P_{C}\rbrack\nonumber\\[0.1cm]
  & = & \frac{1}{2}\bigl(\hat{R}^{AB} + \frac{1}{l^{2}}\hat{e}^{A}\wedge \hat{e}^{B}\bigr)M_{AB} + \frac{1}{l}\hat{T}^{A}P_{A}. 
\end{eqnarray}
onde $\hat{R}^{AB} := d\hat{\omega}^{AB} + \hat{\omega}^{A}\,_{C}\wedge \hat{\omega}^{CB}$ é a curvatura de Riemann $2$-forma em $5$D e $\hat{T}^{A} = D\hat{e}^{A} = d\hat{e}^{A} + \hat{\omega}^{A}\,_{B}\wedge \hat{e}^{B}$ é a torção $2$-forma que é a derivada covariante em $5$D do fünfbein. Note que, a torção $\hat{T}^{A}$ aparece como uma componente desse novo \textit{field strength} obtido pela álgebra (a)dS$_{6}$. Temos ainda como separar as variações de gauge da conexão $\hat{A}^{MN}$ em suas componentes de Lorentz,

\begin{eqnarray*}
\delta_{\epsilon^{AB}}\hat{A}^{MN}=\left\{
	\begin{array}{c}
	\delta\hat{A}^{AB} = \delta\hat{\omega}^{AB} = D\epsilon^{AB},\\
	\delta\hat{A}^{A5}=\dfrac{1}{l}\delta\hat{e}^{A} = - \dfrac{1}{l}\epsilon^{A}\,_{B}\hat{e}^{B} \hspace{30pt}
	 \end{array}
	      \right.
\end{eqnarray*}
mais ``translações''
\begin{eqnarray*}
\delta_{\epsilon^{A5}}\hat{A}^{MN}=\left\{
	\begin{array}{c}
	\delta\hat{A}^{AB} = \delta\hat{\omega}^{AB} = -\frac{s}{l}(\epsilon^{A5}\hat{e}^{B} - \hat{e}^{A}\epsilon^{B5}),\\
	\delta\hat{A}^{A5} = \delta\hat{e}^{A} = l D\epsilon^{A5} \hspace{30pt}
	 \end{array}
	      \right.
\end{eqnarray*}
 
 Portanto, com esse novo grupo de simetria de gauge obtemos uma definição para a conexão donde uma teoria de puro gauge poderá ser definida através de uma ação cuja nova conexão contém os campos dinâmicos da gravitação. Poderíamos nos perguntar: o que aconteceria com a ação de Lovelock em $5$D se exigirmos que ela seja invariante sob essa simetria estendida? A resposta  a essa pergunta nos levará, na verdade, a ação de Chern-Simons em $5$D como veremos a seguir. Lembrando que a família de Lovelock é a generalização natural da ação de Einstein-Hilbert que produz equações de campo sem derivadas superiores. Naturalmente, a ação é Lorentz invariante, ou seja, $\delta_{\epsilon^{AB}}S_{\textrm{Lovelock}} = 0$. Entretanto, ao exigirmos que ela seja invariante também pelas ``translações'' iremos poder fixar as contantes de expansão da série de Lovelock, isto é,
 \begin{equation}
 S_{\textrm{Lovelock}} = \kappa\int_{\mathcal{M}}\sum_{p = 0}^{n = [D/2]}\alpha_{p}L_{p},\label{lovelock1} 
 \end{equation}
 onde $[D/2]$ denota o menor inteiro da divisão e 
 \begin{equation*}
 L_{p} = \varepsilon_{a_{1}....a_{D}}\underbrace{R^{a_{1}a_{2}}\wedge....\wedge R^{a_{2p-1 a_{2p}}}}_{p \, \textrm{vezes}}\wedge \underbrace{e^{a_{2p + 1}}\wedge...\wedge e^{a_{D}}}_{D - 2p \,\textrm{vezes}}. 
 \end{equation*}
 Os coeficientes $\alpha_{p}$ são constantes arbitrárias de dimensão
 \begin{equation*}
 \lbrack\alpha_{p}\rbrack = \lbrack l \rbrack^{-(D - 2p)},
 \end{equation*}
 e $\kappa$ é uma constante com unidade da ação.
 
 Variando a ação de Lovelock em respeito ao viellbein $e^{A}$ produz  equações de campo, que são mais complicadas que as equações de campo de Einstein-Hilbert, envolvendo potências mais altas da curvatura. Essas equações de campo, entretanto, são ainda de segunda-ordem na métrica. A variação com respeito a conexão de spin $\omega^{AB}$ é identicamente nula devido a hipótese de torção nula. As densidades de Lovelock, foram primeiramente estudas no formalismo métrico usual \cite{Lovelock}, a ação de Lovelock é construída sob os mesmo requerimentos que Einstein-Hilbert: covariância sob o grupo de difeomorfismos e equações de campo de segunda-ordem na métrica. Na linguagem das formas diferenciais, (\ref{lovelock1}) é obtida via o requerimento que a densidade de lagrangiano seja invariante sob o grupo de simetria de Lorentz local, isto é, que a lagrangiana seja um $D$-forma, construída inteiramente dos campos fundamentais: viellbein $e$ e conexão de spin $\omega$ e suas derivadas exteriores. 
 
 Essa demanda, contundo, não possibilita a restrição dos valores dos coeficientes $\alpha_{p}$. A fim de obtermos esses coeficientes iremos considerar uma imersão do grupo de Lorentz SO$(1,D-1)$ em um grupo maior. A extensão mínima  para SO$(1,D-1)$  seria os grupos de (anti)-de Sitter SO$(2,D-1)$. Em dimensões \emph{ímpares} é possível construír uma lagrangiana invariante sob o grupo (A)dS, de modo que, através da exigência de invariância da ação sob esse grupo maior pode-se fixar os parâmetros de Lovelock $\alpha_{p}$. A lagrangiana obtida através desse requerimento é a lagrangiana de \emph{Chern-Simons} associada com a densidade de Euler \cite{Za1} para uma dimensão acima de $D$. Vamos mostrar que Lovelock em $5$D sob o grupo  de simetria AdS$_{6}$ nos fornecerá uma teoria de Chern-Simons em $5$D. A ação de Lovelock $5$-dimensional assume a seguinte forma
 \begin{eqnarray}
 S_{\textrm{Lovelock}} = & &\kappa \varepsilon_{ABCDE} \int_{\mathcal{M}_{5}}\biggl(\alpha_{2}\hat{R}^{AB}\wedge \hat{R}^{CD}\wedge \hat{e}^{E} + \alpha_{1}\hat{R}^{AB}\wedge \hat{e}^{C}\wedge \hat{e}^{D} \wedge \hat{e}^{E} \nonumber\\[0.1cm]
 & &  + \,\alpha_{0}\hat{e}^{A}\wedge \hat{e}^{B} \wedge \hat{e}^{C} \wedge \hat{e}^{D}\wedge \hat{e}^{E}\biggr),
 \end{eqnarray}
 as dimensões dos parâmetros são:
 \begin{equation*}
 \lbrack\alpha_{0}\rbrack = \lbrack l\rbrack^{-5},\quad\lbrack\alpha_{1}\rbrack = \lbrack l \rbrack^{-3},\quad \lbrack\alpha_{2}\rbrack = \lbrack l \rbrack^{-1}.
 \end{equation*}
 Portanto, a variação da ação $\delta_{\epsilon^{A5}}S$ produzirá\footnote{Lembrando que a variação da curvatura $2$-forma é: $\delta\hat{R}^{AB} = D\delta\hat{\omega}^{AB}$}:
 \begin{eqnarray*}
 \delta_{\epsilon}S_{\textrm{Lovelock}}  =   \kappa\varepsilon_{ABCDE} &&\int_{\mathcal{M}_{5}}\biggl( 2\alpha_{2}\delta\hat{R}^{AB}\wedge\hat{R}^{CD}\wedge \hat{e}^{E} + \alpha_{2}\hat{R}^{AB}\wedge \hat{R}^{CD}\wedge \delta\hat{e}^{E} +\\[0.1cm]
&& + \alpha_{1}\delta\hat{R}^{AB}\wedge \hat{e}^{C}\wedge \hat{e}^{D} \wedge \hat{e}^{E} + 3 \alpha_{1}\hat{R}^{AB}\wedge \hat{e}^{C}\wedge \hat{e}^{D} \wedge \delta\hat{e}^{E} +\\[0.1cm] && + 5\alpha_{0}\hat{e}^{A}\wedge \hat{e}^{B} \wedge \hat{e}^{C} \wedge \hat{e}^{D}\wedge \delta\hat{e}^{E}\biggr)
 \end{eqnarray*}
 Donde, fazendo-se uso das simetrias do Levi-Civita e substituindo as variações sob ``translações'', obtemos:
 \begin{eqnarray*} 
 \delta_{\epsilon}S_{\textrm{Lovelock}} =\kappa\varepsilon_{ABCDE} &&\int_{\mathcal{M}_{5}} \biggl(5\alpha_{0}\hat{e}^{A}\wedge \hat{e}^{B}\wedge\hat{e}^{C}\wedge(D\epsilon^{E5}) - 6s\alpha_{1}\epsilon^{A5}\hat{e}^{B}\wedge\hat{e}^{C}\wedge\hat{e}^{D}\wedge D\hat{e}^{E} +\\[0.1cm]
 & & + 3\alpha_{1}\hat{R}^{AB}\wedge\hat{e}^{C}\wedge\hat{e}^{D}\wedge D\epsilon^{E5} - 4s\alpha_{2}\epsilon^{A5}\hat{e}^{B}\hat{R}^{CD}\wedge D\hat{e}^{E} + \\[0.1cm]  &&+\alpha_{2}\hat{R}^{AB}\wedge\hat{R}^{CD}\wedge D\hat{e}^{E}\biggr).  
 \end{eqnarray*}
 Fazendo-se uma integração por partes e usando a identidade de Bianchi $D\hat{R}^{AB} = 0$,
 \begin{eqnarray}
  \delta_{\epsilon}S_{\textrm{Lovelock}} =\kappa\varepsilon_{ABCDE} &&\int_{\mathcal{M}_{5}} (-20\alpha_{0}- 6s\alpha_{1})\hat{e}^{A}\wedge\hat{e}^{B}\wedge\hat{e}^{C}\wedge D\hat{e}^{D}\epsilon^{E5} + \nonumber\\[0.1cm]
  && +\, (-6\alpha_{1} - 4s\alpha_{2})\hat{R}^{AB}\wedge\hat{e}^{C}\wedge D\hat{e}^{D}\epsilon^{E5},
 \end{eqnarray}
  Assim, exigindo-se que a ação de Lovelock seja invariante sob essa variação, isto é $\delta_{\epsilon}S_{\textrm{Lovelock}} \stackrel{!}{=} 0$, as constantes $\alpha_{p}$ serão vínculadas da seguinte forma:
\begin{eqnarray}
-20\alpha_{0} - 6s\alpha_{1} = 0, && \quad -6s\alpha_{1} - 4s^{2}\alpha_{2} = 0 \nonumber\\[0.1cm]
\Rightarrow \alpha_{1} = -\frac{2s}{3l^{2}}\alpha_{2}, && \quad \alpha_{0} = \frac{1}{5l^{4}}\alpha_{2}
\end{eqnarray}

Com efeito, a ação de Lovelock em $5$D com um grupo de gauge (A)dS$_{6}$, lê-se:
\begin{eqnarray}
 S_{\textrm{Lovelock}} = \frac{\bar{\alpha_{2}}\kappa}{l}\varepsilon_{ABCDE} &&\int_{\mathcal{M}_{5}}\biggl(\hat{R}^{AB}\wedge \hat{R}^{CD}\wedge \hat{e}^{E} - \frac{2s}{3l^{2}}\hat{R}^{AB}\wedge \hat{e}^{C}\wedge \hat{e}^{D} \wedge \hat{e}^{E} + \nonumber\\[0.1cm]
 & &  + \,\frac{1}{5l^{4}}\hat{e}^{A}\wedge \hat{e}^{B} \wedge \hat{e}^{C} \wedge \hat{e}^{D}\wedge \hat{e}^{E}\biggr),
 \end{eqnarray}
 onde $\bar{\alpha_{2}}$ é uma constante adimensional que pode ser absorvida na constante $\kappa$.
   Essa é justamente a ação de Chern-Simons em $5$D sob o grupo de (A)dS$_{6}$. De fato, podemos notar que Lovelock em dimensões ímpares $D = 2n + 1$, fazendo-se essa extensão da álgebra, sempre nos produzirá como resultado uma teoria de Chern-Simons \cite{Banados2}. A ação de Chern-Simons em $5$D  para uma conexão $A = \frac{1}{2}A^{MN}M_{MN}$ tomando valores numa álgebra de Lie (a)ds$_{6}$ é da forma
 \begin{eqnarray}
 S_{\textrm{CS}_{5}} = \frac{1}{24}\varepsilon_{MNPQRS} && \int_{\mathcal{M}_{5}}\biggl(\hat{A}^{MN}\wedge d\hat{A}^{PQ}\wedge d\hat{A}^{RS} + \frac{3}{2}d\hat{A}^{MN}\wedge (\hat{A}^{2})^{PQ}\wedge d\hat{A}^{RS} + \nonumber\\[0.1cm]
 && + \frac{3}{5}\hat{A}^{MN}\wedge (\hat{A}^{2})^{PQ}\wedge (\hat{A}^{2})^{RS}\biggr).
 \end{eqnarray}
 Cujas equações de campo obtidas pela variação da conexão $\hat{A}$ são da forma
 \begin{equation}
 \frac{1}{4}\varepsilon_{MNPQRS}\hat{F}^{PQ}\wedge \hat{F}^{RS} = 0, \quad \textrm{onde} \quad \hat{F}^{PQ} = d\hat{A}^{PQ} + (\hat{A}^{2})^{PQ}, \label{Chern-Simons5}
 \end{equation}
 com $\hat{F}^{MN} = d\hat{A}^{MN} + \hat{A}^{N}\,_{P}\wedge \hat{A}^{PN}$ é a curvatura de Yang-Mills.
 
 Da mesma maneira que podemos interpretar a teoria de Chern-Simons em $3$D para um grupo de gauge pseudo-ortogonal SO$(1,3)$ ou SO$(2,2)$ como uma teoria de gravitação com constante cosmológica \cite{Witten}, podemos fazer o mesmo com Chern-Simons em $5$D identificando os geradores com Lorentz mais ``translações'' e as componentes da conexão $\hat{A}$ com a conexão de spin em  $5$D e o fünfbein. Com efeito, a ação de Chern-Simons, após uma integração por partes e supondo $\partial\mathcal{M}_{5} = \emptyset$ ou que os campos vão a zero na borda, assume a seguinte forma:
 
 \begin{eqnarray}
 S_{\textrm{CS}_{5D}} = \frac{1}{8l}\varepsilon_{ABCDE} &&\int_{\mathcal{M}_{5}}\biggl(\hat{R}^{AB}\wedge \hat{R}^{CD}\wedge \hat{e}^{E} - \frac{2s}{3l^{2}}\hat{R}^{AB}\wedge \hat{e}^{C}\wedge \hat{e}^{D} \wedge \hat{e}^{E} + \nonumber\\[0.1cm]
 & &  + \,\frac{1}{5l^{4}}\hat{e}^{A}\wedge \hat{e}^{B} \wedge \hat{e}^{C} \wedge \hat{e}^{D}\wedge \hat{e}^{E}\biggr). \label{CS5}
 \end{eqnarray}

 Podemos identificar no segundo e o terceiro termos as densidades de Einstein-Hilbert e constante cosmológica para a gravitação em $5$D, respectivamente. A novidade da ação de Chern-Simons é justamente a presença do primeiro termo, que é da forma $\hat{R}_{5}\wedge \hat{R}_{5}$; como vimos esse termo não entra com uma constante de acoplamento abitrária, pelo contrário, ele vem acopanhado de um número racional que é fixado via a exigência da teoria de Lovelock ser invariante sob AdS$_{6}$, embora a ação esteja escrita em forma manifestamente Lorentz invariante SO$(1,4)$.

 Finalmente, vemos que uma teoria de Lovelock em dimensão ímpar $D = 2n + 1$ cuja ação é invariante sob o grupo de simetria local SO$(2n -2, 2)$ ou SO$(2n-1,1)$ $(n = 0, 1,  2,...)$ representa, de fato, uma teoria de Chern-Simons que pode ser interpretada como uma teoria de gravitação com constante cosmológica. De fato, existe um procedimento em dimensão ímpar que implementado possibilita a obtenção de todos os coeficientes da família de Lovelock de forma geral \cite{Za1}. Esse procedimento produz diretamente uma densidade de lagrangiano $(2n + 1)$-dimensinal 
 \begin{equation*}
 L_{2n + 1}^{\textrm{(A)dS}} = \sum_{p = 0}^{n}\alpha_{p}L_{p}^{2n + 1},
 \end{equation*}
 onde $L_{p}$ são os elementos da série de Lovelock. Nesse caso particular de Lovelock em dimensão ímpar sob (A)dS, todos os coeficientes $\alpha_{p}$ são fixados e assumem os valores dados por
 \begin{equation}
 \alpha_{p} = \kappa \frac{(\pm 1)^{p + l} l^{2p - D}}{(D- 2p)}\binom{n}{p}
\end{equation}  
onde $\kappa$ é uma constante arbitrária adimensional. Portanto, o que veremos a seguir é como essa teoria de Chern-Simons em $5$D nos leva ao modelo de Chamseddine em $4$D via uma redução dimensional mais truncação. Nosso objetivo será mostrar que, de fato, recupera-se o modelo tipo-topológico de Chamseddine \cite{Cham} e que as equações de campo deste são soluções da teoria de Chern-Simons completa sem truncação.
\section{O modelo de Chamseddine}
A relatividade geral é uma teoria invariante de fundo (``background invariant''), o que significa que não existe estrutura geométrica dada \emph{a priori} à variedade espaço-temporal onde a teoria é definida: a métrica pertence aos campos dinâmicos. Uma outra classe de teorias que são independentes de fundo são as teorias topológicas \cite{Blau} tais como as teorias de  Chern-Simons\footnote{Podemos mencionar também as teorias BF como exemplo de teoria topológica. No entanto, a menos que alguns vínculos sejam aplicados à teoria, elas podem endereçar graus de liberdade locais em espaço-tempo de quaisquer dimensões, para mais detalhes vide \cite{Freidel,Piguet3} e as referências contidas.}. De maneira extraordinária \cite{Witten}, a gravitação em um espaço-tempo $3$-dimensional pode ser escrita como uma teoria de Chern-Simons cujos grupos de gauge locais são Poincaré ISO$(1,2)$, mas também para SO$(1,3)$ ou SO$(2,2)$ se existir uma constante cosmológica positiva ou negativa respectivamente. Da mesma forma, podemos descrever teorias de Chern-Simons em dimensões mais altas, na verdade apenas em dimensões ímpares\cite{Za1}. Uma diferença essencial entre gravitação em $3$D e em dimensões maiores que três é que aquela não possui graus de liberdade local enquanto esta possui. O mesmo acontece com as teorias de Chern-Simons em $3$D e em dimensões mais altas. Portanto, a primeira teoria de Chern-Simons que possui graus de liberdade locaias é a em $5$D, como vimos na seção anterior.

Como estamos interessados em  gravitação em $4$D, a pergunta natural que podemos nos fazer é: poderíamos encontrar uma teoria de cunho topológico similar em um espaço-tempo $4$-dimensional? Um resposta a essa pergunta foi dada por Chamseddine \cite{Cham}: uma teoria que além de conter os campos associados à gravitação deve trazer consigo um campo escalar tipo-dilaton\cite{Banados}. Pode-se obtê-la de uma teoria de Chern-Simons em $5$D através de uma redução dimensional e truncação de algumas componentes dos campos fundamentais. Como iremos ver, o conjunto de soluções do modelo de Chamseddine é um subconjunto das soluções da teoria completa, a saber Chern-Simons $5$D reduzida a $4$D. De maneira mais precisa, a proposta do presente trabalho será investigar a dinâmica desse modelo e seus aspectos no nível clássico e compará-las com as da teoria padrão de Einstein.

Portanto, a partir de agora iremos fazer uma redução dimensional da ação (\ref{CS5}) de $5$D para $4$D. Com efeito, iremos fazer algumas decomposições do grupo (A)dS$_{6}$ em termos das representações do grupo de Lorentz em $4$D SO$(1,3)$. As componentes da conexão serão decompostas como $\hat{\omega}^{AB} \,=\, \{\hat{\omega}^{IJ},\,\hat{\omega}^{I4}:= \,\frac{1}{l}\hat{b}^{I}\}$ e $\hat{e}^{A} = \{\hat{e}^{I},\,\hat{e}^{4}\}$. Devemos notar que, via a definição $\varepsilon_{IJKL4} := \varepsilon_{IJKL}$ e utilizando-se das propriedades de anti-simetria do Levi-Civita bem como das quantidades em (\ref{CS5}) a ação lê-se:
\begin{eqnarray}
S_{\textrm{CS}_{5}} = \frac{1}{8l}\varepsilon_{IJKL} && \int_{\mathcal{M}_{5}}\biggl[\underline{\hat{e}^{4}\wedge\hat{R}^{IJ}_{(5)}\wedge\hat{R}^{KL}_{(5)} }+ 4\hat{e}^{I}\wedge\hat{R}^{J4}_{(5)}\wedge\hat{R}^{KL}_{(5)} + \nonumber\\[0.1cm] 
&& - \frac{2s}{3l^{2}}\biggl(\underline{3\hat{e}^{4}\wedge\hat{e}^{I}\wedge\hat{e}^{J}\wedge\hat{R}^{KL}_{(5)} }+ 2 \hat{e}^{I}\wedge\hat{e}^{J}\wedge\hat{e}^{K}\wedge\hat{R}^{L4}_{(5)}\biggr) + \nonumber\\[0.1cm] && + 
\frac{1}{l^{4}}\underline{\hat{e}^{4}\wedge\hat{e}^{I}\wedge\hat{e}^{J}\wedge\hat{e}^{K}\wedge\hat{e}^{L}}\biggr],
\label{CS52}
\end{eqnarray}
onde 
\begin{equation*}
\hat{R}^{IJ}_{(5)} = \hat{R}^{IJ} - \frac{1}{l^{2}}\hat{b}^{I}\wedge\hat{b}^{J}, \quad \hat{R}^{I4}_{(5)} = \frac{1}{l}D\hat{b}^{I};
\end{equation*}
com
\begin{equation*}
\hat{R}^{IJ} = d\hat{\omega}^{IJ} + \hat{\omega}^{I}\,_{K}\wedge\hat{\omega}^{KJ}.
\end{equation*}
Juntando-se os termos em destaque em (\ref{CS52}) a ação assume a seguinte forma
\begin{eqnarray*}
S_{\textrm{CS}_{5}} = \frac{1}{8l}\varepsilon_{IJKL} && \int_{\mathcal{M}_{5}}\hat{e}^{4}\wedge\biggl(\hat{R}^{IJ} - \frac{1}{l^{2}}(\hat{b}^{I}\wedge\hat{b}^{J} + s\hat{e}^{I}\wedge\hat{e}^{J})\biggr)\wedge\biggl(\hat{R}^{KL} - \frac{1}{l^{2}}(\hat{b}^{K}\wedge\hat{b}^{L} + s\hat{e}^{K}\wedge\hat{e}^{L})\biggr) + \\[0.1cm]
&& + \frac{4}{l}\hat{e}^{I}\wedge D\hat{b}^{J}\wedge\hat{R}^{KL} -\frac{4}{l^{3}}\underline{\hat{e}^{I}\wedge D\hat{b}^{J}\wedge\hat{b}^{K}\wedge\hat{b}^{L}} -\frac{4s}{3l^{3}}\hat{e}^{I}\wedge D\hat{b}^{J}\wedge\hat{e}^{K}\wedge\hat{e}^{L},
\end{eqnarray*}
note que o termo em destaque pode ser integrado por partes, usando
\begin{equation*}
 D(\hat{b}^{J}\wedge\hat{b}^{K}\wedge\hat{b}^{L}\wedge\hat{e}^{I}) = 3D\hat{b}^{J}\wedge\hat{b}^{K}\wedge\hat{b}^{L}\wedge\hat{e}^{I} + \hat{b}^{J}\wedge\hat{b}^{K}\wedge\hat{b}^{L}\wedge D\hat{e}^{I} .
\end{equation*} 
 
 Considerando que a integral de uma derivada total vai a zero na borda da variedade $\partial\mathcal{M}_{5}$, a ação assume a forma
 
 \begin{eqnarray}
S_{\textrm{CS}_{5}} = &&\frac{1}{8l}\varepsilon_{IJKL}  \int_{\mathcal{M}_{5}}\hat{e}^{4}\wedge\biggl(\hat{R}^{IJ} - \frac{1}{l^{2}}(\hat{b}^{I}\wedge\hat{b}^{J} + s\hat{e}^{I}\wedge\hat{e}^{J})\biggr)\wedge\biggl(\hat{R}^{KL} - \frac{1}{l^{2}}(\hat{b}^{K}\wedge\hat{b}^{L} + s\hat{e}^{K}\wedge\hat{e}^{L})\biggr) + \nonumber\\[0.1cm]
&& + \frac{2}{l}D\hat{e}^{I}\wedge\hat{b}^{J}\wedge\biggl(\hat{R}^{KL}- \frac{2}{3l^{2}}\hat{b}^{K}\wedge\hat{b}^{L}\biggr) - \frac{2}{l}D\hat{b}^{I}\wedge\hat{e}^{J}\biggl(\hat{R}^{KL} - \frac{2s}{3l^{2}}\hat{e}^{K}\wedge\hat{e}^{L}\biggr).\label{CS53}
\end{eqnarray}
Podemos notar que os campos $\hat{e}^{I}$ e $\hat{b}^{I}$ assumem um papel simétrico na ação (\ref{CS53}), de modo que em princípio poderíamos usar quaisquer um dos dois ou mesmo uma combinação linear de $\hat{e}$ e $\hat{b}$ para definir a forma do vierbein em $4$D. Contudo, uma diferença qualitativa entre essas duas quantidades irá manifestar-se após uma truncação conveniente. A teoria $4$-dimensional é obtida através de uma redução dimensional tipo Kaluza-Klein onde os campos ``tipo-matéria'' são codificados na dimensão extra, ou seja, na $5^{\textrm{a}}$ dimensão. Portanto, assumiremos que uma dimensão espacial representará uma dimensão ``microscópica'' e compacta. Com efeito, faremos um desacoplamento das coordenadas do espaço-tempo em D $=\,5$ nas coordenadas de D $=\,4$ do espaço-tempo $x^{\mu}$, $\mu = 0,...,3$ e a quinta coordenada $\chi:= x^{4}$. Desse modo os campos separam-se,
\begin{eqnarray}
\hat{e}^{I} & = & e^{I}_{\mu}dx^{\mu} + e^{I}_{\chi}d\chi, \nonumber\\[0.1cm]
\hat{b}^{I} & = & b^{I}_{\mu}dx^{\mu} + b^{I}_{\chi}d\chi, \nonumber\\[0.1cm]
\hat{e}^{4} & = & e^{4}_{\mu}dx^{\mu} + e^{4}_{\chi}d\chi, \label{Split} \\[0.1cm] 
\hat{\omega}^{IJ} & = & \omega^{IJ}_{\mu}dx^{\mu} + \omega^{IJ}_{\chi}d\chi. \nonumber
\end{eqnarray}

O desacoplamento das componentes da curvatura de Yang-Mills lêem-se
\begin{equation}
\hat{F}^{MN} = F^{MN} + F^{MN}_{\chi}d\chi \label{YM}
\end{equation}
onde
\begin{equation*}
F^{MN} = \frac{1}{2}F^{MN}_{\mu\nu}dx^{\mu}\wedge dx^{\nu}, \quad F^{MN}_{\chi} = F^{MN}_{\mu\chi}dx^{\mu}.
\end{equation*}
Naturalmente, devemos separar as componentes de cada uma dessas formas em termos de suas componentes de Lorentz SO$(1,3)$,
\begin{eqnarray*}
F^{MN} & = & (F^{IJ}, F^{4I}, F^{5I}, F^{45}),\\[0.1cm]
F^{MN}_{\chi} & = & (F_{\chi}^{IJ},F^{4I}_{\chi}, F^{5I}_{\chi}, F^{45}_{\chi}).
\end{eqnarray*}
Consequentemente, as componentes da curvatura assumem a seguinte forma após as separações feitas em (\ref{Split}) e (\ref{YM})
\begin{eqnarray}
F^{IJ}  & = & R^{IJ}-\frac{1}{l^{2}}b^{I}\wedge b^{J}-\frac{s}{l^{2}}e^{I}\wedge e^{J},\nonumber\\[0.1cm]
F^{I4}  & = & \frac{1}{l} Db^{I} - \frac{s}{l^{2}} e^{I} e^{4}, \nonumber \\[0.1cm]
F^{I5}  & = & \frac{1}{l} De^{I} + \frac{1}{l^{2}} b^{I}\wedge e^{4},\nonumber\\[0.1cm]
F^{45}  & = & \frac{1}{l} d e^4 - \frac{1}{l^{2}} b_{I}\wedge e^{I},\\[0.1cm] 
F_{\chi}^{IJ}  & = & R_\chi^{IJ}+\frac{1}{l^{2}}(b_{\chi}^{I}\wedge b^{J} - b^{I} \wedge b_{\chi}^{J}) 
+\frac{s}{l^{2}}(e_{\chi}^{I}\wedge e^{J} - e^{I}\wedge e_{\chi}^{J}), \nonumber\\[0.1cm]
F_{\chi}^{I4} & = & \frac{1}{l}( D b_{\chi}^{I} + \omega_{\chi}^{I}\,_{J}\wedge b^{J})
+  \frac{s}{l^{2}} (e^{I}_{\chi}\wedge e^{4} - e^{I}\wedge e_{\chi}^{4}), \nonumber\\[0.1cm]
F_{\chi}^{I5} & = & \frac{1}{l}( D e_{\chi}^{I} + \omega{_\chi}^{I}\,_J\wedge e^{J})
- \frac{1}{l^{2}} (b^{I}_{\chi}\wedge e^{4} - b^{I}\wedge e_{\chi}^{4}),\nonumber\\[0.1cm]
F_{\chi}^{45} & = & \frac{1}{l} d e_{\chi}^{4} + \frac{1}{l^{2}} (b^{I}_{\chi}\wedge e^{I} - b^{I}\wedge e_{\chi}^{I}),\label{eqs} \nonumber 
\end{eqnarray}
onde $R^{IJ}$ representa a curvatura $2$-forma associada a conexão de Lorentz $\omega^{IJ}$. As equações de campo (\ref{Chern-Simons5}) são separadas nos seguintes conjuntos de equações
\begin{eqnarray}
\varepsilon_{IJKL}(F^{45}\wedge F^{KL}-2F^{K4}\wedge F^{L5}) & = & 0\nonumber\\[0.1cm]
\varepsilon_{IJKL}F^{J5}\wedge F^{KL} & = & 0,\nonumber\\[0.1cm]
\varepsilon_{IJKL}F^{J4}\wedge F^{KL} & = & 0,\label{eq:auxiliar-4}\\[0.1cm]
\varepsilon_{IJKL}F^{IJ}\wedge F^{KL} & = & 0, \nonumber
\end{eqnarray}
que são $4$-formas e as demais equações correspondendo a $3$-formas, ou seja, as componentes $\chi$
\begin{eqnarray}
\varepsilon_{IJKL}(F_{\chi}^{45}\wedge F^{KL}+F^{45}\wedge F_{\chi}^{KL}
-2F_{\chi}^{K4}\wedge F^{L5}-2F^{K4}\wedge F_{\chi}^{L5}) & = & 0,\nonumber\\[0.1cm]
\varepsilon_{IJKL}(F_{\chi}^{J5}\wedge F^{KL}+F^{J5}\wedge F_{\chi}^{KL}) & = & 0,\label{eq:auxiliar-5}\\[0.1cm]
\varepsilon_{IJKL}(F_{\chi}^{J4}\wedge F^{KL}+F^{J4}\wedge F_{\chi}^{KL}) & = & 0,\nonumber\\[0.1cm]
\varepsilon_{IJKL}F^{IJ}\wedge F_{\chi}^{KL} & = & 0,\nonumber
\end{eqnarray}
com as componentes da curvatura dadas por (\ref{eqs}). A teoria, mesmo com esse desacoplamento continua sendo invariante sob as transformações completas de (A)dS$_{6}$, que agora podem ser identificadas em termos das quantidades $4$D:
\begin{eqnarray}
\delta\hat{\omega}^{IJ} =  D\epsilon^{IJ}
+  \frac{s}{l} ( \epsilon^{I5}\hat{ e}^{J} - \epsilon^{J5}\hat{ e}^{I} ) 
+\frac{1}{l}  (\epsilon^{I4}\hat {b}^{J} - \epsilon^{J4}\hat{b}^{I}),\nonumber\\[0.1cm]
\delta \hat{e}^{I} =   l D\epsilon^{I5}
+ \hat{e}^{J}\epsilon_{J}\,^{I}  + \hat{b}^{I}\epsilon^{45} 
- {\hat{e}^4_\chi}\epsilon^{I4}, \nonumber\\[0.1cm]
\delta \hat{b}^{I} = l D\epsilon^{I4}
+ \hat{b}^{J}\epsilon_{J}\,^{I}  - s\hat{e}^{I}\epsilon^{45} + s\hat{e}_{\chi}^{4}\epsilon^{I5}, \label{eq:auxiliar-6} \\[0.1cm]
\delta\hat{e}^{4} = l d\epsilon^{45}
-\hat{b}_{I}\epsilon^{I5} + \hat{e}_{I}\epsilon^{I4}.\nonumber
\end{eqnarray}
\subsection{Fixação de Gauge e a Ação de Chamseddine}
A ação (\ref{CS53}) e as equações de movimento obtidas (\ref{eq:auxiliar-4}) e (\ref{eq:auxiliar-5}) podem ser simplificadas via uma fixação parcial de gauge que consiste de oito condições
\begin{equation}
b^{I}_{\chi} = 0, \quad e^{I}_{\chi} = 0, \quad I = 0,...,3,
\label{partialgauge1}
\end{equation}
que fixa as simetrias de gauge geradas por $M_{I5} = lP_{I}$ e $M_{I4} = lQ_{I}$, respectivamente, como poder ser visto pelas leis de transformação (\ref{eq:auxiliar-6}) para as componentes $\chi$ de $b^{I}$ e $e^{I}$, onde assumimos que o campo $e^{4}_{\chi} \neq 0$. Essa fixação reduz a simetria de gauge ao grupo SO$(1,3)$ $\times$ (A)dS$_{2}$, onde SO$(1,3)$ é o grupo de Lorentz $4$D e (A)dS$_{2}$ $=$ U$(1)$ se $s > 0$ (teoria com constante cosmológica positiva) ou o grupo de dilatação se $s < 0$ (teoria com constante cosmológica negativa). Naturalmente (\ref{partialgauge1}) é apenas uma fixação de gauge: a teoria continua sendo uma teoria com simetria de gauge sobre o (A)dS$_{6}$ completo.

A teoria em $4$D pode ser obtida através de uma compactificação ou redução dimensional tipo Kaluza-Klein, onde os campos de matéria são codificados nas componentes pentadimensionais dos campos dinâmicos. A abordagem do processo de redução dimensional proposto por Kaluza-Klein é obtida baseada na hipótese de que a variedade $5$-dimensional adimita um decomposição topológica com a seguinte estrutura $\mathcal{M}_{5} = \mathcal{M}_{4}\otimes S_{1}$. $S_{1}$ é um espaço topológico unidimensional e compacto e ``microscópico'', topologicamente equivalente a um círculo de raio $r_{c}$ e, portanto, parametrizado pela coordenada $x^{4} = \chi$ tal que $0 \leq \chi \leq 2\pi r_{c}$. Com efeito, quaisquer campos definidos em $\mathcal{M}_{5}$ são periódicos em $\chi$, e podem ser expandidos como uma série de Fourier como
\begin{equation*}
f(x) = \sum_{n = -\infty}^{\infty}f^{(n)}(x)e^{in\chi /r_{c}}, \quad \forall\,\textrm{campo} \,f,
\end{equation*}  
onde todas as componentes de Fourier satisfazem a condição de realidade, ou seja, $(f^{(n)}(x))^{*} = f^{(-n)}(x)$.

Uma vez que a dependência na coordenada $\chi$ seja conhecida, a redução dimensional é obtida inserindo os campos com suas respectivas expansões em modos de Fourier na ação e integrando sobre a quinta coordenada. O resultado será uma ação efetiva em $4$D envolvendo todos o modos de Fourier e suas interações com essa expansão  em série infinita, a qual, pelo menos no regime perturbativo e espaço-tempo plano, são caracterizadas por um parâmetro de massa que cresce com $n$, isto é, $m_{n} = n/r_{c}$. Esse valor da massa para um regime de baixa-energia pode ser facilmente observado através de uma expansão dos campos dinâmicos sob o vácuo de Minkowski, ou seja, se consideramos que os campos possam ser expandidos como $f(x) = \mathring{f}(x) + h(x)$, onde $\mathring{f}(x)$ é o valor do campo no vácuo considerado e $h(x)$ a flutuação em relação ao background em questão. Obtemos que as flutuações em relação ao vácuo $h(x)$ satisfazem a equação de d'Alembert em $5$D,
\begin{equation*}
(\partial_{0}^{2} - \nabla^{2} - \partial_{\chi}^{2})h = 0,
\end{equation*} 
e suas componentes de Fourier, levando em consideração a condição de periodicidade, assumem a forma de onda plana, isto é, $h \sim \exp(-ik_{\mu}x^{\mu} + in\chi/r_{c})$. Naturalmente, eles satisfazem a relação de dispersão 
\begin{equation}
- \omega^{2} + k^{2} + \frac{n^{2}}{r_{c}^{2}} = 0,
\end{equation}
típica de modos massivos com $m^{2} = n^{2}/r_{c}^{2}$.

Como estamos assumindo que $r_{c}$ é pequeno o suficiente, de maneira que não seja possível  esquadrinhar experimentalmente no regime de energia disponível hoje, segue que os modos massivos com $n \neq 0 $ devem ser muito pesados. Portanto, no regime de baixa-energia podemos nos limitar, pelo menos em primeira aproximação, a apenas os setores de modo zero da expansão dos campos, que significa dizer que todos os campos sejam constantes em $\chi$. Isso significa que
\begin{equation}
\partial_{\chi}f(x) = 0, \quad \forall\,\textrm{campo} \,f.
\label{chi condition}
\end{equation}
O modelo de Chamseddine é obtido \cite{Cham} por uma truncação que consiste em tomar alguns campos a zero:
\begin{equation}
e^{I}_{\chi} = 0,\quad \omega_{\chi}^{IJ} = 0,\quad e^{4}_{\mu} = 0,\quad b^{I}_{\mu} = 0.
\label{truncation}
\end{equation}
Note que a primeira condição é, de fato, nada mais que uma condição de fixação de gauge, a segunda de (\ref{partialgauge1}). As outras três truncações quebram aparentemente a simetria (A)dS$_{6}$ em SO$(1,3)$. Contudo, Chamseddine mostrou que através de uma reordenação dos campos remanecentes em novos multipletos permite mostrar que a teoria residual esconde uma simetria de gauge (A)dS$_{5}$. Para vermos isso, não deveríamos aplicar direto a primeira das condições de fixação de gauge (\ref{partialgauge1}) mas reordenando os campos em multipletos de (A)dS$_{5}$:
\begin{align}
\mathbb{A}^{AB} & =\{\mathbb{A}^{IJ},\mathbb{A}^{4I}\}
:=\{\omega^{IJ},\frac{1}{l} e^{I}\},\label{eq:SO5-connec}\\
\Phi^{A} & =\{\Phi^{I},\Phi^{4}\}:=\{-b_{\chi}^{I},e_{\chi}^{4}\}.\label{eq:SO5-Phi}
\end{align}
Usando essas definições junto com as condições de truncação (\ref{truncation}), a ação (\ref{CS53}) reduz-se a expressão invariante sob (A)dS$_{5}$
\begin{equation}
S^{(4D)} =  \frac{1}{8}\int_{\mathcal{M}_{4}}\!\varepsilon_{ABCDE}\Phi^{A}\mathbb{F}^{BC}\wedge\mathbb{F}^{DE},\label{eq:phi-FF-action}
\end{equation}
onde o campo $\Phi^{A}$ é um multipleto na representação adjunta do grupo de simetria e uma zero-forma, ou seja, um escalar sob difeomorfismos. Note que nenhum parâmetro é necessário em frente a ação, pois qualquer parâmetro poderia ser absorvido em uma redefinição do campo escalar $\Phi^{A}$. A curvature de (A)dS$_{5}$
\begin{equation}
\mathbb{F}^{AB}= d\mathbb{A}^{AB}+\mathbb{A}^{A}\!_{C}\wedge\mathbb{A}^{CB}.
\end{equation}
 Em termos das componentes de SO$(1,3)$ lê-se
\begin{align*}
\mathbb{F}^{IJ} & = R^{IJ}-\frac{1}{l^{2}}e^{I}\wedge e^{J},\quad(R^{IJ}=d\omega^{IJ}+\omega^{I}\!_{K}\wedge\omega^{KJ})\\
\mathbb{\mathbb{F}}^{I4} & = \frac{1}{l}D e^{I},\quad\quad\quad\quad\quad(D e^{I}=d e^{I}+\omega^{I}\!_{J}\wedge e^{J}).
\end{align*}

As transformações de gauge infinitesimais de (A)dS$_{5}$ que deixam a ação de Chamseddine invariante podem ser escritas como
\begin{equation}
\delta  \mathbb{A}^{AB} =  d \epsilon^{AB} 
+ \mathbb{A}^A{}_C \, \epsilon^{CB}-\mathbb{A}^B{}_C \,\epsilon^{CA},\quad
\delta \Phi^{A} = \epsilon^{A}\,_{B}\Phi^{B}, \label{Ads-transf}
\end{equation}
onde $\epsilon^{AB} = -\epsilon^{BA}$ é o parâmetro infinitesimal da transformação. As equações de movimento que se seguem de (\ref{eq:phi-FF-action}) via a variação em relação a conexão $\mathbb{A}$ e ao campo escalar $\Phi$ são
\begin{align}
\frac{\delta S^{(4D)}}{\delta\Phi^{A}} = \frac{1}{8}\varepsilon_{ABCDE}\mathbb{F}^{BC}\wedge\mathbb{F}^{DE} & =0,\label{eq:so5-eom1}\\[0.1cm]
\frac{\delta S^{(4D)} }{\delta\mathbb{A}^{AB}} = \frac{1}{2}\varepsilon_{ABCDE}\mathbb{D}\Phi^{C}\wedge\mathbb{F}^{DE} & =0,\label{eq:so5-eom2}
\end{align}
onde $\mathbb{D}$ é a derivada covariante em respeito a conexão $\mathbb{A}$ que age da seguinte forma: $\mathbb{D} = d + \lbrack \mathbb{A},\quad\rbrack$. Portanto, a atuação nas componentes de $\Phi^{A}$ 
\begin{eqnarray*}
\mathbb{D}\Phi^{I} = D\Phi^{I}+\frac{s}{l}\Phi^4e^I, && \mathbb{D}\Phi^{4} = d\Phi^{4}-\frac{1}{l}e^I\Phi_I.\\
\end{eqnarray*}
As equações de movimento em termos das componentes SO$(1,3)$, assumem a forma
\begin{eqnarray}
\frac{\delta S}{\delta e^{I}} & = & -\frac{1}{2l}\varepsilon_{IJKL}(D\Phi^{J} + \frac{s}{l}e^{J}\Phi^{4})(R^{KL} -\frac{s}{l^{2}}e^{K}\wedge e^{L}) = 0 \nonumber\\[0.1cm]
\frac{\delta S}{\delta \omega^{IJ}} & = & \frac{1}{2}\varepsilon_{IJKL}\biggr((d\Phi^{4} - \frac{1}{l}e_{I'}\Phi^{I'})(R^{KL} - \frac{s}{l^{2}}e^{K}\wedge e^{L}) + \frac{1}{l}(D\Phi^{K} + \frac{s}{l}e^{K}\Phi^{4})De^{L}\biggl) = 0,\nonumber\\[0.1cm]
\frac{\delta S}{\delta \Phi^{4}} & = & \frac{1}{8}\varepsilon_{IJKL}(R^{IJ} - \frac{s}{l^{2}}e^{I} \wedge e^{J})(R^{KL} - \frac{s}{l^{2}}e^{K}\wedge e^{L}) = 0,\label{eqs-componentes}\\[0.1cm]
\frac{\delta S}{\delta \Phi^{I}} & = & \frac{1}{2l}\varepsilon_{IJKL}De^{J}(R^{KL} - \frac{s}{l^{2}}e^{K}\wedge e^{L}) = 0.\nonumber
\end{eqnarray}

Podemos introduzir matéria adicionando a ação (\ref{eq:phi-FF-action}) um termo S$_{m}$ o qual iremos supor que admita invariância sob o grupo de gauge do modelo, a saber (A)dS$_{5}$, e independente do campo escalar $\Phi^{A}$. A invariância sob (A)dS$_{5}$ da ação completa,
\begin{equation}
S = S^{4D} + S_{m}[e,\omega] \label{ação completa}
\end{equation}
pode ser expressada através de uma ``identidade de Ward'' local, que pode ser obtida ao considerarmos uma ação $S = S[\varphi]$, onde $\varphi$ são campos em geral, que possui uma invariância sob certo grupo de transformações, isto é, $\delta\varphi = \epsilon(x)P(\varphi)$ $\Rightarrow$ $\delta S[\varphi] = 0$, onde $P(\varphi)$ é uma função dos campos. Portanto, a identidade de Ward surge quando exigimos a invariância da ação sob a variação dos campos dinâmicos
\begin{equation*}
\delta S = \int dx\,\epsilon(x)\delta\varphi(x)\frac{\delta S[\varphi]}{\delta \varphi} = 0.
\end{equation*}

No nosso caso os campos dinâmicos são a conexão $\mathbb{A}^{AB}$ e o campo escalar $\Phi^{A}$, onde
\begin{equation*}
\delta\mathbb{A}^{AB} = \mathbb{D}\epsilon^{AB},\quad \delta\Phi^{A} = \epsilon^{A}\,_{B}\Phi^{B}.
\end{equation*}
Com efeito, a variação da ação
\begin{equation*}
\delta S[\mathbb{A}, \Phi] = \int dx \epsilon^{[AB]}(x)\biggr(-\frac{1}{2}\mathbb{D}\frac{\delta S}{\delta \mathbb{A^{AB}}} + \Phi_{B}\frac{\delta S}{\delta \Phi^{A}}\biggl) = 0, \quad \forall \epsilon^{AB}(x);
\end{equation*}
onde fizemos uma integração por partes no termo com derivada exterior, e lembrando da antisimetria do parâmetro $\epsilon^{AB}$ devemos antisimetrizar o termo de variação de $\Phi$. Dessa forma, a identidade de Ward assume a forma
\begin{equation*}
W_{AB} S := -\mathbb{D}\frac{\delta S}{\delta\mathbb{A}^{AB}} + \Phi_{A}\frac{\delta S}{\delta \Phi^{B}} - \Phi_{B}\frac{\delta S}{\delta \Phi^{A}} = 0.
\end{equation*}
Estaremos interessados particularmente na identidade de Ward associada à invariância sob os geradores $M_{I4}$:
\begin{equation}
W_{I}S := -\frac{1}{l}e^{J}\frac{\delta S}{\omega^{IJ}} -lD\frac{\delta S}{\delta e^{I}} + \Phi_{I}\frac{\delta S}{\delta \Phi^{4}} - s\Phi^{4}\frac{\delta S}{\delta \Phi^{I}} = 0.
\label{Ward}
\end{equation}
Podemos observar que essa identidade é assegura separadamente para ambas as ações $S^{4D}$ e $S_{m}$. Definindo as seguintes quantidades
\begin{equation*}
\mathcal{T}_{I} := \frac{\delta S_{m}}{\delta e^{I}}, \quad \mathcal{T}_{IJ} := \frac{\delta S_{m}}{\delta \omega^{IJ}},
\end{equation*}
Podemos reescrever (\ref{Ward}) como
\begin{eqnarray}
\frac{1}{l}e^{J}\mathcal{T}_{IJ} + lD\mathcal{T}_{I}& =  & -\frac{1}{l}e^{J} \frac{\delta S^{4D}}{\omega^{IJ}} - lD\frac{\delta S^{4D}}{\delta e^{I}}  + \Phi_{I}\frac{\delta S^{4D}}{\delta \Phi^{4}} \nonumber\\[0.1cm]  &&- s\Phi^{4}\frac{\delta S^{4D}}{\delta \Phi^{I}}.
\end{eqnarray}
onde a última igualdade expressa a invariância de $S^{(4D)}$. Essa identidade nos leva a uma equação de continuidade geral
\begin{equation}
\frac{2}{l}e^{J}\mathcal{T}_{IJ} + D\mathcal{T}_{I} = 0.
\label{continuity eq.}
\end{equation}
A $3$-forma $\mathcal{T}_{I}$ está relacionada com as componentes  do tensor energia-momento $\mathcal{T}^{N}\,_{I}$ na base de vierbein por
\begin{equation}
\mathcal{T}_{I} = \frac{1}{6}\varepsilon_{NJKL}\mathcal{T}^{N}\,_{I}e^{J}\wedge e^{K}\wedge e^{L}.
\label{energy-momentum}
\end{equation}
Podemos notar que $\mathcal{T}_{IJ} = 0$ se a ação da matéria for independente da conexão de spin $\omega$ (\ref{continuity eq.}) é interpretada então como equação de continuidade para energia e momento ($D\mathcal{T}_{I} = 0 \Longleftrightarrow \nabla_{\mu}T^{\mu}_{\nu} = 0$ no formalismo métrico usual da relatividade geral.)
\subsection{Acessibilidade e fixação de gauge}
O modelo de Chamseddine (\ref{eq:phi-FF-action}), estudado na seção anterior, é invariante   sob as transformações de gauge (\ref{Ads-transf}). Portanto, podemos ver a possibilidade de uma fixação parcial de gauge dada por quatro condições que nos permitem levar a zero quatro das componentes do campo escalar $\Phi^{A}$, ou seja,
\begin{equation}
\Phi^{I} = 0, \quad I = 0,...,3. \label{gauge fixing Phi-FF}
\end{equation}
Essa fixação deixa a invriância local de Lorentz explícita. Contudo, devemos analisar se essa transformação, que tem o papel de fixação parcial de gauge, é acessível. Em analogia com a eletrodinâmica onde o campo de gauge é o $4$-potencial $A^{\mu}$ temos que, se $\partial_{\mu}A^{\mu} \neq 0$, é possível encontrar uma transformação de gauge tal que a $4$-divergência do novo potencial $A'^{\mu}$ seja nula, ou seja,  $\partial_{\mu}A'^{\mu}\stackrel{!}{=} 0$. Portanto, sabendo que o potencial $A^{\mu}$ se transforma como
\begin{equation*}
A^{\mu} \longmapsto A'^{\mu} = A^{\mu} + \partial^{\mu}\alpha(x),
\end{equation*}
a transformação de gauge é acessível contanto que
\begin{equation*}
\partial_{\mu}\partial^{\mu}\alpha = 0 \quad \Leftrightarrow\quad \square\alpha = 0.
\end{equation*}
No nosso caso, temos uma conexão $\mathbb{A}$ que se transforma segundo um elemento $g$ do grupo de simetria como
\begin{equation*}
\mathbb{A}' = g^{-1} \mathbb{A} g + g^{-1} d\mathbb{A}.
\end{equation*}
A atuação de uma transformação finita sob o campo $\Phi$ 
\begin{equation*}
\Phi' = g \Phi,
\end{equation*}
onde podemos dividir a matriz elemento do grupo em blocos
$$ g = \left(
\begin{array}{c|cccc}
(g^{IJ})_{4\times 4} & g^{I4} \\
\hline
g^{4I}  & g^{44} \\
\end{array}
\right)$$
Além disso, 
\begin{equation*}
(M_{I4})^{C}\,_{D} = -(\delta^{C}_{I}\eta_{4D} - \delta^{C}_{4}\eta_{ID}) = (P_{I})^{C}\,_{D}.
\end{equation*}
A fim de entendermos como construir uma transformação finita iremos começar com a matriz $P_{I=0}$, assim

\begin{equation*}
\Rightarrow (P_{0})^{C}\,_{D} = -(\delta^{C}_{0}\eta_{4D} - \delta^{C}_{4}\eta_{0D})
\end{equation*}

Dessa forma, $P_{0}$ possui apenas os elementos  $C,D = 0,4$ e a matriz possui a forma

$$ P_{0} = \left[
\begin{array}{cccc}
0 & \cdots & -s \\
\vdots & \ddots & \vdots \\
\sigma & \cdots & 0 \\
\end{array}
\right]$$

\begin{align*}
(P_{0}^{2})^{C}\,_{D}  & = (P_{0})^{C}\,_{K} (P_{0})^{K}\,_{D} \\[0.1cm]
             & = \bigg(\delta^{C}_{0}\eta_{4K} - \delta^{C}_{4}\eta_{0K}\bigg)\bigg(\delta^{K}_{0}\eta_{4D} - \delta^{K}_{4}\eta_{0D}\bigg)\\[0.1cm]
            & = -s\delta^{C}_{0}\eta_{0D} - \sigma\delta^{C}_{4}\eta_{4D}\\[0.1cm]
            & = (-\sigma s)Q_{0} 
\end{align*}
onde podemos ver que $Q_{0}$ é um tipo de matriz ``identidade'' que contém apenas $1$ e $0$ na diagonal o que garante que todas as potências de $Q$ reproduzem seu valor inicial, isto é, $Q^{n} \equiv Q$.

$$ Q_{0} =  \left(
\begin{array}{ccccc}
1 & 0 & 0 & 0 & 0 \\
0 & 0 & 0 & 0 & 0 \\
0 & 0 & 0 & 0 & 0 \\
0 & 0 & 0 & 0 & 0 \\
0 & 0 & 0 & 0 & 1 \\
\end{array}
\right)$$

Por conseguinte, as potências pares de $P_{0}$ são
\begin{equation}
P_{0}^{2n} = (-\sigma s)^{n} Q_{0},
\end{equation}
analogamente, as potências ímpares são
\begin{align}
(P_{0})^{2n + 1} &= P_{0}P_{0}^{2n} \nonumber\\[0.1cm]
                 &= (-\sigma s)^{n}P_{0}.
\end{align}
Seguindo a mesma metodologia feita acima iremos obter a $I$-ésima matriz $P$
\begin{equation*}
(P_{I}^{2})^{C}\,_{D} = (P_{I})^{C}\,_{K} (P_{I})^{K}\,_{D}
\end{equation*}
Donde,
\begin{equation}
(P_{I}^{2})^{C}\,_{D} = -\eta_{II}\delta^{c}_{4}\eta_{4D} - s\delta^{C}_{I}\eta_{ID}
\end{equation}
lembrando que para cada $I$ o elemento $\eta_{II}$ assume apenas os valores $1$ ou $-1$, ou segundo as nossas notações $\sigma = \pm 1$. Dessa forma, podemos definir para cada matriz $Q_{I}$ tal como no caso $Q_{0}$ onde suas potências retornam o valor original, ou seja, 
\begin{equation*}
P_{I}^{2}  = (-\eta_{II} s)Q_{I}.
\end{equation*}
Portanto, as potências pares da $I$-ésima matriz $P$ seguem a fórmula recursiva
\begin{equation}
(P_{I})^{2n} = (-\eta_{II} s)^{n}Q_{I}\quad ; n \geq 1,
\end{equation}
e, as potências ímpares
\begin{equation}
(P_{I})^{2n + 1} = (-\eta_{II} s)^{n}P_{I}.
\end{equation}

Finalmente, temos que a transformação assume a seguinte forma
\begin{align*}
g_{I} & = e ^{\lambda P_{I}} = \displaystyle\sum_{k=0}^{\infty} \frac{\lambda^{k} P_{I}^{k}}{k!}\\[0.2cm]
      & = \mathds{1} + \displaystyle\sum_{k=0}^{\infty}\frac{\lambda^{2k +1}(-\eta_{II}s)^{k}P_{I}}{(2k +1)!} + \displaystyle\sum_{k=1}^{\infty}\frac{\lambda^{2k}(-\eta_{II}s)^{k}Q_{I}}{(2k)!}.\\
\end{align*}
Devemos analisar dois casos, o primeiro ao considerarmos $\eta_{II}s = 1$ o que irá produzir uma função oscilatória nos parâmetros
\begin{equation}
g_{I} = \mathds{1} + \sin\lambda P_{I} + (\cos\lambda - 1)Q_{I}.
\end{equation}
Nesse caso, não temos nenhum problema ou vínculo nos parâmetros da transformação quando aplicada sobre o campo $(\Phi^{I})$, porque as funções não assumem nenhum ponto singular em seu domínio de definição. Em outras palavras, não teremos problemas em definir $\tan\lambda$ como será mostrado nos cálculos a seguir. O segundo caso, é obtido ao considerarmos $\eta_{II}s = - 1$ o que nos leva a funções hiperbólicas nos parâmetros
\begin{equation}
g_{I} = \mathds{1} + \sinh\lambda P_{I} + (\cosh\lambda - 1)Q_{I}.
\end{equation} 

Contudo, a transformaçao finita sobre o campo $\Phi^{A}$, gerada pelo elemento do grupo $g$, é dada por
\begin{equation*}
(g_{I})^{A}\,_{B} \Phi^{B} = \Phi'^{A}
\end{equation*}
Assim, quando $A = I$ temos
\begin{equation}
\Phi'^{I}  =  \cosh\lambda \Phi^{I} - s\sinh\lambda \Phi^{4}
\end{equation}
Finalmente, exigiremos que as componentes desse novo campo $\Phi'^{I} \stackrel{!}{=} 0 $, donde obtemos a seguinte condição sobre o parâmetro
\begin{equation}
\tanh\lambda = s\frac{\Phi^{I}}{\Phi^{4}}\quad ; \mid\tanh\lambda\mid \leq 1.\label{gauge condition on phi1}
\end{equation}
Analogamente, no caso oscilatório, a única diferença é que não necessitamos de vincular o domínio de definição do parâmetro de transformação como no caso hiperbólico, ou seja,
\begin{equation}
\tan\lambda = s\frac{\Phi^{I}}{\Phi^{4}}. \label{gauge on phi 2}
\end{equation}
Em conclusão, vemos que de fato a fixação parcial de gauge é acessível logo as condições (\ref{gauge fixing Phi-FF}) são garantidas contanto que (\ref{gauge condition on phi1}) ou (\ref{gauge on phi 2}) sejam respeitadas.
  
A ação completa, incluindo a matéria, se reduz depois dessa fixação
\begin{eqnarray}
\bar{S} & = & \frac{1}{8}\int_{\mathcal{M}_{4}}\varepsilon_{IJKL}\Phi^{4}\mathbb{F}^{IJ}\wedge \mathbb{F}^{KL} + S_{m}\nonumber\\[0.1cm]
& = & \frac{1}{8}\int_{\mathcal{M}_{4}}\varepsilon_{IJKL}\Phi^{4}(R^{IJ} - \frac{s}{l^{2}}e^{I}\wedge e^{J})\wedge (R^{KL} - \frac{s}{l^{2}}e^{K}\wedge e^{L}) + S_{m} \label{acao com gauge fixing},
\end{eqnarray}
onde a ação de matéria $S_{m}$ é supostamente independente de $\Phi^{A}$ e assumiremos sua independência da conexão de spin daqui em diante. As equações de campo que são derivadas da ação acima\footnote{A qual estamos adicionando a ação da matéria $S_{m}$ que por hipótese deve obedecer a mesma invariância de gauge do modelo de Chamseddine puro, a saber (A)dS$_{5}$.} são
\begin{eqnarray}
\frac{\delta \bar{S}}{\delta e^{I}} & = & -\frac{1}{2}\frac{s}{l^{2}}\Phi^{4}\varepsilon_{IJKL}(e^{J}\wedge R^{KL} - \frac{s}{l^{2}}e^{J}\wedge e^{K} \wedge e^{L}) + \mathcal{T}_{I}  =  0,\nonumber\\[0.1cm]
\frac{\delta \bar{S}}{\delta \omega^{IJ}} & = & \frac{1}{2}\varepsilon_{IJKL}\biggr(d\Phi^{4}(R^{KL} - \frac{s}{l^{2}}e^{K}\wedge e^{L}) + \frac{2s}{l^{2}}\Phi^{4}e^{K}\wedge De^{L}\biggl) =  0, \label{eqs com gauge fixing} \\[0.1cm]
\frac{\delta \bar{S}}{\delta \Phi^{4}} & = & \frac{1}{8}\varepsilon_{IJKL}(R^{IJ} - \frac{s}{l^{2}}e^{I}\wedge e^{J})\wedge (R^{KL} - \frac{s}{l^{2}}e^{K}\wedge e^{L}) = 0,\nonumber
\end{eqnarray} 
onde a Torção $2$-forma é definida por $T^{I} := De^{I}$ e $\mathcal{T}_{I}$ é a energia-momento $3$-forma (\ref{energy-momentum}).

Observe que ao compararmos a primeira das Eqs. (\ref{eqs com gauge fixing}) com as equações padrões de Einstein com constante cosmológica no formalismo de primeira ordem,
\begin{eqnarray}
 & &\varepsilon_{IJKL}\biggr(e^{J}\wedge R^{KL} - \frac{\Lambda}{3}e^{J}\wedge e^{K}\wedge e^{L}\biggl) = -8\pi G\mathcal{T}_{I},\label{einstein first order} \\[0.1cm]
 & & T^{I} = De^{I} = 0,\nonumber
\end{eqnarray}
nos leva a identificação do fator $\dfrac{3s}{l^{2}}$ com a constante cosmológica,
\begin{equation}
\Lambda := \frac{3s}{l^{2}}, \label{cosmological constant}
\end{equation}
e definir a função
\begin{equation}
G(x) := \frac{3}{4\pi \Lambda\Phi^{4}(x)} \label{Newton function}
\end{equation}
como sendo o ``parâmetro'' de Newton, que é inversamente proporcional ao campo escalar $\Phi^{4}$. Dessa forma, as equações de campo (\ref{eqs com gauge fixing}) assumem a forma

\begin{align}
\varepsilon_{IJKL}\biggr( e^{J}\wedge R^{KL}
- \dfrac{\Lambda}{3}e^{J}\wedge e^{K}\wedge e^{L}\biggl) & = -8\pi G(x) \mathcal{T}_{I}\,,\nonumber\\
\varepsilon_{IJKL}\biggr( dG(x)\bigr( R^{KL}-\dfrac{\Lambda}{3}e^{K}\wedge e^{L}\bigl) 
- 2\dfrac{\Lambda}{3}G(x)e^{K}\wedge D e^{L}\biggl)& =0\,,
\label{definitive-PhiFF-field-eq}\\
\varepsilon_{IJKL}\biggr( R^{IJ}-\dfrac{\Lambda}{3}e^{I}\wedge e^{J}\biggl)
\biggr( R^{KL} - \dfrac{\Lambda}{3}e^{K}\wedge e^{L}\biggl) 
& =0\nonumber\,,
\end{align}
onde devemos notar a dependência em $x$ do parâmetro de Newton $G(x)$.

Podemos notar que a teoria é claramente singular quando $\Lambda = 0$ pois esse valor corresponderia um $s$ nulo na métrica de (A)dS$_{5}$ que naturalmente tornar-se-ia singular. A primeira das equações de campo (\ref{eqs com gauge fixing}) possuem a forma das equações de campo de Einstein usuais no formalismo de primeira ordem. Contudo, nosso parâmetro de acoplamento de Newton $G(x)$ é uma função dependente do campo $\Phi^{4}$. A segunda equação determina a torção $T^{I}$ em termos dos campos dinâmicos construtores da teoria $\omega^{IJ}$, $e^{I}$ e $G$. 

Na ausência de matéria uma solução natural de vácuo é a de curvatura constante e livre de torção que é a solução do espaço de (anti-)de Sitter: $R^{IJ} = \dfrac{\Lambda}{3}e^{I}\wedge e^{J}$. A última equação, que claramente admite a solução de curvatura constante de (anti-)de Sitter, é também compatível como soluções não triviais, como veremos. É interessante notar que a condição de fixação de gauge (\ref{gauge fixing Phi-FF}) é equivalente a primeira das  fixações de gauge em (\ref{partialgauge1}). 

Na teoria de Einstein a equação de continuidade do tensor de energia-momento lê-se, no formalismo de primeira-ordem,
\begin{equation}
D\mathcal{T}_{I} = 0
\end{equation}
onde $D$ é  a derivada covariante exterior em respeito a conexão de spin $\omega^{IJ}$ e a $3$-forma $\mathcal{T}_{I}$ está relacionanda ao tensor de energia-momento (\ref{energy-momentum}). A equação de continuidade acima segue das equações de Einstein (\ref{einstein first order}) e da identidade de Bianchi $DR^{IJ} = 0$. Como vimos anteriormente, temos que a equação de continuidade ainda se mantém no nosso caso como uma consequência da invariância sob (A)dS$_{5}$ e da identidade (\ref{Ward}) e com a hipótese de que a ação da matéria é independente do campo escalar $\Phi$ e também da conexão de spin $\omega$. 

Podemos ainda notar a identidade de Ward (\ref{Ward}) aplicando-se a condição de fixação de gauge $\Phi^{I} = 0$. Com efeito, levando em consideração que a ação da matéria $S_{m}$ dependa apenas do vierbein $e^{I}$, isso nos leva a identidade
\begin{equation*}
D\mathcal{T}_{I} = \frac{\Lambda}{3}e^{J}\frac{\delta \bar{S}}{\delta \omega^{IJ}} + \sqrt{\frac{\mid\Lambda\mid}{3}}\Phi^{4}\frac{\delta S}{\delta \Phi^{I}}\bigg|_{\Phi^{I} = 0},
\end{equation*}
onde $\bar{S}$ é a ação completa após a fixação de gauge (\ref{acao com gauge fixing}), e $S$ é a ação completa antes da fixação (\ref{ação completa}). Como $D\mathcal{T}_{I} = 0$, a identidade acima nos mostra que a equação
\begin{equation}
\frac{\delta S}{\delta \Phi^{I}} \bigg|_{\Phi^{I} = 0} = 0 
\label{onshell}
\end{equation}
é válida ``on-shell'', isto é, se as equações de movimento (\ref{eqs com gauge fixing}) da teoria com fixação de gauge são satisfeitas. Em outras palavras, isso é equivalente as equações  da teoria sem fixação obtida através da variação da ação em função de $\Phi^{I}$, avaliadas em $\Phi^{I} = 0$. De fato, essa identidade `` on-shell'' em (\ref{onshell}) pode ser obtida através da identidade de Ward (\ref{Ward}) ao tomar $\Phi^{I} = 0$.

Para finalizar essa seção poderíamos nos perguntar se as equações de movimento obtidas através da truncação da teoria, a saber as equações do modelo de Chamseddine (\ref{eq:so5-eom1}) e (\ref{eq:so5-eom2}), junto com as condições de truncação  (\ref{truncation}) e as condições sobre a dependência na coordenada $\chi$ (\ref{chi condition}), são também soluções das equações de movimento (\ref{Chern-Simons5}) da teoria completa de Chern-Simons sob (A)dS$_{6}$. De fato, as equações da teoria de Chern-Simons completa em $5$D reduzida a quatro dimensões são dadas por (\ref{eqs}), (\ref{eq:auxiliar-4}) e (\ref{eq:auxiliar-5}). Após a imposição das condições de truncação (\ref{truncation}) junto com a restrição (\ref{chi condition}) e renomeando (\ref{eq:SO5-connec}), (\ref{eq:SO5-Phi}), as componentes da curvatura assumem a forma
\begin{eqnarray}
& & F^{IJ} = R^{IJ} -\frac{s}{l^{2}}e^{I}\wedge e^{J},\,F^{I4} = 0,\quad  F^{I5} = \frac{1}{l}De^{I},\, F^{45} = 0, \nonumber\\[0.1cm]
&& F^{IJ}_{\chi} = 0,\, F^{I4}_{\chi} = -\frac{1}{l}D\Phi^{I} - \frac{s}{l^{2}}e^{I}\Phi^{4},  F^{I5}_{\chi} = 0 \label{eqs2}\\[0.1cm]
 & & F^{45}_{\chi} = \frac{1}{l}d\Phi^{4} - \frac{1}{l^{2}}e_{I}\Phi^{I}. \nonumber
\end{eqnarray}
Inserindo as expressões (\ref{eqs2}) nas oito equações (\ref{eq:auxiliar-4}) e (\ref{eq:auxiliar-5}), obtemos equações triviais $0 = 0$, e quatro equações não triviais que são identicas àquelas obtidas através da ação do modelo de Chamseddine, equação (\ref{eqs-componentes}). Concluimos que o conjunto de soluções das equações de movimento de (A)dS$_{5}$ é um subconjunto particular das soluções da teoria (A)dS$_{6}$ de Chern-Simons completa. É interessante notar que as quatro equações não triviais são derivadas da ação de Chern-Simons através da variação dos quatro campos ``destinados'' à truncação. Se tivessimos feito algum outro tipo de truncação nas equações de campo, obteríamos mais equações independentes do que obtem-se da ação truncada diretamente como fora feito. 
\section{Formalismo de Dirac-Bergmann}

Considere um sistema mecânico usual com número de graus de liberdade finitos, descritos por uma lagrangiana que é função das coordenadas generalizadas $q^{I}$ e das velocidades generalizadas $\dot{q}^{I}$ ($I= 1,..., N$) e $N \in \mathbb{N}$, isto é, $L = L(q,\dot{q})$, e a ação que descreve o sistema dinâmico dada pelo funcional $S[q,\dot{q}] :=  \int dt L(q,\dot{q})$. O princípio variacional nos leva as equações de movimento de Euler-Lagrange,
\begin{equation}
\frac{d}{dt}\biggl(\frac{\partial L}{\partial \dot{q}^{I}}\biggr)  =  \frac{\partial L}{\partial q^{I}}, \quad I = 1, 2, ..., N,
\end{equation}
onde definimos a matriz hessiana $M_{IJ}$ por:
\begin{equation}
\biggl(\frac{\partial^{2}L}{\partial\dot{q}^{I}\partial\dot{q}^{J}}\biggr)\ddot{q}^{J}:= M_{IJ}(q,\dot{q})\ddot{q}^{J} = \frac{\partial L}{\partial q^{I}} - \frac{\partial^{2}L}{\partial\dot{q}^{I}\partial\dot{q}^{J}}\dot{q}^{J}:= Q_{I}(q,\dot{q}).
\end{equation}
Em geral, esta matriz hessiana $M_{IJ}$, construída através das derivadas segundas da lagrangiana em relação as velocidades generalizadas, é invertível o que permite resolver as acelerações, $\ddot{q}^{I} = (M^{-1})^{IJ}Q_{J}$. Consequentemente, dadas a posição e velocidade iniciais, pode-se sempre determinar univocamente a trajetória dinâmica, no espaço de configuração, em cada instante. Por outro lado, se a matriz $M_{IJ}$ não for invertível então, se $1 \leq r \leq N$ denota o rank de $M$, haverá $(N - r)$ vetores independentes $u^{I}_{m}$ $(m = 1, ... (N -r))$, satisfazendo a equação de auto-valor $M u = 0$ e, portanto, não podemos resolver para todas as acelerações univocamente. Naturalmente, as equações de movimento não são univocamente determinadas e obtemos $(N - r)$ relações entre as $2N$ variáveis $q, \dot{q}$.
Portanto, podemos dividir as teorias em dois casos associados ao determinante dessa matriz
$$
M = \textrm{det}\bigg|\frac{\partial^{2}L}{\partial\dot{q}^{I}\partial\dot{q}^{J}}\bigg| = \left\{\begin{array}{rcl}
\neq 0, & \textrm{não singular}\\
= 0, & \textrm{singular}
\end{array}
\right. 
$$
a importância que reside nessa classificação está ligada ao processo de passagem para o formalismo hamiltoniano, que tem implicações diretas nos métodos de quantização canônica \cite{Gitman, Hans, Dir,Arnold}. De fato, com a finalidade de transitarmos para o formalismo citado, devemos introduzir os momentos canonicamente conjugados às coordenadas generalizadas $q^{I}$ como
\begin{equation}
\pi_{I} := \frac{\partial L}{\partial\dot{q}^{I}}.
\label{momenta definition}
\end{equation}
Assim, as velocidades generalizadas $\dot{q}^{I}$ deveriam ser invertidas em termos de funções de $q$ e $\pi$, isto é, 
\begin{equation*}
\dot{q}^{I} = v^{I}(q,\pi).
\end{equation*}
Contudo, a invertibilidade só se processa de maneira natural no caso de teorias não-singulares. Esse é a grosso modo a essencia do \emph{teorema da função implícita}, que não serão abordados nessa seção de revisão. Sendo este o caso, define-se a hamiltoniana canônica como uma transformação de Legendre
\begin{equation}
H_{C}(q, \pi) = \sum_{I = 0}^{N}\pi_{I}\dot{q}^{I} - L(q,\dot{q}),\label{hamiltoniana canonica}
\end{equation}
e considerando-se, sua variação funcional,
\begin{equation}
\delta H = \dot{q}^{I}\delta \pi_{I} + \biggl(\pi_{I} - \frac{\partial L}{\partial \dot{q}^{I}}\biggr)\delta \dot{q}^{I} - \frac{\partial L}{\partial q^{I}}\delta q^{I}.
\label{hamilton variation}
\end{equation}
Pela definição (\ref{momenta definition}), o termo do meio é cancelado e
a hamiltoniana só envolve variações dos $q$ e $\pi$ e não envolve variação das velocidades $\dot{q}^I$. O espaço $2N$-dimensional de todos os pares $\pi_{I}, q^{I}$ é chamado de espaço de fase. A dinâmica do sistema é estabelecidade pelas equações de Hamilton,
\begin{equation}
\dot{q}^{I} = \frac{\partial H}{\partial\pi_{I}}, \quad \dot{\pi}_{I} = - \frac{\partial H}{\partial q^{I}}.
\label{eqs de Hamilton}
\end{equation}

Portanto, nos sistemas dinâmicos não-singulares assume-se sempre duas hipóteses: $(1)$ as variações $\delta \pi_{I}, \delta q^{I}$ são completamente independentes e $(2)$ a hamiltoniana é uma função exclusivamente de $\pi_{I}, q^{I}$. Contudo, essas duas hipóteses falham nos casos singulares. De fato, note que a variação da hamiltoniana, sendo dada em termos dos momenta e das posições, é dependente da valoração desta na subvariedade definida por $\pi_{I} = \frac{\partial L}{\partial \dot{q}^{I}}$. Com efeito, se exigimos que as variações das variáveis canônicas respeitem também essas condições, devemos ter
\begin{eqnarray}
\delta \pi_{I} & = & M_{IJ}\delta\dot{q}^{J} + \frac{\partial^{2}L}{\partial\dot{q}^{I}\partial q^{J}}\delta q^{J}, \\[0.1cm]
u^{I}_{m}\delta \pi_{I} & = &  0 + u^{I}_{m}\frac{\partial^{2}L}{\partial\dot{q}^{I}\partial q^{J}}\delta q^{J},
\end{eqnarray}
 que mostra imediatamente que as variações $\delta \pi_{I}, \delta q^{I}$ não são completamente independentes. Com efeito, não podemos obter univocamente, no caso singular, as equações de movimento de Hamilton. As variações não sendo independentes, ou seja, a impossibilidade de eliminarmos todas as velocidades generalizadas em termos de $\pi_{I}$ e $q^{I}$, introduz relações entre as variáveis do espaço de fase, isto é, $\phi_{m}(q,\pi) = 0$ $(m = 1,..., M \leq N)$ chamadas de \emph{vínculos primários}\footnote{Devido ao fato de $\phi(q,\pi)$ ser zero, pela própria definição de momento, usualmente chamamos esses tipos de vínculos de vínculos primários.}. Portanto, a hamiltoniana definida por (\ref{hamiltoniana canonica}), chamada de \emph{hamiltoniana canônica} $H_{C}$, produziria a mesma dinâmica da hamiltoniana canônica acrescida de uma combinação linear dos $\phi$'s
 \begin{equation}
 H_{T} = H_{C} + \lambda_{m}\phi_{m}, \quad (m = 1,2...,M \leq N), \label{hamiltoniana total}
\end{equation}  
onde as quantidades $\lambda_{m}$ são coeficientes arbitrários dos $q$'s e dos $\pi$'s, conhecidos por \emph{multiplicadores de Lagrange}. A teoria física não é capaz de distinguir entre $H_{C}$ e $H_{T}$ pois a hamiltoniana não é mais univocamente determinada.

Vimos acima que a variação da hamiltoniana $H_{C}$ é dada por
\begin{equation*}
\delta H = \dot{q}^{I}\delta\pi_{I} - \biggl(\frac{\partial L}{\partial \dot{q}^{I}}\biggr).
\end{equation*}
 Essa equação é assegurada para quaisquer variações de $q$ e $p$ que são sujeitas a condição dos vínculos primários a serem preservados. Como vimos, a variação dos $q$'s e $\pi$'s não são independetes pois são restritas pelos vínculos e devem ser sujeitas a preservação destes. Desse modo, aplicando-se os métodos de variação levando-se em conta os vínculos primários (\ref{hamiltoniana total}) obtemos as seguintes equações de movimento
 \begin{eqnarray}
 \dot{q}^{I} & = & \frac{\partial H_{C}}{\partial \pi_{I}} + \lambda_{m}\frac{\partial \phi_{m}}{\partial \pi_{I}}, \label{variations of canonical variab}\\[0.1cm]
 \dot{\pi}_{I} & = & - \frac{\partial H_{C}}{\partial q^{I}} - \lambda_{m}\frac{\partial \phi_{m}}{\partial q^{I}}\nonumber.
 \end{eqnarray}
 Temos, desse modo, as equações de movimento de Hamilton, descrevendo como as variáveis $q$ e $\pi$ evoluem no tempo, no entanto, essas equações envolvem coeficientes indeterminados $\lambda_{m}$. 
 
 É conveniente introduzir um formalismo, que faremos uso recorrente, que nos permite escrever essas equações de maneira mais compacta, conhecido como o formalismo dos parenteses de Poisson. É uma operação entre duas funções no espaço de fase, $C(\pi,q)$ e $B(\pi, q)$, definida por
\begin{equation}
\bigr\{C, B\bigl\} := \frac{\partial C}{\partial q^{I}}\frac{\partial B}{\partial\pi_{I}} - \frac{\partial C}{\partial\pi_{I}}\frac{\partial B}{\partial q^{I}}.
\end{equation}
Em termos dos parenteses de Poisson temos que
\begin{equation}
\bigr\{q_{I},\pi^{J}\bigl\} = \delta^{J}_{I},
\end{equation}
os parênteses de Poisson possuem certas propriedades que seguem imediatamente de sua definição que são, antisimetria:
\begin{equation*}
\bigr\{C, B\bigl\} = - \bigr\{B, C\bigl\},
\end{equation*}
linearidade
\begin{equation*}
\bigr\{C_{1} + C_{2}, B\bigl\} = C_{1}\bigr\{C_{2}, B\bigl\} + C_{2}\bigr\{C_{1}, B\bigl\},
\end{equation*}
temos ainda a regra do produto ou de Leibniz
\begin{equation*}
\bigr\{C_{1}C_{2}, B\bigl\} = C_{1}\bigr\{C_{2}, B\bigl\} + \bigr\{C_{1}, B\bigl\}C_{2}.
\end{equation*}
Finalmente, existe uma relação, conhecida como identidade de Jacobi, conectando três variáveis dinâmicas
\begin{equation*}
\bigr\{C_{1},\bigr\{C_{2},C_{3}\bigl\}\bigl\} + \bigr\{C_{2},\bigr\{C_{3},C_{1}\bigl\}\bigl\} + \bigr\{C_{3},\bigr\{C_{1},C_{2}\bigl\}\bigl\} = 0.
\end{equation*}

 As equações de Hamilton, ou melhor, a evolução dinâmica de uma variável $C$ no espaço de fase é dada por
\begin{equation}
\dot{C} = \frac{\partial C}{\partial q^{I}}\dot{q}^{I} + \frac{\partial C}{\partial \pi_{I}}\dot{\pi}_{I}. 
\label{evolução dinâmica}
\end{equation}
Substituindo-se as variações de $q$ e $\pi$ dadas por (\ref{variations of canonical variab}), encontramos que (\ref{evolução dinâmica}) assume a forma
\begin{equation}
\dot{C} = \bigr\{C, H_{C}\bigl\} + \lambda_{m} \bigr\{C, \phi_{m}\bigl\}.\label{ev. dinâmica 2}
\end{equation}
Consequentemente, as equações (\ref{eqs de Hamilton}) podem ser reescritas como
\begin{equation}
\dot{q}^{I} = \bigr\{q^{I}, H_{C}\bigr\} + \lambda_{m}\bigr\{q^{I}, \phi_{m}\bigl\}, \quad \dot{\pi}_{I} = \bigr\{\pi_{I}, H_{C}\bigr\} + \lambda_{m}\bigr\{\pi_{I}, \phi_{m}\bigl\}.
\end{equation}
  Temos que ter alguns cuidados ao trabalharmos com sistemas não-singulares. Como estamos interessados na dinâmica, que é dada pelos colchêtes de Poisson, ao trabalhar com vínculos não podemos utilizá-los antes de resolvermos os colchêtes. Caso contrário obteríamos um resultado contraditório. Para nos lembrar dessa regra no formalismo de Dirac-Bergmann, escrevemos os vínculos com um sinal de ``igualdade fraca'' $\approx$. Por definição, os $M$ vínculos são funções independentes entre si. A subvariedade definida por $\phi_{m} \approx 0$ é chamada de superfície dos vínculos possuindo dimensão $2N - M$ em relação ao espaço de fase. Assim, os vínculos primários são escritos como
  \begin{equation}
  \phi_{m} \approx 0, \quad (m = 1,...,M). \label{primary constraints}
  \end{equation}
  Devemos fazer uso de (\ref{primary constraints}) como igualdade estrita somente depois de termos calculados todos os parênteses de Poisson. Sujeita a essa regra, a evolução de uma variável dinâmica, dada pela hamiltoniana total (\ref{hamiltoniana total}), assume a seguinte forma mais concisa
  \begin{equation}
  \dot{C} \approx \bigr\{C, H_{T}\bigl\} \label{ev. dinâmica 3}
  \end{equation}
  Agora, devemos analisar as consequências dessas equações de movimento. Naturalmente, teremos algumas condições de consistência, pois temos as quantidades $\phi_{m}$ que devem ser zero a todo instante. Temos de garantir que as trajetórias no espaço de fase, que estejam sobre a superfície de vínculos em um dado instante inicial, permaneçam sobre esta na evolução do sistema. Podemos aplicar as equações de movimento (\ref{ev. dinâmica 2}) ou (\ref{ev. dinâmica 3}) tomando a variável dinâmica $C$ sendo um dos vínculos primários da teoria (\ref{primary constraints}). Contudo, sabemos que $\dot{\phi}$ deverá ser zero por consistência e, portanto, obtemos algumas condições de consistência. 
\begin{equation}
\varphi (q, \pi) = 0.
\label{secundary constraints}
\end{equation}
Finalmente, algumas condições podem acontecer através de (\ref{secundary constraints}: um tipo de equação reduz-se a $0 = 0$. De outro modo, podemos obter apenas algumas restrições sobre os multiplicadores de Lagrange $\lambda_{m}$.

O primeiro tipo de equação não precisamos nos preocupar pois a imposição de consitência se processa identicamente. No entanto, o segundo tipo que produz novas relações entre as variáveis canônicas, significa que temos novos vínculos sobre as variáveis $q, \pi$ dados por (\ref{secundary constraints}). Esses novos vínculos advindos da imposição de consistência da evolução dinâmica são chamados de \emph{vínculos secundários}. Eles diferem dos vínculos primários pois esses são consequências diretas de (\ref{momenta definition}) que definem os momenta, enquanto para os vínculos secundários, devemos usar as equações de movimento.

Havendo a existência de vínculos secundários na nossa teoria devemos impor, novamente, as condições de consistência, ou seja, o vínculo secundário produzido deve ser estável e, portanto, exigimos que $\dot{\varphi} \approx 0$. Assim, obtemos uma nova equação
\begin{equation}
\bigr\{\varphi, H_{C}\bigl\} + \lambda_{m}\bigr\{\varphi, \phi_{m}\bigl\} \approx 0.
\label{estabilidade1}
\end{equation}
Essa equação deve ser tratada no mesmo pé de igualdade que as demais, isto é, devemos notar qual dos três tipos descritos ela irá produzir. Se obtemos mais uma relação entre as variáveis canônicas isso indica a preseça de mais um vínculo secundário, logo o processo deve ser realizado mais uma vez devido ao novo vínculo gerado. Devemos aplicar esse algorítmo de condições de consistência até não produzirmos mais relações entre as variáveis canônicas, e no final termos obtido um conjunto de vínculo secundários do tipo (\ref{secundary constraints}) mais um certo número de condições sobre os coeficientes $\lambda$.

Os vínculos secundários serão tratados no mesmo pé de igualdade que os vínculos primários, veremos mais adiante o porquê dessa consideração. Com efeito, é conveniente redefinir o conjunto dos vínculos primários e secundários como sendo rotulados por um índice seguindo a notação
\begin{equation}
\phi_{k} \approx 0, \quad k = M + 1, ..., M + K,
\end{equation}  
onde $K$ é o número total de vínculos secundários. Como eles são escritos como igualdades fracas implica que eles também representam equações que não devemos usar antes de calcularmos o parêntese de Poisson. Portanto, todos os vínculos, ou seja, o número completo de vínculos primários e secundários podem ser escritos de uma maneira mais compacta como
\begin{equation}
\phi_{j} \approx 0, \quad j = 1,..., M + K := J.
\end{equation}

Finalmente, precisamos compreender quais são os tipos de restrições que o terceiro tipo de condição, associada a estabilidade, produzirá sobre os multiplicadores de Lagrange. Lembrando que na hamiltoniana total (\ref{hamiltoniana total}) só entra, em princípio, os vínculos primários,  ou seja $(m = 1,..., M)$. Com efeito, ao analisarmos a estabilidade dos vínculos $\phi_{j}$, onde $(j = 1,..., J)$ e como $J \geq M$, o sistema de equações que produziremos será, em geral, mais que completo. De fato,
\begin{equation}
\bigr\{\phi_{j}, H_{C}\bigl\} + \lambda_{m}\bigr\{\phi_{j}, \phi_{m}\bigl\} \approx 0, \quad m = 1, ..., M, \, j = 1,..., J = M + K,
\label{condições sobre lambda}
\end{equation}
nos proporciona condições sobre os $\lambda$'s, pois nesse caso elas não se reduzem a equações de vínculos, contudo, obtemos um conjunto completo de equações lineares não-homogêneas para $\lambda_{m}$. Uma maneira de fixar os $\lambda$'s é buscar por soluções que nos dê $\lambda$'s como função dos $q$'s e dos $\pi$'s, digamos
\begin{equation}
\lambda_{m} = \mathcal{U}_{m}(q,\pi).
\end{equation} 
Entretanto, essa solução não é única, pois se temos uma solução particular podemos adicionar a ela uma combinação linear das soluções ($\mathcal{V}_{m}(q,\pi)$) associada a parte homogênea de (\ref{condições sobre lambda}):
\begin{equation}
\mathcal{V}_{m} \bigr\{\phi_{j}, \phi_{m}\bigl\} \approx 0,
\label{sol. inomogenea}
\end{equation}
o que nos produzirá a solução geral da equação inomogênea. Desejamos a solução mais geral para (\ref{condições sobre lambda}) e, portanto, devemos considerar todas as soluções independentes de (\ref{sol. inomogenea}), a qual denotaremos por $\mathcal{V}_{am}$, $a = 1, ..., A$. Logo, a solução mais geral de (\ref{condições sobre lambda}) assume a forma
\begin{equation}
\lambda_{m} = \mathcal{U}_{m} + v_{a}\mathcal{V}_{am},
\end{equation}
em termos dos coeficientes $v_{a}$ que são completamente arbitrários.

Substituindo esse expressão para os $\lambda$'s na hamiltoniana total (\ref{hamiltoniana total}) obtemos
\begin{equation*}
H_{T} = H_{C} + \mathcal{U}_{m}\phi_{m} + v_{a}\mathcal{V}_{am}\phi_{m}.
\end{equation*}
Podemos reescrever a equação acima como
\begin{equation}
H_{T} = \bar{H} + v_{a}\phi_{a}, \label{1st class hamiltonian}
\end{equation}
onde $\bar{H} := H_{C} + \mathcal{U}_{m}\phi_{m}$ e $\phi_{a} := \mathcal{V}_{am}\phi_{m}$. Assim, vemos que mesmo após aplicarmos toda a análise de consistência da teoria ainda ficamos com coeficientes completamente arbitrários $v$. Essa é uma diferença crucial  das formulações hamiltonianas nos casos regulares. Isto é, temos funções arbitrárias do tempo na solução das equações de movimento dados as condições iniciais. 

Obviamente que isso é um sinal da presença de algum tipo de simetria na teoria. Efetivamente, muitas vezes em sistemas físicos usamos mais variáveis que o necessário para sua descrição. Constroi-se modelos sem sequer saber, \textit{a priori}, quais são os verdadeiros graus de liberdade. Um exemplo nítido é a eletrodinâmica. O campo eletromagnético no vácuo possui apenas dois graus de liberdade associados às polarizações da onda eletromagnética. No entanto, nós usualmente descrevemos a teoria em termos do vetor potencial $A^{\mu} = (A_{0}, \vec{A})$ o qual possui quatro componentes, ou em termos do tensor $F_{\mu\nu}$ que guarda a informação dos campos eletromagnéticos. 

A relatividade geral possui também dois graus de liberdade, no entanto, sua descrição é feita através do tensor métrico $g_{\mu\nu}$ que possui dez componentes ou via o vierbein, no formalismo de primeira ordem, $e^{I}_{\mu}$ que possui $16$ componentes. Naturalmente, existem relações entre as variáveis chamadas de vínculos. Os vínculos, como vimos acima, são equações envolvendo as variáveis do espaço de fase que devem ser asseguradas a cada instante da evolução do sistema $\lbrace\phi,H\rbrace \approx 0$. A fim de compreender melhor essas relações devemos inserir uma nova classificação dos vínculos que nos permitirá relacioná-los com os geradores infinitesimais de alguma simetria. 
 \begin{definition}
 Uma variável dinâmica $\mathcal{A} (q,\pi)$,  é chamada de primeira-classe caso os seus parênteses de Poisson com todos os vínculos $\phi$'s sejam nulos:
 \begin{equation}
 \lbrace \mathcal{A},\phi_{j}\rbrace \approx 0, \quad 1,..., J. \label{1st-class}
 \end{equation}
 Caso contrário, $\mathcal{A}(q,\pi)$ é chamada de segunda-classe.
\end{definition}

Com efeito, se $\mathcal{A}$ é de primeira-classe, segue que $\lbrace \mathcal{A},\phi_{j}\rbrace$ deve ser alguma combinação linear dos $\phi$'s, pois eles são as únicas entidades, no presente formalismo, que são, por definição, fracamente zero. Portanto, temos
\begin{equation}
\lbrace \mathcal{A},\phi_{j}\rbrace = c_{jk}\phi_{k}.
\end{equation}
Além disso, um resultado interessante que pode ser demonstrado facilmente é: se temos duas variáveis dinâmicas $\mathcal{A}(q,\pi)$ e $\mathcal{B}(q,\pi)$ de primeira-classe, então o parêntese de Poisson dessas duas quantidades, ou seja, $\lbrace\mathcal{A},\mathcal{B} \rbrace$, é também de primeira-classe.

De uma forma matricial, podemos obter essas classificações dos vínculos de primeira e segunda-classe, que em muitos casos pode mostrar-se mais prática. Dessa forma, podemos ainda perder a distinção entre vínculos primários e secundários e ver que essa distinção não é tão importante. De fato, considerando em (\ref{hamiltoniana total}) a soma de todos os vínculos primários e secundários temos a equação para os $\lambda$'s
\begin{equation}
\lbrace H_{C},\phi_{j}\rbrace + \lambda_{i}\lbrace \phi_{j},\phi_{i}\rbrace \approx 0, \quad i,j = 1,..., J. \label{equação de vínculos}
\end{equation}
Se os $\lambda$ são ou não parâmetros completamente arbitrários dependerá crucialmente das propriedades de invertibilidade da matriz antisimétrica 

\begin{equation}
\vartriangle_{ij} := \lbrace \phi_{i}, \phi_{j} \rbrace. \label{matriz dos vínculos}
\end{equation}

Naturalmente, se (\ref{matriz dos vínculos}) é invertível\footnote{Lembrando que apenas matrizes antisimétricas de dimensão par são invertíveis devido a necessidade do determinante ser diferente de zero como condição de invertibilidade.}, podemos resolver (\ref{equação de vínculos}), o que nos daria uma solução única para todos os $\lambda$, e assim uma hamiltoniana única e consequentemente uma evolução temporal univocamente determinada. Contudo, suponha que a matriz não seja invertível. Consequentemente, haverá alguns autovetores nulos. Como somos livre para fazer mudanças de base sob transformações lineares nos vínculos, podemos assumir que exista um subconjunto dos vínculos $\phi_{a}$,  $a = 1,...,A$,cujos parênteses de Poisson com todos os demais vínculos se anulam fracamente, isto é,
\begin{equation}
\lbrace \phi_{a},\phi_{j}\rbrace \approx 0.
\end{equation} 
Nesse caso, uma parte só dos $\lambda$ em (\ref{equação de vínculos}) é determinada, os $\lambda_{a}$, $a = 1,..., A$ permanecendo arbitrários. E portanto, os vínculos $\phi_{a}$ tendo parênteses de Poisson fracamente zero com todos os vínculos,  são os vínculos de \emph{primeira-classe}. Por outro lado, se denotarmos os demais vínculos por $\psi_{m}$, $m = 1,..., J - A$, são chamados de \emph{segunda-classe}\footnote{Dessa forma, vemos que o rank da matriz $\vartriangle_{mn}$, igual a $J - A$, nos informa sobre o número de vínculos de segunda-classe. Consequentemente, o rank, nesse contexto, será sempre uma quantidade par.}\cite{Hen1,Dir}.

Finalmente, vamos tentar dar uma compreensão física sobre esses conceitos de vínculos de primeira-classe. Considere um sistema dinâmico, descrito por um variável $\mathcal{A}(q,\pi)$ que evolui de um dado estado inicial. Assumindo que no instante $t = 0$ nossa variável dinâmica assume o valor $\mathcal{A}_{0}$. Daí, seu valor após um intervalo de tempo infinitesimal $\delta t$ será, após uma expansão em Taylor
\begin{equation*}
\mathcal{A}(\delta t) = \mathcal{A}_{0} + \dot{\mathcal{A}}\delta t. 
\end{equation*}
Por outro lado, $\dot{\mathcal{A}}$ é dada através do parêntese de Poisson com a hamiltoniana (\ref{1st class hamiltonian}), ou seja,
\begin{eqnarray*}
 \mathcal{A} & = & \mathcal{A}_{0} + \lbrace \mathcal{A}, H_{T} \rbrace\delta t, \\[0.1cm]
                   & = & \mathcal{A}_{0} + \delta t \biggl(\lbrace\mathcal{A},\bar{H} \rbrace + v_{a}\lbrace\mathcal{A},\phi_{a}\rbrace\biggr).
\end{eqnarray*}
 Os coeficientes $v$ são completamente arbitrários e, portanto, para um valor distinto de $v$ teremos um $\mathcal{A}$ diferente no mesmo instante de tempo $\delta t$, cuja diferença funcional é dada por
 \begin{eqnarray}
 \delta \mathcal{A} & = & \delta t (v_{a} - \bar{v}_{a})\lbrace \mathcal{A},\phi_{a}\rbrace, \\[0.1cm]
                    & = & \epsilon_{a} \lbrace \mathcal{A},\phi_{a}\rbrace, \quad \epsilon_{a}:= \delta t (v_{a} - \bar{v}_{a}). \label{1st-class as gauge}
 \end{eqnarray}
 As variações das variáveis dinâmicas do nossso sistema, dadas por (\ref{1st-class as gauge}), devem descrever o mesmo estado físico. Essa mudança nas variáveis hamiltonianas consiste em aplicar uma transformação que se processa no espaço das variáveis dinâmicas e, não necessáriamente no espaço de fase, capaz de mudar as variáveis sem alterar o estado físico do sistema. Esse tipo de transformação é conhecido como transformações de gauge cuja  função geradora é dada por $\epsilon_{a}\phi_{a}$. Portanto, chegamos a conclusão que os vínculos de \emph{primeira-classe}, tem o seguinte significado: são geradores de transformações de gauge que alteram as variáveis dinâmicas sem alterarem o estado físico do sistema. 
 
 Além disso, podemos fazer a contagem dos graus de liberdade da teoria, pois a presença de vínculos reduz a dimensão do espaço de fase ($\Gamma_{\textrm{fase}}$). De fato, o número de graus de liberdade é dado por
\begin{equation}
\mathcal{N} = \frac{1}{2}\bigl(\textrm{dim}\Gamma_{\textrm{fase}} - 2\times \textrm{F} - \textrm{S}\bigr), \label{degrees of freedom count}
\end{equation}
onde $F$ representa o número de vínculos de primeira classe e $S$ o número de vínculos de segunda classe. A contagem dos graus de liberdade é feita segundo (\ref{degrees of freedom count}) pois cada vínculo implica em uma condição sobre as coordenadas do espaço de fase. Com efeito, a presença dos vínculos produz uma subvariedade, digamos $\Sigma \subset \Gamma_{\textrm{fase}}$, onde a dinâmica do sistema deve ser avaliada. Entretando, os vínculos de primeira-classe além de proporcionarem a condição citada, eles somam mais um grau de liberdade associado a invariâcia de gauge. Em outras palavras, os estados físicos não são representados por pontos pertencentes à superfície $\Sigma$, mas são representados por órbitas em $\Sigma$, isto é, trajetórias que conectam pontos da subvariedade através dos vínculos de primeira-classe. Dessa forma, os vínculos de primeira-classe produzem uma classe de equivalência entre pontos da subvariedade. Os pontos das classes de equivalência estabelecem as órbitas de gauge ao longo das quais podemos modificar as variáveis dinâmicas sem alterar o estado físico do sistema. Consequentemente, temos de contabilizar esse fato e por isso o fator $2$F para vínculos de primeira-classe. Os vínculos de segunda-classe são apenas condições entre as variáveis do espaço de fase e nada tem a ver com alguma simetria, dessa forma a contabiliza-se apenas uma vez cada S.
 
Como exemplo vamos considerar a aplicação do algorítimo de Dirac-Bergmann no caso da eletrodinâmica de Maxwell livre, definida pela ação
\begin{equation}
S[A(x)] = \int dt\int d^{3}x \biggr(-\frac{1}{4}F_{\mu\nu}F^{\mu\nu}\biggl):=\int d^{4}x \mathcal{L}(A_{\mu},\partial_{\nu}A_{\mu}).
\label{ação de Maxwell}
\end{equation}
Através do princípio da mínima ação obtemos as equações de Maxwell no vácuo expressas em função de $A_{\mu}$. Esse é um exemplo de um sistema de campos, ou seja, a dinâmica está associada com infinitos graus de liberdade, $A_{\mu}(t, \vec{x})$. Portanto, derivadas da lagrangiana serão denotadas por $\frac{\delta L}{\delta A_{\mu}(x)}$, lembrando:
\begin{equation}
\frac{\delta A_{\mu}(x)}{\delta A_{\nu}(y)} = \delta^{\nu}_{\mu}\delta^{3}(x - y), \quad \frac{\delta\partial_{\nu}A_{\mu}(x)}{\delta A_{\kappa}(y)} = \delta^{\kappa}_{\mu}\partial_{\nu}\delta^{3}(x - y).
\label{propr fundamentais no EM}
\end{equation}
A fim de passarmos para descrição hamiltoniana devemos obter o momento canônicamente conjugado a $A_{\mu}(t,x)$ que é obtido por uma generalização de (\ref{momenta definition}) para o caso contínuo\footnote{Como todas as relações, parênteses de Poisson, etc., estão a tempo fixo $x^{0} = t$, só as coordenadas expaciais $x$ estarão explícitas.}
\begin{equation}
\pi^{\mu} : = \frac{\delta L(x)}{\delta\partial_{0}A_{\mu}(y)} =  F^{\mu 0}(x).
\label{momento canonico EM}
\end{equation} 
Naturalmente, $\pi^{0} = 0$ é um vínculo primário da teoria, devido a propriedade de antisimetria de $F^{\mu\nu}$, por outro lado, $\pi^{i} =  F^{i 0} = \partial_{0}A_{i} - \partial_{i}A_{0} = -\pi_{i}$\footnote{Aqui estamos usando a notação de índices covariantes e contra-variantes levando em consideração a métrica $\eta_{\mu\nu} = \textrm{diag}(1,-1,-1,-1)$. Portanto, $A^{\mu} = (A^{0}, \vec{A})$ e $A_{\mu} = \eta_{\mu\nu}A^{\nu} = (A^{0}, -\vec{A})$.}. Isso nos permite expressar as ``velocidades'' $\partial_{0}A_{i} = - \pi_{i} + \partial_{i}A_{0}$ que é necessário para obtermos a forma hamiltoniana. Note que, não podemos inverter para $\partial_{0}A_{0}$ o que caracteriza a teoria de Maxwell como uma teoria singular, det $W = 0$. Assim, a hamiltoniana canônica é escrita como,
\begin{eqnarray}
H_{C} & = & \int d^{3}x \biggr(\pi^{\mu}\partial_{0}A_{\mu} + \frac{1}{4}F_{\mu\nu}F^{\mu\nu}\biggl)\nonumber\\[0.1cm]
      & = & \int d^{3}x\,\pi^{i}(-\pi_{i} + \partial_{i}A_{0}) + \frac{1}{2}F_{i0}F^{i0} + \frac{1}{4}F_{ij}F^{ij}\nonumber\\[0.1cm]
      & = & \int d^{3}x \,\frac{1}{2}(\vec{E}^{2} + \vec{B}^{2}) + \vec{E}^{i}\partial_{i}A_{0} \label{hamiltoniana EM}\\[0.1cm]
      & = & \int d^{3}x \,\frac{1}{2}(\vec{E}^{2} + \vec{B}^{2}) - A_{0}\vec{\nabla} \cdot \vec{E},\nonumber
\end{eqnarray} 
onde fizemos uma integração por partes na última equação. O primeiro termo está claramente associado a densidade de energia armazenada nos campos $\vec{E}$ e $\vec{B}$,  lembrando que $\pi^{i} = F^{i0} = \vec{E}^{i}$ e que $\vec{B}^{i} = \frac{1}{2}\varepsilon^{ijk}F_{jk}$. Portanto, as variáveis canônicas são $A_{i}(x)$ e $E^{i}(x)$, um para para cada ponto do espaço. Os parênteses de Poissson dessas variáveis canônicas são dados por
\begin{equation*}
\bigr\{A_{\mu}(x),\pi^{\nu}(y)\bigl\}_{x^{0} = y^{0}} = \delta^{\mu}_{\nu}\delta^{3}(x - y).
\end{equation*}
Além disso, o colchête de Poisson associado a duas funções $F$ e $G$ no espaço de fase $(A_{\mu}, \pi^{\mu})$ assume a forma
\begin{equation*}
\bigl\{F(x), G(y)\bigr\}_{x^{0} = y^{0}} = \biggl(\frac{\delta F}{\delta A_{\mu}}\frac{\delta G}{\delta \pi^{\mu}} - \frac{\delta G}{\delta A_{\mu}}\frac{\delta F}{\delta \pi^{\mu}}\biggr)\delta^{3}(x - y).
\end{equation*}

Utilizando-se das equações de Hamilton podemos encontrar as equações de movimento para $\pi^{0}$, notando a presença do termo $A_{0}$ em \ref{hamiltoniana EM}, obtemos

\begin{equation}
\dot{\pi}^{0} = \bigr\{\pi^{0},H_{C}\bigl\} = \vec{\nabla}\cdot\vec{E} := \phi_{2}\label{Gauss}
\end{equation}
No entanto, o vínculo primário $\pi^{0}$ deve ser preservado em todo instante. Consequentemente, sua evolução temporal deverá anular-se também para consistência da teoria. Donde resulta o vínculo secundário
\begin{equation}
\phi_{2}:=  \vec{\nabla}\cdot\vec{E} \approx 0,
\end{equation} 
  que é a lei de Gauss no vácuo. Esse vínculo  implica que a variável canônica $E^{i}$ não pode assumir quaisquer valores, mas apenas os valores cuja sua divergência seja nula a todo instante.  
Como a condição de estabilidade (\ref{Gauss}) produz um vínculo secundário, o algorítmo de Dirac-Bergmann exige que analisemos a consistência desse novo vínculo, ou seja, devemos calcular a evolução dinâmica desse vínculo e exigir que seja nulo. O algorítimo continua até que tenhamos obtido uma evolução identicamente zero caracterizando a condição de consistência. No caso eletromagnético o processo termina já na segunda etapa da análise da estabilidade, pois claramente vemos que $\dot{\phi_{2}} \approx 0$. Naturalmente, sabendo que os vínculos de primeira-classe são geradores de simetrias, como discutido anteriormente, iremos fazer uso dessa estrutura. No entanto, devemos ser caltelosos pois como estamos trabalhando com um teoria de campos, as expressões são locais ou mehor são funções dos pontos o que nos leva a ideia de distribuição. A fim de evitarmos complicações como derivações de distribuições como podem ser vistas em (\ref{propr fundamentais no EM}) é bom introduzir a ideia de vínculos ``ponderados'' (\textit{smeared}). Por exemplo,
\begin{equation}
\mathcal{G}_{1}(\eta) := \int d^{3}x \eta(x)\pi^{0}, \quad \mathcal{G}_{2}(\alpha) := \int d^{3}x \alpha(x)\partial_{i}E^{i}
\label{smeared}
\end{equation}
onde $\eta$ e $\alpha$ é uma função suave e arbitrária das coordenadas do espaço $x$, de maneira que a integral seja bem definida. O vínculo ponderado é agora um número o que torna os cálculos bem mais diretos e melhor definidos. Considerando o parêntese de Poisson dos vínculos (\ref{smeared}) com a hamiltoniana, chega-se, com um pouco de álgebra, à conclusão de que são nulos, o que nos mostra que as órbitas geradas pelos vínculos deixa a teoria invariante. É de suma importância explorarmos o que essas órbitas tem as nos dizer. Para tanto, temos

\begin{equation*}
\bigr\{\mathcal{G}_{1}(\eta), \pi^{\mu}\bigl\} = \biggr\{\int d^{3}y\,\eta(y) \pi^{0}, \pi^{\mu}(x)\biggl\} = 0, 
\end{equation*}

\begin{equation*}
\bigr\{\mathcal{G}_{1}(\eta), A_{0}\bigl\} =  \biggr\{\int d^{3}y\,\eta(y) \pi^{0}, A_{0}(x)\biggl\} = -\eta,
\end{equation*} 
\begin{eqnarray}
&&\bigr\{\mathcal{G}_{2}(\alpha), E^{i}\bigl\}  =  -\biggr\{\int d^{3}y\partial_{i}\alpha E^{i},E^{i}\biggl\} = 0,\nonumber \\[0.1cm]
\bigr\{\mathcal{G}_{2}(\alpha), A_{j}\bigl\} & =  &  -\biggr\{\int d^{3}x\partial_{i}\alpha(y) E^{i},A_{j}(y)\biggl\} = \int d^{3}x \partial_{i}\alpha \bigl\{E^{i}(x), A_{j}(y)\bigr\}\nonumber\\[0.1cm]
      & = & \int d^{3}x \partial_{i}\alpha \delta^{i}_{j}\delta^{3}(x-y) = \partial_{i}\alpha,\nonumber
\end{eqnarray}
assim, ao longo das órbitas geradas por $\mathcal{G}_{1}$ e $\mathcal{G}_{2}$ os campos eletromagnéticos permanecem invariantes enquanto o potencial vetor $\vec{A}$ e $A_{0}$   mudam por o gradiente de uma função e por uma função arbitrária $\eta$, respectivamente. Sabemos da teoria de Maxwell que o potencial vetor é definido a menos do gradiente de uma função, caracterizando a invariância de gauge da eletrodinâmica. Vemos que essa mesma estrutura emerge na análise canônica como uma consequência da presença dos vínculos $\mathcal{G}_{1}$ e $\mathcal{G}_{2}$. Nesse contexto, a lei de Gauss é chamada de gerador das transformações de gauge espaciais. Verifica-se de maneira imediata que 
\begin{equation}
\bigr\{\mathcal{G}_{1}, \mathcal{G}_{2}\bigl\} \approx 0,
\end{equation}
 implicando que a teoria de Maxwell apresenta, nesse contexto, apenas vínculos de primeira-classe.

  Os graus de liberdade da teoria são obtidos segundo (\ref{degrees of freedom count}). Temos que o espaço de fase da eletrodinâmica possui dimensão $D = 8$, temos dois vínculos de primeira-classe o que implica que o espaço de fase reduzido, ou seja, o espaço de fase físico, onde se processam os verdadeiros graus de liberdade, será
\begin{equation*} 
  \mathcal{N} = \frac{1}{2}(8 - 2\times 2) = 2.
\end{equation*}  
   Deve ficar claro que o tópico de estudos sobre a análise de sistemas vinculados é muito mais profundo do que está sendo abordado nesssa seção de maneira pragmática e o leitor é convidado a consultar os textos clássicos para maiores detalhes \cite{Hen1, Dir}. 
\section{Análise Canônica do Modelo de Chamseddine}
A fim de estudarmos o conteúdo dinâmico da teoria de Chamseddine  descrita na Seção $2.2$, bem como identificar os graus de liberdade, iremos aplicar a análise hamiltoniana \textit{à la} Dirac \cite{Andreas,Gitman,Date, Hans, Hen1, Dir} que nos possibilitará identificar os vínculos da teoria e suas clasificações. Para realizarmos uma análise hamiltoniana iremos assumir que a variedade espaço-temporal $\mathcal{M}_{4}$ admite topologia $\mathbb{R}\times \Sigma$, onde $\Sigma$ é uma superfície $3$-dimensional tipo-espaço, e iremos decompor a conexão $1$-forma em
\begin{equation}
\mathbb{A}^{AB}_{\mu}dx^{\mu} = \mathbb{A}^{AB}_{t}dt + \mathbb{A}^{AB}_{a}dx^{a}, \quad (a = 1,2,3).
\end{equation}
Consequentemente, a ação decompõe-se da seguinte forma
\begin{equation}
S = \int_{\mathbb{R}}\int_{\Sigma}dtd^{3}x\biggr(l^{a}_{AB}(\Phi, \mathbb{A})\dot{\mathbb{A}}^{AB}_{a} + \mathbb{A}_{t}^{AB}K_{AB}(\Phi,\mathbb{A})\biggl)
\label{splitting da ação}
\end{equation}
onde defimos as quantidades 
\begin{equation}
l^{a}_{AB} := \varepsilon_{ABCDE}\varepsilon^{abc}\mathbb{F}^{CD}_{bc}\Phi^{E}, \quad K_{AB} := \varepsilon_{ABCDE}\varepsilon^{abc}\mathbb{F}^{CD}_{bc}\mathbb{D}_{a}\Phi^{E}.
\end{equation}
Podemos notar ainda que, via o uso da identidade de Bianchi
\begin{equation*}
\mathbb{D}_{[\nu}\mathbb{F}_{\rho\sigma]} = \varepsilon^{\mu\nu\rho\sigma}\mathbb{D}_{\nu}\mathbb{F}_{\rho\sigma} = 0,
\end{equation*}
 pode-se reescrever o termo $K_{AB}$ como uma derivada total de $l^{a}_{AB}$, isto é,
\begin{equation*}
K_{AB} = \mathbb{D}_{a}l^{a}_{AB}.
\end{equation*}

A ação (\ref{splitting da ação}) é de primeira ordem nas derivadas temporais consequentemente nos leva a uma matriz hessiana de determinante nulo na passagem da lagrangiana para hamiltoniana, ou seja, uma teoria singular como vimos na seção anterior. Com efeito, não podemos inverter todos os momentos canonicamente conjugados (\ref{momenta definition}) aos campos dinâmicos, o que caracteriza a presença de vínculos. Em outras palavras, os momentos canônicos não são todos independentes das velocidades. Nesse caso, existem certas relações (vínculos) que conectam as variáveis de momento com os campos dinâmicos e, naturalmente, produz uma dinâmica que não é univocamente determinada pelas equações de movimento. Para tanto, devemos fazer uso do  mecanismo construído por Dirac e Bergmann\footnote{Para um tratamento mais aprofundado do tema uma outra ótima referência é o livro de Kurt Sundermeyer: \textit{Constrained Dynamics}.} para tratar nosso modelo. Nessa seção  de análise de vínculos do modelo de Chamseddine, estaremos seguindo de perto a discussão desenvolvida na seção anterior.

Como a ação é linear na derivada temporal de $\mathbb{A}^{AB}_{a}$ e não contém derivada temporal do campo escalar $\Phi^{A}$, calculando os momentos conjugados, obtemos os vínculos primários\footnote{Usaremos $\frac{\delta T^{AB}}{\delta T^{CD}} = \delta^{AB}_{CD} = (\delta_{C}^{A}\delta_{D}^{B}-\delta_{D}^{A}\delta_{C}^{B})$, para $T^{AB} = - T^{BA}$.}: 

\begin{eqnarray}
 \phi^{a}_{AB} & := & \Pi^{a}_{AB} - l^{a}_{AB} \approx 0,  \nonumber\\[0.1cm]
\phi_{A} & := & \Pi^{\Phi}_{A}  \approx 0, \label{vínculos primários}\\[0.1cm]
\phi^{t}_{AB} & := & \Pi^{t}_{AB}  \approx  0. \nonumber \\[0.1cm]
\mathit{l}^{a}_{BC} &:= &-\frac{1}{4\kappa} \varepsilon_{ABCDE}\varepsilon^{abc}\Phi^A\mathbb{F}^{DE}_{bc},\nonumber
\end{eqnarray}
onde $\Pi^{a}_{AB}$, $\Pi^{\Phi}_{A}$ e $\Pi^{t}_{AB}$ são os momentos canonicamente conjugados a $\mathbb{A}_{a}^{AB}$, $\Phi^{A}$ e $\mathbb{A}_{t}^{AB}$, respectivamente. 
\begin{eqnarray*}
&&\Pi^{t}_{BC}:=\frac{\delta L}{\delta\partial_t\mathbb{A}_t^{BC}}=0,\\
&&\Pi^{a}_{BC}:=\frac{\delta L}{\delta\partial_t\mathbb{A}_a^{BC}}=\frac{1}{4\kappa} \varepsilon_{ABCDE}\varepsilon^{abc}\Phi^A\mathbb{F}^{DE}_{bc},\\
&&\Pi_{A}:=\frac{\delta L}{\delta\partial_t\Phi^A}=0,\\
\end{eqnarray*}
que são os vínculos primários do modelo. Os parênteses de Poisson canônicos são:
 \begin{eqnarray*}
 \lbrace \mathbb{A}_{t}^{AB}(x), \Pi^{t}_{CD}(y)\rbrace&=&\delta_{CD}^{AB}\delta^3(x-y),\\
  \lbrace \mathbb{A}_{a}^{AB}(x), \Pi^{b}_{CD}(y)\rbrace&=&\delta_{a}^{b}\delta_{CD}^{AB}\delta^3(x-y),\\
  \lbrace \Phi^{A}(x), \Pi_{B}(y)\rbrace & = &\delta^{A}_{B}\delta^3(x-y),
 \end{eqnarray*}
 onde $\delta^{AB}_{CD} = (\delta_{C}^{A}\delta_{D}^{B}-\delta_{D}^{A}\delta_{C}^{B})$.
 A hamiltoniana canônica é
\begin{eqnarray*}
H_{C} & = &\int_\Sigma d^3x(p_{i}\dot{q}^{i} -\mathcal{L})\\
&=&\int_\Sigma d^3x(\frac{1}{2}\Pi^{a}_{AB}\dot{\mathbb{A}}_{a}^{AB}-\mathcal{L})=\frac{1}{2}\int_\Sigma d^3x\mathbb{A}_{t}^{BC}\mathbb{D}_a\mathit{l}^{a}_{BC}.\\
\end{eqnarray*}
 Consequentemente, obtemos a hamiltoniana total acrescentando-se a $H_{C}$ os vínculos primários 
 \begin{eqnarray*}
H_{T} & = & H_{C} + \int_\Sigma \lambda^i\phi_i\\
& = &\int_\Sigma d^3x(\frac{1}{2}\mathbb{A}_{t}^{BC}\mathbb{D}_a\mathit{l}^{a}_{BC}+\frac{1}{2}\lambda_{t}^{BC}\phi^{t}_{BC}+\frac{1}{2}\lambda_{a}^{BC}\phi_{BC}^{a}+\lambda^{A}\phi_{A}).
\end{eqnarray*}
Da análise de estabilidade de $\phi^{t}_{BC}$ temos
\begin{eqnarray*}
\dot{\phi}^{t}_{BC}(x)& = &\lbrace \phi^{t}_{BC}(x), H_{T}\rbrace\\
& = &-\mathbb{D}_{a}\mathit{l}^{a}_{BC}=:-K_{BC}\approx 0, 
\end{eqnarray*}
 o que resulta no vínculo secundário $K_{BC}$. É interessante substituir o vínculo  $K_{BC}$ por
\begin{eqnarray*}
&& K_{BC}\to G_{BC} = \mathbb{D}_{a}\Pi^{a}_{BC} + (\Pi_{B}\Phi_{C}-\Pi_{C}\Phi_{B}), 
\end{eqnarray*}
Essa substituição é possível pois as superfícies geradas no espaço de fase por ambos são equivalentes. O algoritmo de Dirac-Bergmann, posto aqui, nos leva a $\lbrace G_{BC}, H_{T}\rbrace \approx 0$ como é fácil de verificar.  Assim podemos ver que  a hamiltoniana total $H_{T}$ é completamente vinculada, isto é, ela é composta exclusivamente de vínculos e pode ser escrita como
\begin{eqnarray}
\label{h1}
H_{T} & = &\int_\Sigma d^3x(\frac{1}{2}\mathbb{A}_{t}^{BC}G_{BC} + \frac{1}{2}\lambda_{t}^{BC}\phi^{t}_{BC} + \frac{1}{2}\lambda_{a}^{BC}\phi_{BC}^{a}+\lambda^{A}\phi_{A}).
\end{eqnarray}
onde
\begin{eqnarray}
\label{constraints1}
\phi^{t}_{BC} & := &\Pi^{t}_{BC}\approx 0,\nonumber\\
\phi_{BC}^{a} & := &\Pi^{a}_{BC} - \mathit{l}^{a}_{BC}\approx 0,\\
\phi_{A} & := &\Pi_{A} \approx 0,\nonumber\\
G_{BC} & := & \mathbb{D}_a\Pi^{a}_{BC}+(\Pi_{B}\Phi_{C}-\Pi_{C}\Phi_{B}).\nonumber
\end{eqnarray}
Escrevendo-os na forma ``ponderada'' temos
\begin{eqnarray}
\label{constraints2}
&&\phi_{1}(\epsilon):= \frac{1}{2}\int d^3z\,\epsilon^{AB}(z)\phi^{t}_{AB}(z),\hspace{30pt} \phi_{2}(\eta) := \frac{1}{2}\int d^3z\,\eta_{a}^{AB}(z)\phi^{a}_{AB}(z),\nonumber\\
&&  \phi_{3}(\zeta) := \int d^3z\,\zeta^{A}(z)\phi_{A}(z), \hspace{55pt} \phi_{4}(\xi):=\frac{1}{2}\int d^3z\,\xi^{AB}(z)G_{AB}(z).\nonumber
\end{eqnarray}
Portanto, a hamiltoniana total assume a forma
\begin{eqnarray}
\label{h2}
H_{T} & = &\phi_{4}(\xi) + \phi_{1}(\epsilon) + \phi_{2}(\eta) + \phi_{3}(\zeta).
\end{eqnarray}
A redefinição do vínculo $K_{BC}$ por $G_{BC}$ foi conveniente pois estes são os \emph{geradores das transformações de gauge}, como podemos ver abaixo,

\begin{eqnarray}
\delta \mathbb{A}^{AB}_{a}  & = & \biggl\{\mathbb{A}^{AB}_{a}, \int_{\Sigma}\epsilon^{CD}G_{CD}\biggr\} = -\mathbb{D}_{a}\epsilon^{AB} \nonumber\\[0.1cm]
\delta \Phi^{A} & = & \biggl\{\Phi^{A}, \int_{\Sigma}\epsilon^{BC}G_{BC}\biggr\} = \epsilon^{A}\,_{B}\Phi^{B} 
\end{eqnarray}
% SÃO AS MESMAS EQUAÇÕES SE PRECISAR EU COLOCO
%\begin{eqnarray*}
%\lbrace \phi_{4}(\epsilon^{..}),\mathbb{A}_{a}^{AB}(x)\rbrace & = &\mathbb{D}_{a}%\epsilon^{AB},\\
%\lbrace \phi_{4}(\epsilon^{..}),\Pi^{a}_{AB}(x)\rbrace & = &\lbrack\Pi^{a},\epsilon^{..}\rbrack_{AB},\\
%\lbrace \phi_{4}(\epsilon^{..}),\Phi^{A}(x)\rbrace & = &-\epsilon^{A}\,_{B}\Phi^{B}=-\lbrack \epsilon^{..},\Phi\rbrack^{A},\\
%\lbrace \phi_{4}(\epsilon^{..}),\Pi_{A}(x)\rbrace & = &-\epsilon_{A}\,^{B}\Pi_{B}=-\lbrack \epsilon^{..},\Pi\rbrack_{A},
%\end{eqnarray*}

Segue das definições acima que $G_{AB}$ é um vínculo de primeira-classe, ou seja, os parênteses de Poisson com todos os demais vínculos ou são fracamente zero ou uma combinação linear de outros vínculos. A álgebra completa dos colchêtes de Poisson dos demais vínculos é\footnote{Definindo: $\lbrack \epsilon^{..},\xi^{.}\rbrack^{A} = \epsilon^{A}\!_{B}\xi^{B}$}
\begin{eqnarray*}
\lbrace \phi_{1}(\epsilon),\phi_{\alpha}(\xi)\rbrace & \approx &0, \qquad(\alpha=1, 2, 3, 4)\\
\lbrace \phi_{4}(\epsilon),\phi_{4}(\xi)\rbrace & \approx &\frac{1}{2}\int d^{3}z\,\lbrack \epsilon,\xi\rbrack^{AB} \phi_{4AB}=\phi_{4}(\lbrack\epsilon,\xi\rbrack),\\
\lbrace \phi_{4}(\epsilon),\phi_{3}(\xi)\rbrace & \approx &\int d^{3}z\, \epsilon^B\,_A\xi^A\phi_{3B} = \int d^3\,z \lbrack\epsilon,\xi\rbrack^B\phi_{3B} = \phi_{3}(\lbrack \epsilon,\xi\rbrack),\\
\lbrace \phi_{4}(\epsilon),\phi_{2}(\xi)\rbrace & \approx &\frac{1}{2}\int d^3z\, \biggl(\lbrace \phi_{4}(\epsilon),\Phi^{a}_{BC}\rbrace \xi_{a}^{BC}-\frac{1}{4\kappa}\varepsilon_{ABCDE}\varepsilon^{abc}\lbrace \phi_{4}(\epsilon),\Phi^{A}\mathbb{F}_{bc}^{DE}\rbrace \xi_{a}^{BC}\biggr)\\
& \approx &\frac{1}{2}\int d^3z\, \biggl(\lbrack \epsilon,\xi_{a} \rbrack^{BC}\Pi^{a}_{BC}-\frac{1}{4\kappa}\varepsilon_{ABCDE}\varepsilon^{abc}(-\lbrack\epsilon,\Phi\rbrack^{A} \xi_{a}^{BC}\mathbb{F}_{bc}^{DE}+\lbrack \epsilon,\Phi\rbrack^{A}\xi_{a}^{BC} \mathbb{F}_{bc}^{DE}\\&&+\Phi^{A}\lbrack \epsilon,\xi_{a} \rbrack^{BC} \mathbb{F}_{bc}^{DE})\biggr) = \frac{1}{2}\int d^3z\, \lbrack \epsilon,\xi_{a} \rbrack^{BC}(\Pi^{a}_{BC} + \mathit{l}^{a}_{BC})\\
& \approx &\phi_2(\lbrack \epsilon,\xi \rbrack),\\
\lbrace \phi_{3}(\epsilon),\phi_{3}(\xi)\rbrace & \approx & 0,\\
\lbrace \phi_{3}(\epsilon),\phi_{2}(\xi)\rbrace & \approx &\frac{1}{8\kappa}\int d^3 z\, \varepsilon_{ABCDE}\varepsilon^{abc}\epsilon^{A}\xi_{a}^{BC}\mathbb{F}_{bc}^{DE},\\
\lbrace \phi_{2}(\epsilon),\phi_{2}(\xi)\rbrace & \approx & -\frac{1}{4\kappa}\int d^3 z\, \varepsilon_{ABCDE}\varepsilon^{abc}\epsilon_a^{AB}\xi_{b}^{CD}\mathbb{D}_{c}\Phi^{E}.  
\end{eqnarray*}
\subsection{Estabilidade e Graus de Liberdade}
 
 Vimos que os vínculos $G_{AB}$ são de primeira-classe. A fim de investigarmos a natureza dos vínculos $\phi^{a}_{AB}$ e $\phi_{A}$, devemos considerar a matriz $\Omega_{\alpha\beta}(x,y)$ formada pelos parêntes de Poisson desses vínculos, onde $\alpha, \beta$ denotam os índices $(A, B, a)$, que surgem na análise da estabilidade. Devemos ter cuidado pois $\Omega_{\alpha\beta}$ é, de fato uma matriz infinita, de modo que devemos analisá-la localmente. Como a hamiltoniana total é uma soma de vínculos
\begin{equation*}
H_{T} = \hat{\lambda}^{\alpha}\phi_{\alpha}, \quad (\alpha = (A,B,a), (AB), \textrm{etc}),
\end{equation*}
onde subentende-se a convenção de De Witt, ou seja, $X^{\alpha}Y_{\alpha} = \sum_{\alpha}\int d^{3}x X^{\alpha}(\vec{x})Y_{\alpha}(\vec{x})$.
A evolução dinâmica dos vínculos
\begin{eqnarray}
\dot{\phi}_{\alpha} & = & \lbrace\phi_{\alpha},H_{T}\rbrace\nonumber\\[0.1cm]
& = & \Omega_{\alpha\beta}\hat{\lambda}^{\beta} \approx 0, \quad \textrm{onde} \quad \Omega_{\alpha\beta} = \lbrace\phi_{\alpha},\phi_{\beta}\rbrace\label{produto}
\end{eqnarray}
Sendo $\Omega(x,y)$ uma matriz infinita, para podermos extrair informação do seu rank devemos definí-la de modo mais preciso, isto é, 
\begin{equation*}
\Omega_{\alpha\beta}(x,y) = \omega_{\alpha\beta}\delta^{3}(x - y).
\end{equation*}
Nesse caso, passamos a informação da estrutura matricial para a matriz finita $\omega_{\alpha\beta}$ e assim podemos definir o rank de $\Omega$ localmente. Seja $r = \textrm{rank}(\omega)$, podemos dividir a matriz $\omega$ e o vetor $\hat{\lambda}$  da seguinte forma

$$ \hat{\lambda} = \left[\begin{array}{c}
\mu_{i}\\
\nu_{a}\\
\end{array}
\right] $$

 e $$ \omega = \left[\begin{array}{c|c} P_{ij} & Q_{ia} \\ \hline
                                                       -Q^{t} & R_{ab} \\
                              \end{array}               
                             \right] $$ 
onde $i = 1,\,...,\,r$, $a = r + 1,\, ...,\, A$ e, onde $\textrm{dim}P = \textrm{rank}(\omega)$, $P$ sendo invertível, ou seja, det$P$ $\neq$ $0$. Consequentemente, a dimensão de $P$ nos informará sobre os vínculos de segunda-classe, isto é, $\phi_{i} (i = 1,..., r)$ são os vínculos de segunda-classe. Dessa forma, (\ref{produto}) produz as seguintes equações

\begin{equation*}
P\mu + Q\nu = 0; \qquad -Q^{t}\mu + R\nu = 0 \label{eq:auxiliar}
\end{equation*}
Resolvendo em componentes para $\mu$ obtemos
\begin{equation}
\mu_{i} = - \bigl( P^{-1}Q \bigr)_{i}\,^{a}\nu_{a}; \qquad \bigl(\underbrace{Q^{t}P^{-1}Q + R}_{\equiv 0}\bigr)\nu = 0,
\end{equation}
 onde $Q^{t}P^{-1}Q + R \equiv 0$ segue, naturalmente, pois como os $\mu$'s são arbitrários e sendo escritos como função dos $\nu$'s. Com efeito, temos que os $\nu_{a}$ são também completamente arbitrários e, de fato, serão associados a simetria de gauge assim como vimos na descrição do algoritmo de Dirac-Bergmann. Dessa forma, a hamiltoniana total terá funções completamente arbitrárias e  assume a forma  

\begin{align*}
H_{T} & = \nu_{a}\phi^{a} - \bigl(P^{-1}Q\bigr)_{i}\,^{a}\nu_{a}\phi^{i}\\[0.1cm]
                & = \nu_{a}\hat{\phi}^{a}\,; \qquad \hat{\phi}^{a} := \phi^{a} - \phi^{i}\bigl(P^{-1}Q\bigr)_{i}\,^{a},
\end{align*}
onde a redefinição $\phi \to \hat{\phi}$ é, de fato, uma mudança de base. Os $\hat{\phi}$ são os vínculos de primeira-classe, isto é,
\begin{equation}
\left\lbrace \hat{\phi}^{a},\hat{\phi}^{b} \right\rbrace \approx 0, \quad \left\lbrace \hat{\phi}^{a},\phi^{i} \right\rbrace \approx 0 \label{redefinição dos vínculos}
\end{equation}
\textit{Demonstração}:

\begin{eqnarray}
\left\lbrace \hat{\phi}^{a},\hat{\phi}^{b} \right\rbrace & = & \left\lbrace\phi^{a} - \phi^{i}\bigl(P^{-1}Q\bigr)_{i}\,^{a},\phi^{b} - \phi^{j}\bigl(P^{-1}Q\bigr)_{j}\,^{b}\right\rbrace \nonumber\\
                                                         & = & \left\lbrace\phi^{a},\phi^{b}\right\rbrace - \left\lbrace\phi^{a},\phi^{j}\bigl(P^{-1}Q\bigr)_{j}\,^{b}\right\rbrace - \left\lbrace\phi^{i}\bigl(P^{-1}Q\bigr)_{i}\,^{a},\phi^{b}\right\rbrace + \nonumber \\
                                                         &  & +  \left\lbrace\phi^{i}\bigl(P^{-1}Q\bigr)_{i}\,^{a},\phi^{j}\bigl(P^{-1}Q\bigr)_{j}\,^{b}\right\rbrace \nonumber \\
                                                         & = & \left\lbrace\phi^{a},\phi^{b}\right\rbrace - \left\lbrace\phi^{a},\phi^{j}\right\rbrace\bigl(P^{-1}Q\bigr)_{j}\,^{b} - \left\lbrace\phi^{i},\phi^{b}\right\rbrace\bigl(P^{-1}Q\bigr)_{i}\,^{a} + \nonumber \\
                                                         & & + \left\lbrace\phi^{i},\phi^{j}\right\rbrace\bigl(P^{-1}Q\bigr)_{i}\,^{a}\bigl(P^{-1}Q\bigr)_{j}\,^{b}.\nonumber \\
                                                         & = & R^{ab} - \bigl(Q^{t}\bigr)^{aj}\bigl(P^{-1}Q\bigr)_{j}\,^{b} - \underbrace{Q^{ib}}\bigl(P^{-1}Q\bigr)_{i}\,^{a} + P^{ij}\bigl(P^{-1}Q\bigr)_{i}\,^{a}\bigl(P^{-1}Q\bigr)_{j}\,^{b}. \nonumber \\
                                                         & = & R^{ab} - \bigl(Q^{t}P^{-1}Q\bigr)^{ab} - \bigl(Q^{t}P^{-1}Q\bigr)^{ba} - \bigl(\underbrace{PP^{-1}}_{\mathds{1}}Q\bigr)^{ja}\bigl(P^{-1}Q\bigr)_{j}\,^{b}.\nonumber\\ 
                                                         & = &  R^{ab} - \bigl(Q^{t}P^{-1}Q\bigr)^{ab} - \bigl(Q^{t}P^{-1}Q\bigr)^{ba} + \bigl(Q^{t}P^{-1}Q\bigr)^{ab}. \nonumber                                                         
\end{eqnarray}
Lembrando que,  $Q^{t}P^{-1}Q\,+\,R\,=\,0$, obtemos o primeiro resultado de (\ref{redefinição dos vínculos}). A demonstração do segundo resultado segue de maneira análoga
\begin{eqnarray*}
\left\lbrace \hat{\phi}^{a},\phi^{i} \right\rbrace & \approx & \left\lbrace \phi^{a} - \phi^{j}(P^{-1}Q)j\,^{a}, \phi^{i} \right\rbrace\\[0.1cm]
 & \approx & \Omega^{ai} - (P^{-1}Q)_{j}\,^{a}\Omega^{ji}\\[0.1cm]
 &\approx & -Q^{t}_{ai} + P_{i}\,^{j}(P^{-1}Q)_{j}\,^{a}\\[0.1cm]
 & \approx & 0 \qquad \Box.
\end{eqnarray*}

 A natureza dos vínculos $\phi_{\alpha}$ é determinada pela equação de autovalor (\ref{produto}). Como vimos acima, segue-se que algumas combinações lineares de $\phi_{\alpha}$ são de primeira-classe, portanto, a matriz não é invertível na superfície gerada pelos vínculos no espaço de fase. Entretanto, existe uma submatriz de dimensão igual ao rank de $\Omega$ invertível, logo nem todos os $\phi'$s são de primeira-classe. Em outras palavras, a determinação do rank de $\Omega$ nos dará a quantidade exata de vínculos de segunda-classe e consequentemente os de primeira. De fato, estamos interessados nas soluções não-triviais de (\ref{produto}), ou seja, será que existe soluções não-triviais para $\hat{\lambda}$ ? Se formos capazes de encontrar $\hat{\lambda}$ estaremos encontrando os multiplicadores de Lagrange de modo a satisfazer a condição de estabilidade dos vínculos e teremos uma ideia do número de vínculos de primeira-classe. Aplicando os resultados obtidos acima no caso do modelo de Chamseddine, podemos obter os autovetores nulos (modo-$0$) $\hat{\lambda}_{\beta}$ de $\Omega_{\alpha\beta}$ associados a equação de autovalor. Lembrando que
\begin{equation*}
\Omega^{ab}_{AB,\,CD} := \lbrace\phi^{a}_{AB},\phi^{b}_{CD}\rbrace = - 2\varepsilon_{ABCDE}\varepsilon^{abc}D_{c}\Phi^{E},
\end{equation*}
\begin{equation*}
\Omega^{a}_{AB,\,C} := \lbrace \phi^{a}_{AB},\phi_{C}\rbrace = \varepsilon_{ABCDE}\varepsilon^{abc}\mathbb{F}^{DE}_{bc}, 
\end{equation*}
e 
\begin{equation*}
\Omega_{AB} :=\ \lbrace\phi_{A},\phi_{B}\rbrace = 0.
\end{equation*}
Portanto, a equação de autovalor (\ref{produto}) assume a seguinte forma
$$
\Omega_{\alpha\beta}\hat{\lambda}_{\beta} = \left(\begin{array}{cc}
\Omega^{ab}_{AB,\,CD} & \Omega^{a}_{AB,\,F}\\
-\Omega^{b}_{CD,\,E} & 0 \end{array}\right) \left(\begin{array}{c} \hat{\lambda}^{CD}_{b}\\
\hat{\lambda}^{F}\end{array}\right) = 0.
$$
Naturalmente, temos o seguinte conjunto de equações
\begin{eqnarray}
-\Omega^{b}_{CD,\,E}\hat{\lambda}^{CD}_{(l)b} & =& \varepsilon_{ABCDE}\varepsilon^{abc}\mathbb{F}^{AB}_{bc}\hat{\lambda}^{CD}_{(l)a} \approx 0,\label{1vínculo} \\[0.2cm]
\Omega^{ab}_{AB,\,CD}\hat{\lambda}^{CD}_{(l)b} + \Omega^{a}_{AB,\,F}\hat{\lambda}^{F}_{(l)} & = & \varepsilon_{ABCDE}\varepsilon^{abc}\biggl(2 \hat{\lambda}^{CD}_{(l)b} D_{c}\Phi^{E} + \mathbb{F}^{CD}_{bc}\hat{\lambda}^{E}_{(l)}\biggr) \approx 0. \label{2vínculo}
\end{eqnarray}
Temos que (\ref{1vínculo}) adimite três soluções $\hat{\lambda}^{CD}_{(l)a} = \mathbb{F}^{CD}_{la}$, com $ l = 1, 2, 3$. Substituindo em (\ref{2vínculo}), obtemos que $\hat{\lambda}^{E}_{(l)} = D_{l}\Phi^{E}$, que é, de fato, uma consequência da identidade
\begin{equation}
\varepsilon_{ABCDE}\varepsilon^{abc}\biggl(2\mathbb{F}^{CD}_{lb}D_{c}\Phi^{E} + \mathbb{F}^{CD}_{bc}D_{l}\Phi^{E}\biggr) = \delta^{a}_{l}K_{AB} \approx 0.
\end{equation}

Consequentemente, a matriz $\Omega$, de dimensão $35$, possui no mínimo três autovetores nulos: $\hat{\lambda}_{(l)\beta} = \bigl(\mathbb{F}^{CD}_{la}, D_{l}\Phi^{E}\bigr)$ que correspondem a vínculos de primeira classe. Que são dados explicitamente por
\begin{equation}
H_{a} =\phi^{b}_{AB}\mathbb{F}^{AB}_{ab} + D_{a}\Phi^{A}\phi_{A}, \label{spatialdiff}
\end{equation}
os quais são responsáveis por gerar os difeomorfismos espaciais ``melhorados'' (\textit{improved}) que iremos definir a seguir. A ação do nosso modelo, além da invariância de gauge associada ao grupo de simetria local, admite invariância sob difeomorfismos em relação a um parâmetro $\eta^{\mu}$: $\delta_{\eta}\mathbb{A}^{AB}_{\mu} = \pounds_{\eta}\mathbb{A}^{AB}_{\mu}$ e $\delta_{\eta}\Phi^{A} = \pounds_{\eta}\Phi^{A}$, que pode ser representada por:
\begin{eqnarray*}
\delta_{\eta}\mathbb{A} & = & i_{\eta}d\mathbb{A} + d(i_{\eta}\mathbb{A})\\[0.1cm]
                             & = & i_{\eta}\mathbb{F} - i_{\eta}\mathbb{A}^{2} + d(i_{\eta}\mathbb{A})\\[0.1cm]
                            & = & i_{\eta}\mathbb{F} + \delta_{(i_{\eta}\mathbb{A})}\mathbb{A}.
\end{eqnarray*}
Portanto, define-se os difeomorfismos melhorados como diferindo dos difeomorfismo por uma transformação de gauge com parâmetro $\epsilon := i_{\eta}\mathbb{A}$, ou seja, 
\begin{equation}
\delta_{I}\mathbb{A}^{AB}_{\mu} = \eta^{\nu}\mathbb{F}^{AB}_{\nu\mu}, \quad \delta_{I}\Phi^{A} = \eta^{\mu}D_{\mu}\Phi^{A}. \label{improved diff}
\end{equation}
Verificamos que (\ref{spatialdiff}) é o gerador dos difeomorfismos espaciais melhorados
\begin{eqnarray}
\delta\mathbb{A}^{AB}_{a}  & := & \biggl\{\mathbb{A}^{AB}_{a},\int_{\Sigma}H_{b}\eta^{b}\biggr\} = \eta^{b}\mathbb{F}^{AB}_{ba}, \\[0.2cm]
\delta \Phi^{A} & := & \biggl\{\Phi^{A},\int_{\Sigma} H_{b}\eta^{b}\biggr\} = \eta^{b}D_{b}\Phi^{A}.
\end{eqnarray}
Assim, os vínculos de primeira-classe deveriam ser apenas $G_{AB}$ $=$ $0$ e $H_{a}$ $=$ $0$, se a teoria fosse \emph{genérica}\cite{Banados1} como veremos adiante, implicando que o vínculo associado ao gerador dos difeomorfismos temporais não é independente, ou seja, é consequência dos demais vínculos. Se fosse esse o caso, o vínculo escalar hamiltoniano seria evitado e uma quantização via laços seria mais simples e seguiria os passos dos trabalhos \cite{Piguet1,Piguet2, Piguet3}. Podemos verificar esse fato explicitamente escrevendo a atuação dos difeomorfismos temporais com a seguinte parametrização: $\eta$ $=$ $(\eta^{0},\,0,\,0,\,0)$
\begin{equation*}
\delta_{\eta}\mathbb{A}^{AB}_{\mu} = \eta^{0}\mathbb{F}^{AB}_{t\mu} + D_{\mu}(\eta^{0}\mathbb{A}^{AB}_{t})
\end{equation*}
e
\begin{equation}
\delta_{\eta}\Phi^{A} = \eta^{0}D_{t}\Phi^{A} - \eta^{0}\mathbb{A}^{A}_{t}\,_{B}\Phi^{B}.
\end{equation}
Os modos-$0$ de $\Omega$ : $\bigl(\mathbb{F}^{AB}_{ab},\,D_{a}\Phi^{A}\bigr)$ $:=$ $\hat{\lambda}_{\alpha}$, e as equações de movimento 
\begin{equation*}
\frac{1}{2}\Omega^{b}_{AB},^{c}_{CD}\mathbb{F}^{CD}_{tc} + \Omega^{b}_{AB},_{C}\Phi^{C} \stackrel{*}{=} 0,
\end{equation*}
\begin{equation}
\frac{1}{2}\Omega^{c}_{A},_{BC}\mathbb{F}^{BC}_{tc} \stackrel{*}{=}  0
\end{equation}
isto é, a validade é assegurada \textit{on-shell}. Portanto,
\begin{equation*}
\bigl(\mathbb{F}^{CD}_{tc}, D_{t}\Phi^{C}\bigr)= \textrm{modo}-0 := \hat{W}_{t} = \xi^{\alpha}\hat{V}_{\alpha}
\end{equation*}
$\Rightarrow$
\begin{equation}
\delta_{\eta}\bigl(\mathbb{A}^{AB}_{\mu},\Phi^{A}\bigr) = \eta^{0}\xi^{\alpha}\hat{V}_{\alpha} = \delta_{\xi}\bigl(\mathbb{A}^{AB}_{\mu},\Phi^{A}\bigr),
\end{equation}
que é um difeomorfismo melhorado com o parâmtero dado por $\xi$ $=$ $(0,\,\eta^{0}\xi^{a})$. Contudo, para nos certificarmos do número de vínculos de primeira-classe é fundamental que calculemos o rank da matriz $\Omega$.
\subsection{Teorias Genéricas e Contagem dos Graus de Liberdade}
Iremos agora analisar o conceito de teoria genérica em modelos tipo-topológicos e fazer a contagem dos graus de liberdade propagados pelo modelo. O conceito de generalidade (\emph{genericallity}) foi introduzido em \cite{Banados1} e é de grande relevância para a quantização na representação de laços de teorias (tipo-) topológicas.  Na série de trabalhos \cite{Piguet1, Piguet2, Piguet3}, estudou-se a quantização de laços em teorias topológicas em dimensões mais baixas. Portanto, no nosso caso, temos de analisar se o modelo de Chamseddine é compatível com a seguinte
\begin{definition}
 Dizemos que a teoria é genérica, quando ela satisfaz a seguinte condição: as transformações de gauge e os difeomorfismos espaciais formam uma base para todos os vínculos de primeira-classe da teoria.
 \end{definition}
 Essa é uma generalização da definição original dada em \cite{Banados1}. Além  disso, se a teoria é genérica, com $n$ sendo a dimensão da base dos vínculos de primeira-classe, então existem $n$ soluções não-triviais da equação (\ref{produto}), digamos $\hat{\lambda}_{l}$ $(l = 1,..., n)$. Isso implica que a matriz $\Omega$ de dimensão $N$ tenha rank máximo ($r_{\textrm{máx}}$) compatível com a definição de generalidade acima. Em outras palavras, o número de vínculos de segunda-classe (rank $(\omega) = r$) será 
 \begin{equation}
 \textrm{rank}(\omega) = N - n := r_{\textrm{máx}},\label{rank-maximo}
 \end{equation}
 que será nossa definição de rank máximo.
 
 O significado físico dessa condição algébrica que define o conceito de generalidade é bem objetivo. Essas condições, simplesmente expressam que as transformações de gauge (\ref{Ads-transf}) e os difeomorfismos melhorados (\ref{improved diff}) são independentes e que não existem outros vínculos de primeira-classe entre os $\phi$'s além de $G_{AB}$ e $H_{a}$. Essa condição é importante para a proposta de quantização na representação de laços pois garante que a invariância sob os difeomorfismos temporais não é independente  daquela sob os difeomorfismos espaciais e das demais simetrias de gauge. Em outras palavras, os difeomorfismos temporais podem ser escritos em termos dos difeomorfismos espaciais e das transformações de gauge internas,  o que facilita muito a aplicação das técnicas de quantização via laços pois nesse caso o vínculo hamiltoniano ou escalar é consequência dos demais. Com efeito, algumas perguntas surgem naturalmente:
\begin{itemize}
\item Será que o modelo de Chamseddine caracteriza-se pela condição de generalidade?\vspace{0.1cm}
\item Qual o rank($\omega$) e o número de graus de liberdade que são propagados pelo modelo?
\end{itemize}

Portanto, podemos calcular facilmente qual seria o rank máximo de $\omega$, lembrando que o rank dessa matriz nos fornece o número de vínculos de segunda-classe. Assim, como a matriz $\omega$ possui dimensão $35$ e, só temos $3$ vínculos de primeira classe associados a $H_{a}$, o rank máximo seria
\begin{equation*}
r_{\textrm{max}} = 35 - 3 \quad \Rightarrow \quad \textrm{rank}(\omega)\big|_{\textrm{max}} = 32.
\end{equation*}
Para que nosso modelo seja genérico deveríamos encontrar o rank$(\omega)$ $=$ $r_{\textrm{max}} = 32$. Contudo, calculamos o rank dessa matriz, via um programa feito no \emph{Mathematica}, donde obtemos 
\begin{equation*}
\textrm{rank}(\omega) = 26 < r_{\textrm{max}}.
\end{equation*}
Consequentemente, não temos um modelo genérico, o que significa que existem mais vínculos de primeira-classe que os $G_{AB}$ e $H_{a}$ que ainda não fomos capazes de identificar. Na verdade, é possível saber o número de vínculos de primeira-classe $\#$ F$_{0}$ que ainda estão faltando: o número total de vínculos é  $n = 55$, o rank  $r = 26$ e $G_{AB}$ e $H_{a}$ somam F $= 23$ logo:
\begin{equation*}
\# \textrm{F}_{0} =  (n - r) - F = 6,
\end{equation*}
ou seja, ainda existem $6$ vínculos de primeira-classe que precisam ser identificados e analisados.

 Portanto, precisamos diagonalizar a matriz $\omega$ em blocos de maneira a identificar uma submatriz, digamos $\omega_{1}$ de dimensão $26$ que comporte, de fato, todos os vínculos de segunda-classe. Em outras palavras, devemos ser capazes de separar por completo os vínculos de primeira e segunda-classe. Além disso, precisamos buscar por uma interpretação física do que poderia ser essas novas simetrias associadas aos vínculos de primeira-classe que ainda não identificamos. Essa é uma tarefa árdua que necessita ser feita a fim de se buscar uma possível quantização de laços do modelo. Embora no trabalho \cite{Mignemi} uma análise canônica tenha sido feita, levando-se em consideração o caso genérico, não foi calculado explicitamente o valor do rank que é capaz de demonstrar a não-generalidade do modelo de Chamseddine. 

Finalmente, a contagem dos graus de liberdade da teoria é feita ao computarmos a dimensão do espaço de fase reduzido. Para tanto, temos que o número total de vínculos do modelo é $ n = 55$, distribuídos como se segue:
\begin{eqnarray*}
\phi^{t}_{AB}  \rightarrow  10 & & \phi^{a}_{AB}  \rightarrow  30  \\[0.1cm]
\phi_{A}  \rightarrow  5 & & G_{AB}  \rightarrow  10. 
\end{eqnarray*} 
O espaço de fase $(\Gamma)$ é formado pelas variáveis dinâmicas $(\mathbb{A}^{AB}_{\mu}, \Phi^{A})$ e seus momentos canonicamente conjugados. Levando-se em conta as simetrias da conexão, verifica-se que a dimensão do espaço de fase é $ N = 90$. O número de vínculos de segunda-classe S $= r$, naturalmente, o número de vínculos de primeira-classe será $\textrm{F} = n - r$. Portanto, a dimensão do espaço de fase reduzido que nos dará informação sobre a contagem dos graus de liberdade é dado por (\ref{degrees of freedom count}) e assim:
\begin{eqnarray*}
\mathcal{N} & = & \frac{1}{2}\bigl(N - r - 2(n - r)\bigr)\\[0.1cm]
            & = & \frac{r}{2} - 10,
\end{eqnarray*}  
como $r = 26$, segue que $\mathcal{N} = 3$. Dessa forma, a análise canônica nos mostra que o número de graus de liberdade da teoria é $3$: isso corresponde aos dois graus de liberdade propagados pelo campo gravitacional mais um correspondendo ao campo escalar tipo diláton $\Phi^{4}$ - o parâmetro de acoplamento de Newton $G(x)$.

\chapter{Consequências Físicas do Modelo}
\pagestyle{fancy}
\lhead{\bfseries 3. Aproximação Linear e Soluções Cosmológicas}
\rhead{}
\section{Aproximação Linear e Ondas Gravitacionais}
A fim de investigar o limite newtoniano do modelo de Chamseddine ou ainda olhar para a presença de soluções tipo ondas gravitacionais no vácuo, realiza-se uma linearização do modelo, de maneira que dividiremos as variáveis dinâmicas do modelo entre as variáveis de fundo, que serão marcadas com um superescrito ``\,$\mathring{}$\,'', e as variáveis de perturbação. A tarefa consiste em fazer perturbações em torno desse fundo fixo. Em princípio, podemos considerar qualquer tipo de espaço curvo como sendo o fundo. Contudo, os campos deverão ser descritos da seguinte forma

\begin{equation}
\omega^{IJ} = \mathring{\omega}^{IJ} + a^{IJ},\quad e^{I} = \mathring{e}^{I} + h^{I},\quad G = \mathring{G} + \phi, 
\end{equation}
onde $\mathring{\omega}$, $\mathring{e}$ e $\mathring{G}$ são a conexão, o vierbein e o campo escalar (função de Newton), respectivamente, descrevendo o fundo.
A menos de termos de ordens mais altas que um na perturbação, a curvatura $R = d\omega + \omega^{2}$ e a torção $T^{I} = De^{I}$ assumem a forma
\begin{eqnarray*}
R^{IJ} =  \mathring{R}^{IJ} + \mathring{D}a^{IJ}, && T^{I} = \mathring{T} + \mathring{D}h + a^{I}\,_{J}\wedge \mathring{e}^{J},
\end{eqnarray*}
onde $\mathring{D}$ é a derivada covariante correspondendo a conexão de fundo $\mathring{\omega}$. Dessa forma, aplicando esse divisão nos campos, $\mathbb{F}^{IJ}$ torna-se 
\begin{equation}
\mathbb{F}^{IJ} = \underbrace{\mathring{R}^{IJ} - \frac{\Lambda}{3}\mathring{e}^{I}\wedge\mathring{e}^{J}}_{\textrm{fundo}} + \underbrace{\mathring{D}a^{IJ} -\frac{\Lambda}{3}\mathring{e}^{I}\wedge h^{J} -\frac{\Lambda}{3}h^{I}\wedge\mathring{e}^{J}}_{\textrm{perturbação}}
\end{equation}
Além disso, o fundo considerado aqui é um espaço-tempo de curvatura constante, de de Sitter, solução da equação
\begin{equation*}
\mathring{\mathbb{F}}^{IJ} = \mathring{R}^{IJ} - \frac{\Lambda}{3}\mathring{e}^{I}\wedge \mathring{e}^{J} = 0.
\end{equation*}
Então: $ \mathbb{F}^{IJ} = \mathcal{F}^{IJ}$ $=$ $\mathring{D} a^{IJ} -\dfrac{\Lambda}{3}\mathring{e}^{I}\wedge h^{J} -\dfrac{\Lambda}{3}h^{I}\wedge\mathring{e}^{J} (+ \textrm{ordens} > 1)$  e a ação do modelo lê-se
\begin{equation}
S_{\textrm{pert}} = \int\varepsilon_{I\,J\,K\,L}\,\mathcal{F}^{IJ}\wedge\mathcal{F}^{KL}\mathring{G} + S_{\textrm{matéria}}.\label{ação-linear}
\end{equation}
Iremos assumir que a ordem zero do parâmetro de Newton $\mathring{G}$ é uma constante diferente de zero denotada por $G_{0}$ e interpretada como a constante de Newton no momento presente, que tem uma dinâmica ao longo evolução cósmica. 

Fazendo-se a variação da ação (\ref{ação-linear}) a respeito de $h^{I}$, $a^{IJ}$ e $\phi$, obtemos as equações de movimento em primeira ordem
\begin{eqnarray}
\varepsilon_{I\,J\,K\,L}\mathring{e}^{K}\wedge\mathcal{F}^{IJ} & = &  - 8\pi G_{0}\mathcal{T}_{I}, \label{eq. de perturbação} \\[0.1cm] 
\varepsilon_{I\,J\,K\,L} G_{0}\mathring{e}^{K}\wedge\biggl(\mathring{D}h^{I} + a^{I}\,_{J}\mathring{e}^{J}\biggr) & = & 0, \nonumber \\[0.1cm]
 0 & = & 0 \nonumber
\end{eqnarray}
Note que a terceira equação é trivial na primeira ordem. A torção em ordem zero é nula: $\mathring{T} = 0$, pois nosso vácuo é o espaço-tempo de de Sitter cuja torção é, de fato, nula. A primeira equação de movimento nos mostra que a energia-momentum $3$-forma $\mathcal{T}_{I}$ deve ser considerada como primeira ordem. Além disso, a segunda equação nos mostra que a torção é nula também em primeira ordem:

\begin{equation*}
\mathring{D}h^{I} + a^{I}\,_{J}\wedge \mathring{e}^{J} = 0.
\end{equation*}
Consequentemente, ficamos com as primeiras equações de campo (\ref{eq. de perturbação}), onde a perturbação da conexão em primeira ordem $a^{IJ}$ deverá ser resolvida em termos das perturbações do vierbein $h^{I}_{\mu}$ e suas derivadas de acordo com a condição de torção nula. Dessa forma, vemos que o modelo de Chamseddine, no limite linear, tendo como fundo fixo o espaço-tempo de de Sitter, assume o mesmo conteúdo físico e dinâmico que as equações de Einstein da relatividade geral com constante cosmológica.

Podemos mostrar que o conteúdo da equação (\ref{ação-linear}) é o mesmo que o do regime linear da ação de Palatini com constante cosmológica. Em outras palavras, podemos partir da ação de Palatini com constante cosmológica, que abarca o conteúdo de Einstein mais constante cosmológica, fazer uma linearização e mostrar que obtemos as mesmas equações de movimento. Assim,
\begin{equation*}
S = \kappa\int \varepsilon_{IJKL}\bigl(e^{I}\wedge e^{J} \wedge R^{KL} - \frac{\Lambda}{6}  e^{I}\wedge e^{J} \wedge e^{K} \wedge e^{L}\bigr) + S_{\textrm{matéria}},
\end{equation*} 
introduzindo $e^{I} \,=\, \mathring{e}^{I} + h^{I}$ e $\omega^{IJ} = \mathring{\omega}^{IJ} + a^{IJ}$ na ação, obtemos
\begin{eqnarray*}
S & = & \kappa\int \varepsilon_{IJKL}\bigl(\mathring{e}^{I}\wedge \mathring{e}^{J}\wedge  \mathring{R}^{KL} - \frac{\Lambda}{6}  \mathring{e}^{I} \wedge \mathring{e}^{J} \wedge \mathring{e}^{K} \wedge \mathring{e}^{L}\bigr) +  \\[0.1cm]
 & & + \kappa\int \varepsilon_{IJKL} \bigl(\mathring{e}^{I} \wedge \mathring{e}^{J}\wedge\mathring{D}a^{KL} + \mathring{e}^{I}\wedge h^{J}\wedge \mathring{D}a^{KL} + 2h^{I}\wedge\mathring{e}^{J}\wedge\mathring{R}^{KL} + \\[0.1cm]
 && -\frac{2}{3}\Lambda\mathring{e}^{I}\wedge\mathring{e}^{J}\wedge h^{K}\wedge h^{L} - \frac{2}{3}\Lambda \mathring{e}^{I}\wedge \mathring{e}^{J}\wedge \mathring{e}^{K}\wedge h^ {L}\bigr) + S_{\textrm{matéria}}.
\end{eqnarray*}

Como estamos considerando perturbações em torno do fundo de Sitter ($\Lambda > 0$) a curvatura de Riemann $2$-forma é dada $\mathring{R}^{IJ} \,=\, \dfrac{\Lambda}{3}\mathring{e}^{I}\wedge\mathring{e}^{J}$  e a ação lê-se 
\begin{eqnarray}
S_{\textrm{pert}} & = & \kappa\int \varepsilon_{IJKL}\bigl(\mathring{e}^{I}\wedge\mathring{e}^{J}\wedge\mathring{D}a^{KL} + \mathring{e}^{I}\wedge h^{J}\wedge \mathring{D}a^{KL} + \nonumber \\[0.1cm] 
& & - \frac{2}{3}\Lambda \mathring{e}^{I}\wedge \mathring{e}^{J}\wedge h^{K}\wedge h^{L}\bigr) + S_{\textrm{matéria}}.
\label{Palatine-pertubação}
\end{eqnarray}

Através da variação da ação (\ref{Palatine-pertubação}) em relação ao vierbein perturbado  ($\delta S_{\textrm{pert}} / \delta h^{L}$) obtemos as seguintes equações de movimento.
\begin{equation}
\varepsilon_{IJKL}\mathring{e}^{J}\wedge\biggl( \mathring{D}a^{KL} - \frac{\Lambda}{3}\bigl(\mathring{e}^{K}\wedge h^{L} - \mathring{e}^{L}\wedge h^{K}\bigr)\biggr) = -8\pi G\mathcal{T}_{I},
\label{Eqm-Palatini}
\end{equation}
que são as mesmas equações de movimento obtidas no regime linear de Chamseddine descritos na equação (\ref{eq. de perturbação}). Uma primeira implicação é que a teoria admite um limite newtoniano assim como as equações de Einstein. Uma segunda implicação é a existência de solução de ondas gravitacionais mesmo com constante cosmológica. Com efeito, como no regime linear o modelo de Chamseddine coincide com Einstein com constante cosmológica, podemos nos basear nos resultados descritos de maneira extensa nos trabalhos feitos em \cite{Ash5, Bernabeu1, Bernabeu2}, onde está mostrado que, além da solução de curvatura constante, existe, de fato, propagação de soluções tipo ondas. Nos referimos aos artigo deles para mais detalhes e no Apêndice $3$ da tese.
 
\section{Soluções Cosmológicas} 

Existem dois aspectos na cosmologia hoje que a torna mais encantadora do que nunca. Primeiro, existe uma quantidade enorme de dados e observações acuradas \cite{Supernova, Planck} abrangendo um espectro das várias escalas de aplicabilidade das leis da Natureza. O outro aspecto da cosmologia moderna, que a distingue das tentativas anteriores de se compreender o universo, é o fato de termos desenvolvido um arcabouço teórico que concorda quantitativamente, de maneira espetacular, com os dados. Essas duas características formam a base para o entusiasmo despendido na Cosmologia Moderna. Temos uma teoria que faz predições e estas podem ser testadas.

A partir da formulação einsteiniana da gravitação, uma das consequências imediatas das lições conceituais trazidas é a de que o espaço-tempo, ao contrário do que se imaginavam, é uma entidade dinâmica e ``flexível''. O espaço-tempo ganha o \emph{status} de um campo dinâmico capaz de se curvar e de interagir na presença de matéria-energia. A gravitação seria interpretada como a capacidade ``elástica'' do espaço-tempo, e sua dinâmica regidas pelas equações de campo da relatividade geral.

Na década de $1920$, o matemático russo Alexander Friedmann (Para uma bela introdução aos desenvolvimentos de Friedmann consulte \cite{jean}) e o padre e astrônomo belga Lemaître analisaram, independentemente, as implicações das equações de Einstein ao universo como um todo e obtiveram algo que revolucionou e trouxe grande precisão a nova era das ciências cosmológicas. Assim como a gravitação da Terra opera sobre uma pedra, lançada para o alto que, ou estará sempre subindo, ou estará descendo, mas nunca parada (exceto no exato momento em que ela alcança sua altura máxima), Friedmann e Lemaître perceberam que o mesmo acontece com a matéria e radiação espalhadas por todo o universo. Com efeito, teríamos que o tecido do espaço ou tem de estar se expandindo ou se contraíndo mas não pode estar parado e inerte. Eles obtiveram equações que governariam a evolução do universo a partir das equações de Einstein que não se diferenciam muito do movimento de uma pedra sendo lançada para cima.

Apresento aqui uma situação física bem heurística e ilustrativa \cite{Rov2} que nos leva a ter uma intuição física por trás da descrição da cosmlógica de Friedmann. O modelo é bem simples e realístico. No contexto da física newtoniana, consideremos um universo formado por uma distribuição esférica de galáxias. Assumindo que essas galáxias distribuem-se de maneira uniforme no espaço e mantêm-se  uniformimente distribuídas no decorrer do tempo. Seja $\rho(t)$ a densidade, tempo-dependente, de galáxias no universo e que elas se atraem gravitacionalmente. Seja $O$ o centro dessa distribuição e $O'$ uma galáxia, digamos a nossa, a uma distância $r(t)$ do centro. Como já é bem conhecido, pelo teorema das cascas de Newton e a lei de Gauss, a força gravitacional em $O'$ devido as galáxias fora da esfera de raio $r(t)$ é nula. A força gravitacional no interior dessa superfície esférica, segundo o teorema, é a mesma que se considerássemos toda a massa concentrada no centro $O$. Com efeito, a força gravitacional percebida em $O'$ é

\begin{equation}
\frac{d^{2}r(t)}{dt^{2}} = -\frac{4\pi G}{3} r(t) \rho(t). \label{friedmann-newton}
\end{equation} 
Considerando que a densidade permanece espacialmente constante, ou seja, toda a variação temporal é incorporada no nosso fator de escala $r(t)$, isto é, ela se escala uniformemente com $r^{-3}$. Assim, $\rho(t) = \rho_{0}r^{-3}(t)$, onde $\rho_{0}$ é uma constante associada a densidade quando $r(t) = 1$, ou seja, o fator de escala $r(t)$ é normalizado de maneira que, no tempo presente, tenhamos por definição $r(t) = 1$. Naturalmente, temos

\begin{equation}
\ddot{r} = - \frac{4\pi}{3} \rho_{0}\frac{1}{r^{2}}, \label{cosmologia-newton}
\end{equation}
que representa o correspondente newtoniano das equações cosmológicas de Friedmann \cite{Padmanabhan2}. O mais interessante que são as mesmas equações obtidas considerando-se a relatividade geral no caso de curvatura espacial nula.

Essa ``plasticidade'' do espaço-tempo nos fornece as chaves para a interpretação da descoberta de Hubble \cite{Mac}. Em vez da interpretação antropocêntrica do movimento centrífugo das galáxias através de uma versão cósmica da explosão, a relatividade geral nos fornece um novo paradigma, durante bilhões de anos o espaço está se expandindo. Nesse processo de expansão, ele leva as galáxias a separar-se cada vez mais umas das outras. Portanto, a origem do movimento de recessão das galáxias não é uma explosão que aconteceu num certo lugar no espaço, como muitos indoutos acreditam, ao contrário, deriva-se da expansão do próprio espaço. 

Se o universo está se expandindo, obviamente que ele era menor no passado (Mais precisamente, as galáxias eram mais próximas umas das outras e  o Universo era mais denso e consequentemente mais quente). Usando as equações da Relatividade Geral, e algumas hipóteses sobre os tipos de matéria que compõe o Universo, é possível rebobinar o filme cósmico para reconstruírmos o passado histórico do cosmos. Eventualmente - algo em torno de $14$ bilhões de anos atrás, de acordo com nossas melhores estimativas- podemos atingir um momento de densidade infinita e consequentemente de curvatura infinita. Essa singularidade é popularmente conhecida como \textit{Big Bang}. 

O modelo $\Lambda$CDM (Lambda-Cold Dark Matter), o qual é hoje o chamado modelo padrão da cosmologia, assume que a evolução dinâmica do universo seja regida pelas equações do modelo cosmológico de Friedmann-Robertson-Walker-Lemaître (FRWL), também conhecido como a cosmologia do \emph{Big Bang}. Os ingredientes fundamentais nesse modelo são:

\begin{itemize}
\item \textbf{Simetria}. Espaço-tempo é descrito como uma variedade diferenciável sem torção que localmente apresenta simetria sob o grupo de Lorentz SO$(1,3)$. \vspace{0.1cm}
\item \textbf{Dinâmica}. Através do princípio variacional, aplicado à ação de Einstein-Hilbert com constante cosmológica, 
\end{itemize}

\begin{equation*}
S_{\textrm{EH}}[g] = \frac{1}{16\pi G} \int_{\mathcal{M}}\,d^{4}x \sqrt{-g}\bigl(R - 2 \Lambda\bigr) + S_{\textrm{matéria}},
\end{equation*} 
chega-se às equações e campo de Einstein,
\begin{equation}
R_{\mu\nu} - \frac{1}{2}R - \Lambda g_{\mu\nu} = 8\pi G T_{\mu\nu}, \quad T_{\mu\nu}:=\frac{\delta S_{\textrm{matéria}}}{\delta g^{\mu\nu}}. \label{einstein-cosmologia}
\end{equation}
\begin{itemize}
\item \textbf{Princípio Cosmológico}. A métrica do espaço-tempo é solução do sistema de equações de Einstein e este é por hipótese \emph{globalmente hiperbólico}\footnote{Essa hipótese de um espaço-tempo globalmente hiperbólico nos possibilita a descrição da RG como um problema de Cauchy, ou seja, um problema de valor inicial\cite{Yvonne}.}. Além disso, o espaço-tempo quando observado em escalas acima de $100$ Mpc\footnote{A unidade de medida tradicionalmente utilizada em astronomia é o parsec (pc), cuja definição está associada à distância de uma estrela que possui um paralaxe de um arco de segundo para uma linha de base igual à distância entre a Terra e o Sol (chamada de AU, unidade astronômica). Assim  $1$ pc $=$ ($1''$ em radianos)$^{-1}$ $\times$ AU $=\, 3.1\,\times\,10^{16}$ m $=\,3.26$ anos-luz.} admite uma estrutura de folheações em superfícies $3$-dimensionais, do tipo-espaço, que são supostamente \emph{homogêneas e isotrópicas}.
\end{itemize}
Assume-se também que a matéria é descrita como um fluído perfeito de densidade $\rho(t)$ e pressão $p(t)$.

Com essas hipóteses sobre a evolução do universo as equações (\ref{einstein-cosmologia}) reduzem-se as equações de Friedmann mais gerais
\begin{eqnarray}
 H^{2} & = &  \frac{8\pi}{3}\rho + \frac{\Lambda}{3} - \frac{k}{a^{2}}, \label{FRWL-eqn}\\[0.1cm]
 2 \dot{H}  +  3H^{2} & = &  -8\pi G p, \nonumber
\end{eqnarray}
onde $a$ é o fator de escala, $H = \dot{a}/a$ é o parâmetro de Hubble e, $k = 1, 0,-1$ nos informa sobre a geometria espacial para universo fechado ( $k = 1$, curvatura espacial positiva), espacialmente plano ($k = 0$), e universo aberto ( $k = -1$, curvatura espacial negativa), respectivamente. A métrica é a métrica de FRWL dada por
\begin{equation}
ds^{2} = -dt^2 + a^2(t)\lbrack \frac{dr^2}{1-kr^2}+r^2d\theta^2+r^2\sin^2\theta d\varphi^2\rbrack, 
\end{equation}
As duas equações de Friedmann (\ref{FRWL-eqn}) combinadas, no caso de um fluído perfeito, pode ser substituída pela equação de continuidade
 \begin{equation}
 \frac{d}{dt}\bigl(\rho a^{3}\bigr) + p\frac{d}{dt}(a^{3}) = 0. \label{eq-continuidade FRWL}
 \end{equation}

 A fim de se comparar a magnitude dos três termos do lado direito da equação (\ref{FRWL-eqn}) podemos dividir toda a equação pelo parâmetro de Hubble ao quadrado e definir

\begin{equation}
\Omega_{\textrm{m}}:= \frac{8\pi G \rho}{3H^{2}},\quad \Omega_{\Lambda}:= \frac{\Lambda}{3H^{2}}, \quad \Omega_{k}:= -\frac{k}{a^{2}H^{2}}, \label{densidades}
\end{equation}
tal que
\begin{equation}
\Omega_{\textrm{m}} + \Omega_{\Lambda} + \Omega_{k} = 1.
\end{equation}
Evidências observacionais da radiação cósmica de fundo (\textit{CMB}) \cite{Supernova} apontam para um universo com geometria espacial plana, ou seja, $k = 0$, o que implica $\Omega_{\textrm{m}} + \Omega_{\Lambda} \sim 1$. Poderíamos também levar em consideração o fator da radiação $\Omega_{\textrm{radiação}}$ que nos primórdios da evolução cósmica (na era cosmológica chamada de \emph{radiation-dominated in ultra-hot matter}) era dominante no universo. Dessa forma, a matéria-energia era basicamente composta de radiação, devido as altas temperaturas, e se escalava com $\rho_{\textrm{radiação}} \sim \dfrac{1}{a^{4}}$. Contudo, sua contribuição no tempo presente é negligenciável $\Omega_{\textrm{radiação}}(t_{0})\sim 0$ e, de acordo com o modelo $\Lambda$CDM,

\begin{equation}
\Omega_{\textrm{m}} = \Omega_{\textrm{bariônica}} + \Omega_{\textrm{matéria escura}}
\end{equation}\label{matéria}
 e as observações nos indicam que, atualmente, 
 
 \begin{eqnarray}
 \Omega_{\textrm{bariônica}} & \sim & 0.0227 \pm 0.0006,\nonumber\\[0.1cm]
 \Omega_{\Lambda} & \sim & 0.74 \pm 0.03.
 \end{eqnarray}
 Portanto, $\Omega_{\Lambda}$ representando a ``energia escura'' é basicamente $30$ vezes maior que a matéria escura observada. Esses são os fatos observados \cite{WMAP,Planck}.
 
 O desafio tem sido trazer um conteúdo teórico que possa abarcar e interpretar, de maneira matematicamente precisa, essas observações e esses novos efeitos. O modelo $\Lambda$CDM que é amplamente aceito pela comunidade científica, exibe uma tentativa de se explicar esses efeitos considerando o universo com geometria espacial plana e contendo apenas dois componentes que competem na evolução cósmica: matéria (bariônica e escura) como definidas em (\ref{matéria}) sem pressão  e a constante cosmológica que atribui à energia escura a responsabilidade pela expansão acelerada do universo. Modelos alternativos tem sido propostos que levam em conta modificações da ação de Einstein-Hilbert substituíndo-a por alguma densidade escalar que seja uma função arbitrária da curvatura e da torção e/ou apenas levando-se em consideração apenas potências do escalar de curvatura nos chamados modelos $f(R)$\footnote{Esses modelos $f(R)$ tem sido estudados e trazem, de certa forma, graus de liberdade novos em relação a RG usual. De fato, leva-se toda a discussão das observações e hipóteses de existência de matéria e energia escura a uma reformulação da dinâmica do espaço-tempo. Entretanto, um problema sério enfrentado é sua incapacidade, até o momento, de obter um princípio preciso através do qual essa função $f(R)$ possa ser obtida. Esse é um desafio muito complexo pois existe uma infinidade de parâmetros arbitrários que poderiam produzir a mesma dinâmica. Com efeito, a escolha desses parâmetros que é feita nos modelos $f(R)$ a fim de ``fitar'' com os dados observacionais, e não sua predição teórica, parece ser algo pouco razoável.}\cite{Felice, Capo}.
 \subsection{Homogeneidade e isotropia}
 A fim de explorar o conteúdo físico do modelo de Chamseddine, buscaremos nessa seção por soluções cosmológicas para compará-las com os resultados do modelo $\Lambda$CDM \cite{Planck} descrito na sub-seção precedente. Estaremos analisando as soluções das equações de campo (\ref{eqs com gauge fixing}) considerando-se que o espaço-tempo possa ser folheado através de uma família de superfícies $3$-dimensionais do tipo espaço, seguindo os requisitos do Princípio Cosmológico que foram especificados na seção anterior. A métrica que descreve a  cosmologia do \emph{Big Bang} é a métrica de Friedmann-Robertson-Walker-Lamaître (FRWL), dada por
 
\begin{eqnarray*}
ds^2=-dt^2+a^2(t)\lbrack \frac{dr^2}{1-kr^2}+r^2d\theta^2+r^2\sin^2\theta d\varphi^2\rbrack, 
\end{eqnarray*}
dependendo do fator de escala $a(t)$ e do parâmetro de curvatura espacial $k = 0, \pm 1$. As coordenadas do espaço-tempo são a coordenada temporal $t$ e a parte espacial descritas em coordenadas esféricas $r, \theta, \varphi$. A métrica de FRWL admite seis isometrias geradas por seis vetores de Killing associados com as três invarinâncias sob translações espaciais $\xi_{(a)}$ e três rotações $\xi_{[ab]}$ tais que
\begin{equation}
\xi_{(a)} = \sqrt{1 - kr^{2}}\partial_{a}, \quad \xi_{[ab]} = x_{a}\partial_{b} - x_{b}\partial_{a}. \label{killing-cosmology}
\end{equation}
Assumiremos que a torção e o campo escalar (o parâmetro de Newton $G$) possuem as mesmas isometrias geradas que a métrica, isto é, $\pounds T^{\alpha}_{\mu\nu} = 0$ e $\pounds G = 0$, onde $\pounds$ denota a derivada de Lie correspondendo aos seis vetores \ref{killing-cosmology}. Essas condições implicam que $G = G(t)$ e após resolver as equações diferenciais da derivada de Lie, chega-se que as componentes não nulas da torção como definida em (\ref{Torção-cap1}) são

\begin{eqnarray*}
 &&T^{r}\!_{\theta\varphi}=2f(t)a(t)r^2\sqrt{1-kr^2}\sin\theta, \hspace{20pt}  T^{\varphi}\!_{r\theta}=\frac{2f(t)a(t)}{\sqrt{1-kr^2}\sin\theta},\\
  &&T^{\varphi}\!_{r\theta}=-\frac{2f(t)a(t)\sin\theta}{\sqrt{1-kr^2}}, \hspace{70pt} T^{r}\!_{rt}=T^{\theta}\!_{\theta t}=T^{\varphi}\!_{\varphi t}=h(t)
 \end{eqnarray*}
 onde $f(t)$ e $h(t)$ são funções arbitrárias do tempo a serem determinadas pelas equações de movimento. Trabalhando no formalismo de primeira-ordem, podemos escolher uma parametrização diagonal\footnote{Essa parametrização é sempre possível, via uma fixação de gauge, desde que a métrica seja diagonal. Caso a métrica não seja diagonal é possível ao menos transformar a matriz dos vierbein em uma matriz triangular. } para o vierbein\footnote{Lembrando que a torção $T^{I}_{\mu\nu}$ é a $2$-forma definida por $T^{I} = De^{I}$, cujas componentes estão relacionadas através do inverso do vierbein $T^{\alpha}_{\mu\nu} = e^{\alpha}_{I}T^{I}_{\mu\nu}$.},
 
 \begin{eqnarray*}
&&e^0=dt, \hspace{50 pt} e^1=\frac{a(t)}{\sqrt{1-kr^2}}dr,\\
&&e^2=a(t)rd\theta, \hspace{23 pt} e^3=a(t)r\sin\theta d\varphi
\end{eqnarray*} 
Nessa base os elementos da torção $2$-forma $T^{I} = \frac{1}{2}e^{I}_{\alpha}T^{\alpha}_{\mu\nu}dx^{\mu}\wedge dx^{\nu}$ podem ser calculados. A componente $T^{0}$ é dada por
\begin{equation*}
T^{0} = \frac{1}{2}e^{0}_{t}T^{t}_{\mu\nu}dx^{\mu} \wedge dx^{\nu} = 0,
\end{equation*}
pois todas as componentes $T^{t}_{\mu\nu}$ são nulas. As componentes espaciais serão

\begin{eqnarray*}
T^{1} & = & \frac{1}{2}e^{1}_{r}T^{r}_{kl}dx^{k}\wedge dx^{l}\\[0.1cm]
      & = & \frac{a(t)}{\sqrt{1 -kr^{2}}}T^{r}_{rt}dr\wedge dt + \frac{a(t)}{\sqrt{1 - kr^{2}}} T^{r}_{\theta\varphi}d\theta \wedge d\varphi \\[0.1cm]
      & = & h(t) e^{1}\wedge e^{0} + f(t)e^{2}\wedge e^{3}.
\end{eqnarray*}
Analogamente, as demais componentes são
\begin{equation*}
T^{2} = h(t) e^{2} \wedge e^{0} - f(t) e^{1}\wedge e^{3},\quad T^{3} = h(t)e^{3}\wedge e^{0} + f(t) e^{1}\wedge e^{2},
\end{equation*}
de maneira mais compacta temos
\begin{equation}
T^{0} = 0, \quad T^{i} = h(t)e^{i}\wedge e^{0} + \frac{1}{2}f(t)\varepsilon^{i}_{jk} e^{j}\wedge e^{k}, \quad i, j, k = 1, 2, 3.\label{Componentes da Torção}
\end{equation}
Da equação $T^{I} = \mathcal{C}^{I}\,_{J}\wedge e^{J}$ que relaciona a torção com a contorção, podemos obter a contorção $1$-forma
\begin{eqnarray}
T^{I} & = & \mathcal{C}^{I}\,_{J}\wedge e^{J} \quad \Rightarrow \quad \frac{1}{2} T^{I}\,_{JK}e^{J}\wedge e^{K} = - \mathcal{C}^{I}\,_{JK} e^{J}\wedge e^{K}\nonumber\\[0.1cm]
 T^{I}\,_{JK}e^{J} & = & \bigl(\mathcal{C}^{I}\,_{KJ} - \mathcal{C}^{I}\,_{JK}\bigr)e^{J}\label{Contorção}
\end{eqnarray}  
Daí, somando-se e subtraíndo-se as permutações cíclicas dos índices da equação (\ref{Contorção}), encontra-se a expressão explícita da contorção como função dos elementos da torção
\begin{equation}
\mathcal{C}_{[IJ]} = \frac{1}{2}\biggl(T_{IJK} - T_{JIK} - T_{KIJ}\biggr)e^{I}. \label{contorção-torção}
\end{equation}
As componentes não nulas da torção em termos dos índices $I, J, K$ do espaço tangente são
\begin{eqnarray}
T_{123} = T^{r}_{\theta\varphi}, && T_{321} = T^{\varphi}_{r\theta}\nonumber\\[0.1cm]
T_{213} = T^{\theta}_{r \varphi}, && T_{110} = T_{220} = T_{330} = h(t), \label{Torção 2-forma} 
\end{eqnarray}
dessa forma, é bem simples mostrar que as componentes da contorção leem-se
\begin{equation}
\mathcal{C}^{0i} = h(t)e^{i},\quad \mathcal{C}^{ij} = - f(t)\varepsilon^{ij}_{k}e^{k}
\label{Contorção 1-forma}
\end{equation}
Lembrando que a conexão de spin $\omega$ é decomposta em sua parte livre de torção mais contorção como vimos em (\ref{conexão-completa}). Segundo (\ref{xi-eqn}) temos que as componentes não nulas de $\xi_{IJK}$ serão
\begin{equation*}
\xi_{101} = \xi_{202} = \xi_{303} = H(t)= \frac{\dot{a}}{a}, \quad \xi_{212} = \xi_{313} = \frac{\sqrt{1-kr^{2}}}{ar},\quad \xi_{323} = \frac{\cot\theta}{a r},     
\end{equation*}
com efeito, as componentes da conexão livre de torção como definido em  (\ref{conexão-sem torção}) não nulas serão
\begin{eqnarray*}
\bar{\omega}^{0i} = H(t)e^{i}, && \bar{\omega}^{12} = - \frac{\sqrt{1-kr^{2}}}{ar}e^{2}\\[0.1cm]
\bar{\omega}^{13} = -\frac{\sqrt{1 - kr^{2}}}{ar} e^{3}, && \bar{\omega}^{23} = -\frac{\cot\theta}{ar}e^{3}.  
\end{eqnarray*} 
Dessa forma a conexão de spin que gera essa torção lê-se
\begin{eqnarray*}
&&\omega^{0i}=(H+h)e^i, \hspace{60 pt} \omega^{12}=-\frac{\sqrt{1-kr^2}}{ar}e^2-fe^3,\\
&&\omega^{31}=\frac{\sqrt{1-kr^2}}{ar}e^3-fe^2, \hspace{20 pt} \omega^{23}=-\frac{\mbox{cot}\theta}{ar}e^3-fe^1, 
\end{eqnarray*}
onde $H:= \dot{a}(t)/a(t)$ é o parâmetro de Hubble. A curvatura de Riemann é dada por
\begin{eqnarray*}
&&R^{0i}=\lbrack (\dot{H}+\dot{h})+H(H+h)\rbrack e^0\wedge e^i+f(H+h)\varepsilon^{i}\!_{jk}e^j \wedge e^k,\\
&&R^{ij}=\lbrack (H+h)^2+\frac{k}{a^2}-f^2\rbrack e^i\wedge e^j+(\dot{f}+Hf)\varepsilon^{ij}\!_{k}e^k \wedge e^0,
\end{eqnarray*}

Consequentemente, 
\begin{eqnarray*}
&&\mathbb{F}^{0i}=\lbrack (\dot{H}+\dot{h})+H(H+h)-\frac{\Lambda}{3}\rbrack e^0 \wedge e^i+f(H+h)\varepsilon^{i}\!_{jk}e^j\wedge e^k,\\
&&\mathbb{F}^{ij}=\lbrack (H+h)^2+\frac{k}{a^2}-f^2-\frac{\Lambda}{3}\rbrack e^i\wedge e^j+(\dot{f}+Hf)\varepsilon^{ij}\!_{k}e^k\wedge e^0.
\end{eqnarray*}
\subsection{Equações de campo}
Assumiremos que a matéria constitui-se de um fluído perfeito de densidade $\rho_{\textrm{m}}$ e pressão $p_{\textrm{m}}$, com o tensor de energia-momento $\mathcal{T}^{I}\,_{J} = \textrm{diag}(-\rho_{\textrm{m}},p_{\textrm{m}},p_{\textrm{m}},p_{\textrm{m}} )$. Substituindo nas equações de campo (\ref{eqs com gauge fixing}), com $dG = \dot{G}e^{0}$, obtemos um sistema de equações diferenciais
\begin{eqnarray}
\label{ce1}
&&U^2+\frac{k}{a^2}-f^2-\frac{\Lambda}{3}
=\frac{4\pi G}{3}\rho_{\rm m} \,,\\
\label{ce2}
&&  U^2+\frac{k}{a^2}-f^2-\Lambda+2(\dot{U}+HU)\
= -4\pi G \,p_{\rm m},\\
\label{ce3}
&&\dot{\Phi}^4(U^2+\frac{k}{a^2}-f^2-\frac{\Lambda}{3})+\frac{2\Lambda}{3}\Phi^4h=0,\\
\label{ce4}
&&f(\dot{\Phi}^4U+\frac{\Lambda}{3}\Phi^4)=0,\\
\label{ce5}
&&(U^2+\frac{k}{a^2}-f^2-\frac{\Lambda}{3})(\dot{U}+HU-\frac{\Lambda}{3})-2fU(\dot{f}+Hf)=0,
\end{eqnarray}
onde $U := H + h$ e $G = G(t)$ é o parâmetro de acoplamento de Newton.
\subsection{Equações de continuidade}
Uma primeira equação de continuidade para a energia e pressão segue naturalmente da equação de continuidade para o tensor de energia-momento (\ref{continuity eq.}) que resume-se em $D\mathcal{T}_{I} = 0$. Calculando as componentes da energia-momento $3$-forma, de (\ref{energy-momentum}), obtemos
\begin{eqnarray*}
\mathcal{T}_{0}  =  \frac{\rho_{\textrm{m}}(t)}{6}\varepsilon_{ijk}e^{i}\wedge e^{j} \wedge e^{k}, &&
\mathcal{T}_{i}   =  -\frac{p_{\textrm{m}}}{2}\varepsilon_{ijk}e^{0}\wedge e^{j}\wedge e^{k},
\end{eqnarray*}
consequentemente,
\begin{eqnarray*}
\mathcal{T}_{0} & = & -\rho_{\textrm{m}}e^{1}\wedge e^{2} \wedge e^{3} = -\frac{\rho_{\textrm{m}}(t)a^{3}(t)r^{2}\sin\theta}{\sqrt{1-kr^{2}}}dr \wedge d\theta \wedge d\varphi \\[0.1cm]
\mathcal{T}_{1} & = & -p_{\textrm{m}}e^{0}\wedge e^{2} \wedge e^{3} = -p_{\textrm{m}} a^{2}(t) r^{2}\sin\theta dt\wedge d\theta \wedge d\varphi,\\[0.1cm]
\mathcal{T}_{2} & = & -p_{\textrm{m}}e^{0}\wedge e^{3} \wedge e^{1} = -p_{\textrm{m}}\frac{a^{2}(t)r\sin\theta}{\sqrt{1 - kr^{2}}} dt \wedge d\varphi \wedge dr,\\[0.1cm]
\mathcal{T}_{3} & = &-p_{\textrm{m}}e^{0}\wedge e^{1} \wedge e^{2} =   -p_{\textrm{m}}\frac{a^{2}(t)r}{\sqrt{1 - kr^{2}}}dt \wedge dr \wedge d\theta.
\end{eqnarray*}
A equação $D\mathcal{T}_{0} = 0$ produz a equação de continuidade da densidade-pressão-torção
\begin{equation}
\dot{\rho}_{\textrm{m}} + 3H(p_{\textrm{m}} + \rho_{\textrm{m}}) + 3h p_{\textrm{m}} = 0.
\label{continuidade1}
\end{equation}
As equações $D\mathcal{T}_{i} = 0$ para $i = 1, 2, 3$ são trivialmente satisfeitas, sendo da forma $0 = 0$.

Note a dependência nas funções da torção na equação (\ref{continuidade1}). Contudo, para matéria sem pressão (cold matter), ou seja, poeira, essa equação de continuidade assume a forma usual\cite{Padmanabhan2}:
\begin{equation}
\frac{d}{dt}\biggl(\rho_{\textrm{m}}a^{3}\biggr) = 0, \quad \textrm{se},\, p_{\textrm{m}} = 0.\label{continuidade2}
\end{equation}
Uma segunda equação de continuidade pode ser encontrada da seguinte forma: Notando que ao substituirmos $U = H + h$ nas equações (\ref{ce1},\ref{ce2}) nos leva a uma equação de Friedmann sem constante cosmológica usual:
 \begin{eqnarray}
H^{2} = \frac{8\pi G_{0}}{3}\rho_{\textrm{tot}}, && 2\dot{H} + 3H^{2} = -8\pi G_{0}p_{\textrm{tot}}, \label{friedmann-like}
\end{eqnarray}
onde $G_{0}$ é a constante de Newton, avaliada no tempo presente de $G(t)$, e $\rho_{\textrm{tot}}$, $p_{\textrm{tot}}$ são as densidade e pressão ``totais'',
\begin{eqnarray*}
\rho_{\textrm{tot}} & = & \frac{G}{G_{0}}\bigl(\rho_{\textrm{m}} + \rho_{k} + \rho_{T} + \rho_{\Lambda}\bigr),\\[0.1cm]
p_{\textrm{tot}} & = &  \frac{G}{G_{0}}\bigl(p_{\textrm{m}} + p_{k} + p_{T} + p_{\Lambda}\bigr),
\end{eqnarray*}
com
\begin{equation*}
\rho_{k} = \frac{3}{8\pi G}\frac{k}{a^{2}}, \quad \rho_{T} = \frac{3}{8\pi G}(f^{2} - 2Hh - h^{2}), \quad \rho_{\Lambda} = \frac{\Lambda}{8\pi G}
\end{equation*}
\begin{equation*}
p_{k} = - \frac{\rho_{k}}{3}, \quad p_{T} = \frac{1}{8\pi G} (2\dot{h} + 4Hh + h^{2} - f^{2}), \quad p_{\Lambda} = - \rho_{\Lambda}.
\end{equation*}
$\rho_{T}$ e $p_{T}$ são interpretadas como as contribuições da torção à densidade e pressão $\rho_{\textrm{tot}}$ e $p_{\textrm{tot}}$. Como consequência das equações tipo Friedmann (\ref{friedmann-like}), a densidade total e pressão satisfazem a equação de continuidade
\begin{equation*}
\dot{\rho}_{\textrm{tot}} + 3H(\rho_{\textrm{tot}} + p_{\textrm{tot}}) = 0.
\end{equation*}
\subsection{Poeira com $\Lambda > 0$ e $k = 0$}
Nessão seção apresentamos as soluções gerais das equações (\ref{ce1})-(\ref{ce5}) considerando-se o caso de matéria na ausência de pressão (poeira), com $p_{\textrm{m}} = 0$, considerando uma constante cosmológica positiva e uma geometria espacial plana, como nos indica os resultados experimentais \cite{Supernova,Planck}. Da equação \ref{ce4} segue que
\begin{equation}
\textrm{ou} \quad f(t) = 0, \quad \textrm{ou}\quad \dot{G}U - \frac{\Lambda}{3}G = 0.
\label{sol.1}
\end{equation}
Verificamos que a primeira condição nos leva a solução trivial do espaço de de Sitter com constante cosmológica $\Lambda$, com o vierbein ou a métrica sendo definidas pelo fator de escala $a(t) = e^{\sqrt{\frac{\Lambda}{3}}t}$.

Dessa forma, assumimos a função $f(t)$ como sendo diferente de zero. As equações a serem resolvidas  são (\ref{ce1})-(\ref{ce3}), (\ref{ce5}) e a segunda de (\ref{sol.1}), junto com o parâmetro de Hubble definido em termos do fator de escala $a(t)$. A solução geral é dada pelas seguintes expressões obtidas pelo uso do programa Mathematica\cite{Mathematica}, onde a coordenada temporal foi redefinida por
\begin{equation*}
\tau(t):=\sqrt{\frac{\Lambda}{3}}t.
\end{equation*}
\begin{itemize}
\item \textit{Fator de escala}:
\end{itemize}
\begin{equation}
a(t) = C_{4}\bigl(3e^{\tau} + C_{3}e^{-\tau}\bigr)^{\frac{1}{3}} \bigl(\cosh(\tau - C_{1})\bigr)^{\frac{2}{3}}. \label{fator de escala}
\end{equation}
\begin{itemize}
\item \textit{Parâmetro de torção} $f(t)$:
\end{itemize}
\begin{eqnarray}
f(t) & = & \frac{\sqrt{\Lambda}}{3}\biggl(\bigl(-9e^{2\tau} - 3C_{3} + (6 e^{2\tau} - 2C_{3})\bigr)\tanh(\tau - C_{1}) + \nonumber\\[0.1cm] 
&& + \bigl(3e^{2\tau} +  C_{3}\bigr)\tanh^{2}(\tau - C_{1})\bigr)\bigg/\bigl(3 e^{2\tau} + C_{3}\bigr)\biggr)^{\frac{1}{2}}.
\label{fator f}
\end{eqnarray}

\begin{itemize}
\item \textit{Parâmetro de torção} $h(t)$:
\end{itemize}
\begin{equation}
h(t) = \sqrt{\frac{\Lambda}{3}}\frac{\bigl(-3e^{2\tau} + C_{3} + (3e^{2\tau}+ C_{3})\tanh(\tau - C_{1})\bigr)}{9e^{2\tau} + 3C_{3}}.
\label{fator h}
\end{equation}
\begin{itemize}
\item \textit{Parâmetro de Hubble} $H(t)= \dot{a}/a$:
\end{itemize}
\begin{equation}
H(t) = \sqrt{\frac{\Lambda}{3}}\tanh(\tau - C_{1}) - h(t).
\end{equation}
\begin{itemize}
\item \textit{Parâmetro de Newton} $G(t) = -3/(8\pi\Lambda\Phi(t))$:
\end{itemize}
\begin{equation}
G(t) = C_{2} \sinh(\tau - C_{1}).
\end{equation}
\begin{itemize}
\item \textit{Densidade de matéria} $\rho_{\textrm{m}}(t)$:
\end{itemize}
\begin{equation}
\rho_{\textrm{m}} = \frac{3}{8\pi G(t)} \biggl(\bigl(H(t) + h(t)\bigr)^{2} - f^{2}(t) - \frac{\Lambda}{3}\biggr).
\end{equation}
As quatro constantes de integração $C_{1}$, $C_{2}$, $C_{3}$, $C_{4}$ e a constante cosmológica $\Lambda$ devem ser determinadas por cinco condições físicas, as quais escolhemos sendo:
\begin{eqnarray}
a(0) & = & 0 : \textrm{hipótese do Big Bang},\nonumber\\[0.1cm]
a(t_{0}) & = & 1: t_{0} = \textrm{idade atual do universo},\nonumber\\[0.1cm]
H(t_{0}) & = & H_{0}: \textrm{valor atual do parâmetro de Hubble},\\[0.1cm]
G(t_{0}) & = & G_{0}: \textrm{valor atual do parâmetro de Newton},\nonumber\\[0.1cm]
\rho_{\textrm{m}} & = & \rho_{0}: \textrm{valor atual da densidade de matéria},\nonumber
\end{eqnarray}
com os dados experimentais e observacionais \cite{Planck} dados por 
\begin{eqnarray*}
t_{0} & = & 13.8\times 10^{9}\, \textrm{anos} \quad (1\textrm{Gy} = 10^{9} \textrm{anos}),\\[0.1cm]
H_{0} & = & 0.0693\, \textrm{Gy}^{-1},\\[0.1cm]
\rho_{0} & = & 2.664 \times 10^{-27}\, \textrm{Kg}\,\textrm{m}^{-3}, \\[0.1cm]
  G_{0} & = & 6.674 \times 10^{-11}\textrm{m}^{3}\,\textrm{s}^{-2}\,\textrm{Kg}^{-1}.
\end{eqnarray*}
Para uma comparação com os resultados do modelo padrão $\Lambda$CDM, necessitamos da fórmula do fator de escala $a(t)$, em um universo dominado por matéria escura fria (poeira) da densidade relativa \cite{Planck} $\Omega_{\textrm{m}} = 0.309$. Para comparação, negligenciando a contribuição da radiação, temos o fator de escala normalizado de $\Lambda$CDM é \cite{Padmanabhan2}
\begin{equation}
a_{\Lambda \textrm{CDM}} (t) = \biggl(\frac{\sinh\bigl(\frac{3}{2}H_{0}\sqrt{1- \Omega_{\textrm{m}}}t\bigr)}{\sinh\bigl(\frac{3}{2}H_{0}\sqrt{1 - \Omega_{\textrm{m}}}t_{0}\bigr)}\biggr)^{\frac{2}{3}}
\end{equation}
A Figura $3.1$ mostra a evolução temporal do fator de escala $a$, do parâmetro de Hubble $H$,  do parâmetro de desaceleração $q:= - \ddot{a}a/(\dot{a}a^{2})$, da densidade de massa $\rho_{\textrm{m}}$, e do parâmetro de Newton normalizado $G/G_{0}$. Cada uma das quantidades foi comparada com seu correspondente no modelo $\Lambda$CDM . Com excessão da desaceleração $q$, as demais quantidades apresentam um desvio bem pequeno. O parâmetro de Newton que tem de ser igual ao valor atual da constante de acoplamento de Newton $G_{0}$ no tempo presente, mostra-se com uma ligeira diminuição no passado da evolução cósmica, aumentando  cerca de $85\%$ desde seu valor nos primórdios do Big Bang. A desaceleração $q$ difere notavelmente do modelo padrão da cosmológica, no entanto, o tempo de transição entre as eras de aceleração e desaceleração é praticamente coincidente. O valor atual de $q(t_{0}) = -0.25$ é, contudo, apenas a metade do valor previsto por $\Lambda$CDM.

\begin{figure}[!htb]
\centering
\includegraphics[scale=0.58]{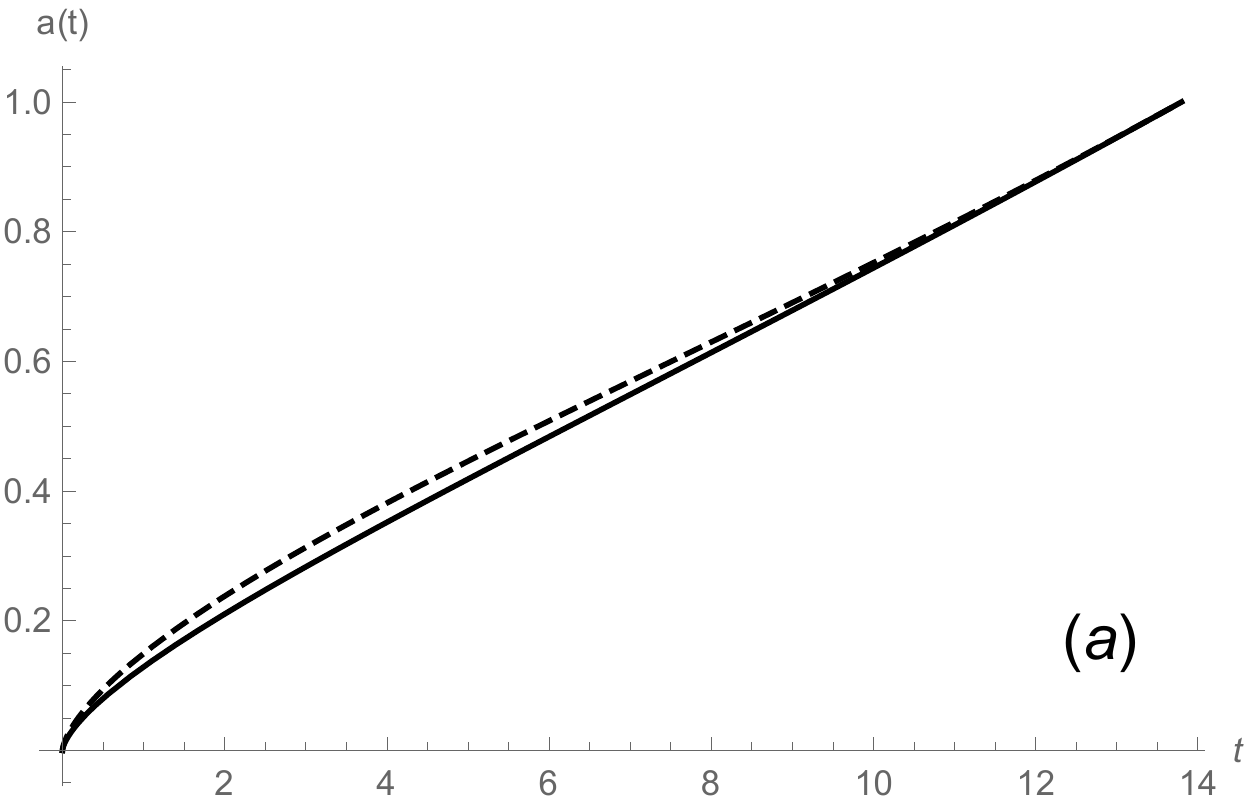}
\includegraphics[scale=0.58]{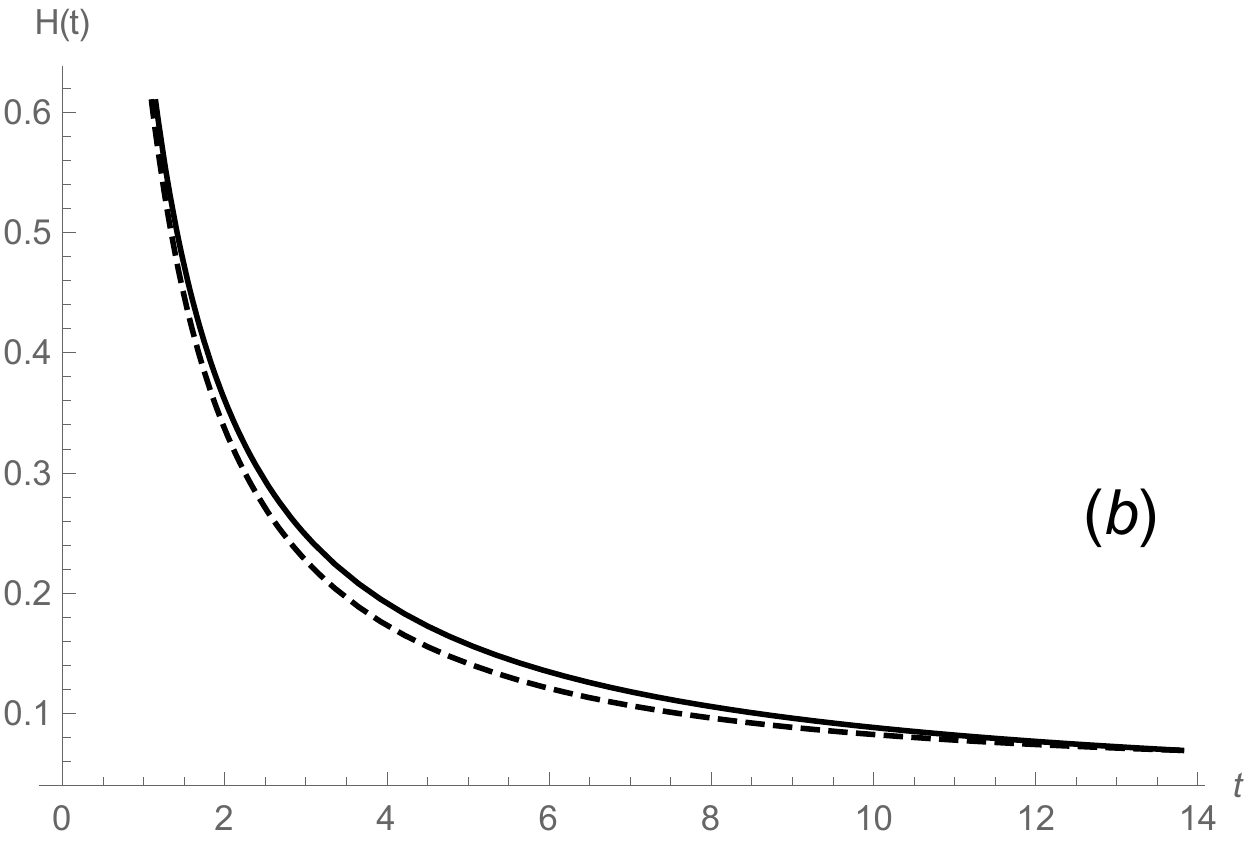}
\includegraphics[scale=0.58]{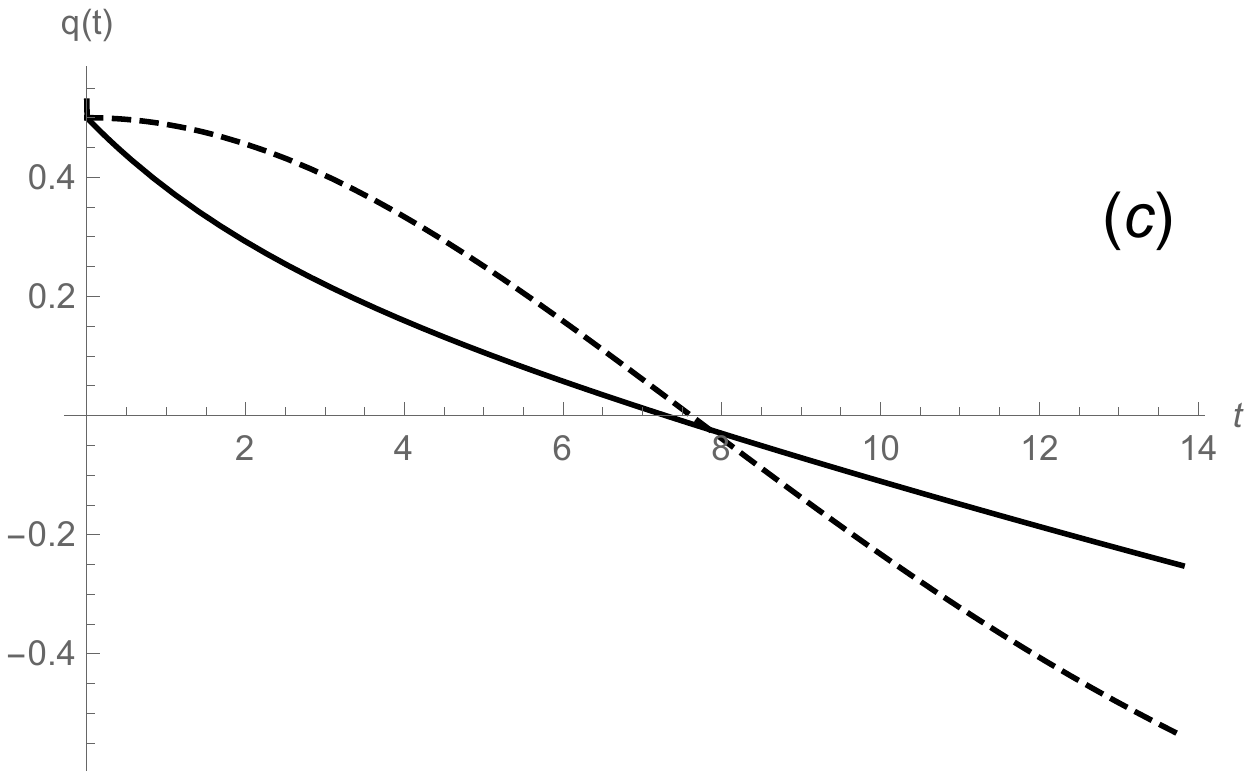}

\includegraphics[scale=0.58]{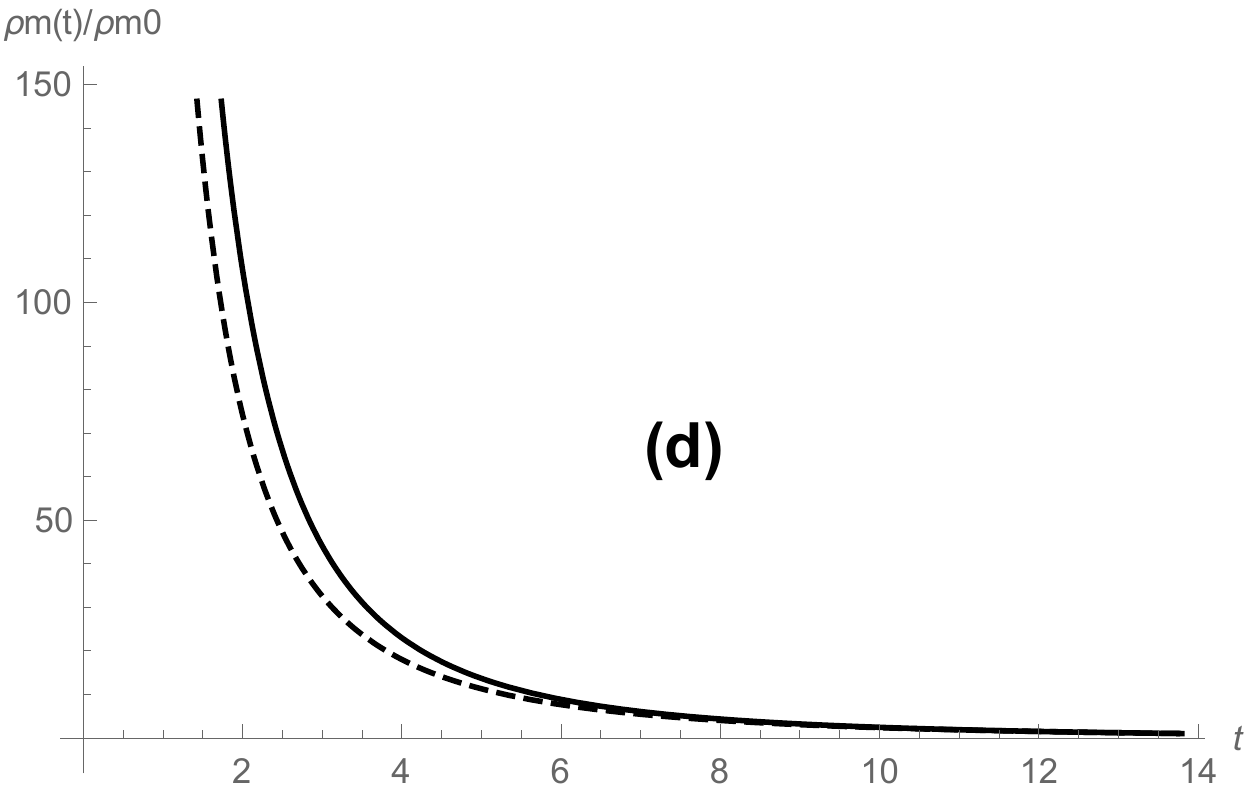}
\includegraphics[scale=0.58]{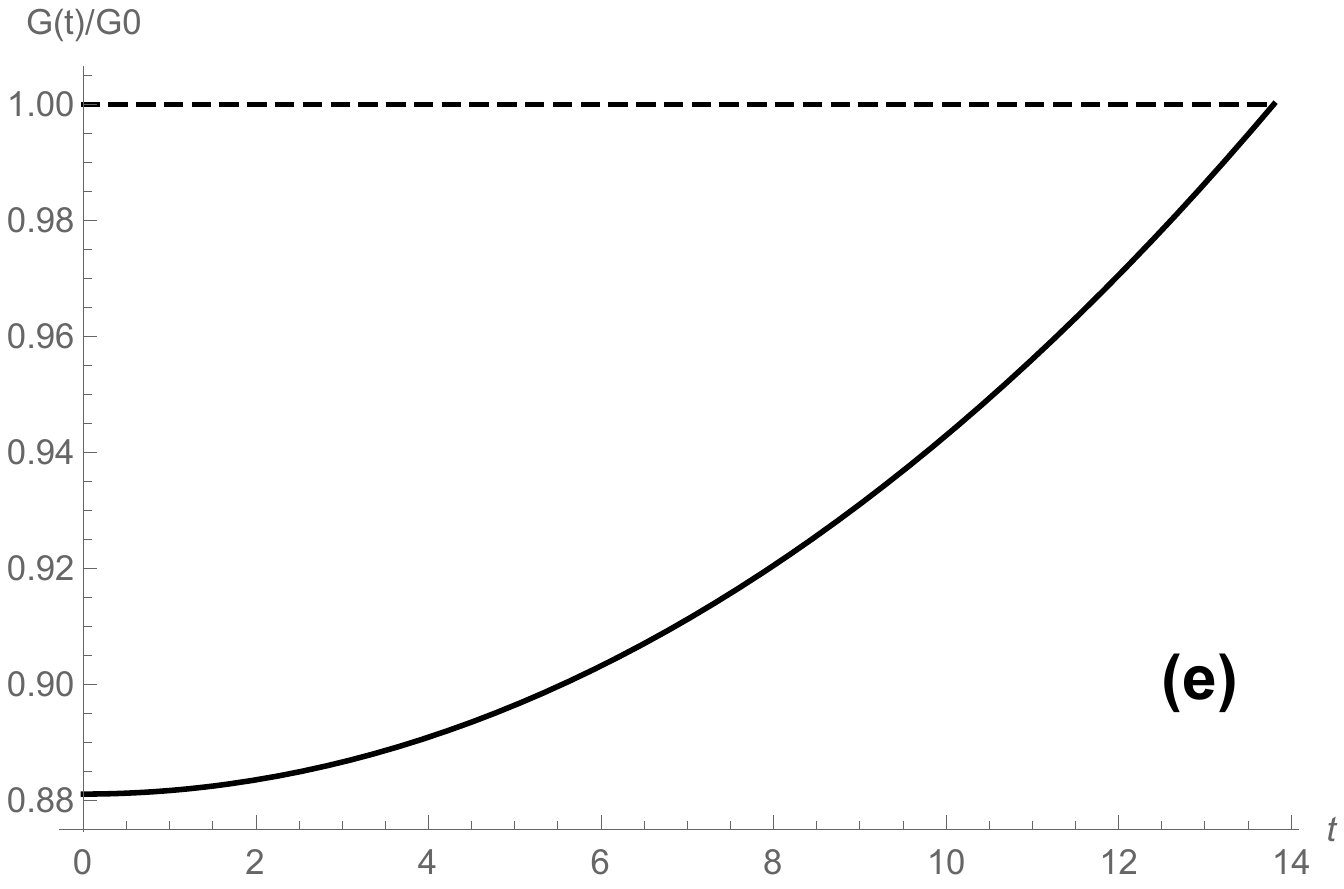}
\label{fig31}
\caption{\it\small (a) Fator de escala normalizado $a(t)$; 
(b) Parâmetro de Hubble $H(t)$;
(c) Parêmetro de desaceleração $q(t)$;
(d) densidade de matéria escura (fria)  $\rho_m(t)$;
(e) Parâmetro de acolplamento gravitacional tempo-dependente $G(t)$;
Linhas sólidas: predições do nosso modelo; linhas tracejadas: resultados padrões de $\Lambda$CDM . }
\end{figure} 
A evolução temporal dos parâmetros da torção $h$ e $f$, bem como das densidades relativas $\Omega_{\textrm{m}}$, $\Omega_{\Lambda}$ e $\Omega_{T}$ para matéria, constante cosmológica e torção, respectivamente, são mostradas na Figura $3.2$(a-b).

\begin{figure}[!htb]
\centering
\includegraphics[scale=0.58]{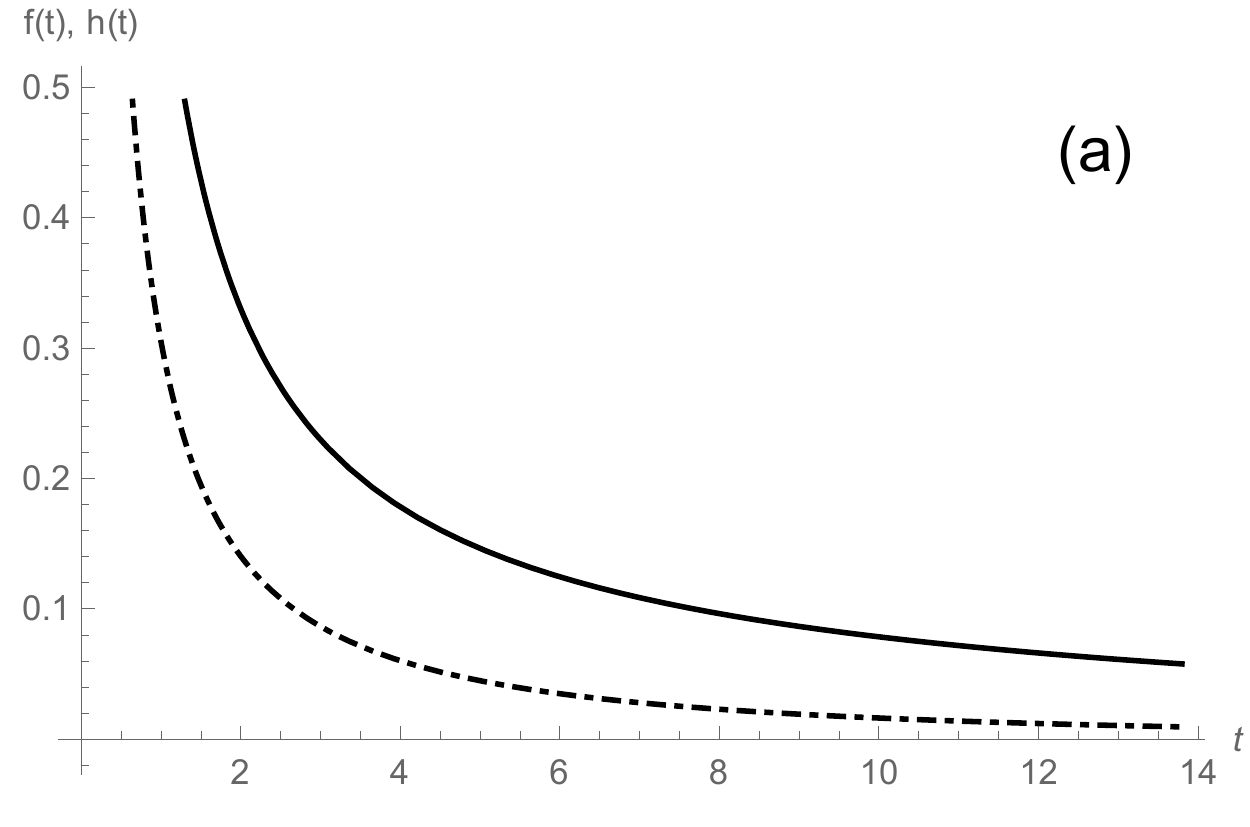}
\includegraphics[scale=0.58]{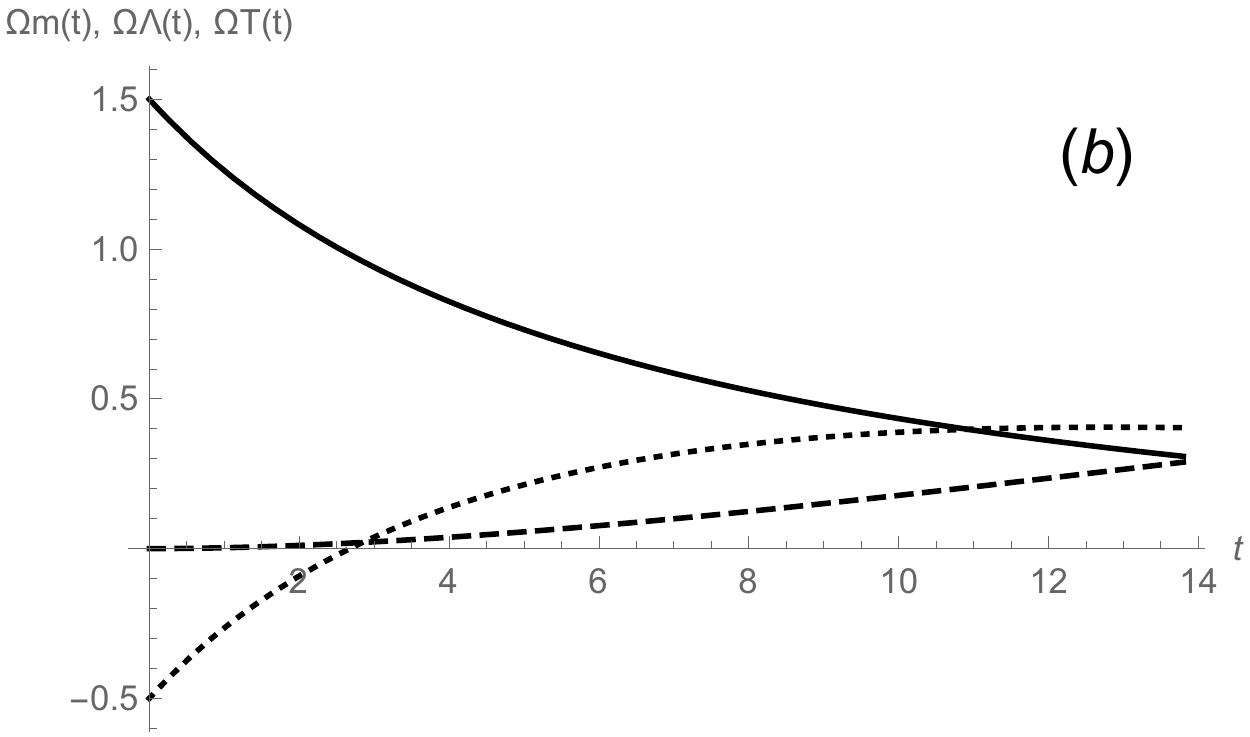}
\includegraphics[scale=0.50]{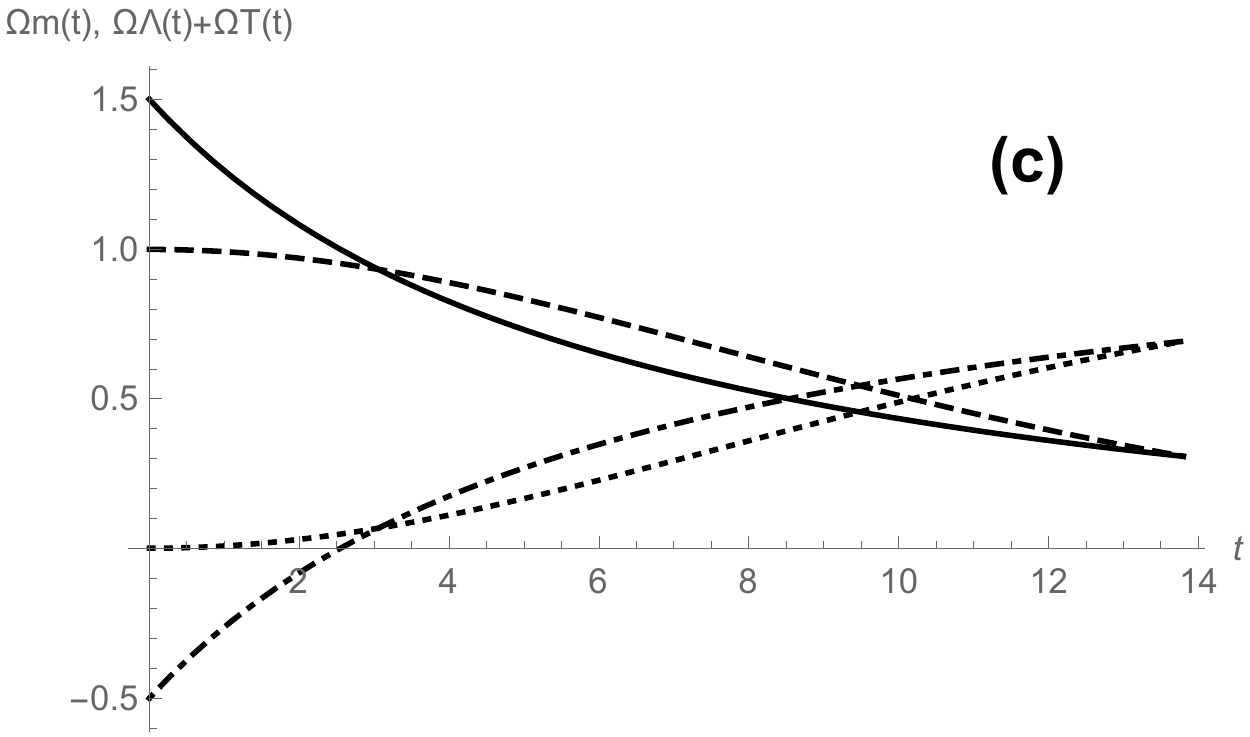}
\label{figura2}
\caption{\it\small(a) Parâmetro da torção $f(t)$ (linha sólida) e $h(t)$ (linha tracejada); 
(b) Densidades relativas $\Omega_m(t)$ (linha sólida), $\Omega_{\Lambda}(t)$ (linha tracejada) e  $\Omega_T(t)$ (linha pontilhada);
(c)  $\Omega_m(t)$ (linha sólida) e $\Omega_{\Lambda}(t)$ + $\Omega_T(t)$ 
(linha tracejada-pontilhada);  resultados de $\Lambda$CDM   são mostrados para $\Omega_m$ (linha tracejada) e $\Omega_{\Lambda}$ (linha pontilhada).}
\end{figure}

Observa-se da figura $3.2$(c) que o fim do domínio da era de matéria fria, ocorreu $t \sim 10.2$ Gy para $\Lambda$CDM, e  $t \sim 8.5$ Gy para o nosso modelo. Domínio da matéria sendo definido, no nosso caso, como o domínio de $\Omega_{\textrm{m}}$ sobre a soma $\Omega_{\Lambda} + \Omega_{T}$. Finalmente, os valores atuais das densidades relativas do nosso modelo são:
\begin{itemize}
\item $\Omega_{\textrm{m}}(t_{0}) = 0.308$,\vspace{0.1cm}
\item $\Omega_{\Lambda}(t_{0}) = 0.289$,\vspace{0.1cm}
\item $\Omega_{T}(t_{0}) = 0.403$
\end{itemize} 
Esses valores devem ser colocados em comparação com os valores do modelo $\Lambda$CDM\footnote{Não foi levado em conta em nossos cálculos as atualizações feitas nesse ano para a nova determinação de $H_{0}$. Em junho de $2016$ foi reportado pela NASA, através do WMAP\cite{WMAP16}, uma série de medidas mais acuradas para o valor atual da constante de Hubble. A melhor estimativa obtida foi: $H_{0} = 73.24 \pm 1.74$ Km s$^{-1}$ Mpc$^{-1}$. Esse valor produz uma taxa de expansão do universo maior do que era previsto pelo $\Lambda$CDM.} que são $\Omega_{\textrm{m}}(t_{0}) = 0.308$ e $\Omega_{\Lambda}(t_{0}) = 0.692$: observamos que no nosso modelo a torção contribui junto com a constante cosmológica para a aceleração. Finalmente, como uma questão de análise de consistência, verificamos que nossa solução das equações de campo, de fato, satisfazem a equação de continuidade (\ref{continuidade2}).

Além disso, fizemos uma busca por outras soluções, pois como a torção pode contribuir para a aceleração, poderia-se esperar soluções que apresentem uma aceleração positiva mesmo considerando-se o caso de constante cosmológica negativa. Isso ocorre, por exemplo, para a classe de modelos investigadas em \cite{Za}. No nosso caso, checamos que não há solução com $\Lambda < 0$ e aceleração positiva que seja capaz de satisfazer as condições de contorno representadas pelos valores atuais da densidade de matéria e dos parâmetros de Hubble e Newton. Outro tipo de classe de soluções que envolvem ricochete (bounce) que houve em algum tempo no passado existe, mas nenhum deles são compatíveis nem de longe com as condições de contorno físicas.

\chapter*{Conclusões e Perspectivas}
\addcontentsline{toc}{chapter}{Conclusões e Perspectivas}
\pagestyle{fancy}
\lhead{\bfseries Conclusões e Perspectivas}
\rhead{}
Vimos, com grandes detalhes, como a redução dimensional e a truncação de um teoria de Chern-Simons em $5$D considerando-se como grupo de gauge o (A)dS$_{6}$ nos leva ao modelo de Chamseddine em $4$D. O modelo de Chamseddine envolve um campo escalar de tipo-dilaton o qual interpretamos como sendo o parâmetro de acoplamento gravitacional de Newton que, como vimos pelas nossas predições, estaria variando ao longo da evolução cósmica. Exploramos as soluções das equações de campo, de modo que: mostramos que, no limite linear do modelo, existe a presença de ondas gravitacionais, assim como as previstas pela RG padrão e detectadas esse ano \cite{ligo1,ligo2}, e da existência de um limite newtoniano. O parâmetro de Newton é considerado constante em ordem zero, contudo, permanece indeterminado em primeira ordem.
\medskip

As soluções de ondas gravitacionais corroboram os resultados obtidos pela análise canônica do modelo, onde obtivemos os graus de liberdade propagados: dois para o ``gráviton'' e um associado ao campo escalar, ou seja, o parâmetro de Newton. Exploramos as soluções cosmológicas do tipo FRWL, onde obtemos soluções obedecendo a condições de contorno físicas, a saber, os valores atuais dos parâmetros físicos: parâmetros de Newton e Hubble e a densidade de matéria fria. Mostrando um comportamento que se assemelha bem, pelo menos qualitativamente, ao modelo padrão da cosmologia $\Lambda$CDM.

\medskip

A constante cosmológica das nossas soluções deve ser positiva, contudo, com um valor menor do que o reportado no modelo $\Lambda$CDM pois, no nosso modelo, a torção contribui de maneira expressiva para a aceleração atual do universo. Um modelo similar mas bem diferente em estrutura foi estudado em \cite{Za}. A maior diferença, é que a ação possui  o campo escalar  apenas como um fator suplementar à parte quadrática da curvatura de Riemann. No nosso caso, o campo escalar está completamente conectado com toda a densidade de lagrangiana.
\medskip

 Apesar ter termos obtido bons resultados da análise clássica no âmbito cosmológico em \cite{Top-Grav}, os desafios atualmente tem sido o estudo e a compreensão física da análise canônica do modelo de Chamseddine, que mostrou-se ser não genérico o que dificultou a quantização via laços. O estudo da teoria complera de Chern-Simons em $5$D está em progresso, o que nos permitirá explorar um domínio de soluções muito maior. Embora a contagem dos graus de liberdade de Chern-Simons em  D $=$ $5$ tenha sido feita em \cite{Hen2,Miskovic}, possuindo $13$ graus de liberdade, a separação dos vínculos de primeira e segunda classe ainda não foi feita. Portanto, um dos objetivos gerais para futuros trabalhos é separar esses vínculos de primeira e segunda classe e buscar pela quantização do modelo. Além disso, em paralelo, tem sido feito um estudo aprofundado das  soluções do modelo com simetria esférica e a procura de soluções  tipo-Schwarzschild tanto da teoria completa de Chern-Simons em $5$D quanto o correspondente modelo em $4$D, bem como identificar os efeitos, na escala de compactificação da dimensão extra, considerando modos de Kaluza-Klein além do modo zero.

\renewcommand{\theequation}{A.\arabic{equation}}
\chapter*{Derivada de Lie e Aplicações}\setcounter{equation}{0}
\addcontentsline{toc}{chapter}{Apêndice 1: Derivada de Lie}
\pagestyle{fancy}
\lhead{\bfseries Apêndice 1: Derivada de Lie}
\rhead{}
\label{apendice}
Esse apêndice tem a finalidade pedagógico-instrumental no sentido de fornecer as ferramentas matemáticas necessárias para os cálculos apresentados na tese. Trataremos principalmente de ideias e conceitos sobre geometria diferencial que são largamente utilizadas na gravitação bem como em teoria geral de campos. As referências clássicas\cite{Schutz,Felsager,Isham,Tu,Crampin,report} podem ser consultadas para algum detalhe de rigor matemático.

\section*{Definições e Aplicações}
Seja $C\,:\, I \subset \mathbb{R}\, \rightarrow \,\mathcal{M}$ uma curva sobre a variedade
$\mathcal{M}$ cujos vetores tangentes, $\xi\, =\, \dfrac{d}{d\lambda}$ são descritos em termos da base  de coordenadas $\bigl\{\partial_{\mu}\bigr\}$, como 
\begin{equation}
\xi = \xi^{\mu}\frac{\partial}{\partial x^{\mu}} = \frac{dx^ {\mu}}{d\lambda}\frac{\partial}{\partial x^{\mu}}.
\label{eq1}
\end{equation}
onde $\lambda$ é o parâmetro que descreve a curva sobre a variedade. Com efeito, o vetor tangente nos informa sobre a evolução cinemática sobre a curva definida na variedade, em outras palavras, $\xi$ nos dá uma informação sobre a velocidade ao longo dessa curva, suposta regular, isto é, $\dfrac{dx^{\mu}}{d\lambda}\, \neq \, 0$ em todo seu domínio. Quando estamos lhe dando com variedades não euclidianas a pergunta que automaticamente fazemos é: como comparar vetores em pontos distintos na variedade? Se desejamos obter conceitos como os de uma derivada direcional necessitamos, em princípio, comparar vetores
e de certa forma encontrar algum mecanismo que faça o transporte de vetores para compará-los em um mesmo ponto. O que geralmente se faz é trazer uma estrutura afim para variedade introduzindo o conceito de conexão e derivação que se comporte de maneira covariante sob transformações gerais de coordenadas (difeomorfismos). 

Uma alternativa ao conceito de derivada covariante reside na ideia de uma transformação ativa dos pontos da curva (\textit{Lie dragging}) levando a comparação dos vetores no ponto desejado. Esse conceito nos leva a definição da derivada de Lie que está intimamente relacionada, como veremos, com a variação funcional dos campos sobre difeomorfismos. Assim, consideremos um vetor $v^{\mu}$ avaliado no ponto $x$. Desejamos comparar esse vetor porém em um ponto $x'$, infinitesimalmente próximo, de maneira que
\begin{equation}
x^{\mu} \longmapsto x'^{\mu} = x^{\mu} + \xi^{\mu}(x)
\label{eqA2}
\end{equation}
exigindo que sobre a transformação (\ref{eqA2}) o vetor exiba covariância, isto é,

\begin{equation}
v'^{\mu}(x' = x + \xi) = \frac{\partial x'^{\mu}}{\partial x^{\nu}} v^{\nu}(x),
\label{eqA3}
\end{equation} 
donde obtemos,
\begin{equation}
 \frac{\partial x'^{\mu}}{\partial x^{\nu}} = \delta^{\mu}_{\nu} + \partial_{\nu}\xi^{\mu}.
 \label{eqA4}
\end{equation}
Introduzindo (\ref{eqA4}) em (\ref{eqA3}), segue que

\begin{equation}
v'^{\mu}(x + \xi) = v^{\mu}(x) + \partial_{\nu}\xi^{\mu}v^{\nu}(x) 
\label{eqA5}
\end{equation}
fazendo-se uma expansão em Taylor no lado esquerdo da Eq. (\ref{eqA5}) e, guardando informação apenas em primeira ordem, temos que $v'^{\mu}(x + \xi)\, \simeq\, v'^{\mu}(x) \, + \xi^{\nu}\partial_{\nu}v'^{\mu}$, como $\xi$  já é infinitesimal, multiplicado por $v'^{\mu}$, já é  de primeira ordem daí, podemos tomar, nesse termo, $v'^{\mu} \simeq v^{\mu}$, pois qualquer variação já produziria termos de segunda ordem.  

\begin{eqnarray*}
v'^{\mu}(x) + \xi^{\nu}\partial_{\nu}v^{\mu} & = &  v^{\mu}(x) + \partial_{\nu}\xi^{\mu}v^{\nu}(x)\\ [0.1cm]
v'^{\mu}(x) - v^{\mu}(x) & = &  \partial_{\nu}\xi^{\mu}v^{\nu} - \xi^{\nu}\partial_{\nu}v^{\mu} \\[0.1cm]
\delta^{(dif)} v^{\mu} & = & \bigl(\bigl[\xi, v\bigr]\bigr)^{\mu}
\end{eqnarray*}
onde,
\begin{equation}
\bigl(\pounds_{\xi}v\bigr)^{\mu}:= \partial_{\nu}\xi^{\mu}v^{\nu} - \xi^{\nu}\partial_{\nu}v^{\mu},
\end{equation}
é o que chamamos da componente $\mu$ da derivada de Lie de um vetor contravariante na direção do vetor $\xi$. Portanto, vemos que a variação funcional de um vetor sobre difeomorfimos nos leva a definição de uma nova maneira de computar diferenças, sem a necessidade de recorrer a uma conexão. Um dos aspectos interessantes da derivada de Lie é que sua ação em um vetor preserva seu rank e é dada pelo comutador (colchêtes de Lie) entre o vetor e a direção desejada. Explicitamente, temos
 
\begin{eqnarray}
\pounds_{\xi}v & = & \bigl[\xi, v\bigr]\nonumber\\[0.1cm]
& = & \bigl[\xi^{\mu}\partial_{\mu}, v^{\nu}\partial_{\nu}\bigr]\nonumber\\[0.1cm]
& = & \xi^{\mu}\partial_{\mu}v^{\nu}\partial_{\nu} - v^{\mu}\partial_{\nu}\xi^{\mu}\partial_{\mu} \nonumber\\[0.1cm]
& = & \underbrace{\bigl( \xi^{\nu}\partial_{\nu}v^{\mu} - v^{\nu}\partial_{\nu}\xi^{\mu}\bigr)}_{\textrm{componentes da derivada de Lie}}\partial_{\mu} 
\end{eqnarray} 
A generalização para tensores de rank maiores é imediata devido a regra de Leibniz que toda derivada deve satisfazer. E dessa forma se tivermos um tensor, digamos de rank $2$, $T^{\alpha\beta}\,=\, u^{\alpha}v^{\beta}$, segue que
\begin{equation}
\pounds_{\xi}T^{\alpha\beta} = \xi^{\mu}\partial_{\mu}T^{\alpha\beta} - T^{\mu\beta}\partial_{\mu}\xi^{\alpha} - T^{\alpha\mu}\partial_{\mu}\xi^{\beta}.
\end{equation} 

A fim de obtermos a forma de atuação da derivada de Lie em vetores covariantes bem como em formas diferenciais, podemos fazer o uso da aplicação da derivada em um campo escalar $\varphi$, isto é, sendo que $\pounds_{\xi}\varphi\,=\, \xi^{\mu}\partial_{\mu}\varphi$. Para tanto, basta tomarmos esse campo escalar como sendo $\varphi\,=\, v^{\mu}\omega_{\mu}$ e, fazendo-se uso da regra de Leibniz mais uma vez temos
\begin{eqnarray}
\pounds_{\xi}\varphi & = & (\pounds_{\xi}v^{\mu})\omega_{\mu} + v^{\mu}(\pounds_{\xi}\omega_{\mu})\nonumber \\[0.1cm]
 & = & \bigl(\xi^{\nu}\partial_{\nu}v^{\mu} - v^{\nu}\partial_{\nu}\xi^{\mu}\bigr)\omega_{\mu} + v^{\mu}(\pounds_{\xi}\omega_{\mu})\nonumber \\[0.1cm]
 \xi^{\nu}\partial_{\nu}(v^{\mu}\omega_{\mu}) & = & \xi^{\nu}\partial_{\nu}v^{\mu}\omega_{\mu} - v^{\nu}\partial_{\nu}\xi^{\mu}\omega_{\mu} + v^{\mu}\pounds_{\xi}\omega_{\mu}\nonumber \\
 \xi^{\nu}\partial_{\nu}v^{\mu}\omega_{\mu} + v^{\mu}\xi^{\nu}\partial_{\nu}\omega_{\mu} & = & \xi^{\nu}\partial_{\nu}v^{\mu}\omega_{\mu} - v^{\nu}\partial_{\nu}\xi^{\mu}\omega_{\mu} + v^{\mu}\pounds_{\xi}\omega_{\mu}\nonumber \\[0.1cm]
 v^{\mu}\bigl(\pounds_{\xi}\omega_{\mu}\bigr) & = & v^{\mu}\bigl(\xi^{\nu}\partial_{\nu}\omega_{\mu} + \omega_{\mu}\partial_{\nu}\xi^{\nu}\bigr)\nonumber
\end{eqnarray}
portanto, a derivada de Lie de uma $1$-forma, por exemplo, bem como de vetores covariantes, lê-se como
\begin{equation}
\pounds_{\xi}\omega_{\mu} = \xi^{\nu}\partial_{\nu}\omega_{\mu} + \omega_{\mu}\partial_{\nu}\xi^{\nu},
\end{equation}
e dessa forma podemos identificar como, por exemplo, a métrica do espaço-tempo muda funcionalmente quando fazemos uma transformação de difeomorfismos. Além do mais, estamos de posse de uma maneira de caracterizar simetrias no espaço-tempo. Como as transformações de difeomorfismos são transformações quaisquer das coordenadas, automaticamente já englobam transformações lineares tais como: rotações, translações, \textit{boosts}, etc. Com efeito, podemos fazer uma investigação criteriosa das simetrias do espaço-tempo através dessa ferramenta.
\section*{Simetrias}
A busca por simetrias faz parte da espinha dorsal de toda teoria física moderna. Dessa forma, um princípio de simetria, ou seja, um conjunto de transformações que deixa invariante certas quantidades tem guiado de maneira precisa, bem como trazendo a tona predições dantes impensadas, na busca pela compreensão da Natureza. A física de partículas, bem como a construção dos inúmeros aceleradores de partículas como o grande colisor de hadrons (LHC), tem por base um princípio de simetria que alinha teoria-experimento. A detecção em $2012$, pela colaboração ATLAS e CMS no CERN, da partícula que apresenta as propriedades demandas pelo bóson de Higgs, necessária no mecanismo de geração de massa das partículas \cite{englert, higgs} (rendendo o prêmio Nobel de $2013$ par os físicos F. Englert e P. Higgs por suas contribuições teóricas), é um dos maiores exemplos do poder que reside nesse conceito. Nas escalas cósmicas a aplicabilidade desse ferramental é extremamente relevante. Os modelos de cosmologia moderna são profundamente alicerçados em simetrias. O universo, quando observado em escalas acima de $100$ Mpc\footnote{A unidade de medida tradicionalmente utilizada em astronomia é o parsec (pc), cuja definição está associada a uma estrela que possui um paralaxe de um arco de segundo para uma linha de base igual a metade da distância entre a Terra e o Sol (chamada de AU, unidade astronômica). Assim pc $=$ ($1''$ em radianos)$^{-1}$ $\times$ AU $=\, 3.1\,\times\,10^{16}$ m $=\,3.26$ anos-luz.} é homogêneo e isotrópico. E o espaço-tempo satisfazendo esse princípio cosmológico é descrito pela métrica de Friedmann-Lemaître-Robertson-Walker (FRLW) nas coordenadas em co-movimento (o referencial de repouso cósmico). 

De fato, a pergunta que se faz é uma via de mão dupla: dada uma métrica do espaço-tempo desejamos saber quais sãos as possíveis direções de isometria. De outra forma, é possível que nosso interesse agora seja obter a métrica de maneira a satisfazer certas condições de simetria dadas a priori. A derivada de Lie é justamente a ferramenta matemática que nos possibilita responder tais perguntas. Assim, para encontrarmos simetrias de uma métrica é equivalente a condição 
\begin{equation}
\pounds_{\xi}g_{\alpha \beta} = \xi^{\mu}\partial_{\mu}g_{\alpha \beta} + g_{\mu \beta}\partial_{\alpha}\xi^{\mu} + g_{\alpha \mu}\partial_{\beta}\xi^{\mu} = 0.
\label{eqA6}
\end{equation}
assim, da Eq.(\ref{eqA6}) o que nos interessará é encontrar o campo de vetores de Killing que nos darão informações sobre as direções de isometria da métrica. Para tanto, consideremos um conjunto de curvas congruentes, de modo que
os vetores tangentes as curvas formem um campo vetorial. Suponha ainda que escolhamos $\lambda$ como parâmetro de evolução de modo que $\lambda$ seja uma das coordenadas do espaço-tempo, por exemplo $ \lambda = x^{\alpha}$. Daí, como o vetor tangente é dado por $\xi^{\mu} \,=\, \dfrac{dx^{\mu}}{d\lambda}$ e, nesse caso, estamos escolhendo uma coordenada em particular para o parâmetro de evolução da curva, ou seja, uma única direção fixa. Portanto, as componentes de $\xi^{\mu}$ são constantes
\begin{equation}
\xi^{\mu} = \frac{dx^{\mu}}{d\lambda} = \frac{\partial x^{\mu}}{\partial x^{\alpha}} = \delta^{\mu}_{\alpha},
\label{eqA7}
\end{equation}
por exemplo, se a curva é tipo-tempo, logo escolhemos $\lambda \,=\, t$ e obtemos o chamado vetor de Killing dado por $\xi^{\mu} \,=\, (1, 0, 0, 0)$. Para tal escolha, $\partial_{\beta}\xi^{\mu}\,=\, 0$ e a derivada de Lie da métrica é apenas,
\begin{eqnarray}
\pounds_{\xi} g_{\alpha\beta} & = & \xi^{\mu}\partial_{\mu}g_{\alpha\beta} \nonumber\\[0.1cm]
                       & = & \frac{dx^{\mu}}{d\lambda}\frac{\partial g_{\alpha\beta}}{\partial x^{\mu}} = \frac{\partial g_{\alpha\beta}}{\partial \lambda},\nonumber
\end{eqnarray}
isso significa que se a derivada de Lie se anula, a métrica é independente da coordenada $\lambda$. Com efeito, a métrica não apresenta variação ao longo da congruência de curvas geradas por $\dfrac{d}{d\lambda}$, e temos então uma simetria do espaço-tempo associada ao vetor de Killing. Quaisquer direções nas quais a métrica permanece invariante são chamadas de isometrias. De fato, podemos encontrar equações diferenciais que descrevam essas direções de simetria do espaço-tempo. Tomando a derivada de Lie $\pounds_{\xi}g\,=\,0$, de (\ref{eqA6}), temos
\begin{equation}
 \xi^{\mu}\partial_{\mu}g_{\alpha \beta} + g_{\mu \beta}\partial_{\alpha}\xi^{\mu} + g_{\alpha \mu}\partial_{\beta}\xi^{\mu} = 0,
 \label{eqA8}
\end{equation}
observando que os termos em deriva do parâmetro ($g_{\mu\beta}\partial_{\alpha}\xi^{\mu}$) podem ser reescritos como uma derivada total, isto é, $g_{\mu\beta}\partial_{\alpha}\xi^{\mu}\,= \, \partial_{\alpha}(g_{\mu\beta}\xi^{\mu}) - \partial_{\alpha}g_{\mu\beta}\xi^{\mu}$, que podemos introduzir na Eq. (\ref{eqA8}), 
\begin{equation}
0 = \xi^{\mu}\partial_{\mu}g_{\alpha\beta} + \partial_{\alpha}(g_{\mu\beta}\xi^{\mu}) - \partial_{\alpha}g_{\mu\beta}\xi^{\mu} + \partial_{\beta}(g_{\alpha\mu}\xi^{\mu}) - \partial_{\beta}g_{\alpha\mu}\xi^{\mu}.
\label{eqA9}
\end{equation}
Lembrando que a ação da métrica sobre um vetor contravariante o transforma em um vetor covariante, ou seja, $g_{\mu\beta}\xi^{\mu}\,=\,\xi_{\beta}$ e selecionando os termos em derivadas da métrica a Eq. (\ref{eqA9}) torna-se
\begin{equation}
 0 = -\xi^{\mu}\underbrace{\bigl(\partial_{\alpha}g_{\mu\beta} + \partial_{\beta}g_{\alpha\mu} - \partial_{\mu}g_{\alpha\beta}\bigr)} + \partial_{\alpha}\xi_{\beta} + \partial_{\beta}\xi_{\alpha}.
 \label{eqA10}
\end{equation}
O termo em destaque em (\ref{eqA10}) é claramente identificado com a conexão de Levi-Civita $\Gamma^{\alpha}_{(\mu\nu)}$ que é amplamente utilizada na RG usual no formalismo métrico (formalismo de $2^{a}$ ordem)\cite{Trautman1,Zee,Bryce,Moshe,Inverno,Weinberg}, onde a parte anti-simétrica, isto é, a Torção ($\Gamma^{\alpha}_{[\mu\nu]}\,\equiv\,0$) é tomada por hipótese como sendo zero. Além disso, a derivada covariante agindo em um vetor $A_{\mu}$ é da forma: $\nabla_{\mu}A_{\nu}\,=\, \partial_{\mu}A_{\nu} - \Gamma^{\lambda}_{\mu\nu}A_{\lambda}$, com  $\Gamma^{\lambda}_{\mu\nu}\,=\, \dfrac{1}{2}g^{\lambda\alpha}\bigl(\partial_{\mu}g_{\alpha\nu} + \partial_{\nu}g_{\mu\alpha} - \partial_{\alpha}g_{\mu\nu}\bigr)$ $\Longrightarrow$ $2 g_{\lambda\beta}\Gamma^{\lambda}_{\mu\nu}\,=\,\bigl(\partial_{\mu}g_{\beta\nu} + \partial_{\nu}g_{\mu\beta} - \partial_{\beta}g_{\mu\nu}\bigr)$. Daí a Equação (\ref{eqA10}) assume a seguinte forma
\begin{equation*}
0 = -2\xi^{\mu}g_{\mu\lambda}\Gamma^{\lambda}_{\alpha\beta} + \partial_{\alpha}\xi_{\beta} + \partial_{\beta}\xi_{\alpha},
\label{eqA11}
\end{equation*}
ou ainda, como $\xi^{\mu}g_{\mu\lambda}\,=\,\xi_{\lambda}$, segue que

\begin{equation*}
\partial_{\alpha}\xi_{\beta} -\Gamma^{\lambda}_{\alpha\beta}\xi_{\lambda} + \partial_{\beta}\xi_{\alpha} - \Gamma^{\lambda}_{\beta\alpha}\xi_{\lambda} = 0,
\label{eqA12}
\end{equation*}
Finalmente, obtemos a chamada Equação de Killing
\begin{equation}
\nabla_{\alpha}\xi_{\beta} + \nabla_{\beta}\xi_{\alpha} = 0, 
\label{eqA13}
\end{equation}
dada uma métrica ($g_{\mu\nu}$), podemos perguntar por todas as soluções da Equação (\ref{eqA13}) cuja resposta nos dará as direções de simetria do espaço-tempo.

\section*{Simetrias do Espaço de Minkowski}

Iremos agora aplicar esses conceitos desenvolvidos sobre derivação de Lie, e suas consequentes conexões com as simetrias espaço-temporais, ao caso particular de uma métrica de cunho hiperbólica, isto é, com um padrão de medição não necessariamente positivo-definido. Em outras palavras, estaremos considerando o padrão de medição introduzido pela propagação da luz, a saber, espaço da relatividade restrita, cujo elemento de linha é
 
\begin{equation}
ds^{2} = - c^{2}dt^{2} + d\vec{x}^{2}. 
\end{equation} 

Assim estaremos considerando um espaço-tempo plano no qual a métrica $\eta_{\mu\nu}$, em coordenadas cartesianas, 
\begin{equation}
\eta_{\mu\nu} = \textrm{diag}(-1, 1, 1, 1)
\end{equation}
e a conexão de Christoffel se anula, isto é, $\Gamma^{\alpha}_{\mu\nu} \equiv 0$. Naturalmente, as derivadas covariantes que constituem as equações de Killing são levadas à derivações parciais planas.

\begin{equation}
\partial_{\beta}\xi_{\alpha} + \partial_{\alpha}\xi_{\beta} = 0
\label{killing}
\end{equation}
aplicando uma nova derivação à equação (\ref{killing}), obtemos
\begin{equation}
\partial_{\mu}\partial_{\beta}\xi_{\alpha} + \partial_{\mu}\partial_{\alpha}\xi_{\beta} = 0.
\label{killing1}
\end{equation}

Agora, premutando ciclicamente os índices da equação (\ref{killing1}), 

\begin{eqnarray}
\partial_{\mu}\partial_{\beta}\xi_{\alpha} + \partial_{\mu}\partial_{\alpha}\xi_{\beta} & = & 0 ,\\[0.1cm] \label{killing2}
\partial_{\alpha}\partial_{\mu}\xi_{\beta} + \partial_{\alpha}\partial_{\beta}\xi_{\mu}& =& 0, \\[0.1cm] \label{killing3}
\partial_{\beta}\partial_{\alpha}\xi_{\mu} + \partial_{\beta}\partial_{\mu}\xi_{\alpha} & =& 0. \label{killing4}
\end{eqnarray}

Somando-se as equações (\ref{killing2}) e (\ref{killing3}) e subtraindo (\ref{killing4}), supondo, por simplicidade, que as componentes de $\xi$ são de classe $C^{\infty}$ ou apresentem diferenciabilidade o suficiente, para evitarmos maiores dificuldades, segue que 

\begin{equation}
\partial_{\mu}\partial_{\alpha}\xi_{\beta} = 0 \label{killing5},
\end{equation}
ou seja, a segunda derivada de $\xi_{\beta}$ se anula. Isso significa que $\xi_{\beta}$ tem de ser uma função linear das coordenadas. Portanto, a forma mais geral que $\xi_{\beta}$ assume pode ser lida como
\begin{equation}
\xi_{\alpha} = a_{\alpha} + \Sigma_{\alpha\beta} x^{\beta}. \label{xi}
\end{equation}
Substituindo (\ref{xi}) na equação (\ref{killing}), obtem-se
\begin{eqnarray}
 0 & = & \Sigma_{\alpha\beta} + \Sigma_{\beta\alpha},
\end{eqnarray}
de modo que $a_{\alpha}$ fica completamente arbitrário enquanto $\Sigma_{\alpha\beta}$ deve ser uma matriz anti-simétrica. Temos, portanto, dez campos vetoriais independentes, cada um assumindo a forma dada pela equação (\ref{xi}) para cada escolha independente das constantes $a_{\alpha}$ e $\Sigma_{\alpha\beta} = - \Sigma_{\beta\alpha}$.

A escolha mais simples dentre os $10$ campos vetoriais seria assumir apenas uma das constantes $a_{\alpha}$ diferente de zero. Tomando-se $\Sigma_{\alpha\beta} = 0$ e uma das componentes (digamos, $n$, com $n = 0,1,2,3$) de $a_{\alpha}$ obtemos quatro campos vetoriais constantes,
\begin{equation*}
\xi^{\alpha}_{(n)} = \delta^{\alpha}_{n}.
\end{equation*}
Representando vetores unitários de cada uma das direções das coordenadas espaço-temporais. Agora, se colocarmos essas contantes a zero $a_{\alpha} = 0$ e escolhendo, a título pedagógico, apenas um dos seis $\Sigma_{[\alpha\beta]}$, iremos obter ou rotações espaciais ou \textit{boosts} de Lorentz. Com efeito, iremos mostrar que na verdade esses vetores de Killing correspondem aos elementos da álgebra do grupo de Lorentz, ou seja, são os geradores das transformações do grupo $SO(1,3)$.

Por exemplo, com $\Sigma_{21} =  -\Sigma_{12} = 1$ e todos os demais zeros, o campo de Killing será

\begin{eqnarray}
\xi & = & \xi^{\alpha}\partial_{\alpha} \\[0.1cm]
    & = & \bigl(\eta^{\alpha\beta}\Sigma_{\beta\mu}x^{\mu}\bigr)\partial_{\alpha} \\[0.1cm]
    & = & x\partial_{y} - y\partial_{x}. \label{geradorx}
\end{eqnarray}
Esse é o gerador das rotações no plano-$xy$ ou em torno do eixo $z$. Analogamente, para $\Sigma_{23}$ e $\Sigma_{31}$ que está associado aos geradores de rotação em torno do eixo $x$ e $y$, respectivamente. Juntos eles formam os geradores do subgrupo especial e ortogonal SO$(3)$ $\subset$ SO$(1,3)$. Por outro lado, se os índices não nulos forem associados à componente temporal teremos, nesse caso, um \textit{boost}. Para $\Sigma_{10} = -\Sigma_{01} = 1$ temos
\begin{equation}
\xi = x \partial_{t} + t \partial_{x}, \label{geradort}
\end{equation}
esse é o gerador de um \textit{boost} de Lorentz. Para vermos isso de maneira mais clara, vamos exponenciar\footnote{Estamos assumindo aqui que estamos trabalhando com grupos de Lie contínuos e conexos com a identidade que podem ser descritos, a grosso modo, como a exponencial da álgebra $(\textrm{Grupo} = e^{\textrm{álgebra}})$.} o gerador com algum parâmetro a ser interpretado,
\begin{eqnarray}
\Lambda & = & e^{\lambda\bigl(x\partial_{t} + t\partial_{x}\bigr)} \\[0.1cm]
      & = & \sum^{\infty}_{n=0} \frac{\lambda^{n}}{n!}\biggl(x \partial_{t} + t \partial_{x}\biggr)^{n}.
\end{eqnarray}
 O resultado da ação da transformação $\Lambda$ sobre as coordenadas lê-se
 
 \begin{eqnarray}
 t' & = & (\cosh{\lambda}) t + (\sinh{\lambda}) x, \nonumber \\[0.1cm]
 x' & = & (\sinh{\lambda}) t + (\cosh{\lambda}) x,\nonumber \\[0.1cm]
 y' & = & y,\nonumber \\[0.1cm]
 z' & = & z.\nonumber
 \end{eqnarray}
 Note que podemos colocar a função $\cosh{\lambda}$ em evidência 
 
 \begin{eqnarray*}
 t' & = & \cosh{\lambda}( t + \tanh{\lambda} x), \\[0.1cm]
 x' & = & \cosh{\lambda}( x + \tanh{\lambda} t), 
  \end{eqnarray*}
  lembrando que $\cosh{\lambda} \geq 1$ e que $\mid\tanh{\lambda}\mid < 1$. Como pela relatividade especial $v/c < 1$ poderímos identificar a tangente hiperbólica como um parâmetro que mede a razão entre as velocidades relativa dos referenciais e a velocidade da luz. Assim, defina $\tanh{\lambda} := \beta = - v/c$.
  
  Por outro lado, temos da relação $\cosh^{2}{\lambda} - \sinh^{2}{\lambda} = 1$ que
  
\begin{equation}
\cosh^{2}{\lambda}(1 - \tanh^{2}{\lambda}) = 1, \quad \Rightarrow \quad \cosh{\lambda} = \frac{1}{\sqrt{1-\beta^{2}}} \equiv \gamma. 
\end{equation}  
  Finalmente, vemos que através das equações de Killing somos capazes de recuperar um \textit{boost} de Lorentz com velocidade $v$ na direção da coordenada $x$, bem como todo o conjunto de transformações de $SO(1,3)$. Podendo ser reescrito na forma mais familiar
 
 \begin{eqnarray}
 ct' & = & \gamma( ct - \beta x), \\[0.1cm]
 x' & = & \gamma( x  - vt), \\[0.1cm]
 y' & = & y, \\[0.1cm]
 z' & = & z.
 \end{eqnarray}
 
 Portanto, encontramos exatamente as $10$ isometrias do espaço de Minkowski. Esse é o máximo de soluções independentes das equações de Killing. No caso de um espaço-tempo estacionário e esfericamente simétrico (Schwarzchild) temos um campo vetorial de Killing tipo-tempo e outros $3$ campos de Killing associados a rotações espaciais.
 
 \subsection*{Simetria Esférica} 
 
 Vamos agora dar uma noção sobre o significado de um espaço-tempo estacionário e esfericamente simétrico. Ser estacionário implica a existência de um campo vetorial de Killing tipo-tempo. Esfericamente simétrico requer a existência de um conjunto completo de vetores de Killing que sejam geradores das rotações espaciais.
 
 Se desejamos trabalhar em um espaço-tempo estacionário via a existência de um vetor de Killing tipo-tempo, podemos escolher a coordenada temporal para ser o parâmetro $\lambda = t$, e as condições de simetria implicam que
 \begin{equation}
 \pounds_{\xi} g_{\alpha\beta} = \xi^{\mu}\partial_{\mu}g_{\alpha\beta} + g_{\mu\beta}\partial_{\alpha}\xi^{\mu} + g_{\alpha\mu}\partial_{\beta}\xi^{\mu} = 0.
 \end{equation}
 Contudo, com $x^{0} = t = \lambda$, as componentes de $\xi$ são constantes, de modo que $\partial_{\alpha}\xi^{\mu} = 0$. Portanto,
 
 \begin{equation}
 0 = \xi^{\mu}\partial_{\mu}g_{\alpha\beta} \quad \Rightarrow \quad \partial_{t}\bigl(g_{\alpha\beta}\bigr) = 0,
 \end{equation}
e temos um sistema de coordenadas cuja métrica é independente da coordenada temporal.

Para a simetria esférica, sabemos que temos de ter $3$ campos vetoriais de Killing que juntos geram o grupo de rotações espaciais, a saber, SO$(3)$. Podemos tomar duas das coordenadas, mas elas não irão apresentar relações de comutação triviais, de maneira que a métrica não pode ser independente das duas coordenadas  ao mesmo tempo. Tomando a forma, já familiar, em coordenadas Cartesianas
\begin{eqnarray}
\xi_{1} & = & y \partial_{z} - z \partial_{y}, \\[0.1cm]
\xi_{2} & = & z \partial_{x} - x \partial_{z}, \\[0.1cm]
\xi_{3} & = & x \partial_{y} - y \partial_{x}.
\end{eqnarray}  
Nossa tarefa agora será reescrever os vetores de Killing em coordenadas esféricas. Ou seja, devemos fazer uma mudança de coordenadas assumindo que

\begin{eqnarray}
x^{i} \longmapsto x'^{i} = x'^{i}(x) &  & x^{i} = \{x,y,z\}, \quad x'^{i} = \{r,\theta, \phi\}. \nonumber
\end{eqnarray}

Lembrando que 
\begin{eqnarray}
x = r\sin{\theta}\cos{\phi} & & y = r \sin{\theta}\sin{\phi} \\[0.1cm] 
z = r\cos{\theta} & & r = \sqrt{x^{2} + y^{2} + z^{2}}
\end{eqnarray}
a ideia básica será transformar as derivações cartesianas pelas coordenadas curvilíneas, ou seja, sendo $x^{i} = x^{i}(x')$, podemos fazer uso da regra da cadeia
\begin{equation}
\partial_{i} = \frac{\partial x'^{j}}{\partial x^{i}}\partial'_{j}, \label{partial}
\end{equation}
onde estamos fazendo uso da convenção de soma de Einstein, isto é, indíces repetidos indicam soma. De (\ref{partial}), vemos que precisamos calcular os elementos da matriz Jacobiana que conecta as mudanças de coordenadas.

\begin{displaymath}
\biggl(\frac{\partial x'^{j}}{\partial x^{i}}\biggr) = 
\left( \begin{array}{ccc}
\frac{\partial r}{\partial x} & \frac{\partial r}{\partial y} & \frac{\partial r}{\partial z} \\
\frac{\partial \theta}{\partial x} & \frac{\partial \theta}{\partial y} & \frac{\partial \theta}{\partial z} \\
\frac{\partial \phi}{\partial x} & \frac{\partial \phi}{\partial y} & \frac{\partial \phi}{\partial z}
\end{array} \right)
\end{displaymath}

Por exemplo, 
\begin{equation*}
\frac{\partial}{\partial x} = \frac{\partial r}{\partial x}\frac{\partial}{\partial r} + \frac{\partial \theta}{\partial x}\frac{\partial}{\partial \theta} + \frac{\partial \phi}{\partial x}\frac{\partial}{\partial \phi}. 
\end{equation*}
Assim, sendo que $r = \sqrt{x^{2} + y^{2} + z^{2}}$ temos
\begin{equation}
\frac{\partial r}{\partial x} = \frac{2 x}{2\sqrt{x^{2} + y^{2} + z^{2}}} = \frac{x}{r} = \sin{\theta} \cos{\phi},
\end{equation}
analogamente fazendo-se as derivações em relação a $y$ e a $z$, tem-se
\begin{eqnarray}
\frac{\partial r}{\partial y} = \sin{\theta}\sin{\phi}, & & \frac{\partial r}{\partial z} = \cos{\theta}.
\end{eqnarray}

Agora passemos ao cálculo dos elementos $\partial\theta/\partial x^{i}$. Temos que $z = r \cos{\theta}$, portanto segue-se, naturalmente, que $\cos{\theta} = \dfrac{z}{\sqrt{x^{2} + y^{2} + z^{2}}}$.  Fazendo uma derivação implícita temos
\begin{eqnarray}
\frac{\partial}{\partial x}\biggl(\frac{z}{\sqrt{x^{2} + y^{2} + z^{2}}}\biggr) & = & \frac{\partial\theta}{\partial x} \frac{\partial}{\partial \theta}(\cos{\theta})\nonumber \\[0.1cm]
\frac{z x}{r^{3}} & = & \sin{\theta} \frac{\partial \theta}{\partial x} \nonumber.
\end{eqnarray}

Substituindo-se os valores de $z$ e $x$ em suas respectivas representações em coordenadas esféricas
\begin{eqnarray}
\frac{r^{2}\cos{\theta}\sin{\theta}\cos{\phi}}{r^{3}} & = & \sin{\theta} \frac{\partial \theta}{\partial x} \nonumber \\[0.1cm]
\frac{\partial \theta}{\partial x} & = & \frac{\cos{\theta}\cos{\phi}}{r}. 
\end{eqnarray}
Analogamente para $\partial\theta/\partial y$, temos

\begin{equation}
\frac{\partial \theta}{\partial y} = \frac{\cos{\theta}\sin{\phi}}{r}.
\end{equation}

Finalmente, a variação em relação à coordenada $z$ precisa de um pouco mais de cautela
\begin{eqnarray}
\frac{\partial}{\partial z}\biggl(\frac{z}{\sqrt{x^{2} + y^{2} + z^{2}}}\biggr) & = & -\sin{\theta} \frac{\partial\theta}{\partial z} \nonumber \\[0.1cm]
\frac{1}{\sqrt{x^{2} + y^{2} + z^{2}}} - \frac{z^{2}}{r^{3}} & = & - \sin{\theta}\frac{\partial\theta}{\partial z}\nonumber.
\end{eqnarray}
Novamente, fazendo-se as substituições chega-se
\begin{eqnarray}
\frac{1}{r} - \frac{r^{2}\cos^{2}{\theta}}{r^{3}} & = & -\sin{\theta}\frac{\partial \theta}{\partial z}\nonumber \\[0.1cm]
\frac{1 - \cos^{2}{\theta}}{r} & = & -\sin{\theta}\frac{\partial\theta}{\partial z} \nonumber \\[0.1cm]
\frac{\partial \theta}{\partial z} & = & - \frac{1}{r} \sin{\theta}.
\end{eqnarray}

Finalmente as componentes $\partial\phi/\partial x^{i}$. Essas componentes exigem um pouco mais de atenção e cuidado. Nesse caso iremos escolher para derivação implícita $x = r \sin{\theta}\cos{\phi}$. 

\begin{eqnarray}
\frac{\partial}{\partial x}\biggl(\frac{x}{\sqrt{x^{2} + y^{2} + z^{2}}}\biggr) & = & \frac{\partial}{\partial x} (\sin{\theta}\cos{\phi})
\end{eqnarray}
lembrando que tanto $\phi$ quanto $\theta$ são funções das coordenadas antigas, contribuindo para a derivação 

\begin{eqnarray}
\frac{\partial}{\partial x}\biggl(\frac{x}{\sqrt{x^{2} + y^{2} + z^{2}}}\biggr) & = & \frac{\partial \theta}{\partial x}\cos{\phi}\cos{\theta} + \frac{\partial \phi}{\partial x}\sin{\theta}\frac{\partial}{\partial\phi}(\cos{\phi}).
\end{eqnarray}
substituindo os valores funcionais de $x$ e de $\partial\theta/\partial x$ , obtemos

\begin{eqnarray}
\frac{1 - \sin^{2}{\theta}\cos^{2}{\phi}}{r} & = & \frac{\cos^{2}{\theta}\cos^{2}{\phi}}{r} - \sin{\theta}\sin{\phi}\frac{\partial\phi}{\partial x} \nonumber\\[0.1cm]
\frac{1 - \cos^{2}{\phi}(\sin^{2}{\theta} + \cos^{2}{\theta})}{r} & = & -\sin{\theta}\sin{\phi}\frac{\partial\phi}{\partial x} \nonumber\\[0.1cm]
\frac{\sin^{2}{\phi}}{r} & = & -\sin{\theta}\sin{\phi}\frac{\partial\phi}{\partial x} \nonumber\\[0.1cm]
\frac{\partial\phi}{\partial x} & = & - \frac{1}{r}\frac{\sin{\phi}}{\sin{\theta}}.
\end{eqnarray}
Analogamente, obtemos os demais elementos da matriz jacobiana

\begin{eqnarray}
\frac{\partial \phi}{\partial y}  =  \frac{1}{r} \frac{\cos{\phi}}{\sin{\theta}}, & & \quad \frac{\partial \phi}{\partial z} = 0.
\end{eqnarray}
Assim, a matriz jaconiana com os valores dos elementos de transformação para coordenadas esféricas assume a seguinte forma

$$ J = (\partial x^{i}/\partial x'^{j}) = \left(
\begin{array}{ccc}
\sin{\theta} & \sin{\theta}\cos{\phi} & \cos{\theta} \\
\frac{1}{r}\cos{\theta}\cos{\phi} & \frac{1}{r}\cos{\theta}\sin{\phi} & -\frac{1}{r}\sin{\theta} \\
-\frac{1}{r\sin{\theta}}\sin{\phi} & \frac{1}{r\sin{\theta}}\cos{\phi} & 0 \\
\end{array}
\right) $$
Portanto, os vetores de Killing, em coordenadas esféricas, são agora naturalmente reescritos. Consideremos o vetor de Killing responsável por ser o gerador das rotações no plano-$yz$ ou em torno do eixo dos $x$
\begin{equation*}
\xi_{1}  =  y \partial_{z} - z \partial_{y}.
\end{equation*}
Fazendo-se uma expansão das derivadas em termos das correspondentes coordenadas esféricas chega-se
\begin{eqnarray*}
\frac{\partial}{\partial z} & = & \cos{\theta}\frac{\partial}{\partial r} - \frac{1}{r}\sin{\theta}\frac{\partial}{\partial \theta}, \\[0.1cm] \frac{\partial}{\partial y} & = & \sin{\theta}\sin{\phi}\frac{\partial}{\partial r} + \frac{1}{r}\cos{\theta}\sin{\phi}\frac{\partial}{\partial \theta} + \frac{1}{r\sin{\theta}}\cos{\phi}\frac{\partial}{\partial \phi}.
\end{eqnarray*}
\begin{equation}
\therefore \quad \xi_{1} = -\sin{\phi}\partial_{\theta} - \cot{\theta}\cos{\phi}\partial_{\phi}. \label{ki1}
\end{equation}
Daí, os demais vetores de Killing seguem de maneira análoga e assumem a seguinte forma
\begin{eqnarray}
\xi_{2} & = & -\cos{\theta}\partial_{\theta} + \cot{\theta}\sin{\phi}\partial_{\phi}\\[0.1cm]\label{ki2}
\xi_{3} & = & \partial_{\phi}.\label{ki3}
\end{eqnarray}

Ou ainda, podemos escrever os campos vetoriais  na forma 
\begin{eqnarray}
\xi_{1} & = & \left(
0,  0, -\sin{\phi}, -\cot{\theta}\cos{\phi}
\right) \\[0.1cm]
\xi_{2} & = & \left(
0,  0,  -\cos{\phi}, \cot{\theta}\sin{\phi}\right) \\[0.1cm]
\xi_{3} & = & \left(
0,  0,  0, 1\right)
\end{eqnarray}
além disso, esse vetores satisfazem a álgebra de Lie
\begin{eqnarray*}
\lbrack \xi_1,\xi_2\rbrack=-\xi_3,\quad \lbrack \xi_2,\xi_3\rbrack=-\xi_1,\quad \lbrack \xi_3,\xi_1\rbrack=-\xi_2
\end{eqnarray*}
ou de maneira mais compacta
\begin{equation*}
\lbrack \xi_{i},\xi_{j} \rbrack = -\varepsilon_{ijk}\xi_{k},
\end{equation*}
onde $\epsilon_{ijk}$ é o tensor de Levi-Civita completamente anti-simétrico invariante sob o grupo SO$(3)$ definido por
$$
\varepsilon_{ijk} = \begin{cases}
0, & \textrm{se dois dos índices forem iguais,}\\
1, & \textrm{se} \,\, i,j,k \,\, \textrm{forem uma permutação par de}\,\, 1,2,3,\\-1, & \textrm{se}\,\, i,j,k \,\, \textrm{forem uma permutação ímpar de}\,\, 1,2,3.
\end{cases}
$$

\renewcommand{\theequation}{B.\arabic{equation}}
\chapter*{O papel da Torção}\setcounter{equation}{0}
\addcontentsline{toc}{chapter}{Apêndice 2: Torção no Formalismo de Palatini}
\pagestyle{fancy}
\lhead{\bfseries Apêndice 2: Torção}
\rhead{}
\label{apêndice 2}
Nosso objetivo agora será mostrar que os conteúdos das Eqs.(\ref{Palatini1}) e (\ref{Palatine2}) reproduzem as equações de Einstein da Relatividade Geral usual. De fato, considerando-se a Eq. (\ref{Palatini1}) em componentes, temos
\begin{equation}
\underbrace{\varepsilon_{IJKL}e^{K}_{\rho}}\varepsilon^{\mu\nu\rho\sigma}R^{IJ} \,_{\mu\nu} = 0,
\label{Palatine3}
\end{equation}
observando que o termo em destaque podemos fazer uso da Fórmula de Caley para o determinante de uma matriz, isto é,
\begin{equation}
 \varepsilon_{\mu\nu\rho\sigma}\textrm{det}e\,=\, 
 \varepsilon_{IJKL}e^{I}_{\mu}e^{J}_{\nu}e^{K}_{\rho}e^{L}_{\sigma},
 \label{Caley}
\end{equation}
desejamos, na verdade, uma maneira de relacionar o tensor totalmente anti-simétrico - símbolo de Levi-Civita com apenas um dos vierbein como destacado na Eq. (\ref{Palatine3}). Para tanto, iremos utilizar dos inversos dos vierbeins para isolarmos a relação desejada em (\ref{Caley}).
\begin{eqnarray}
\bigl(\varepsilon_{\mu\nu\rho\sigma}\textrm{det}e & = &
\varepsilon_{IJKL}e^{I}_{\mu}e^{J}_{\nu}e^{K}_{\rho}e^{L}_{\sigma}\bigr)e^{\mu}_{A}e^{\nu}_{B}
e^{\sigma}_{C} \\ [0.1cm]
e^{\mu}_{A}e^{\nu}_{B}
e^{\sigma}_{C}\varepsilon_{\mu\nu\rho\sigma}\textrm{det}e & = & \varepsilon_{IJKL}e^{K}_{\rho}\delta^{I}_{A} \delta^{J}_{B} \delta^{L}_{C}\\[0.1cm]
e^{\mu}_{A}e^{\nu}_{B}
e^{\sigma}_{C}\varepsilon_{\mu\nu\rho\sigma}\textrm{det}e & = & \varepsilon_{ABKC}e^{K}_{\rho}
\end{eqnarray}
ou ainda, podemos escrever de maneira equivalente
\begin{equation}
\varepsilon_{IJKL}e^{K}_{\rho} = e^{\alpha}_{I} e^{\beta}_{J} e^{\lambda}_{L} \varepsilon_{\alpha\beta\rho\lambda} \textrm{det}e
\label{eqn6}
\end{equation}
inserindo (\ref{eqn6}) em (\ref{Palatine3}) obtemos
\begin{equation}
e^{\alpha}_{I} e^{\beta}_{J} e^{\lambda}_{L} \varepsilon_{\alpha\beta\rho\lambda}\varepsilon^{\mu\nu\rho\sigma}R^{IJ} \,_{\mu\nu} = 0,
\label{eqn7}
\end{equation}
como os epsilons são completamente anti-simétricos, podemos colocar o índice $\rho$ a frente em ambos sem que se altere o sinal, pois estamos fazendo permutações pares, e lembrando que 
\begin{equation}
\varepsilon_{\rho\alpha\beta\lambda} \varepsilon^{\rho\mu\nu\sigma} = \delta^{[\mu}_{\alpha} \delta^{\nu}_{\beta} \delta^{\sigma]}_{\lambda}
\label{eqn8}
\end{equation}
a equação (\ref{eqn8}) nada mais é que o determinante de uma matriz $3 \,\times\, 3$ formada por deltas de Kronecker, ou seja,

\begin{equation}
\varepsilon_{\rho\alpha\beta\lambda} \varepsilon^{\rho\mu\nu\sigma} = \delta_{\alpha}^{\mu} \delta_{\beta}^{\nu} \delta_{\lambda}^{\sigma} +  \delta_{\alpha}^{\nu} \delta_{\beta}^{\sigma} \delta_{\lambda}^{\mu} +  \delta_{\alpha}^{\sigma} \delta_{\beta}^{\mu} \delta_{\lambda}^{\nu} -  \delta_{\alpha}^{\mu} \delta_{\beta}^{\sigma} \delta_{\lambda}^{\nu} -  \delta_{\alpha}^{\sigma} \delta_{\beta}^{\nu} \delta_{\lambda}^{\mu} - \delta_{\alpha}^{\nu} \delta_{\beta}^{\mu} \delta_{\lambda}^{\sigma} 
\label{eqn9}
\end{equation}
Assim, de (\ref{eqn9}) em (\ref{eqn7}) temos
\begin{eqnarray}
 0 & = & e_{I}^{\mu} e_{J}^{\nu} e_{L}^{\sigma} R^{IJ} \,_{\mu\nu} + e_{I}^{\nu} e_{J}^{\sigma} e_{L}^{\mu} R^{IJ} \,_{\mu\nu} + e_{I}^{\sigma} e_{J}^{\mu} e_{L}^{\nu} R^{IJ} \,_{\mu\nu} - e_{I}^{\mu} e_{J}^{\sigma} e_{L}^{\nu} R^{IJ} \,_{\mu\nu} +  \nonumber \\ [0.1cm]
 & &  - e_{I}^{\sigma} e_{J}^{\nu} e_{L}^{\mu} R^{IJ} \,_{\mu\nu} - e_{I}^{\nu} e_{J}^{\mu} e_{L}^{\sigma} R^{IJ} \,_{\mu\nu}.
\end{eqnarray}
Daí, aplicando as propriedades do vierbein ficamos com

\begin{equation}
e^{\sigma}_{L} R + e^{\mu}_{L}R^{\nu\sigma} \,_{\mu\nu} + e^{\nu}_{L}R^{\sigma\mu} \,_{\mu\nu} - e^{\nu}_{L} R^{\mu\sigma} \,_{\mu\nu} - e^{\mu}_{L}R^{\sigma\nu} \,_{\mu\nu} - e^{\sigma}R^{\nu\mu} \,_{\mu\nu} = 0,
\label{eq10}
\end{equation}
pelas propriedades de anti-simetria do tensor de curvatura de Riemann, temos que $R_{[\mu\nu][\rho\sigma]}$ e da definição do tensor de Ricci:  contração entre o primeiro e o terceiro índice, $R^{\mu\sigma} \,_{\mu\nu} \,\equiv\,R^{\sigma}_{\nu}$. Obtem-se

\begin{equation}
e^{\mu}_{L}R^{\sigma}_{\mu} - \frac{1}{2} e^{\sigma}_{L}R = 0
\label{eqn11}
\end{equation}
vemos que o índice $L$ está livre, daí podemos multiplicar toda a Eq. (\ref{eqn11}) por $e^{L}_{\nu}$ de modo que
\begin{equation}
R^{\sigma}_{\nu} - \frac{1}{2}\delta^{\sigma}_{\nu}R = 0.
\label{eq12}
\end{equation}
Finalmente, lembrando que $R^{\sigma}_{\nu}\,=\, g^{\sigma\mu}R_{\mu\nu}$ e $\delta^{\sigma}_{\nu} = g^{\mu\sigma}g_{\mu\nu}$, obtemos as equações de Einstein no vácuo como desejávamos
\begin{equation}
R_{\mu\nu} - \frac{1}{2}g_{\mu\nu} R = 0.
\label{eq13}
\end{equation}

A fim de que a ação de Palatini reproduza os resultados da Relatividade Geral usual, Einstein-Hilbert, devemos ainda trabalhar a segunda equação de movimento e verificar que seu conteúdo nos mostra, pela dinâmica das equações, que a torção é identicamente nula. Assim, nosso objetivo será mostrar que a Eq. (\ref{Palatine2}) $\Longrightarrow$ $T^{\alpha}\,_{\mu\nu} \,\equiv\, 0$, ou seja, vamos obter via equações de movimento que a parte anti-simétrica da conexão $\Gamma^{\alpha}\,_{[\mu\nu]}$, na ausência de férmions, é de fato a conexão usual de Levi-Civita usada arbitrariamente por Einstein que deliberadamente assume a torção nula. Portanto, uma das grandes vantagens de se usar o formalismo de Palatini é justamente a capacidade de mostrarmos, sem imposição a priori, de que a torção é zero como subproduto das equações de movimento mostrando assim a equivalência\footnote{Obviamente se introduzirmos matéria fermiônica espinorial automaticamente, pela presença da interação gravitacional, ganhamos uma  derivada covariante com um  termo linearmente proporcional a conexão de spin\cite{Gasperini}. Logo, em presença de férmions inevitavelmente trazemos a torção como um componente da teoria.} entre o formalismo de primeira e segunda ordem da RG.

Escrevendo a Eq. (\ref{Palatine2}) em componentes
\begin{equation}
\varepsilon_{IJKL} e^{L}_{\rho} \epsilon^{\mu\nu\rho\sigma}T^{K}\,_{\mu\nu}  = 0
\label{eqn14}
\end{equation}
analogamente, fazendo-se uso da fórmula de Caley
\begin{equation}
\varepsilon^{\rho\mu\nu\sigma}\epsilon_{\rho\alpha\beta\lambda} e^{\alpha}_{I} e^{\beta}_{J} e^{\lambda}_{K} T^{K}\,_{\mu\nu} = 0.
\end{equation}

Fazendo-se uso novamente do determinante (\ref{eqn9}) e desenvolvendo os termos, obtem-se:

\begin{equation}
e^{\sigma}_{K} T^{K}\,_{IJ} + e^{\sigma}_{J} T^{K}\,_{KI} + e^{\sigma}_{I} T^{K}\,_{JK} - e^{\sigma}_{J} T^{K}\,_{IK} - e^{\sigma}_{I} T^{K}\,_{KJ} - e^{\sigma}_{K} T^{K}\,_{JI} = 0,
\label{eqn15}
\end{equation}
como $T^{K}\,_{IJ} \,=\, -\, T^{K}\,_{JI}$, podemos reescrever os termos da Eq.(\ref{eqn15}) como
\begin{equation}
e^{\sigma}_{K}T^{K}\,_{IJ} + e^{\sigma}_{J}T^{K}\,_{KI} + e^{\sigma}_{I}T^{K}\,_{JK} = 0.
\label{eqn16}
\end{equation}
multiplicando toda a Eq.(\ref{eqn16}) por $e^{I}_{\sigma}$, segue-se que
\begin{equation}
T^{K}\,_{KJ} + T^{K}\,_{KJ} - \delta^{I}_{I} T^{K}\,_{KJ} = 0
\label{eqn17}
\end{equation}
lembrando que $\delta^{I}_{I}$ representa o traço da matriz identidade $\mathds{1}_{4\times 4} $ , logo concluímos que o traço da Torção é identicamente nulo, ou seja, $T^{K}\,_{KJ}\,=\, 0$. E substituido esse resultado na Eq.(\ref{eqn16}), sendo o vierbein diferente de zero,
\begin{equation}
e^{\sigma}_{K} T^{K}\,_{IJ} = 0 \Longrightarrow  T^{K}\,_{IJ} = 0.
\end{equation}

Portanto, mostrando que as componentes da torção são todas nulas nos leva a conexão usual de Levi-Civita da RG. Dessa forma, vemos que o formalismo de Palatini onde assumimos tanto a conexão (estrutura afim) quanto o vierbein (estrutura geométrica) como campos independentes; as equações de movimento, na ausência de férmions, reduzem-se exatamente as equações de campo de Einstein.

\subsection*{Torção e curvatura no mesmo pé de igauldade} 

Na descrição de Einstein da gravitação toda dinâmica é estabelecida através do tensor métrico. Einstein utiliza-se da conexão de Levi-Civita \cite{Moshe, Inverno, Wal} que tem por base a hipótese de que a parte anti-simétrica desta é identicamente nula. Em outras palavras, para relatividade geral, no formalismo de segunda ordem, assume-se que a torção é nula. A condição de Torção nula pode ser justificada de uma maneira mais precisa quando estamos no formalismo de primeira ordem \cite{ Gasperini, Nakahara, Bertlmann}. Se a estrutura métrica \textit{vierbein} ($e^{I}$) e a estrutura afim (conexão de Lorentz $\omega^{IJ}$) são considerados como campos dinâmicos independentes, como proposto por Palatini \cite{ Ash1, Rov2, Simone}. Consequentemente, fazendo-se as variações na ação, segue-se que em quatro-dimensões e na ausência de matéria fermiônica, as equações de campo ainda assim nos levam a condição de torção nula. Portanto, $T^{I} = 0$ não é uma restrição necessária da teoria mas uma consequência das equações de movimento.

Se não assumirmos  $T^{I} = 0$ \textit{a priori}, partículas fermiônicas irão poder revelar a presença de torção pois elas se acoplam naturalmente a $T^{I}$. Com efeito, teríamos trajetórias distintas das geodésicas de matéria não-fermiônica. Naturalmente, a trajetetória de um elétron iria diferir das geodésicas de um fóton pois, este não se acoplaria a torção enquanto aquele traria à tona essa nova estrutura do espaço-tempo. Esses efeitos, obviamente, podem não ser muito significantes para atual acuidade experimental, ou mesmo serem muito difíceis de se mensurar ou comparar trajetórias de partículas com difentes \textit{spins} a fim de se separar os efeitos da curvatura e da torção em uma dada região do espaço-tempo.

No entanto, um outro aspecto que daria um novo \textit{status} de relevância para a torção é sua capacidade de agir no mesmo pé de igualdade geométrico que a curvatura para o espaço-tempo. De fato, nosso modelo é uma tentativa de analisar esses efeitos, considerando-se uma modificação da ação usual da relatividade geral. A possibilidade da torção como variável dinâmica pode ter consequências cosmológicas muito importantes. A fim de examinarmos esses efeitos, vale a pena notar como a torção contribui para a curvatura do espaço-tempo.

No formalismo de primeira ordem da gravitação, os campos fundamentais como já mencionado, são o \textit{vierbein} ($e^{I} = e^{I}_{\mu}dx^{\mu}$) e a conexão de \textit{spin} ($\omega^{IJ} = \omega_{\mu}^{IJ}dx^{\mu}$). Essas duas $1$-formas correspondem a diferentes aspectos da geometria, a saber, estrutura métrica e estrutura afim, respectivamente. Com efeito, são consideradas como campos dinâmicos independentes (veja por exemplo as referências \cite{Toloza,Za,Za1, Za2, Carroll}). Nesse formalismo, a ação de Einstein-Hilbert com constante cosmológica lê-se
\begin{equation*}
S_{EH}[e,\omega] = \int_{\mathcal{M}}\varepsilon_{IJKL}\bigl(R^{IJ} - \frac{\Lambda}{6}e^{I}\wedge e^{J}\bigr)\wedge e^{K}\wedge e^{L}.
\end{equation*}
Variando em relação ao \textit{vierbein} e a conexão, obtem-se as equações de campo de Einstein bem como a condição de torção nula
\begin{equation}
\varepsilon_{IJKL}\bigl(R^{IJ} - \frac{\Lambda}{3}e^{I}\wedge e^{J}\bigr)\wedge e^{K} = 0, \quad \varepsilon_{IJKL}T^{I}\wedge e^{J} = 0.
\end{equation}
Em geral, a conexão de \textit{spin} pode ser separada em uma parte sem torção mais a contorção, $\omega^{IJ} = \mathring{\omega}^{IJ} + \mathcal{C}^{IJ}$, onde 

\begin{equation}
 0 = de^{I} + \mathring{\omega}^{I}_{K}\wedge e^{J} \Rightarrow \mathring{\omega}^{I}_{K\mu} = e^{I}_{\nu}(\nabla_{\mu}e^{\nu}_{J}), \label{torsion}
\end{equation}
e a contorção é proporcional a torção,
\begin{equation}
T^{I} = \mathcal{C}^{I}_{J}\wedge e^{J}.
\end{equation}
Em (\ref{torsion}) $\nabla$ é a derivada covariante em relação a conexão de Christoffel \cite{Bertlmann}. Dessa forma, a curvatura $2$-forma, $R^{IJ} = d\omega^{IJ} + \omega^{I}_{K}\wedge \omega^{KJ}$, separa-se na forma usual livre de torção da geometria riemanniana, e a parte remanescente que é dependentende da torção,
\begin{equation}
R^{IJ} = \mathring{R}^{IJ} + \mathring{D}\mathcal{C}^{IJ} + \mathcal{C}^{I}_{K}\wedge\mathcal{C}^{KJ}. \label{curvature}
\end{equation}
Onde, $\mathring{R}^{IJ} = d\mathring{\omega}^{IJ} + \mathring{\omega}^{I}_{K}\wedge\mathring{\omega}^{KJ}$ é a curvatura de Riemann e $\mathring{D}$ é a derivada covariante exterior para a conexão livre de torção $\mathring{\omega}^{IJ}$. Portanto, se a torção for diferente de zero, vemos que os termos remanescentes poderiam contribuir para as equações de campo no mesmo pé de igualdade que o vierbein para curvatura $R^{IJ}$.

\renewcommand{\theequation}{C.\arabic{equation}}
\chapter*{Ondas Gravitacionais}\setcounter{equation}{0}
\addcontentsline{toc}{chapter}{Apêndice 3: Ondas Gravitacionais }
\pagestyle{fancy}
\lhead{\bfseries Apêndice 3: Ondas Gravitacionais}
\rhead{}
\label{apendice}

 Em $1916$, um ano após a formulação final das equações da relatividade geral, Albert Einstein mostrou que suas equações faziam predições da existência de ondas gravitacionais. Ele encontrou que no regime de linearização das suas equações de campo elas  apresentam soluções de onda: ondulações transversais do próprio tecido do espaço que se propagam na velocidade da luz. Einstein percebeu que a amplitude dessas ondas seria extremamente pequena e de difícil detecção. 
 
Duas estrelas de nêutron orbitando ao redor uma da outra devem perder energia devido à  emissão de ondas gravitacionais \cite{Damour}. Essas ondas, assim como as ondas eletromagnéticas, são portadoras de energia e momento. O processo de linearização das equações de Einstein gera, no famoso calibre de De Donder, equações de onda para a geometria do espaço-tempo\cite{Gasperini}. Seria como imaginar um lago sendo o espaço-tempo e, ao soltarmos uma pedra sobre este, vemos ondas que se propagam a partir da fonte emissora. Do mesmo modo, em alguns eventos gravitacionais, como por exemplo a explosão de uma Supernova, é de se esperar pelas equações que haja ondas de geometria, ou seja, flutuações geométricas em torno de um fundo fixo que se tem \textit{a priori}.

 No mesmo ano, um brilhante astrônomo e físico alemão, Karl Schwazschild, enquanto servia na Rússsia na primeira guerra mundial, encontrou soluções exatas para as equações de Einstein. Ele escreveu um trabalho fundamental que conseguia, pela primeira vez, resolver de maneira exata as equações da relatividade geral, para o caso particular de simetria esférica. Sua solução deu origem a uma predição sem precedentes na ciência. O espaço-tempo tendo essa plasticidade e capacidade de se curvar na presença de matéria-energia, seria possível concentrar tanta matéria-energia em uma região do espaço-tempo, que a curvatura, interpretada  campo gravitacional, seja tão intensa que nem mesmo a luz conseguiria escapar. Essa solução foi entendida, mais tarde, através de contribuições importantantíssimas do físico indiano Subrahmanyan Chandrasekhar \cite{Ash6}, como a descrição dos chamados \textit{buracos negros}. Desde então, experimentos a fim de se detectar ondas gravitacionais tem sido propostos e uma das possí­veis fontes para a propagação dessas ondas seria um caso onde dois buracos negros que giram em torno um do outro se fundem. Esse processo de fusão de dois buracos negros seria capaz de gerar ondas gravitacionais. E exatamente $100$ anos depois dessas predições, dantes impensáveis, o observatório LIGO (\textit{Laser Interferometer Gravitational-Wave Observatory}) anunciou, a detecção mais surpreendente de todos os tempos na ciência \cite{ligo1,ligo2}. Um século após as predições fundamentais de Einstein e Schwarzschild foi reportado a primeira detecção direta de ondas gravitacionais, advindas de um sistema binário de buracos negros fundindo em um único. Essas observações proporcionam um acesso singular \cite{louis}, em toda história do desenvolvimento científico, as propriedades fundamentais do espaço-tempo no regime de campo gravitacional intenso e alta velocidade, e confirmando as predições da relatividade geral. A cada vez, a Relatividade Geral tem triunfado em seus testes \cite{Will,Will2}.
 
\section*{Ondas Gravitacionais sobre Espaço-Tempo de de Sitter}
Vimos que o formalismo de Palatini para a relativvidade geral, sem matéria fermiônica, reproduz os mesmos resultados dinâmicos que o formalismo de segunda-ordem\footnote{A única diferença sendo que a torção nula, no formalismo Einstein-Cartan, é consequência das equações de movimento.}. A busca por soluções  tipo-onda reduz-se a uma análise de perturbações em torno de um vácuo com curvatura constante, a saber de Sitter, no formalismo métrico. Com efeito, separando a métrica do espaço-tempo como $g_{\mu\nu} \,=\, \mathring{g}_{\mu\nu} \,+\, h_{\mu\nu}$, com $\mid h_{\mu\nu}\mid \ll 1$ segue, naturalmente, que as quantidades perturbadas em primeira ordem, denotadas com um $1$ sobrescrito,  leem-se 
\begin{equation}
\Gamma^{\alpha (1)}\,_{\mu\nu} = \frac{1}{2}\mathring{g}^{\alpha\beta}\bigl(\mathring{\nabla}_{\mu} h_{\beta\nu} + \mathring{\nabla}_{\nu} h_{\mu\beta} - \mathring{\nabla}_{\beta} h_{\mu\nu}\bigr).
\label{eqq}
\end{equation}

Nesse caso, abaixar e subir os índices e o cálculo de derivadas covariantes são feitas usando-se a métrica de fundo $\mathring{g}_{\mu\nu}$. A expressão para curvatura pode ser obtida da equação(\ref{eqq}), em primeira ordem, notando que termos tipo $\Gamma^{2}$ são negligenciáveis no cálculo da ordem requerida.
\begin{eqnarray}
R^{\alpha (1)}\,_{\kappa\mu\nu} & = & \frac{1}{2}\bigl(\mathring{\nabla}_{\mu}\mathring{\nabla}_{\nu}h^{\alpha}_{\kappa} + \mathring{\nabla}_{\mu}\mathring{\nabla}_{\kappa}h^{\alpha}_{\nu} - \mathring{\nabla}_{\mu}\mathring{\nabla}^{\alpha}h_{\kappa\nu} -\mathring{\nabla}_{\nu}\mathring{\nabla}_{\mu}h^{\alpha}_{\kappa} \nonumber \\ [0.1cm]
& & - \mathring{\nabla}_{\nu}\mathring{\nabla}_{\kappa}h^{\alpha}_{\mu} + \mathring{\nabla}_{\nu}\mathring{\nabla}^{\alpha}h_{\kappa\mu} \bigr).
\label{eq. curvatura}
\end{eqnarray}
Fazendo-se uma contração do tensor de curvatura, obtemos o tensor de Ricci, $R^{(1)}_{\alpha\kappa} \,=\, R^{\mu (1)}\,_{\alpha \mu \kappa}$, que nesse caso assume a seguinte forma
\begin{equation}
R^{(1)}_{\alpha\kappa} = \frac{1}{2}\bigl(\mathring{\nabla}_{\mu}\mathring{\nabla}_{\kappa}h^{\mu}_{\alpha} + \mathring{\nabla}_{\mu}\mathring{\nabla}_{\alpha}h^{\mu}_{\kappa} - \mathring{\nabla}_{\mu}\mathring{\nabla}^{\mu}h_{\alpha\kappa} - \mathring{\nabla}_{\kappa}\mathring{\nabla}_{\alpha}h^{\mu}_{\mu}\bigr).
\label{eq. ricci}
\end{equation}

Sendo nossa região de interesse longe o suficiente da fonte emissora, ou seja, $T_{\mu\nu}\,=\, 0$, \textit{i.e.},  estamos considerando a dinâmica das perturbações na região livre de matéria-energia. Portanto, a equação $R_{\alpha\kappa} \,=\, 0$ deverá ser assegurada em todas as ordens. Demanda-se que $\mathring{R}_{\alpha\kappa} \,=\, 0$, tem-se que a métrica do fundo deve ser solução das equações de Einstein na região livre de matéria-energia. A próxima ordem, $R^{(1)}_{\alpha\kappa}\,=\, 0$ nos fornece as equações de movimento obedecidas pelas perturbações no regime que estamos interessados 
\begin{equation}
\mathring{\nabla}_{\mu}\mathring{\nabla}_{\kappa}h^{\mu}_{\alpha} + \mathring{\nabla}_{\mu}\mathring{\nabla}_{\alpha}h^{\mu}_{\kappa} - \mathring{\nabla}_{\mu}\mathring{\nabla}^{\mu}h_{\alpha\kappa} - \mathring{\nabla}_{\kappa}\mathring{\nabla}_{\alpha}h^{\mu}_{\mu} = 0,
\label{eqs-onda1}
\end{equation}
Desejamos reescrever as equações (\ref{eqs-onda1}) em uma forma tipo-onda em um fundo com curvatura diferente de zero. Denotaremos o traço da métrica $h^{\mu}_{\mu}$ por $h$, e a métrica de traço-reverso  é definida como $\bar{h}_{\mu\nu} \,=\, h_{\mu\nu} \,-\, 1/2 \mathring{g}_{\mu\nu}h$. Além disso, sabemos que sob uma transformação de difeomorfismos infinitesimal,
\begin{equation*}
 x^{\mu} \, \longrightarrow \, x^{\mu} \,+\, \xi^{\mu}, 
 \end{equation*}
 a perturbação da métrica varia funcionalmente como
\begin{equation}
h_{\mu\nu} \longrightarrow h_{\mu\nu} + \mathring{\nabla}_{\mu}\xi_{\nu} + \mathring{\nabla}_{\nu}\xi_{\mu},
\label{gauge-metric}
\end{equation}
usando dessa liberdade de gauge podemos impor, contanto que seja acessível, a condição de divergência nula dessa métrica de traço-reverso,
\begin{equation}
\mathring{\nabla}_{\mu}(h^{\mu}_{\nu} -\frac{1}{2}\delta^{\mu}_{\nu}h) = \mathring{\nabla}_{\mu}\bar{h}^{\mu\nu} = 0,
\label{gauge-deDonder}
\end{equation}
essa é a generalização  do conhecido gauge de de Donder \cite{Buonanno, Flanagan, Padmanabhan}.
Lembrando que
\begin{equation}
[\mathring{\nabla}_{\mu},\mathring{\nabla}_{\nu}]T^{\alpha}\,_{\beta} = \mathring{R}^{\alpha}\,_{\mu\nu\lambda}T^{\lambda}\,_{\beta} + \mathring{R}^{\lambda}\,_{\mu\nu\beta}T^{\alpha}\,_{\lambda},
\label{equ}
\end{equation}
uma vez que essas condições são impostas, podemos manipular (\ref{eqs-onda1}) fazendo uso das relações dadas em (\ref{equ}), e finalmente obtemos 
\begin{equation}
\mathring{\square}\bar{h}_{\mu\nu} + 2\mathring{R}_{\alpha\mu\beta\nu}\bar{h}^{\alpha\beta} = 0,
\label{eqs-onda2}
\end{equation}
onde $\mathring{\square} \equiv \mathring{\nabla}_{\alpha}\mathring{\nabla}^{\alpha}$, é o operador de  D'Alembert covariante.  As equaç?os (\ref{eqs-onda2}) descrevem a propagação de uma onda gravitacional em uma região livre de matéria em um vácuo com curvatura diferente de zero. No caso especial de interesse, esse vácuo seria o fundo de de Sitter, ou seja, $\mathring{R}_{\mu\alpha\beta\nu}\,=\,\dfrac{\Lambda}{3}(\mathring{g}_{\mu\beta}\mathring{g}_{\alpha\nu} - \mathring{g}_{\mu\nu}\mathring{g}_{\alpha\beta})$. As equações de onda assumem a seguinte forma\footnote{Lembrando que a métrica de traço-reverso $\bar{h}_{\mu\nu}$ possui o inverso do traço de $h_{\mu\nu}$. Para ver isso, basta verificar que : $\bar{h}_{\mu\nu} = h_{\mu\nu} -1/2 \mathring{g}_{\mu\nu}h$ $\longmapsto$ $\bar{h} = \mathring{g}^{\mu\nu}\bar{h}_{\mu\nu} = -h$.} 
\begin{equation}
\mathring{\square}\bar{h}_{\mu\nu} + \frac{2\Lambda}{3}(\bar{h}_{\mu\nu} + \mathring{g}_{\mu\nu}h) = 0.
\label{eqs-onda3}
\end{equation}

%-----------------------------------------------
%%%%%%%%%%%%%    FIM DA TESE    %%%%%%%%%%%%%%%%
%-----------------------------------------------

%-----------------------------------------------
%%%%%%%%%%%%%%%%%%BIBLIOGRAFIA%%%%%%%%%%%%%%%%%%
%-----------------------------------------------

\end{document}